| | |
|---|---|
| **Author** | Ahmad Hosney Awad Eid |
| **Title** | Optimized Automatic Code Generation for Geometric Algebra Based Algorithms with Ray Tracing Application |
| **Faculty** | Engineering |
| **Department** | Electrical Engineering, Computer Division |
| **Location** | Port-Said |
| **Degree** | Doctor of Philosophy |
| **Date** | 18 / 4 / 2010 |
| **Language** | English |
| **Supervision Committee** | Prof. Abd El Hay Ahmad Sallam<br>Dr. Mohamed Yousef Farghly<br>Dr. Rawya Yahya Rizk<br>Dr. Randa El Sayed Atta |

**English Abstract**

Automatic code generation for low-dimensional geometric algorithms is capable of producing efficient low-level software code through a high-level geometric domain specific language. Geometric Algebra (GA) is one of the most suitable algebraic systems for being the base for such code generator. This work presents an attempt at realizing such idea in practice. A novel GA-based geometric code generator, called GMac, is proposed. Comparisons to similar GA-based code generators are provided. The possibility of fully benefiting from the symbolic power of GA while obtaining good performance and maintainability of software implementations is illustrated through a ray tracing application.

# ACKNOWLEDGEMENT

I would like to express my gratitude to all the people who tried to help me learn the science and art of engineering that made this work possible. I offer my gratitude to **Prof. Abd El Hay Sallam** for showing me how excellent research should be conducted. I offer my gratitude to **Prof. Sayed Sorour** and **Dr. Mohamed Youssef Farghly** for showing me how to appreciate the beauty and elegance of mathematics. I offer my gratitude to **Dr. Kamel Al Serafi** and **Dr. Mohammed Dessouki** for showing me how engineers should think and work. I offer my gratitude to **Dr. Mohammed Sami** and **Dr. Mohammed Al-Dakkiki** for showing me how to link mathematics and computers to solve engineering problems. I would like to thank **Dr. Rawya Rizk** and **Dr. Randa Atta** for their support and concern. To all the above: I hope I was not a disappointment.

I would like to express my special thanks to **Prof. Amin Shoukry** and **Prof. Taher El Sonni** for their valuable discussions in the proceeding of the ICCTA 2009 that strongly affected the presentation of this work.

Finally, I would like to dedicate this work to the **Faculty of Engineering** in Port-Said; ***for that no gratitude is abundant***.

# SUMMARY

Geometry is an essential part of almost every solution to problems in computer science and engineering. The process of manually creating geometry-enabled software for representing and processing geometric data is tedious and error-prone. The use of automatic code generators for such process is more suitable especially for low-dimensional geometric problems. Such problems can be commonly found in computer graphics, computer vision, robotics, and many other fields of application. The main function of a geometric code generator is to transform a high-level description of the geometric algorithm into an efficient low-level software code capable of processing the relevant geometric information. A high-level geometric Domain Specific Language (DSL) is used for coding the input to the code generator. A mathematical algebraic system with special characteristics must be used for being the base of the DSL and the transformations required for final code generation. Geometric Algebra (GA) is one of the most suitable algebraic systems for such task.

This work illustrates the use of GA as the base for creating a geometric software code generator for low-dimensional geometric problems. The proposed code generator, called GMac, is compared to other similar GA-based code generators. The comparisons are made between the code generators architectures, in addition to the performance of generated code. The proposed code generator is applied in a ray tracing application to enhance its features and performance. The results compare well with traditional approaches based on vector and matrix algebras. The net result is the ability to easily create efficient implementations for GA-based algorithms for solving low-dimensional geometric problems. Such GA-based algorithms require less memory, less processing power, less design effort, and less debugging and maintenance. The reduction of design and maintenance efforts is the key to reducing cost of software development for geometry-aware software systems.

# Table of Contents

# List of Figures

# List of Tables

# List of Abbreviations

| | |
|---|---|
| AP | Automatic Programming |
| API | Application Programming Interface |
| AST | Abstract Syntax Tree |
| BCM | Basis-Change Matrix |
| BIH | Bounding Interval Hierarchy |
| BRDF | Bidirectional Reflectance Distribution Function |
| BTDF | Bidirectional Transmission Distribution Function |
| CAD | Computer Aided Design |
| CAGD | Computer Aided Geometric Design |
| CAS | Computer Algebra System |
| CG | Computer Graphics |
| CGA | Conformal Geometric Algebra |
| CLI | Common Language Interface (a Microsoft .NET technology) |
| CS | Computer Science |
| CSG | Constructive Solid Geometry |
| DBMS | Database Management System |
| DSL | Domain Specific Language |
| GA | Geometric Algebra |
| Gaalop | Geometric Algebra Algorithms Optimizer |
| Gaigen | Geometric Algebra Implementation Generator |
| GGPR | Geometrically Generated Parametric Representation |
| GMac | Geometric Macro (the proposed code generator) |
| GP | Generative Programming |
| GPCM | Generalized Pinhole Camera Model |
| GPGPU | General Purpose GPU |
| GPU | Graphics Processing Unit |
| IDE | Integrated Development Environment |
| iff | If and only if |
| IPM | Inner-Product Matrix |
| LID | Linearly Independent |
| OOP | Object Oriented Programming |
| RSTRI | Rotationally-Symmetric TR Instancing |
| SDL | Scene Description Language |
| TR | (Translate-Rotate) Euclidean transform |
| TSR | (Translate-Scale-Rotate) Euclidean transform |
| TSRI | TSR Instancing |
| w.r.t | With Respect To |

# List of symbols

| | |
|---|---|
| $\mathcal{F}$ | A field of scalars |
| $\mathcal{V}$ | A space of vectors |
| $\oplus$ | sum of vectors, direct sum of subspaces |
| $\otimes$ | Scalar multiplication of a scalar and a vector |
| $(\mathcal{F}, \mathcal{V}, \oplus, \otimes)$ | A vector space over a scalar field with two operations |
| $span(a_1, a_2, \ldots, a_k)$ | The vector subspace spanned by vectors $a_1, a_2, \ldots, a_k$ |
| $ga_span(a_1, a_2, \ldots, a_k)$ | The geometric algebra spanned by vectors $a_1, a_2, \ldots, a_k$ |
| $\langle e_1, e_2, \ldots, e_n \rangle^{\mathcal{V}}$ | The vectors $e_1, e_2, \ldots, e_n$ construct a basis for the vector space $\mathcal{V}$ |
| $\langle e_1, e_2, \ldots, e_n \rangle_{\perp}^{\mathcal{V}}$ | The vectors $e_1, e_2, \ldots, e_n$ construct an orthogonal basis for the vector space $\mathcal{V}$ |
| $\langle e_1, e_2, \ldots, e_n \rangle_{\perp 1}^{\mathcal{V}}$ | The vectors $e_1, e_2, \ldots, e_n$ construct an orthonormal basis for the vector space $\mathcal{V}$ |
| $\dim(\mathcal{V})$ | Dimension of the vector space $\mathcal{V}$ |
| $\mathrm{Rep}(x)_{\mathbf{E}}$ | Representation of a vector $x$ with respect to basis vectors in $\mathbf{E}$ |
| $\mathcal{W} \le \mathcal{V}$ | The vector space $\mathcal{W}$ is a subspace of the vector space $\mathcal{V}$ |
| $\mathcal{W} + \mathcal{U}$ | The subspace sum of subspaces $\mathcal{W}$ and $\mathcal{U}$ |
| $\mathcal{W} \oplus \mathcal{U}$ | The direct subspace sum of subspaces $\mathcal{W}$ and $\mathcal{U}$ |
| $a \parallel b$ | Vector $a$ is parallel to vector $b$ |
| $a \perp b$ | Vector $a$ is orthogonal to vector $b$ |
| $\mathcal{W}^{\perp}$ | The orthogonal complement of a subspace $\mathcal{W}$ |
| $\langle u, v \rangle$ | A bilinear form on vectors $u$ and $v$ |
| $\mathbf{A}^t$ | The transpose of matrix $\mathbf{A}$ |
| $(u_1, u_2, \ldots, u_n)$ | A row vector of components $u_1, u_2, \ldots, u_n$ |
| $(u_1, u_2, \ldots, u_n)^t$ | A column vector of components $u_1, u_2, \ldots, u_n$ |
| $C\ell_{p,q,r}$ | A Clifford Algebra of signature $p, q, r$ |
| $C\ell_{p,q}$ | A non-degenerate Clifford Algebra of signature $p, q, 0$ |
| $C\ell_n$ | A non-degenerate Euclidean Clifford Algebra of signature $n, 0, 0$ |
| $\mathcal{G}$ | A Geometric Algebra |
| $\langle A \rangle$ | The grade of a *k*-vector $A$ |
| $\langle A \rangle_k$ or $A_{\langle k \rangle}$ | The *k*-vector component of a multivector $A$, the $k^{\text{th}}$ grade of a multivector $A$ |
| $A_{\langle \max \rangle}$ | The nonzero *k*-vector component of the multivector $A$ having maximum grade |
| $\langle A \rangle_{e_1 \wedge e_2}$ or $A_{\langle e_1 \wedge e_2 \rangle}$ | The real coefficient associated with the basis blade $e_1 \wedge e_2$ of multivector $A$ |
| $\mathcal{B}^k$ | The set of all *k*-blades of a geometric algebra $\mathcal{G}$ |
| $\mathcal{G}^k$ | The set of all *k*-vectors of a geometric algebra $\mathcal{G}$ |
| $\mathcal{V}^k$ | The set of all *k*-versors of a geometric algebra $\mathcal{G}$ |
| $\mathcal{V}^+$ | The set of all versors with even grades of a geometric algebra $\mathcal{G}$ |
| $\mathcal{V}^-$ | The set of all versors with odd grades of a geometric algebra $\mathcal{G}$ |
| $I$ | The unit pseudo-scalar of the geometric algebra $\mathcal{G}$ |
| $A^{\sim}$ or $\widetilde{A}$ | The reverse of a multivector $A$ |
| $A^{\wedge}$ or $\widehat{A}$ | The grade involution of a multivector $A$ |
| $A^{\dagger}$ | The Clifford conjugate of a multivector $A$ |
| $\lVert A \rVert$ | The norm of a multivector $A$ |
| $[\![ A ]\!]$ | The quasi-norm of a multivector $A$ |
| $\lvert A \rvert$ | The magnitude of a multivector $A$ |
| $\lvert\lambda\rvert$ | The absolute value of a scalar $\lambda$ |
| $A^{-1}$ | The inverse of the versor $A$ |
| $e_i$ | Basis vector $i$ of the 1-vector space $\mathcal{G}^1$ |
| $e_J^k$ | Basis element $J$ of the $k$-vector space $\mathcal{G}^k$ |
| $A^*$ | The dual of a multivector $A$ |
| $A^{\odot}$ | The un-dual of a multivector $A$ |
| $A^{*B}$ | The dual of a blade $A$ w.r.t. blade $B$ |
| $A^{\odot B}$ | The un-dual of a blade $A$ w.r.t. blade $B$ |
| $A \wedge B$ | The outer product of multivectors $A$ and $B$ |
| $A \cdot B$ | The Hestenes inner product of multivectors $A$ and $B$ |
| $A \bullet B$ | The fat-dot product (modified Hestenes inner product) of multivectors $A$ and $B$ |
| $A * B$ | The scalar product of multivectors $A$ and $B$ |
| $A \rfloor B$ | The left-contraction product of multivectors $A$ and $B$ |
| $A \lfloor B$ | The right-contraction product of multivectors $A$ and $B$ |
| $A \underline{\times} B$ | The commutator product of two multivectors |
| $A \overline{\times} B$ | The anti-commutator product of two multivectors |
| $x \leftarrow E$ | Assign the value of expression $E$ to variable $x$ |

# Chapter 1 : Introduction

Automatic Programming (AP) is a type of computer programming methodology in which some mechanism generates a computer program rather than have human programmers write the actual code. One of the software engineering paradigms used for realizing such idea is Generative Programming (GP) [1]. Generative programming is based on modeling whole software families, rather than single special-purpose software systems. In GP given a particular requirements-specification, a highly customized and optimized intermediate or final software product can be automatically manufactured on demand. The manufacturing process is based on elementary, reusable implementation components by means of configuration knowledge. To achieve such goal, GP utilizes many concepts and techniques that include Generic Programming, Aspect-Oriented Programming, Static Meta-programming, and Automatic Code Generators.

## 1.1 Compilers, Interpreters, and Code Generators

In the early days of electronic computers, programming was an act of writing simple assembly code that directly corresponds to the machine language of the targeted hardware system. As the complexity of software and diversity of hardware rose, the need for human-friendly computer languages grew accordingly. Such languages could be executed on a variety of hardware systems without change to the original source code. The main tools used for executing such code were compilers and interpreters.

As illustrated in Figure 1-1, a compiler takes a number of textual high-level code files as input and eventually transforms them into optimized low-level machine code for the targeted platform. Such platform could be a specific hardware system or another layer of intermediate software representation like the .NET intermediate language or the Java byte code. On the other hand, interpreted languages are also common in practice for performing simple, non time-critical tasks. Interpreters are generally much easier to create compared to compilers but suffer from being too slow for many practical applications. Interpreters are common in operating systems shell scripting languages and some scientific software systems such as Matlab and Python. Typical operation of an interpreter is shown in Figure 1-2.

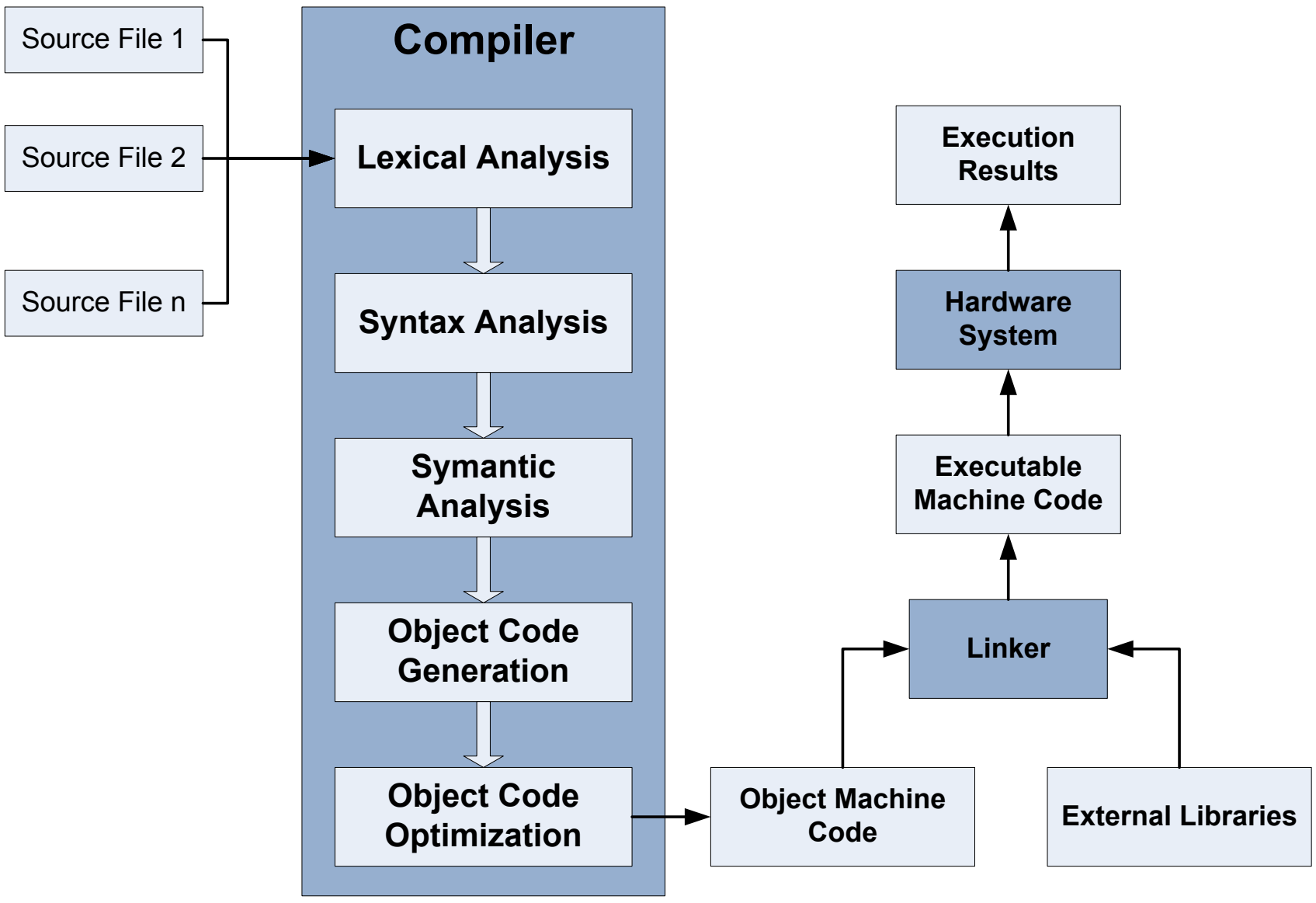

**Figure 1-1: Typical operation of a compiler**

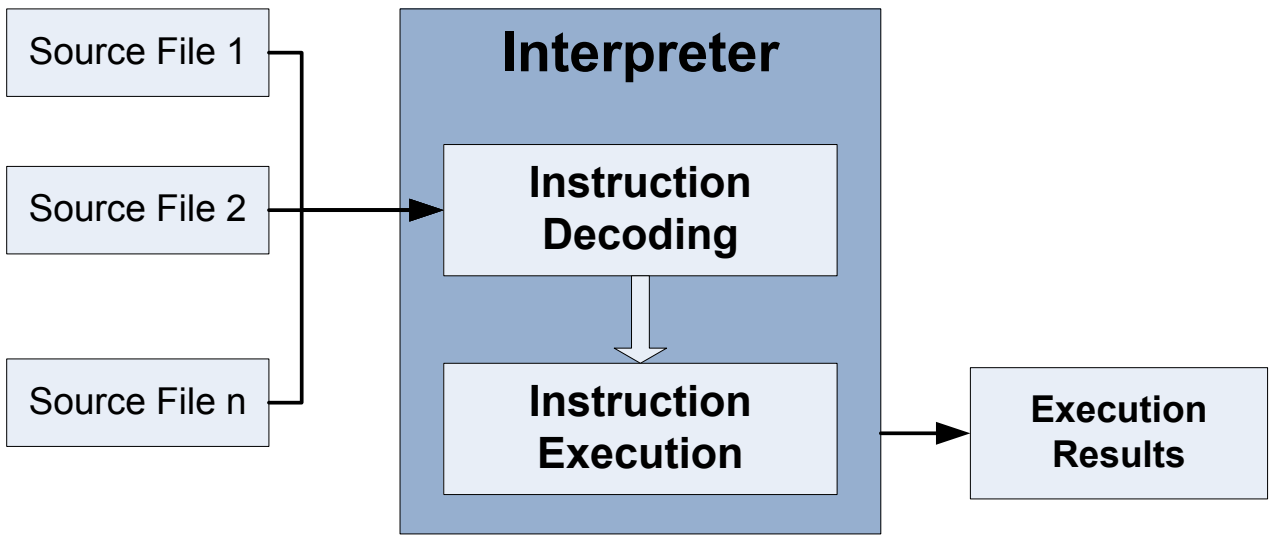

**Figure 1-2: Typical operation of an interpreter**

In recent years, practical software systems have become so complex that common compilers and interpreters are not sufficient by themselves anymore. More complex software development systems and techniques are becoming industry standards. For example the Microsoft .NET platform consists of an advanced Integrated Development Environment (IDE) called Visual Studio. The .NET technology relies on a well structured and mature software framework.

The .NET 3.5 environment integrates many general-purpose language compilers such as VB, C#, C++, and F#. In such framework a third type of software generation system is commonly found; namely the code generator. Many cooperating code generators are present in the .NET platform for performing various coding tasks automatically on behalf of the developer. Such code generators have become essential parts in any modern software development process in the last decade.

Formally, a code generator (also called a source-to-source compiler) can be defined as a program that takes a higher-level specification of a piece of software and produces its implementation in the form of a target language source code or textual data file. The input specifications are usually written in a well designed Domain Specific Language (DSL). Code generators address three important software development issues [1]:

1. ***Raising the intentionality of software system description***: Intentional descriptions directly and explicitly state the system requirements avoiding any cluttering implementation details. Such intentionality is achieved through domain-specific notations and is fully or partially implemented through code generators.
2. ***Computing an efficient implementation***: In practice, there is no one-to-one correspondence between the high-level structure of the input specifications and the final structure of the output implementation. In addition, a small change in the specifications may radically change the final implementation details. Code generators can automate many of the transformations required to obtain an efficient and maintainable implementation from a high-level domain-specific specification. Such automation is indispensible in modern software development as the manual effort required for coding such transformations is becoming unimaginably huge.
3. ***Avoiding the library scaling problem***: The library scaling problem concerns the horizontal scaling of conventional software component libraries. A difficult choice has to be made when adding new features to an existing software system. The first approach is to implement new features as new methods and data members in existing components to preserve performance. Such approach results in exponential growth of code and poor maintainability. The second approach is to re-factor the existing components to preserve system modularity and organization. Such approach usually reduces performance of final software because of increased component-call overhead. Alternatively, a code generator can be used to achieve both effective re-factoring and good performance. Code generators can eliminate call-overhead between software components and can automatically perform various domain-specific optimizations. As a result, near-linear library scaling with very good performance can be achieved with reduced effort.

A practical code generator usually performs the tasks shown in Figure 1-3. When parsing the high-level inputs, the code generator checks the DSL code against its correct syntax and semantics using methods very similar to what traditional compilers apply. Next, the DSL code is usually transformed into an intermediate representation form with default behavior substituted for missing data in the input DSL code. The third step is to perform any possible domain-specific or target-code optimizations. Finally, the code generator outputs the optimized code in the selected target language, usually, in a well formatted textual form.

From the previous description, it is apparent that code generators can cover a very wide spectrum of modern software development tasks. In addition, code generators are very similar to traditional compilers except that compilers usually output very low-level machine\byte code that is usually not understandable by humans. Code generators, on the other hand, can output any textual files ranging from whole programs in a general purpose language to software documentations and test plans. Thus code generators can become very complex software systems by themselves just like traditional compilers, or even more.

One of the widely known uses of code generators is corroborated by the internet. In the last two decades, the enormous connectivity provided by the internet resulted in an explosion in the amount of data required by even the simplest software systems. Such data explosion boosted the development of modern relational Database Management Systems (DBMS) and increased the complexity of modern Object Oriented Software (OOP) systems. Such increased complexity required writing large amounts of very similar code for handling the software interface between applications and DBMS's. Such interface typically contains code for data exchange and modification transactions, database connection maintenance, error reporting, and general system maintenance and archiving activities. The problem is that such processes must be implemented for almost every class in the application code that typically contains hundreds of classes. The only solution to such coding problem is the use of code generators to generate most of the application-DBMS interface code automatically. Currently, there are many successful code generators that take the database structure as input and produce a suitable code interface as output in seconds. Such code generators include MyGeneration [2], LLBLGen [3], and CodeSmeth [4].

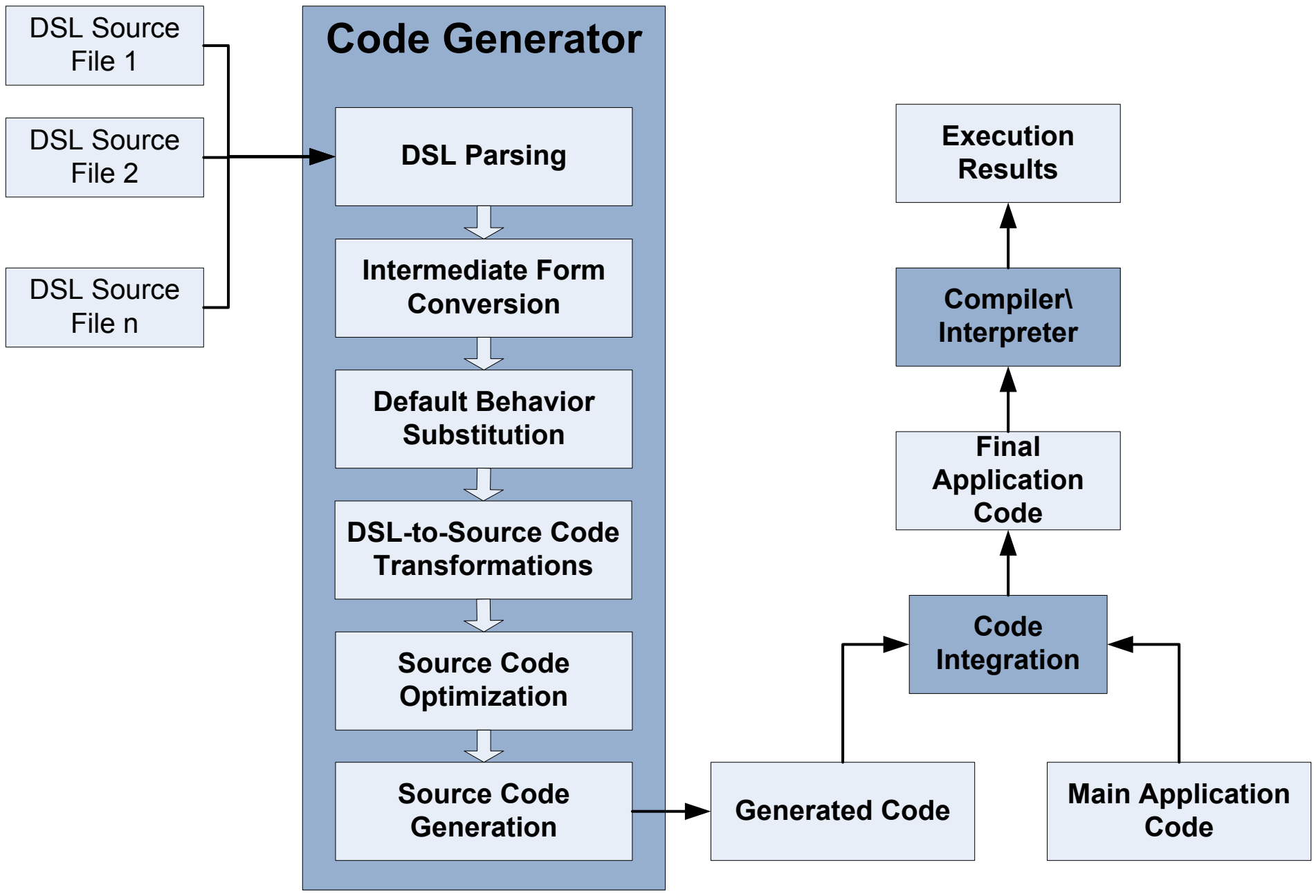

**Figure 1-3: Typical operation of a code generator**

## 1.2 Geometric Processing in Modern Software Systems

The need for processing large amounts of relational data in modern software systems is not the only challenge such system face. Another very important challenge is the need for efficiently processing large amounts of geometric data. Software applications in robotics, computer vision, physical modeling, flight simulation, and computer graphics are typical examples. In addition, commercial software targeting video game development and human-computer interaction applications require a fair amount of geometric processing code.

The current approach dominating the development of such software systems is the use of specialized hand coded geometric software libraries capable of representing the objects and transformations required in the geometric processing activities. Examples of such libraries include parts of the Visualization Toolkit (VTK) library [5], OpenGL [6], DirectX [7], and OpenCV [8] libraries. Such approach has several disadvantages hindering the wide applicability of geometric processing techniques among regular software developers. The first source of disadvantages is the generality of such geometric libraries. Such geometric libraries are bulky and full of unnecessary integration and initialization overhead. Such overhead increases the difficulties of integration with the main application code. In addition, the integration overhead usually degrades the performance of the final software system. Moreover, the learning curve of such libraries is as steep as the generality of the capabilities of the geometric library itself. The second source of disadvantages is the rigidness often found in such libraries. A better situation for a software designer would be to directly represent and process geometric data using the main design elements of the original software system; rather than using less related, predefined concepts specific to a certain geometric processing library. Such approach, if possible, would give greater control over the details, maintainability, and performance of the final software system.

Automatic code generation of geometric computations code from a high-level geometric DSL would be an effective alternative for such general purpose geometric libraries. In addition to all the advantages of using code generators discussed in section 1.1, such approach could enable regular software developers to create efficient geometric processing code with less effort and enhanced control over the generated code. The use of such technique is, unfortunately, not yet realized. The main reason behind this situation is the nature of the mathematical algebraic representations commonly used for representing geometry in practice. Figure 1-4 illustrates the steps required for the two approaches used for creating software implementations for geometric algorithms used in geometric processing applications.

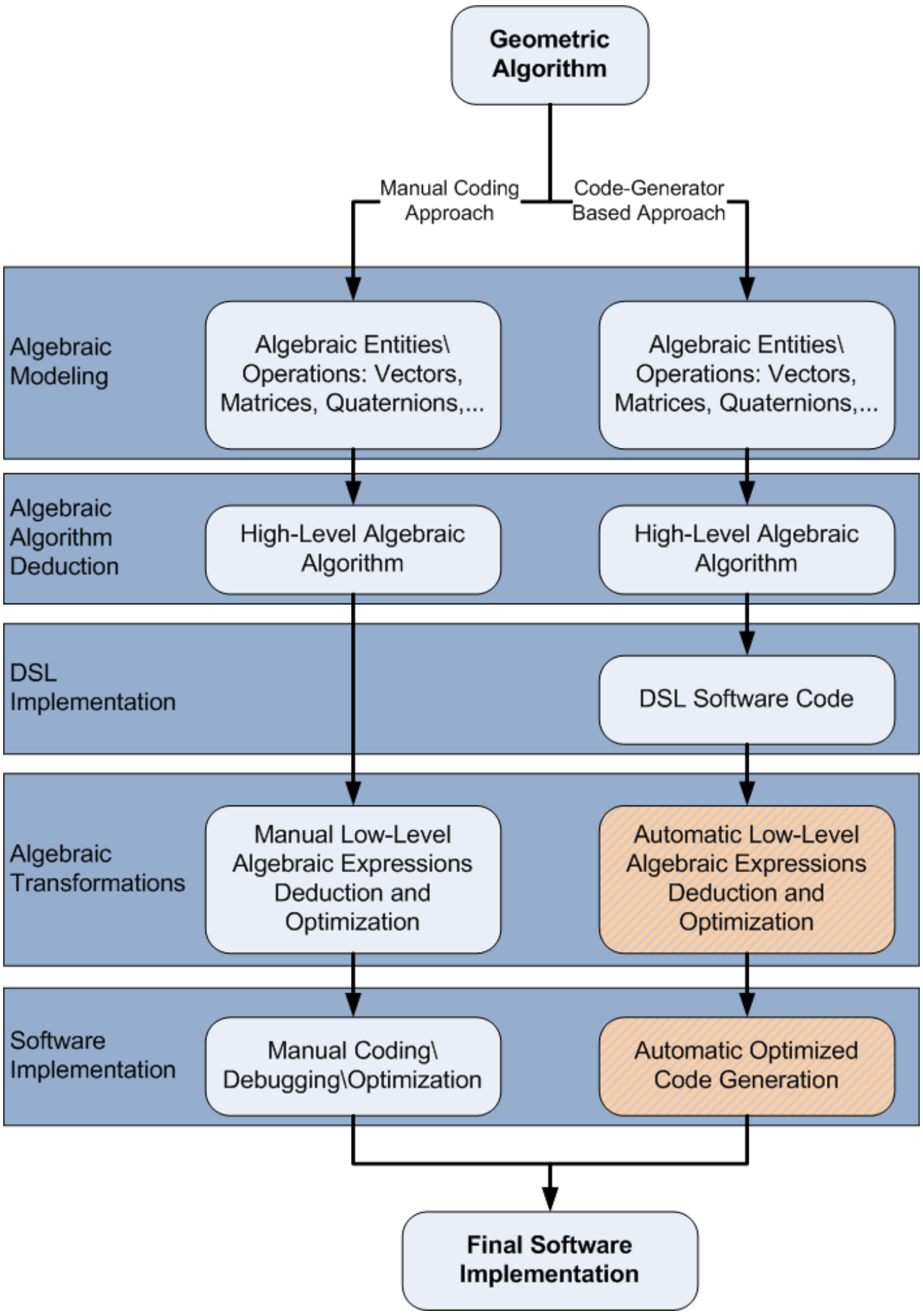

**Figure 1-4: Two approaches for implementing geometric algorithms**

## 1.3 Algebraic Representation of Geometry

A successful code generator for geometric processing problems must be based on a suitable mathematical algebraic system for representing geometric concepts. The geometric DSL of the code generator is based on such algebraic system. All transformations and optimizations required to convert the high-level geometric DSL code into the final implementation code utilizes the rules of the algebraic system as well. Thus, such algebraic system should satisfy a number of requirements in order to be suitable for such task:

1. ***Universality and generality***: The algebraic system should be able to express the widest spectrum of geometric entities, transformations, and concepts possible. The algebraic system should be able to effectively model different types of geometries commonly used in practice. Such geometries include, but are not limited to, Euclidean and non Euclidean geometries (spherical, elliptic, and hyperbolic), affine geometry, projective geometry, and conformal geometry.
2. ***High-level expressive symbolic mathematical language***: The algebraic system should be symbolically compact and efficient in representing geometric entities and transformations. Such compactness and expressiveness is necessary in order to make the geometric DSL suitable for software developers to write high-level code with.
3. ***Uniformity of representation***: The algebraic system should model similar geometric concepts in a uniform manner. If such requirement is violated, it would be necessary to use awkward conversions between different algebraic representations. Such conversions will result in difficulty of modeling and, eventually, in poor performance of final generated code.
4. ***Ease of learning***: The algebraic system should be relatively easy to learn by a wide spectrum of engineers and software developers. The algebraic system should not be more complex than the underlying geometric concepts it models. If the first three requirements are met by the algebraic system, the learning process is likely simplified.

5. ***Efficiency of computation***: The algebraic system should be capable of being used for practical computations, not only for symbolic representations. There should be suitable algebraic rules for transforming the high-level symbolic language into lower-level efficient to compute algebraic expressions. If such rules are not present, there will be no method for creating an efficient software implementation from the high-level DSL.

Many algebraic systems are traditionally used for representing geometric concepts in practice. Such systems include complex and real numbers, vector algebra, matrix algebra, linear algebra, quaternion algebra, Plücker coordinates, Grassmann-Cayley algebra, and tensor algebra. With the exception of tensor algebra, none of the other mathematical systems is general enough to be used by itself to represent a wide-enough spectrum of geometric concepts; which is the first requirement on the algebraic system. All previous systems cannot be qualified as having a high-level symbolic language for expressing geometric concepts in a manner suitable for regular software developers; which contradicts the second requirement. The necessity of using several of such algebraic systems in order to solve a certain geometric problem directly contradicts with the third requirement; namely uniformity of representation. The use of such many algebraic systems in practice effectively hinders the ability of typical engineers and software developers to learn all of them. It is typically required to learn so much mathematical systems, when to use or not to use them, and necessary techniques for transforming mathematical representations using such algebraic systems. The fifth requirement is certainly attainable using current algebraic representations. Unfortunately, in order to achieve efficient computations using such algebraic systems a tedious and awkward process of hand tuning and manual optimization must be made. Such process is usually application specific, and thus cannot be used for creating an automatic code generator.

From the previous discussion, it is not surprising that creating a code generator for solving geometric problems based on such algebraic systems is quite a challenging task. The problem is not that geometry is hard to automate or that code generation is a difficult task. The actual problem is with the algebraic representations commonly used for modeling geometric concepts. ***Geometric Algebra*** (GA) is the one of the closest algebraic systems to fulfilling all five requirements for creating an effective geometric code generator.

## 1.4 Geometric Algebra: Advantages and Obstacles

As will be illustrated in this section, geometric algebra effectively meets most of the requirements of the previous section. The only requirement not currently met by GA-based software is the fifth one: the capability of achieving efficient computations through GA. This work is primarily concerned with attaining such requirement for low-dimensional geometric problems. Such problems are common in many fields in computer science and engineering such as computer vision, computer graphics, 3D geometric modeling, and robotics.

### 1.4.1 Universality and Generality of Geometric Algebra

The first requirement on a suitable algebraic system for automatic code generation is universality and generality. An approximate view of the historical development leading to GA is shown in Figure 1-5. The figure is, unfortunately, incomplete as it neglects the enormous scientific contributions of the scientists of the Islamic civilization during the middle ages to both geometry and algebra [9]. As can be seen from Figure 1-5, geometric algebra is a powerful mathematical system encompassing many mathematical concepts under the same framework. GA is mainly based on the algebraic system called ***Clifford Algebra***, but with a strong emphasis on geometric interpretation. Clifford algebra is mathematically capable of representing all of the following algebraic systems and more:

1. ***Vector and matrix algebras***: A Clifford algebra is constructed on a base vector space [10]. Thus, a multivector is a natural generalization of traditional vectors. In addition, all types of vectors used in practice can be modeled using multivectors. Such vector types include free direction vectors, position vectors, and normal vectors. A special type of multivectors, called a versor, is capable of fully representing all orthogonal linear transform on the base vector space. Versors provide a much better mathematical treatment for orthogonal transform than orthogonal square matrices do. Such treatment is thoroughly illustrated and discussed in [11], [12], and [10].

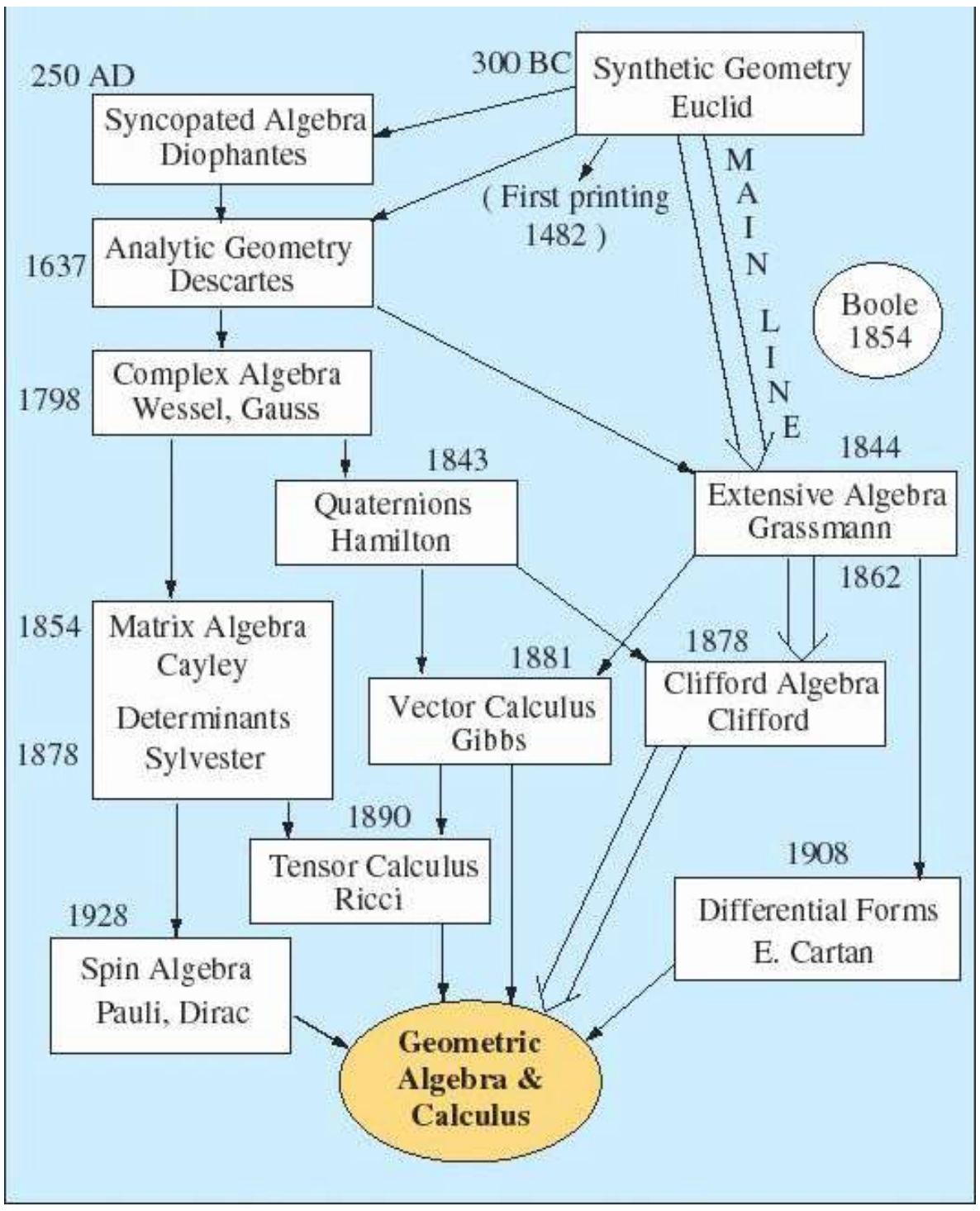

**Figure 1-5: Approximate Family Tree of Geometric Algebra and Calculus [13]**

2. ***Complex numbers***: The use of complex numbers in science and engineering is prevalent. The need for generalizing complex numbers to higher dimensional entities required answering an important question about the meaning of the imaginary unit $i = \sqrt{-1}$. Geometric algebra provides both an answer to the question and a method for generalization as discussed in [11] and [14].
3. ***Quaternions algebra***: Quaternions are efficient representations for rotation operators in 3D Euclidean space. Geometric algebra generalizes quaternions to operate in any dimension. Such generalized rotation operators are special types of multivectors called rotors [11], [14]. In addition, rotors can be applied to any multivector including traditional vectors and other rotors.

4. ***Grassmann–Cayley algebra***: The Grassmann–Cayley algebra is a form of modeling algebra for use in projective geometry. Such algebra uses subspaces as basic elements of computation, a formalism which allows the translation of synthetic geometric statements into invariant algebraic statements. This can create a useful framework for the modeling of conics and quadrics among other forms. Clifford algebra was historically constructed as a generalization to the Grassmann–Cayley algebra and quaternion algebra.
5. ***Plücker coordinates***: Plücker coordinates [11] are used in practice as a compact and computationally efficient technique for representing and processing points, lines, and planes. Geometric algebra gives both an elegant interpretation and good generalization for such system as illustrated in [11]. Thus, using GA can effectively replace and enhance the use of Plücker coordinates in practice.
6. ***Tensor algebra***: As illustrated in [12] and [15], tensor algebra and Clifford algebra are functionally equivalent and share the same level of generality as mathematical algebraic systems. Clifford algebra can be defined as an ideal on tensor algebra. Alternatively, tensor algebra can be defined as multi-linear functions on a Clifford algebra. Having both alternatives, the discussion in [15] strongly recommends using Clifford algebra over tensor algebra in practical applications. The main argument is that the main unit of computation in Clifford algebra is the higher-level more-abstract multivector along with an arsenal of well-defined products and operations. On the other hand, tensor algebra ultimately relies on addition and multiplication of scalars with an extensive use of summations and coordinates. Thus, tensor algebra has a lower-level symbolic representation capability compared to Clifford algebra.

As a natural consequence of being an algebraic generalization for many systems, GA has been proven to be capable of modeling many types of geometries. Such geometries include metric and non-metric geometries alike. Examples of metric geometries successfully modeled by GA include Euclidean,

Spherical, and Hyperbolic geometries [16]. In addition, projective geometry is a non-metric geometry successfully modeled by geometric algebra [11].

### 1.4.2 High-Level Expressiveness of Geometric Algebra

Figure 1-6 illustrates the main elements of geometric algebra. The main algebraic entity in GA is the multivector. In GA, two types of multivectors are used to model geometric concepts. The first type, called a blade, is used for representing oriented subspaces in the base vector space. Thus, blades can be used to compute with subspaces using standard GA products and operations. Blades can be used for representing geometric objects like points, vectors, normals, planes, lines, spheres, circles, tangents, and many others. Being able to perform computations on such objects in a direct manner is a great advantage from a modeling point of view. The second type of multivectors is called a versor. Versors can efficiently and uniformly represent orthogonal transforms, like rotations and reflections, on multivectors. Thus, versors can be used to transform all blades and versors resulting in a unified transformational treatment for diverse geometric entities. Figure 1-6 also shows some of the many operations that can be applied to multivectors. Such rich algebraic language for representing geometry is certainly suitable for creating a high-level DSL for modeling geometric concepts.

### 1.4.3 Uniformity of Representation in Geometric Algebra

The main source of uniformity in geometric algebra is its universality and generality discussed previously. For example, in the 5D Conformal Geometric Algebra (CGA), latter discussed in chapter 3, multivectors can represent many types of 3D Euclidean objects. Such objects include points, lines, planes, spheres, circles, point-pairs, tangent vectors and planes, and normal vectors. In addition, all Euclidean transformations can be represented using versors; a special type of multivectors. The versor product is used to apply any versor to any multivector. As a result, the expression for applying any Euclidean transform to any of the objects mentioned previously is exactly the same. Such example illustrates the main characteristic of geometric algebra: unification of representation resulting in uniformity of geometric transformations.

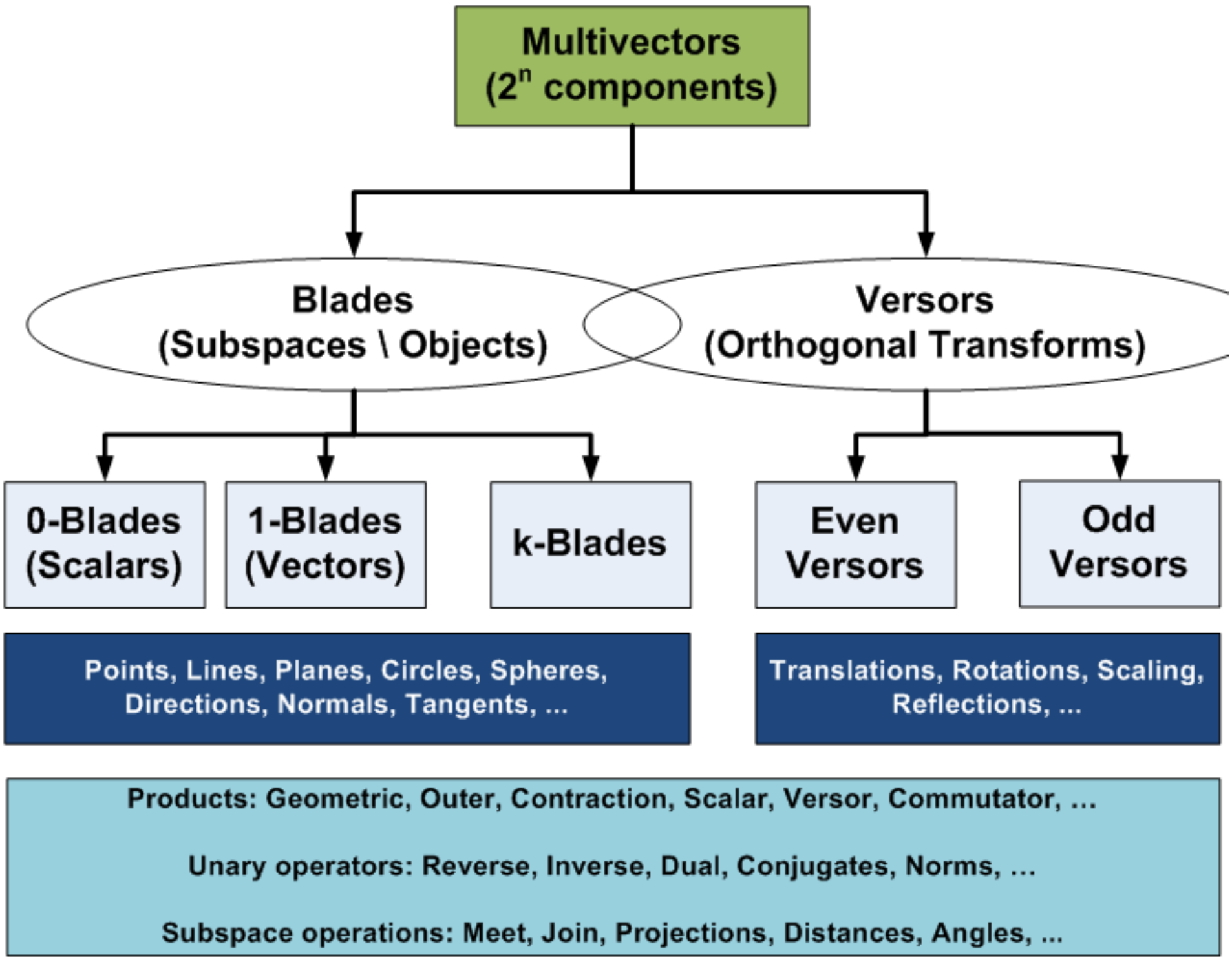

**Figure 1-6: Main elements of geometric algebra**

### 1.4.4 Learning Geometric Algebra

From the personal experience gained through this work, GA is not a hard to learn algebraic system. Given suitable GA references, the learning process of GA could reach a reasonable result in a relatively short time. Many concepts in GA are so simple and geometrically institutive that they can be taught in high-schools. Unfortunately, there is a lack of good teaching foundations of GA in most universities and schools around the world. Teaching GA should replace or go side by side with many mathematical devices currently taught in universities and schools. More effort must be spent in engineering and computer science communities to completely remove this obstacle. Fortunately the awareness of GA is rising rapidly. Very good introductions to GA with emphasis on GA computations are emerging constantly. Good examples of such books directed towards computer scientists and engineers include [11], [10], [17], [18], and [19].

### 1.4.5 Computing with Geometric Algebra

The main algebraic entity of GA, called the multivector, is used to mathematically represent many geometric concepts. It can represent both geometric objects and the linear transforms operating on them. Although this is a powerful feature from a modeling point of view it has a serious drawback. If a direct method for implementing multivectors is applied, huge amounts of memory and processing power are required. Such demand is equivalent to the level of generality that multivectors can provide. A geometric algebra constructed on a vector space of dimension $n$ is itself a $2^n$-dimensional space. Thus, a single multivector in, say, 3D Euclidean space would require 8 real coefficients for its representation. This problem highly forbids GA from being directly used in applications requiring efficient numerical computations. This work is an attempt to eliminate such problem to make GA gain its full potential in engineering and computer science applications.

## 1.5 Problem Statement

This work is concerned with designing an automatic code generator for creating software implementations for low-dimensional geometric algorithms based on geometric algebra. A successful GA-based code generator must have the following characteristics:

1. The software designer should be able to write high-level code that directly corresponds to GA objects and operations describing a certain geometric algorithm in a well designed DSL.
2. The GA-based code generator should be capable of transforming the high-level DSL code into optimized low-level code in a given target language.
3. The generated low-level code should be executed in the most efficient manner possible in regard to memory and processor requirements.
4. The integration between the generated code and the main application code should be simple, efficient, and highly maintainable.

Currently, there are very few GA-based code generators partially supporting the first two points. None of such code generators fully support all points.

## 1.6 Contributions

The contributions of this work include:

- The design of a novel code generator, called GMac, capable of producing optimized low-level code from a high-level description of a geometric algorithm based on geometric algebra.
- The design of a simple Domain Specific Language (DSL) that has enough features to convert many GA capabilities into algorithms that can produce good and efficient code through the proposed code generator; GMac. The code written in the DSL is highly maintainable and efficient as it is directly based on GA operations.
- The implementation of a new ray tracer based on OOP concepts in C#. The ray tracer can act as a flexible testing platform for new geometric modeling and ray tracing techniques.
- The implementation of a GA-based C# library, called Twister, capable of parametrically modeling complex free-form shapes based on the orbit of a point moving under Euclidean transformations.
- The use of the Twister library to define a generalized pinhole camera model that can simulate many types of projection techniques for ray tracing applications within a unified geometric framework.
- The use of GMac and its DSL to implement important geometric algorithms and computations in ray tracing. Unlike previous attempts for the same goal [20], the implementations of this work produce very efficient code. In most cases the code is comparable to or more efficient than hand-tuned code based on traditional mathematical techniques.

## 1.7 Contents of Thesis

This thesis consists of eight chapters and an appendix. Chapter 2 contains a literature review representing the problems facing writing good software implementations based on GA and the attempts to overcome such problems.

Chapter 3 presents a mathematical background to GA. The chapter begins by the full definition of GA emphasizing the geometric interpretations of various elements and operations of GA. An addition the chapter presents the basics of the well-applied Conformal Geometric Algebra (CGA).

Chapter 4 contains a full description for the proposed code generator GMac. The chapter begins by describing GMac design requirements, basic operation principle, and architecture. The chapter then provides a description of the Domain Specific Language (DSL) used in describing GA algorithms for GMac. The chapter concludes with a comparison of the architecture of GMac and two other similar code generators: Gaigen 2 and Gaalop.

Chapter 5 provides a brief introduction to ray tracing and describes the architecture of the base ray tracer implementation used to practically test GMac code performance.

Chapter 6 illustrates some of the possible modeling capabilities of CGA to model 3D free form objects. In addition, the chapter illustrates the enhancements applied to the base ray tracer using GMac in the scene modeling and rendering stages.

Chapter 7 containing some performance comparisons illustrating the advantages of GMac over Gaigen 2 and the advantages of using GA through GMac compared to traditional mathematical tools in ray tracing.

Chapter 8 concludes the thesis and presents some of the possible future work related to geometric algebra, GMac, and ray-tracing.

Finally, appendix A covers some basic mathematical concepts on which the definition of GA is based. Such concepts include fields, vectors, and bilinear forms. The appendix then moves to the algebraic definition of Clifford algebra which is the algebra GA is based upon.

# Chapter 2 : Literature Review

This chapter starts by presenting the main problems facing software implementations based on geometric algebra in section 2.1. The chapter then introduces some efforts made to solve such problems. In section 2.2 the direct software implementations of GA are presented along with their disadvantages. Section 2.3 introduces several generative approaches that include the Clifford library, Gaigen, Gaigen 2, and Gaalop.

## 2.1 Introduction

Since the revival of Clifford algebra and geometric algebra there were many attempts to write related software implementations. These attempts were mainly inspired by the characteristics of geometric algebra as a universal mathematical system. Such characteristics include generality, compactness of representation, robustness, unification of concepts, independence of dimensionality, and a high-level coordinate-free algebraic representation of geometric concepts. Unfortunately, any software implementation must eventually deal with coordinates in the form of basis-blades and their associated coefficients. The exponential nature of the number of basis blades $2^n$ compared to the base vector space of dimension $n$ almost deemed any implementation attempt useless in practice. In [20], the author describes the issues facing any efficient implementation of GA as follows:

- Multivectors are the general elements of computation in geometric algebra, but they are "big" compared to the dimensionality of the base vector space.
- A multivector can represent any multivector type (vector, rotor, circle, and so on) in a geometric algebra. This abstraction is great from a mathematical point of view, but it makes multivectors sparse in practice. This enforces overhead on the representation: either memory is wasted by storing zero coordinates, or processing time is wasted on compression of those coordinates.

- The number of basic geometric algebra operations on multivectors is quite large. Preferably one would like every operation to be implemented for every combination of multivector types, but this leads to a combinatorial explosion when encoded individually for maximum efficiency.
- The metric, encoded in the quadratic form, is a basic feature of a geometric algebra, affecting the most basic products. Many different metrics are useful and need to be allowed. Looking up the metric at run-time (e.g., from a table) is too costly. One should therefore not just implement a general geometric algebra and use it in a particular situation; for efficiency, one needs a special implementation for each metrically-different geometric algebra.
- For geometric algebra with low dimensionality the execution of each individual product or operation requires only a few processor instructions. Therefore any overhead imposed by the implementation (such as a conditional branch due to looping) results in a significant performance degradation.
- Expressions consisting of multiple products and operations can often be executed much more efficiently by folding them into one calculation rather than executing them one by one through a series of function calls.

The only hope for overcoming such problems came from the use of generative programming approaches [1]. Only when such approaches are combined with the rich and well-structured nature of GA as a mathematical system, the problems began to be solved efficiently. This chapter represents the previous attempts at solving the GA implementation problems with both failures and successes. The successful attempts would have never emerged but from the lessons learned from the failed ones.

## 2.2 Direct Implementations

Seveal direct implementations of geometric algebra were created [21], [22], [23]. All these implementations share a common theme of creating a single representation data structure as a software representation for multivectors. Some direct implementations applied some sort of coordinate compression to remove zero coefficients from multivectors. On the other hand, all such representations make heavy use of conditional statements and loops for implementing standard GA operations on multivectors. Although coordinate compression may improve memory and processing requirements, such extensive use of loops and conditionals degraded performance considerably. For example in [20], it is shown that a direct implementation of geometric algebra can often be about 100 times slower than traditional linear algebra-based geometric implementations. The net result is the unfortunate dismissal of geometric algebra from numerical computation applications.

The frustrations associated with direct implementations are apparent in GA literature. For example in [24] it is stated that there is still a lot of work to be done when it concerns implementations and applications. Even though geometric algebra looks promising for computational geometry at first sight, it turns out that a mapping from theory to practice is not as straightforward as one would hope. Obviously the old methods and theories have been used in practice for years and have undergone severe tweaking and tuning. Geometric algebra still has a long way to go before one will see implementations that allow one to benefit from the theoretical expressiveness and elegance without sacrificing performance. As the author of [24] sadly expresses, GA is best used in theory. After which actual implementations give up on the generality of GA and simply provide specialized algorithms for specific problem domains.

## 2.3 Generative Programming Approaches

Faced with the frustrations of the direct implementations, several researchers tried to find a way for GA to attain its true potential in scientific applications. The hope came from the well structured nature of GA itself. Applying concepts from generative programming to obtain good software implementations based on GA models was the alternative those researchers took. Such approach proved to be very successful and promises to deliver GA to its true potential in practical applications.

### 2.3.1 The Clifford Library

This library was a first attempt on applying generative programming concepts to implement a GA library. The author describes the library on his web site [25]. The library avoided storage for and operations on blades in a multivector that are always zero. The library did this at compile time using a combination of C++ template meta-programming, preprocessor macros, and a small amount of offline code generation. Performance was substantial compared to other libraries available at the time; mostly because the library code avoided doing any heap allocations.

However, compile-times were beyond horrid. The author's abuse of the accidentally Turing-complete C++ template compiler led to one 400 line unit-test for the 5 dimensional conformal geometric algebra taking three hours to compile on a typical C++ compiler. More often than not, the compiler crashed. The author concludes that interesting as that library may have been, it was completely unusable for anything practical. The Clifford library will not be investigated further in this work.

### 2.3.2 Gaigen

The second attempt resulted in the Gaigen software library [26]. Gaigen stands for ***G***eometric ***A***lgebra ***I***mplementation ***Gen***erator. The goal of the Gaigen implementation as described in [26] was to create an efficient, general implementation of geometric algebras of relatively low dimension, based on an orthogonal basis of any signature, for use in applications like computer graphics, computer vision, physics and robotics. The approach taken is to let the user specify the properties of the geometric algebra required, and to automatically generate source code accordingly. The resulting source code consists of three layers, of which the lower two are automatically generated. The top layer hides the implementation and optimization details from the user and provides a dimension independent, object oriented interface to using geometric algebra in software, while the lower layers implement the algebra efficiently. Coordinates of multivectors are stored in a compressed form, which does not store coordinates of grade parts that are known to be equal to zero. Optimized implementations of products can be automatically generated according to a profile analysis of the user application.

Gaigen generates C++ source code for a specific geometric algebra, according to the user's needs. The following are some properties of the algebra that the user can specify:

- Dimension.
- Signature of the orthogonal basis vectors {+1, 0, -1}.
- Reciprocal null vectors.
- What products to implement (geometric product, (modified) Hestenes inner product, left and right contraction, outer product, scalar product)
- How to optimize the implementation of these products.
- The storage order of the coordinates.
- What extra functions to implement (e.g. reverse, addition, inverse, projection, rejection, outermorphism).
- What coordinate memory (re-)allocation method should be used.

The way Gaigen implements geometric algebras seemed feasible to the authors. The process of optimizing an algebra implementation for a specific application is slightly annoying but doesn't take much time in practice. The authors saw no other general way to implement low dimensional geometric algebras as efficiently as current linear algebra implementations than to take at least a partial code-generation approach. Implementing the whole algebra using C++ templates causes too much overhead. Writing every algebra implementation or even its product code by hand is too tedious. Gaigen is quite extreme, relative to direct GA implementations, in that almost all code is generated, but the products will always have to be generated and turned into code.

A lot of improvements were still required before reaching the maximum performance software implementations of geometric algebra could achieve. Although Gaigen is a huge step forward compared to the performance of GABLE [22] (1000x faster), and an order of magnitude step forward compared to CLU [23] (20x faster), there is still a factor of 3 to 10 to go before the performance of equivalent linear algebra methods can be obtained.

Something the authors did not consider in Gaigen but thought to be a useful feature is the interoperability between geometric algebras. Assuming someone wants to use both a 3D Euclidean and a 5D conformal geometric algebra in the same application. Although this is entirely possible with Gaigen, there are no functions provided to "map" multivectors from one algebra to the other. One has to go to the coordinate level to do this. It would be nice if Gaigen could generate such translation functions, but what these translation functions should do exactly depends on the interpretation the user assigns to the multivectors from the respective algebras.

Because Gaigen 2 outperforms Gaigen in all aspects, Gaigen will not be considered further in this work.

### 2.3.3 Gaigen 2

Motivated by the partial success of Gaigen, the authors created another version called Gaigen 2. The architecture of Gaigen 2 is fully described in [20] and [27]. Their paper [27] describes their Geometric Algebra Implementation Generator Gaigen 2. Gaigen 2 synthesizes highly efficient GA implementations from the specification of the algebra. Functions over such algebras can be defined in a high-level coordinate-free domain-specific language, and Gaigen 2 transforms these functions into low-level coordinate-based code. This code can be emitted in any target language through a custom back-end. Benchmarks of their implementation show that the combination of GA and Gaigen 2 can rival the performance of standard geometry techniques, despite the greater abstraction and genericity of GA. To obtain this high performance, Gaigen 2 must adapt the generated code to the program that links to it. This is done via a profiling feedback loop. While running, the generated code makes a connection to the code generator. The generated code sends information about functions that should be optimized. The code generator registers this information and sends back new type information. After the program terminates, the code is regenerated according to the recorded profile. This profiling feedback technique may also be useful to implement other types of algebras. The full code generation process of Gaigen 2 is shown in Figure 2-1. Gaigen 2 is currently the only standard software implementation of geometric algebra providing good performance of generated code for general applications.

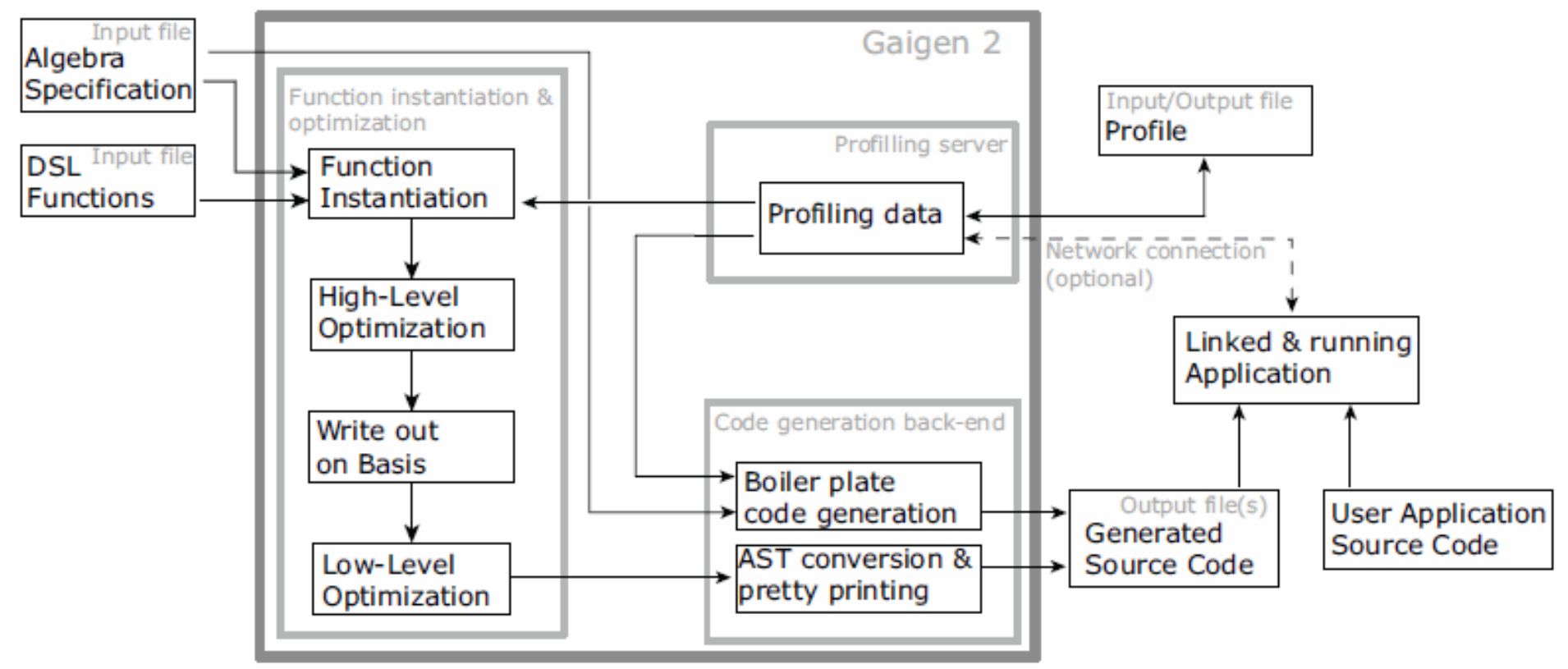

**Figure 2-1: The full Gaigen 2 code generation process [20]**

Although Gaigen 2 is currently the best available GA software implementation generator, it suffers from some disadvantages. First, Gaigen 2 introduces many unnecessary classes and functions to be integrated with the user application source code. Such integration may require manual linking by introducing other adaptor classes or functions. Such behavior may reduce readability of code necessary for code maintenance. Such behavior can also reduce overall application performance because of passing the same data through several application layers.

The second disadvantage is in the multiple profiling stages required for optimal code generation. As described in [20], the choice of which functions to instantiate with which specialized arguments should be determined by profiling the user application. Gaigen 2 cannot perform a static analysis of the user application source code before compilation. Instead, Gaigen 2 must extract the required information by running the application with profiling instrumentation added to it. In each profiling stage, the user application is recompiled and run through a "representative" input. The instrumentation records the usage profile of each DSL function. This usage profile consists of the specialized multivector types used as the arguments, and the number of invocations with those arguments. It is important that the user application "visits" all parts of its geometry code, otherwise the profile may miss information. This architecture may result in producing an over-fitted implementation to the data that was available during code generation. A new set of data used in actual practice may experience degraded performance. This is the most serious problem with Gaigen 2.

### 2.3.4 Gaalop

As described in [28] Gaalop stands for ***G***eometric ***A***lgebra ***Al***gorithms ***Op***timizer. It is mainly intended to produce optimized hardware implementations based on 5D CGA expressions. Although this work is concerned only with GA software implementations, Gaalop is considered here for two reasons. The first is that the author of Gaalop states that it is currently capable of producing C code. In addition the author claims the ability of Gaalop to produce code for other high-level languages, like Java, using the current architecture of Gaalop as can be seen in Figure 2-2. The second reason is that the basic idea behind Gaalop is the same one behind GMac; the proposed code generator system of this work. Namely, the use of a symbolic processing engine to produce optimized assignment statements in a given target language based on high-level GA expressions. The Gaalop project is not yet in a mature form like Gaigen 2 relatively is in.

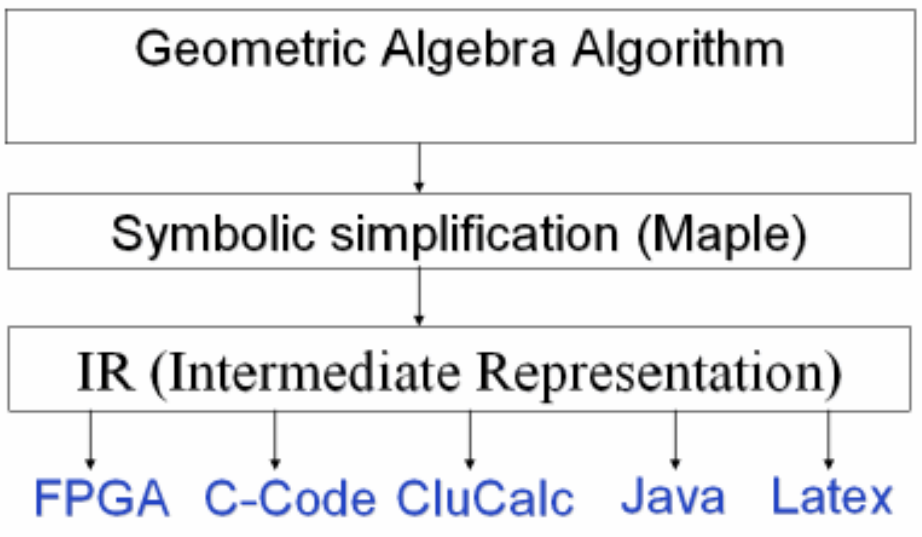

**Figure 2-2: Architecture of Gaalop [28]**

In addition to the disadvantage of being limited to 5D CGA, Gaalop suffers from several other disadvantages. The code generated by Gaalop has to be manually integrated with main application code. In addition, Gaalop has no DSL as a good code generator should. The use of Gaalop in practical applications is thus not a practical task.

# Chapter 3 : Geometric Algebra

This chapter introduces some mathematical concepts associated with the development of geometric algebra. In addition, the chapter briefly presents the conformal model of 3D Euclidean space; the 5D CGA. The 5D CGA is the geometric algebra on which the case study of this work is based.

Section 3.1 is the definition of geometric algebra. Section 3.2 illustrates some common geometric interpretations of GA. Section 3.3 illustrates calculations with basis in GA. Section 3.4 extends the bilinear forms on vectors to multivectors. Section 3.5 introduces some useful geometrically significant operations on multivectors. An introduction to the 5D CGA is presented in section 3.6. Section 3.7 illustrates some of the possible geometric objects that can be represented by CGA blades like lines, circles, planes and spheres. Section 3.8 represents some basic Euclidean transformations represented by CGA versors. Finally, section 3.9 extends such basic transformations to more general ones in an intuitive manner; thanks to the excellent algebraic structure of GA.

## 3.1 Geometric Algebra

Although GA can be defined by assigning an interpretation to a Clifford algebra, provided in Appendix A, the former definition of Clifford algebra hides the rich structure of GA. An alternative definition is thus presented to make this structure explicit. That definition is similar to the definition given in [29] but generalized to GA with any signature. A more extended, mathematically rigor definition can be found in [10].

***Definition***: A geometric algebra over the vector space $(\mathcal{F}, \mathcal{V}, \oplus, \otimes)$ with a quadratic form $\mathrm{Q}$ is a set $\mathcal{G}$ along with two binary operations (geometric sum $a,b \mapsto a+b$ and geometric product $a,b \mapsto ab$ ) and a grade operator ( $\langle \cdot \rangle_k : \mathcal{G} \to \mathcal{G}$ ) for which the following axioms are met:

***Ring Axioms***: The set $\mathcal{G}$ is a non-commutative ring over the two operations.

$$A+B=B+A \quad \forall A,B \in \mathcal{G} \tag{3.1}$$

$$A+(B+C)=(A+B)+C \quad \forall A,B,C \in \mathcal{G} \tag{3.2}$$

$$A(BC)=(AB)C \quad \forall A,B,C \in \mathcal{G} \tag{3.3}$$

$$A(B+C)=AB+AC \quad \forall A,B,C \in \mathcal{G} \tag{3.4}$$

$$(A+B)C=AC+BC \quad \forall A,B,C \in \mathcal{G} \tag{3.5}$$

$$\exists 0 \in \mathcal{G} : A+0=A \quad \forall A \in \mathcal{G} \tag{3.6}$$

$$\exists 1 \in \mathcal{G} : A\,1=1A=A \quad \forall A \in \mathcal{G} \tag{3.7}$$

$$\begin{gathered} \forall A \in \mathcal{G} \quad \exists -A \in \mathcal{G} \text{ such that} \\ A+(-A)=(-A)+A=0\,,0 \in \mathcal{G}^0 \end{gathered} \tag{3.8}$$

***Grade Axioms***: Each element of $\mathcal{G}$ (called a multivector) is composed of the sum of simpler multivectors (called *k*-vectors). The grade operator extracts the *k*-vector part of a multivector.

- The grade operator is a projection operator:

$$\left\langle \langle A \rangle_k \right\rangle_k = \langle A \rangle_k \quad \forall A \in \mathcal{G} \tag{3.9}$$

- A *k*-blade is the geometric product of nonzero orthogonal vectors. The sets of *k*-blades are defined as:

$$\begin{aligned} &\mathcal{B}^0 = \mathcal{F} \quad ,\mathcal{B}^1 = \mathcal{V} \\ &,\mathcal{B}^k = \{A \in \mathcal{G} : A = a_1 a_2 \cdots a_k \ , a_i a_j = -a_j a_i \ \forall i \neq j \\ &\qquad\qquad\qquad\qquad ,a_1, a_2, \ldots, a_k \in \mathcal{V} - \{\mathbf{0}\}\} \end{aligned} \tag{3.10}$$

- A *k*-vector is a linear combination of *k*-blades. The sets of *k*-vectors are defined as:

$$\begin{aligned} &\mathcal{G}^0 = \mathcal{B}^0 = \mathcal{F} \quad ,\mathcal{G}^1 = \mathcal{B}^1 = \mathcal{V} \\ &,\mathcal{G}^k = \{A \in \mathcal{G} : A = \sum_i \lambda_i A_i \quad ; A_i \in \mathcal{B}^k \,, \lambda_i \in \mathcal{F}\} \end{aligned} \tag{3.11}$$

- A multivector is the geometric sum of *k*-vectors and the $k^{\text{th}}$ grade of a multivector is a *k*-vector:

$$\forall A \in \mathcal{G} : A = \sum_k A_k \quad ,\langle A \rangle_k = A_k \in \mathcal{G}^k \tag{3.12}$$

- The grade operator is linear:

$$\left\langle \lambda A + B \right\rangle_k = \lambda \left\langle A \right\rangle_k + \left\langle B \right\rangle_k \quad \forall A, B \in \mathcal{G}, \lambda \in \mathcal{G}^0 \tag{3.13}$$

***Geometric sum and product Axioms***: These define the geometric sum and product of vectors (or 1-vectors) and scalars (or 0-vectors).

- The geometric sum of two scalars:

The geometric sum of two scalars is their ordinary addition (3.14)

- The geometric sum of two vectors is their ordinary addition:

$$a + b = a \oplus b \quad \forall a, b \in \mathcal{G}^1 \tag{3.15}$$

- The geometric product of two scalars is their ordinary multiplication:

$$\alpha\beta = \alpha \cdot \beta \quad \forall \alpha, \beta \in \mathcal{G}^0 \tag{3.16}$$

- The geometric product of a scalar and a vector is their scalar product:

$$\lambda a = \lambda \otimes a \quad \forall \lambda \in \mathcal{G}^0, a \in \mathcal{G}^1 \tag{3.17}$$

- The geometric product of a scalar and a multivector is commutative:

$$\lambda A = A \lambda \quad \forall \lambda \in \mathcal{G}^0, A \in \mathcal{G} \tag{3.18}$$

***Signature Axiom***: This axiom gives the space the desired signature hence defining orthogonality of vectors and possibly associating a metric with the space.

- The geometric product of a vector with itself is a scalar (called the signature of the vector) resulting from the quadratic form $\mathrm{Q}$:

$$aa = a^2 = \mathrm{Q}(a) \in \mathcal{G}^0 \quad \forall a \in \mathcal{G}^1 \tag{3.19}$$

***Remarks***:

- A geometric algebra $\mathcal{G}$ is itself a linear space over the field $\mathcal{G}^0$ with operations $\oplus$ equivalent to geometric summation and $\otimes$ equivalent to geometric multiplication with scalars. Thus a multivector is a vector with additional mathematical structure.

- A multivector $A$ is called a null-multivector if it squares to zero: $A^2 = AA = 0$.

- The zero multivector of the geometric product is the scalar $0 \in \mathcal{G}^0$, and the unity multivector is the scalar $1 \in \mathcal{G}^0$.

- For the definition of geometric algebra to be consistent, only a unique zero multivector is permitted. Therefore, the zero multivector $\mathbf{0} \in \mathcal{G}$ must be considered identical to the scalar $0 \in \mathcal{G}^0$

- In Euclidean space, if two vectors anti-commute then they are orthogonal:

$$ab = -ba \Leftrightarrow a \perp b \quad \forall a, b \in \mathcal{G}^1 \tag{3.20}$$

In addition, if two vectors commute then they are parallel:

$$ab = ba \Leftrightarrow a \parallel b \quad \forall a, b \in \mathcal{G}^1 \tag{3.21}$$

Hence, the geometric product of two vectors defines their mutual relation with respect to direction in Euclidean space.

- The set of *k*-vectors $\mathcal{G}^k$ is a subspace of $\mathcal{G}$:

$$\mathcal{G}^k \leq \mathcal{G} \quad \forall k = 0, 1, 2, \ldots, n \tag{3.22}$$

- The geometric algebra $\mathcal{G}$ is the direct sum of all the sets of *k*-vectors:

$$\mathcal{G} = \mathcal{G}^0 \oplus \mathcal{G}^1 \oplus \cdots \oplus \mathcal{G}^n \tag{3.23}$$

- The set of homogeneous multivectors is defined to be the union of the sets of *k*-vectors:

$$\mathcal{H} = \mathcal{G}^0 \cup \mathcal{G}^1 \cup \cdots \cup \mathcal{G}^n \tag{3.24}$$

- The set of all blades is defined to be the union of the sets of *k*-blades:

$$\mathcal{B} = \mathcal{B}^0 \cup \mathcal{B}^1 \cup \cdots \cup \mathcal{B}^n \tag{3.25}$$

- The set of *k*-versors is defined to be the set of multivectors with maximum grade *k* that are products of 1-vectors having non-zero signatures:

$$\mathcal{V}^k = \{A : A = a_1 a_2 \cdots a_r \text{ and } A_{\langle \max \rangle} = A_{\langle k \rangle} \quad ; a_i \in \mathcal{G}^1 ; \mathrm{Q}(a_i) \neq 0\} \tag{3.26}$$

Thus, every non-null *k*-blade is a *k*-versor but not all versors are blades.

- Every vector is a blade, every *n*-vector is a blade and it can be proven that every (*n-1*)-vector is a blade (called a hyper-blade or pseudo-vector). In addition, a scalar can be considered a 0-blade for convenience:

$$\begin{aligned} &\mathcal{F} = \mathcal{G}^0 = \mathcal{B}^0 \quad , \mathcal{V} = \mathcal{G}^1 = \mathcal{B}^1 \\ &, \mathcal{G}^{n-1} = \mathcal{B}^{n-1} \quad , \mathcal{G}^n = \mathcal{B}^n \end{aligned} \tag{3.27}$$

Scalars are called improper blades while other higher-grade blades are called proper blades.

- Generally, the geometric product of multivectors is neither commutative nor anti-commutative. Nevertheless, some special multivectors commute with all other multivectors. Such multivectors are called central:

$$(AB = BA \quad \forall B \in \mathcal{G}) \Leftrightarrow A \text{ is central} \quad ; \forall A \in \mathcal{G} \tag{3.28}$$

For vector spaces with even dimension, only scalars are central while for vector spaces with odd dimension any multivector on the form (scalar + pseudo-scalar) is central:

$$\begin{aligned} &AB = BA \quad \forall B \in \mathcal{G} \\ &\forall A \in \begin{cases} \mathcal{G}^0 & , \dim(\mathcal{G}^1) \text{ is even} \\ \mathcal{G}^0 \oplus \mathcal{G}^n & , \dim(\mathcal{G}^1) \text{ is odd} \end{cases} \end{aligned} \tag{3.29}$$

The previous definition and remarks only tell half the story. The other more important half is the geometric interpretation associated with the multivectors and operations of the geometric algebra. The following section is an investigation for some of the possible geometric interpretations.

## 3.2 Geometric Interpretations

The construction of a geometric algebra is mainly based on two elements: scalars and vectors. Other multivectors like versors, blades, and *k*-vectors are constructed from scalars and vectors by repeated application of geometric product, the grade operator, and the geometric sum. The geometric product of two vectors can be decomposed into inner and outer products, each of which has a geometric meaning and significance of its own. Other multivectors like rotors and pseudo-scalars are special cases of multivectors that have significant geometric importance in applications. That is the reason for giving them special definitions and investigations in GA literature. The same situation is observed with operations such as multivector inverse, reverse, and grade involution. Such operations have direct geometric interpretations and hence are essential in the study of GA. Thus, the main characteristic of GA is that it will always be open for further additions of multivector types and operations. That process of addition of structures is mainly motivated by the applications and interpretations of GA. As soon as a new consistent geometric interpretation is found, a new GA is created and most of its tools are directly ready for applications.

### 3.2.1 Scalars

The role of scalars in GA is the same in linear algebra. Scalars provide means of constructing linear combinations of multivectors to create new multivectors. Scalars are also means of quantifying relations between multivectors in GA in the form of measurements. The length of a vector, the angle associated with a rotor, and the weight associated with a blade to signify a weighted oriented subspace are some example for scalars in GA. Thus, scalars are very elementary in the study of GA.

### 3.2.2 Vectors

The geometric interpretations of vectors are so many and diverse that they sometimes led to confusion about the nature of the physical quantities modeled by vectors. Some of the possible interpretations of vectors are [30]:

- Abstract depiction of vectors as manipulatable arrows having no physical interpretation, though it can be intuitively helpful in developing an abstract geometric interpretation.
- Vectors as points designate places in a Euclidean space or with respect to a physical reference frame. This interpretation requires designation of a distinguished point (the origin) by the zero vector.
- Position vector $x$ for a particle that can “move” along a particle trajectory $x = x(t)$ must be distinguished from places that remain fixed.

- Kinematic vectors, such as velocity $v = v(t)$ and acceleration are "tied" to particle position $x(t)$. Actually, they are vector fields defined along the whole trajectory.
- Dynamic vectors such as momentum and force representing particle interactions.
- Rigid bodies. It is often convenient to use a vector $a$ as a 1D geometric model for a rigid body like a rod or a ruler. Its magnitude $|a|$ is then equal to the length of the body, and its direction $a/|a|$ represents the body's orientation or, better, its attitude in space.

Whatever interpretation is assigned to vectors an important point must be cleared. Vectors are no more than mathematical models for physical phenomena. When the model (vectors) fails or becomes inefficient, it should be replaced with a more suitable model. The idea behind GA is to provide many other mathematical models for representing physical phenomena including vectors, bivectors, rotors, versors …etc. The choice of which models to use is what is meant by the geometric interpretation associated with the GA.

### 3.2.3 The Geometric Product

Many interpretations can be assigned to the geometric product according to the interpretations assigned to various multivectors of GA. For example, the geometric product of two unit vectors in Euclidean space can be interpreted as a directed arc segment on a unit circle in the plane defined by the two unit vectors as shown in Figure 3-1 [31]. Another interpretation is that the geometric product of two vectors in Euclidean space defines their relative relation in space as it can be expressed according to the famous GA relation $ab = a \rfloor b + a \wedge b \; ; \forall a,b \in \mathcal{G}^1$. The first term $a \rfloor b$ is the usual inner product of the two vectors, which is proportional to cosine the angle between the two vectors. The second term $a \wedge b$ is the outer product between the two vectors that defines the plane in which they both reside. From that relation, if the two vectors are parallel then their geometric product is the same as the product of their magnitudes as $a \wedge b = 0$ and $a \rfloor b = |a||b|\cos(0) = |a||b|$. On the other hand, if the two vectors are orthogonal then their geometric product equals their outer product because $a \rfloor b = |a||b|\cos(\pm\pi) = 0$ and hence their geometric product can be interpreted as the plane they both define. Hence, the geometric product of free vectors in Euclidean space encodes the full geometric relationship between the two vectors.

The previous two interpretations are not the only interpretations for the geometric product. More interpretations are met along the paths to different applications. In addition to such interpretations, the geometric product provides a means for defining the inverse of vectors and blades. The interpretation of such inverse depends on the application. Such a powerful inverse concept results in many other powerful operations in GA like the dual of multivectors, division of multivectors, and projections of blades. The inverse cannot be defined using any single derived product like the inner or outer products. Only through using the full geometric product, the inverse of multivectors can be defined. Hence the geometric product is considered by many GA researchers the main and most elementary product between vectors.

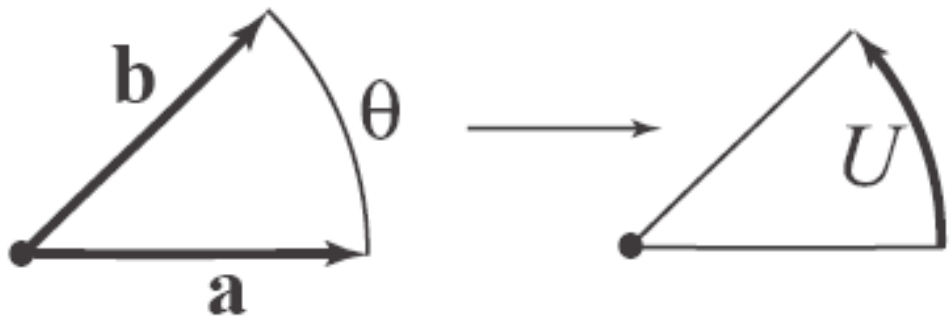

**Figure 3-1: Creating a rotor from the geometric product**

### 3.2.4 The Contraction and Inner Product

For a long time, the inner product of vectors was the only known general product of vectors. The result of the inner product is a scalar proportional to the angle between the two vectors. In GA, the inner product is given many interpretations just like the geometric product. The generalization of the inner product of vectors to other multivectors is best accomplished using the contraction product of multivectors. Although many other generalizations are present, like the Hestenes inner product and the fat-dot product (also called the modified Hestenes inner product), the contraction has the most direct and general geometric interpretation. Hence, many authors advocate its use as the most suitable generalization of inner product between vectors [32], [17]. In 3D Euclidean space, the contraction of a vector $x$ and a 2-blade $B_2$ is the vector $y$ that belongs to the blade $B_2$ and is orthogonal to the original vector $x$. Hence, the contraction is a grade-lowering process compared to the outer product which is a grade-rising process.

### 3.2.5 The Outer Product

The outer product is the second most important product after the geometric product. Even the contraction can be defined using the outer and geometric products. The outer product of two vectors is a new entity called a bivector that represents the extension of a vector in the direction of the other vector. As shown in Figure 3-2, the outer product of vectors $a$ and $b$ returns a directed area element of area $|a||b|\sin(\theta)$.

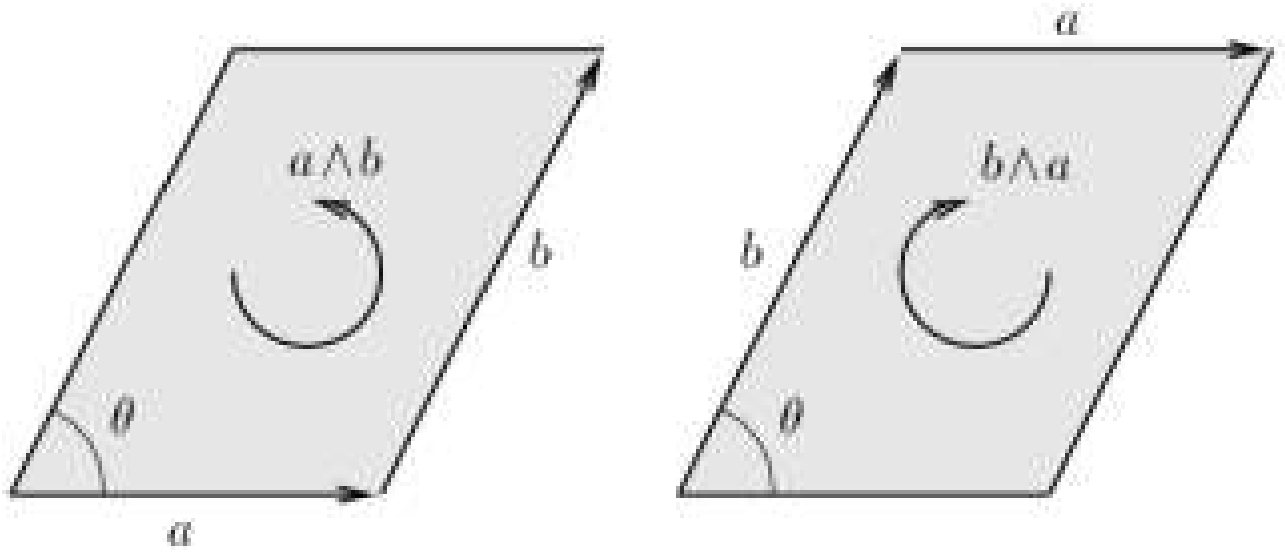

**Figure 3-2: The Outer Product [33].**

Through the outer product it is possible to describe, in a vector space, subspaces of arbitrary dimension; now called *k*-blades in GA literature. The outer product is also important for the extension of linear mappings using what is called outermorphisms [11]. An outermorphism is a generalized linear mapping that can act on multivectors and preserves their outer product.

### 3.2.6 Bivectors

It is common in GA literature to represent bivectors with parallelograms as in Figure 3-2. It is nonetheless important to realize that bivectors do not specify a shape. Given $a \wedge b$, there is no unique way to recover the vectors $a$ and $b$. All that the bivector encodes is the plane itself, together with an area and a handedness (direction). For this reason, it is sometimes better to replace the directed parallelogram with a directed circle (or any other planar shape) with the same area.

Bivectors were rediscovered many times in the history of physics and mathematics. Surprisingly, the most important bivector in mathematics is the unit imaginary number $i$ . Imaginary numbers and complex numbers were a mystery for a long time with no apparent direct physical interpretation. Nevertheless, imaginary and complex numbers are an essential corner for many models in physics and engineering. Geometric algebra provides a direct and clear interpretation of the unit imaginary as a bivector with unit magnitude. A bivector is the outer product of two vectors. Hence, whenever an interpretation is given for the vectors and their geometric product, bivectors are naturally explained with nothing "imaginary" about them. Bivectors are also associated with rotations in 3-dimensions. Quaternions, like complex numbers, were used to perform 3D rotations in many engineering applications without knowing what they are until bivectors were used. A quaternion is simply a scalar plus a bivector in 3D Euclidean space. Since the use of bivectors can be generalized to any dimension, the need for generalizations for imaginary, complex and quaternion numbers was fulfilled using GA.

### 3.2.7 Versors

A versor is simply the geometric product of a number of vectors. Versors are a very important tool that enables the replacement of matrices for the study of orthogonal linear mappings. Matrices were traditionally used to represent orthogonal transformations, like rotations and reflections, for a long time. Using versors, all of the properties of orthogonal transformations can be studied more efficiently and clearly compared to matrices. The reason behind such clarity is that matrices depend on a choice of basis and only act upon vectors. On the other hand, versors are coordinate-free representations that can act on any multivector including vectors, blades, and other versors. An excellent introduction to the use of versors to represent orthogonal linear transforms can be found in [11].

### 3.2.8 Blades

A blade is a special kind of multivectors that is the outer product of linearly independent vectors. Thus, a blade in Euclidean space is simply a special kind of versors. Blades are very important in GA because they are usually given a very important geometric interpretation. A blade represents a weighted oriented subspace in GA. Any subspace in a vector space can be represented by a blade definable up to a scale and sign (orientation). Hence, linear combinations of blades extend the power of vectors to multi-dimensions. Such association of blades and subspaces results in the ability of direct representation of geometric operations on subspaces using blades. Operations like projections of subspaces, rotations of subspaces, intersections and unions of subspaces are all possible using blades. That removed the barriers between points, lines, and planes in 3D geometry and unified their treatments as special grades of blades. For example, the angle between a line and a plane and the angle between two planes has totally different expressions in the traditional representations. In GA, the two angles can be calculated using the same simple expression by using blades as representatives for lines and planes.

### 3.2.9 Rotors

In 3D Euclidean space, a rotor is a scalar plus a bivector. Thus a rotor is essentially a versor. A rotor is mathematically equivalent to a quaternion, which is used to perform arbitrary rotations. The surprising fact is that in the 5D GA called Conformal Geometric Algebra (CGA), a rotor can perform translations, rotations, and dilations (uniform scaling) for elements of an equivalent 3D Euclidean space. Hence, rotors are suitable for many applications in computer graphics when studied within the CGA model. That is because rotors in CGA provide a unified method for performing all the important transformations necessary for rendering real world scenes in computer graphics.

## 3.3 Basis for a Geometric Algebra

Although using basis is not essential in the study of geometric algebra, they can be useful for some kinds of problems. They are essential for any GA software implementation. They are also useful for the proof of some theoretical results.

### 3.3.1 Definition of GA Basis

Having an orthonormal basis for $\mathcal{V}, \dim(\mathcal{V}) = n$ : $\mathbf{E} = \langle e_1, e_2, \ldots, e_n \rangle_{\perp 1}^{\mathcal{V}}$ then the following is true:

$$e_i e_j = \begin{cases} \mathrm{Q}(e_i) = \langle e_i, e_i \rangle & , i = j \\ -e_j e_i & , i \neq j \end{cases} \quad \forall\, 1 \le i \le n, 1 \le j \le n \tag{3.30}$$

Where $\mathrm{Q}(e_i) = \langle e_i, e_i \rangle \in \{0, 1, -1\}$ according to the signature of the unit vector $e_i$ .

For Euclidean spaces, this relation becomes:

$$e_i e_j = \begin{cases} 1 & , i = j \\ -e_j e_i & , i \neq j \end{cases} \quad \forall\, 1 \le i \le n, 1 \le j \le n \tag{3.31}$$

- The set of *k*-vectors $\mathcal{G}^k$ has basis $\mathbf{E}^k = \langle e_1^k, e_2^k, \ldots, e_m^k \rangle^{\mathcal{G}^k}$ where:

$$\begin{aligned} & e_J^k = e_{j_1} e_{j_2} \cdots e_{j_k} \quad , 0 < j_1 < j_2 < \cdots < j_k \le n \\ & \forall J = 1, 2, \ldots, m \quad , m = \binom{n}{k} \end{aligned} \tag{3.32}$$

- A unity pseudo-scalar of $\mathcal{G}$ is the basis for $\mathcal{G}^n$ :

$$I = e_1 e_2 \cdots e_n \tag{3.33}$$

- A basis for $\mathcal{G}$ is the set:

$$\begin{aligned} \mathbf{E}^{\mathcal{G}} &= \mathbf{E}^0 \cup \mathbf{E}^1 \cup \cdots \cup \mathbf{E}^n \\ &= \langle 1, e_1, \ldots, e_n, e_1^2, \ldots, e_{m_2}^2, e_1^3, \ldots, e_{m_3}^3, \ldots, I \rangle^{\mathcal{G}} \\ &, m_k = \binom{n}{k} \quad \forall k = 2, 3, \ldots, n-1 \end{aligned} \tag{3.34}$$

Hence if $n = \dim(\mathcal{V})$ then:

$$\dim(\mathcal{G}^k) = \binom{n}{k} \quad \forall k = 0, 1, 2, \ldots, n \tag{3.35}$$

In addition:

$$\dim(\mathcal{G}) = \sum_{k=0}^{n} \dim(\mathcal{G}^k) = 2^n \qquad (3.36)$$

For example, if $n = \dim(\mathcal{V}) = 4$ then Table 3-1 shows the basis for different *k*-vectors.

The unity pseudo-scalar is: $I = e_1e_2e_3e_4$ and the basis for $\mathcal{G}$ is:

$$\mathbf{E}^{\mathcal{G}} = \left\langle 1, e_1, e_2, e_3, e_4, e_1e_2, e_1e_3, e_1e_4, e_2e_3, e_2e_4, e_3e_4, e_1e_2e_3, e_1e_2e_4, e_1e_3e_4, e_2e_3e_4, I \right\rangle^{\mathcal{G}}$$

and $\dim(\mathcal{G}) = 2^4 = 16$

**Table 3-1: Basis vectors for the geometric algebra $\mathcal{G}_4$**

| $k$ | $\dim(\mathcal{G}^k) = \binom{n}{k}$ | $\mathbf{E}^k = \left\langle e_1^k, e_2^k, \ldots, e_m^k \right\rangle^{\mathcal{G}^k}$ |
|---|---|---|
| 0 | 1 | $\langle 1 \rangle^{\mathcal{G}^0}$ |
| 1 | 4 | $\langle e_1, e_2, e_3, e_4 \rangle^{\mathcal{G}^1}$ |
| 2 | 6 | $\langle e_1e_2, e_1e_3, e_1e_4, e_2e_3, e_2e_4, e_3e_4 \rangle^{\mathcal{G}^2}$ |
| 3 | 4 | $\langle e_1e_2e_3, e_1e_2e_4, e_1e_3e_4, e_2e_3e_4 \rangle^{\mathcal{G}^3}$ |
| 4 | 1 | $\langle e_1e_2e_3e_4 \rangle^{\mathcal{G}^4}$ |

The previous GA axioms and relations are enough by themselves to calculate any expression involving geometric products and summations. For example if $A = 5 + e_1 - 3e_1e_3$ and $B = 2e_1e_4 + e_2e_3e_4$ are two multivectors in a Euclidean space then $AB$ can be calculated as follows:

$$AB \quad = (5 + e_1 - 3e_1e_3)(2e_1e_4 + e_2e_3e_4)$$

Using the ring axioms and axiom (3.18) the expression becomes:

$$AB = 10e_1e_4 + 2e_1e_1e_4 - 6e_1e_3e_1e_4 + 5e_2e_3e_4 + e_1e_2e_3e_4 - 3e_1e_3e_2e_3e_4$$

Using relation (3.31) the expression is reduced to:

$$AB = 10e_1e_4 + 2e_4 + 6e_3e_4 + 5e_2e_3e_4 + e_1e_2e_3e_4 + 3e_1e_2e_4$$

Rearranging the expression into *k*-vectors:

$$\begin{aligned} AB &= (2e_4) + (10e_1e_4 + 6e_3e_4) + (3e_1e_2e_4 + 5e_2e_3e_4) + (e_1e_2e_3e_4) \\ &= \langle AB \rangle_1 + \langle AB \rangle_2 + \langle AB \rangle_3 + \langle AB \rangle_4 \end{aligned}$$

where: $\langle AB \rangle_1 = (2e_4)$, $\langle AB \rangle_2 = (10e_1e_4 + 6e_3e_4)$,

$\langle AB \rangle_3 = (3e_1e_2e_4 + 5e_2e_3e_4)$, $\langle AB \rangle_4 = (e_1e_2e_3e_4)$

For comparison, calculating $BA$ gives the expression:

$$\begin{aligned} BA &= (2e_1e_4 + e_2e_3e_4)(5 + e_1 - 3e_1e_3) \\ &= 10e_1e_4 + 5e_2e_3e_4 - 2e_1e_1e_4 + e_2e_3e_4e_1 - 6e_1e_4e_1e_3 - 3e_2e_3e_4e_1e_3 \\ &= 10e_1e_4 + 5e_2e_3e_4 - 2e_1e_1e_4 + (-1)^3e_1e_2e_3e_4 - (-1)^2 6e_1e_1e_3e_4 - (-1)^4 3e_1e_2e_3e_3e_4 \\ &= 10e_1e_4 + 5e_2e_3e_4 - 2e_4 - e_1e_2e_3e_4 - 6e_3e_4 - 3e_1e_2e_4 \\ &= (-2e_4) + (10e_1e_4 - 6e_3e_4) + (-3e_1e_2e_4 + 5e_2e_3e_4) + (-e_1e_2e_3e_4) \end{aligned}$$

Hence $AB \neq BA$ for general multivectors.

### 3.3.2 Reciprocal Basis

Having the basis $\langle e_1, e_2, \ldots, e_n \rangle^{\nu}$ the reciprocal basis $\langle e^1, e^2, \ldots, e^n \rangle^{\nu}$ can be obtained as [10]:

$$\begin{aligned} &e^k = (-1)^{k+1}(e_1 \wedge e_2 \wedge \cdots \wedge e_{k-1} \wedge e_{k+1} \wedge \cdots \wedge e_n)E^n \\ &\text{Where } E_n = e_1 \wedge e_2 \wedge \cdots \wedge e_n, E^n = e^n \wedge e^{n-1} \wedge \cdots \wedge e^1 \qquad (3.37) \\ &, E_nE^n = 1 \Rightarrow E^n = \frac{1}{(E_n)^2}E_n \end{aligned}$$

The reciprocal basis satisfies the following:

$$e_j \cdot e^k = \delta_j^k = \begin{cases} 1, j = k \\ 0, j \neq k \end{cases} \tag{3.38}$$

$$e_i A_k e^i = (-1)^k (n - 2k) A_k \quad \forall A_k \in \mathcal{B}^k \tag{3.39}$$

$$e_i (e^i \rfloor A_k) = k A_k \quad \forall A_k \in \mathcal{B}^k \tag{3.40}$$

$$e_i (e^i \wedge A_k) = (n - k) A_k \quad \forall A_k \in \mathcal{B}^k \tag{3.41}$$

Reciprocal basis are important for applications when the given basis vectors are non-orthogonal. Examples of such applications can be found in [11] and [10].

## 3.4 Extending the Bilinear Form

One of the main sources of power in GA is its associated bilinear form used to define the geometric product of vectors. Extending the bilinear form to multivectors [17] is essential for the definition of a very important product called the contraction product. The contraction product is a generalization for the inner product of vectors to general multivectors.

***Definition***: Having a geometric algebra $\mathcal{G}$ with an associated bilinear form $\mathrm{B}(u,v) = \langle u,v \rangle \; \forall u,v \in \mathcal{G}^1$ then it can be extended to multivectors as follows:

- For scalars the bilinear form is their ordinary product:

$$\langle \lambda, \mu \rangle = \lambda \mu \quad \forall \lambda, \mu \in \mathcal{G}^0 \tag{3.42}$$

- For $k$-blades of the same grade the bilinear form is defined as:

$$\begin{aligned} &\text{If } A_k = a_1 a_2 \cdots a_k \in \mathcal{G}^k, B_k = b_1 b_2 \cdots b_k \in \mathcal{G}^k \text{ then} \\ &\langle A_k, B_k \rangle = \det(\mathbf{C}) \text{ where } \mathbf{C} = (c_{ij}), c_{ij} = \langle a_i, b_j \rangle \end{aligned} \tag{3.43}$$

- For $k$-blades of different grade the bilinear form is zero:

$$\langle A_r, B_s \rangle = 0 \quad \forall A_r \in \mathcal{G}^r, B_s \in \mathcal{G}^s, r \neq s \tag{3.44}$$

- The bilinear form is linear in both arguments for any multivector:

$$\begin{aligned} &\langle \lambda A + B, C \rangle = \lambda \langle A, C \rangle + \langle B, C \rangle \\ &,\langle A, \lambda B + C \rangle = \lambda \langle A, B \rangle + \langle A, C \rangle \quad \forall A, B, C \in \mathcal{G}, \lambda \in \mathcal{G}^0 \end{aligned} \tag{3.45}$$

In addition, the extended bilinear form can be proven to be symmetric:

$$\langle A, B \rangle = \langle B, A \rangle \quad \forall A, B \in \mathcal{G} \tag{3.46}$$

This definition results in the relation:

$$\begin{aligned} &(\langle A, C \rangle = \langle B, C \rangle \quad \forall C \in \mathcal{G}) \Leftrightarrow A = B \\ &\forall A, B \in \mathcal{G} \end{aligned} \tag{3.47}$$

This is equivalent to a generalized definition of orthogonality on multivectors. Such generalized bilinear form is also called the scalar product of multivectors:

$$A * B = \langle A^{\sim}, B \rangle = \langle AB \rangle_0 = \sum_{r,s} \langle A_{\langle r \rangle} B_{\langle s \rangle} \rangle_0 \quad \forall A, B \in \mathcal{G} \tag{3.48}$$

Where $A^{\sim}$ is the reverse of multivector $A$ as defined in the next section. All other generalizations for the inner product of vectors can be derived from the extended bilinear form. More details on the derivation of the contraction product and the Hestenes inner product can be found in [11], [10], and [12].

## 3.5 Operations on Multivectors

Several operations on multivectors are provided in this section to complete the mathematical discussion required for the following chapters. All such operations are essentially derived from the axioms of GA. More information on such operations can be found in [11] and [10].

### 3.5.1 Bilinear Products and Outermorphisms

In addition to the geometric, outer, and contraction products there are several other useful GA bilinear products. Some of such products are used more than others in practice. All such products can be defined using the geometric product. In addition, any GA bilinear product ($\otimes$) satisfies the relation:

$$\begin{aligned} A \otimes B &= \left( \sum_{r=0}^{2^n-1} a_r E_r \right) \otimes \left( \sum_{r=0}^{2^n-1} b_r E_r \right) \\ &= \left( \sum_{m=0}^{2^n-1} \sum_{k=0}^{2^n-1} a_m b_k (E_m \otimes E_k) \right) \\ \forall A &= \sum_{r=0}^{2^n-1} a_r E_r, \quad B = \sum_{r=0}^{2^n-1} b_r E_r \end{aligned} \tag{3.49}$$

Thus it is sufficient to compute the product of the basis blades $E_m \otimes E_k$ to be able to compute the product of any two multivectors. Table 3-2 illustrates some possible definitions of bilinear products between basis blades $E_m$ where $g_m$ are the integers representing the grade of $E_m$ .

**Table 3-2: GA bilinear products of basis blades**

| **Bilinear Product ($\otimes$)** | $E_m \otimes E_k$ |
|---|---|
| Outer Product ($\wedge$) | $\langle E_m E_k \rangle_{g_m + g_k}$ |
| Left Contraction Product ($\rfloor$) | $\langle E_m E_k \rangle_{g_k - g_m}$ |
| Right Contraction Product ($\lfloor$) | $\langle E_m E_k \rangle_{g_m - g_k}$ |
| Scalar Product ($*$) | $\langle E_m E_k \rangle_0$ |
| Hestenes Inner Product ($\cdot$) | $\begin{cases} 0 & g_m = 0 \; or \; g_k = 0 \\ \langle E_m E_k \rangle_{\lvert g_k - g_m \rvert} & otherwise \end{cases}$ |
| Fat-Dot Product ($\bullet$) | $\langle E_m E_k \rangle_{\lvert g_k - g_m \rvert}$ |
| Commutator Product ($\overline{\times}$) | $\frac{1}{2}(E_m E_k + E_k E_m)$ |
| Anti-Commutator Product ($\underline{\times}$) | $\frac{1}{2}(E_m E_k - E_k E_m)$ |

The case with outermorphisms [11] is similar. An outermorphism is a generalization of a linear transform on vectors that preserves the outer product. Any outermorphism $L[]$ satisfies the following relation:

$$\begin{aligned} L[A] &= L[\sum_{k=0}^{2^n-1} a_k E_k] \\ &= \sum_{k=0}^{2^n-1} a_k L[E_k] \end{aligned} \tag{3.50}$$

Thus it is sufficient to find the values of $L[E_k]$. Such values can be easily evaluated as follows:

$$\begin{aligned} L[E_k] &= L[e_{i_1} \wedge e_{i_2} \wedge \ldots \wedge e_{i_m}] \\ &= L[e_{i_1}] \wedge L[e_{i_2}] \wedge \ldots \wedge L[e_{i_m}] \end{aligned} \tag{3.51}$$

Where $m$ is the grade of $E_k = e_{i_1} \wedge e_{i_2} \wedge \ldots \wedge e_{i_m}$. From the last two relations it is sufficient to find the effect of $L[]$ on the basis vectors $e_i, i = 1, \ldots, n$ in order to compute the effect of $L[]$ on any multivector.

### 3.5.2 Involutions and Conjugates

Three metric-independent linear operations can be defined on *k*-blades and generalized to multivectors through their linearity. Such operations appear in many important GA expressions and equations throughout GA literature.

An involution is an operation that maps an operand to itself when applied twice [10]. The most common involution in GA is the reverse defined on a *k*-blade $A_k = a_1 \wedge a_2 \wedge \ldots \wedge a_k$ as:

$$\widetilde{A_k} = A_k^{\sim} = a_k \wedge a_{k-1} \wedge \ldots \wedge a_1 = (-1)^{k(k-1)/2} A_k \tag{3.52}$$

The reverse is metric-independent because it is defined using the outer product and grade operations only. Extension to general multivectors comes from the reverse being a linear operation:

$$(\lambda A + B)^{\sim} = \lambda A^{\sim} + B^{\sim} \quad \forall A, B \in \mathcal{G}; \lambda \in \mathcal{F} \tag{3.53}$$

In addition, the reverse has the following important properties:

$$(AB)^{\sim} = \widetilde{B}\widetilde{A}$$
$$(A^{\sim})^{\sim} = A \qquad (3.54)$$

The second operation is the grade involution defined on *k*-blades as:

$$\widehat{A_k} = A_k^{\wedge} = (-1)^k A_k \qquad (3.55)$$

The grade involution has the following properties:

$$(\lambda A + B)^{\wedge} = \lambda A^{\wedge} + B^{\wedge},$$
$$(AB)^{\wedge} = \widehat{A}\widehat{B} \qquad (3.56)$$
$$(A^{\wedge})^{\wedge} = A \quad \forall A, B \in \mathcal{G}; \lambda \in \mathcal{F}$$

The third operation is called the Clifford Conjugate. It is a combination of the first two operations:

$$A_k^{\dagger} = (A_k^{\sim})^{\wedge} = (A_k^{\wedge})^{\sim} = (-1)^{k(k+1)/2} A_k \qquad (3.57)$$

The Clifford conjugate satisfy the following properties:

$$(\lambda A + B)^{\dagger} = \lambda A^{\dagger} + B^{\dagger},$$
$$(AB)^{\dagger} = B^{\dagger} A^{\dagger} \qquad (3.58)$$
$$(A^{\dagger})^{\dagger} = A \quad \forall A, B \in \mathcal{G}; \lambda \in \mathcal{F}$$

A fourth operation common in practice is simply called the conjugate of a multivector. The full definition of a conjugate is given in [10]. The definition of a positive-definite norm for arbitrary multivectors, called the magnitude of a multivector, is based on such operation.

### 3.5.3 Norms

The quasi-norm is a bilinear function on multivectors that map two multivectors to a real number. The value of the quasi-norm is not necessarily positive for general spaces; thus the prefix “quasi-”. The quasi-norm of two multivectors is defined using the scalar product as:

$$[\![A]\!] = A * A^{\sim} = \langle AA^{\sim} \rangle_0 \quad \forall A \in \mathcal{G} \qquad (3.59)$$

The quasi-norm is very important in defining the inverse of versors in the next subsection. The inverse of versors is used in the versor product to apply a versor to a multivector to represent orthogonal linear transforms on multivectors. The quasi-norm is positive-definite for Euclidean spaces.

Another norm common in literature is defined as:

$$\begin{aligned} \|A\| &= \sqrt{\sum_k \left[\!\left[ A_{\langle k \rangle} \right]\!\right]} = \sqrt{\sum_k \left| \left\langle A_{\langle k \rangle} \tilde{A}_{\langle k \rangle} \right\rangle_0 \right|} \\ \forall A &= \sum_k A_{\langle k \rangle} \in \mathcal{G} \end{aligned} \tag{3.60}$$

Such norm is always positive but not guaranteed to be positive-definite for non-Euclidean spaces.

In some applications, a positive-definite norm must be defined on all multivectors including null-blades. Such norm is called the magnitude of the multivector $|A|$. A full definition for the magnitude is given in [10]. All three norms are completely equivalent for Euclidean spaces.

### 3.5.4 Inverse, Dual, and Versor Product

An important class of multivectors is called versors. A versor is the geometric product of non-null vectors. Versors can efficiently represent orthogonal transforms on multivectors through the versor product. In order to define the versor product, the inverse of a versor $V$ must first be defined as:

$$V^{-1} = \frac{\tilde{V}}{[\![V]\!]} = \frac{\tilde{V}}{\tilde{V}V} \tag{3.61}$$

Where versors have the following important property:

$$V\tilde{V} = \tilde{V}V = [\![V]\!] \in \mathcal{F} \tag{3.62}$$

In addition, every non-null blade is a versor and the previous equations are applicable to non-null blades as well.

The versor product of a versor $V$ and a multivector $A$ is then defined as:

$$VAV^{-1} \tag{3.63}$$

Versors with unity quasi-norm are usually called rotors in GA literature. The inverse of a rotor is its reverse, as can be seen from (3.61). Thus, the versor product with a rotor $R$ is reduced to:

$$R A \widetilde{R} \tag{3.64}$$

The dual operation is another important operation usually used for blades. Assuming two blades $A_r, B_s, r \leq s$ where the subspace represented by $A_r$ is fully contained in the subspace represented by the non-null blade $B_s$ . Then, the dual of $A_r$ with respect to $B_s$ is defined as the (*s*-*r*)-blade:

$$C_{s-r} = A_r^{*B_s} = A_r \rfloor B_s^{-1} \tag{3.65}$$

The dual of a blade is another blade that represents the orthogonal complement of its associated subspace. If the dual is with respect to the space pseudo-scalar $I$ , it is simply written as $A^*$.

The dual operation is not an involution since for general blades $A, B$ : $(A^{*B})^{*B} = \pm A$ . An inverse operation, called the un-dual [11], is thus used for obtaining the original blade. The un-dual is defined as:

$$A^{\odot B} = A \rfloor B \tag{3.66}$$

When taking the un-dual with respect to the space pseudo-scalar $I$ , the un-dual is simply written as $A^{\odot}$.

The dual can be used to define the projection of a blade $A$ into another non-null blade $B$ as:

$$\begin{aligned} P_B(A) &= (A^{*B})^{\odot B} \\ &= \frac{1}{[\![B]\!]}(A \rfloor B) \rfloor B \end{aligned} \tag{3.67}$$

Thus, whatever geometric entities the two blades represent, the projection of the two entities is defined using the same equation.

## 3.6 Introduction to Conformal Geometric Algebra

The conformal model is one of the most remarkable successes of geometric algebra. It is based on a geometric algebra called Conformal Geometric Algebra (CGA). It is simply an embedding of the traditional n-dimensional Euclidean space into a (n+2)-dimensional space having a Minkowski metric; where one of the basis has a negative signature. This chapter will focus on the special case of 3D Euclidean space and its corresponding 5D CGA. The material presented in this chapter is mainly based on [11].

Assuming an orthonormal basis for 3D Euclidean space $\langle e_1, e_2, e_3 \rangle_{\perp 1}^{\mathbf{R}^3}$ with all basis vectors having signature 1, the corresponding 5D CGA can be defined by adding two LID vectors to the basis. There are two common choices for the two vectors in GA literature. The first choice is to add two orthonormal vectors $e$ and $\bar{e}$ having signatures 1 and -1 respectively. The second choice is by adding two null vectors $e_0$ and $e_\infty$. The relations between the four vectors are as follows [11]:

$$\begin{aligned} e_0 \cdot e_\infty &= -1, \quad e_0 \cdot e_0 = e_\infty \cdot e_\infty = 0, \\ e \cdot \bar{e} &= 0, \quad e \cdot e = 1, \quad \bar{e} \cdot \bar{e} = -1, \\ e_0 &= \tfrac{1}{2}(\bar{e} + e), \quad e_\infty = \bar{e} - e \\ \bar{e} &= e_0 + \tfrac{1}{2} e_\infty, \quad e = e_0 - \tfrac{1}{2} e_\infty \end{aligned} \tag{3.68}$$

Although the orthogonal basis is useful for simplifying some computations, the other basis is more suitable for most applications. Hence, this chapter will be using the second basis $\langle e_0, e_1, e_2, e_3, e_\infty \rangle$ exclusively.

The following sections introduce the basics of the conformal model. The blade $I_3 = e_1 \wedge e_2 \wedge e_3$ will be called the Euclidean pseudo-scalar. The blade $I = e_o \wedge I_3 \wedge e_\infty = e_o \wedge e_1 \wedge e_2 \wedge e_3 \wedge e_\infty$ will be called the CGA pseudo-scalar. The inverses of two blades are related through:

$$I^{-1} = e_o \wedge I_3^{-1} \wedge e_\infty \tag{3.69}$$

## 3.7 Representing Objects of Euclidean Space

The power of the CGA comes from the use of different GA subspaces to represent different classes of objects in Euclidean space. For example, a direction vector, a position vector, and a normal vector are all represented by 3-tupples in traditional linear algebra. The three entities model very different mathematical and physical concepts of 3D Euclidean space. In addition, the three entities behave very differently under geometric operations. A direction vector is translation-invariant; it should not be affected by translations, only by rotations. A position vector represents a point in Euclidean space and is affected by all Euclidean transforms. The inverse of a Euclidean transform cannot be directly applied to a normal vector to obtain a correctly transformed normal vector. CGA provide an alternative representation for all three entities using distinct sets of basis blades. In addition, all Euclidean transforms can be represented using a unified method through CGA versors. Such method of representation is compact, efficient, and universal in that it acts on all multivectors in the same way to produce correct results with correct geometric interpretations regardless of dimension or selection of basis.

### 3.7.1 Representing Points

In traditional linear algebra, a position vector $p_E = p_1 e_1 + p_2 e_2 + p_3 e_3$ is used to represent Euclidean points in 3D space. In CGA an alternative more powerful representation is used:

$$\begin{aligned} p &= e_0 + p_E + \tfrac{1}{2} p_E^2 e_\infty \\ &= e_0 + p_1 e_1 + p_2 e_2 + p_3 e_3 + \tfrac{1}{2}(p_1^2 + p_2^2 + p_3^2) e_\infty \end{aligned} \qquad (3.70)$$

The multivector $p$ is a CGA vector satisfying the relations:

$$p \rfloor e_\infty = -1, \quad p^2 = 0 \qquad (3.71)$$

This multivector is called a normalized point representation or a normalized point for short. The Euclidean distance of any two points $p_E, q_E$ having normalized points $p, q$ can be obtained through their inner product as follows:

$$d_E^2(p_E, q_E) = -2 p \rfloor q \qquad (3.72)$$

Hence, the Euclidean distances of points can be directly obtained through a simple GA operation; the contraction product of normalized points.

The origin of Euclidean space is traditionally represented by a zero position vector. Substituting into (3.70) the corresponding normalized point is equal to $e_o$. Thus the basis blade $e_o$ is a representation of the Euclidean origin in CGA. In addition, as illustrated in [11] the basis blade $e_\infty$ represents the point at infinity; a very important extension to Euclidean space with many useful implications.

### 3.7.2 Representing Spheres

A sphere in 3D Euclidean space can be fully defined using two different but related methods. The first is by defining its center point $c_E$ and radius $r$. The second is by defining 4 non-planar points $p_{1E}, p_{2E}, p_{3E}, p_{4E}$ belonging to the sphere. 5D CGA has two corresponding representations of the sphere. Both methods rely on representing a sphere using a CGA blade. The first representation is called the direct sphere representation or direct sphere for short. It is a 4-blade $\sigma$ created by taking the outer product of the normalized points corresponding to 4 points on the sphere:

$$\sigma = p_1 \wedge p_2 \wedge p_3 \wedge p_4 \tag{3.73}$$

The second representation is called the dual sphere representation or a dual sphere for short. It is a CGA vector $\omega$ constructed from the normalized point of the center $c$ and the radius:

$$\omega = c - \tfrac{1}{2} r^2 e_\infty \tag{3.74}$$

Interestingly, the two blades are related through:

$$\omega = \sigma^* = \sigma \rfloor I^{-1} \tag{3.75}$$

Thus the center and radius of a sphere passing through four points in 3D Euclidean space can be obtained from the multivector:

$$\omega = (p_1 \wedge p_2 \wedge p_3 \wedge p_4) \rfloor I^{-1} \tag{3.76}$$

The last equation is a clear example of the algorithmic importance of GA in practice. If a program can be written in a compact form similar to (3.76) and then transformed into its basic coordinate expressions automatically, a very powerful geometrically based software engine is created. The aim of this work is to create such engine.

A clear observation resulting from (3.74) is that a normalized point is a dual sphere with radius zero. This observation is important to numeric calculations based on CGA. If a vector on the form (3.74) results from a GA computation, it might be misinterpreted as a dual sphere with very small radius due to round off errors. Care must be taken to distinguish true spheres from normalized points with very small radii as stated in [20].

### 3.7.3 Representing Planes

As in the case of the sphere, a plane in 3D Euclidean space has blades for direct and dual plane representations related by the dual operation. Assuming a plane having a normal vector $n_E$ with distance $\delta$ from the origin and points $p_{1E}, p_{2E}, p_{3E}$ on the plane; the blade of the direct plane $\Pi$ is the outer product of the three normalized points and the point at infinity:

$$\Pi = p_1 \wedge p_2 \wedge p_3 \wedge e_\infty \tag{3.77}$$

While the dual plane blade $\Theta$ is the vector:

$$\Theta = n_E + \delta e_\infty \tag{3.78}$$

The two blades are related through:

$$\Theta = \Pi^* = \Pi \rfloor I^{-1} \tag{3.79}$$

Thus the normal and origin distance of a plane passing through three points can be obtained by the relation:

$$\Theta = (p_1 \wedge p_2 \wedge p_3 \wedge e_\infty) \rfloor I^{-1} \tag{3.80}$$

A quick study of the plane blades reveals the nature of plane representation in the conformal model. A plane is very similar to a sphere except that its fourth point is the point at infinity represented by $e_\infty$. In addition a plane can be given by one point $p_E$ and a direction bivector $A_E = v_{1E} \wedge v_{2E} = n_E \rfloor I_3$ where $n_E$ is the plane's normal and $v_{1E}, v_{2E}$ are two LID Euclidean direction vectors parallel to the plane. In this case the direct plane is constructed as:

$$\begin{aligned}\Pi &= p \wedge A_E \wedge e_\infty \\ &= p \wedge v_{1E} \wedge v_{2E} \wedge e_\infty\end{aligned} \tag{3.81}$$

This form of constructing planes can be used to completely replace the representation of planes by normal vectors. The main reason behind the importance of such representation is that there is no unique normal vector to a plane in dimensions higher than three. On the other hand, a unique 2-blade parallel to the plane can be defined in any dimension.

### 3.7.4 Representing Circles

Having a circle in a plane with direction bivector $A$ passing through the circle's center $c_E$ and assuming the circle passes through three points with position vectors $p_{1E}, p_{2E}, p_{3E}$. The 3-blade of the direct circle $\phi$ and the 2-blade of the dual circle $\varphi$ are:

$$\begin{aligned}\phi &= p_1 \wedge p_2 \wedge p_3,\\ \varphi &= (c + \tfrac{1}{2} r^2 e_\infty) \wedge (-c \rfloor \hat{A} e_\infty)\\ &= \phi^*\end{aligned} \tag{3.82}$$

### 3.7.5 Representing Lines

A line passing through two points $p_{1E}, p_{2E}$ parallel to direction vector $v_E = p_{2E} - p_{1E}$ can have a direct line representation $L$ constructed by:

$$\begin{aligned}L &= p_1 \wedge p_2 \wedge e_\infty\\ &= p_1 \wedge v_E \wedge e_\infty\end{aligned} \tag{3.83}$$

The dual line representation is simply:

$$l = L^* = L \rfloor I^{-1} \tag{3.84}$$

From the previous relations, a line is a special circle passing through two points and the point at infinity $e_\infty$.

### 3.7.6 Representing Point Pairs

A point pair is a new geometric construct in CGA. It results from intersecting a line with a sphere where the line passes through the sphere's center. A point pair having the two points $p_{1E}, p_{2E}$ has a direct point pair representation constructed by:

$$P = p_1 \wedge p_2 \tag{3.85}$$

The dual point pair representation is simply:

$$p = P^* = P \rfloor I^{-1} \tag{3.86}$$

A point pair can be used to model a line segment or two coupled points with fixed distance and orientation in Euclidean space.

### 3.7.7 Representing Direction Vectors and Bivectors

Direction vectors have different behavior from position vectors. While a direction vector is translation-invariant, a position vector is not. The same translation-invariance goes for direction bivectors; which are the outer product of two LID direction vectors. The direct direction vector having coordinates $v_E = v_1 e_1 + v_2 e_2 + v_3 e_3$ is:

$$v = v_E \wedge e_\infty \tag{3.87}$$

While the direct direction bivector constructed from direction vectors $v_{1E}, v_{2E}$ and having coordinates $A_E = v_{1E} \wedge v_{2E} = a_1 e_1 \wedge e_2 + a_2 e_2 \wedge e_3 + a_3 e_1 \wedge e_3$ is:

$$A = A_E \wedge e_\infty \tag{3.88}$$

In 5D CGA, a normal vector $n_E$ and the corresponding direction bivector $A_E$ are related through the relations:

$$\begin{aligned} n_E &= A_E^* = A_E \rfloor I_3^{-1}, \\ A_E &= n_E^{\odot} = n_E \rfloor I_3 \end{aligned} \tag{3.89}$$

Thus, a direction bivector can completely replace a normal vector in any application using CGA as the representation for 3D Euclidean space.

### 3.7.8 Representing Tangent Vectors and Bivectors

A tangent vector is a direction vector "glued" to a fixed point in Euclidean space. It results from intersection operations between touching elements like two touching circles or a line touching a circle. Assuming a direction vector $v_E$ fixed at a point $p_E$ the direct tangent vector is:

$$v_t = p \wedge (-p \rfloor v) \tag{3.90}$$

The same is true for a tangent bivector that has a direction bivector $A_E$ fixed at a point $p_E$:

$$A_t = p \wedge (p \rfloor A) \tag{3.91}$$

Where $A = A_E \wedge e_\infty$ and $v = v_E \wedge e_\infty$.

## 3.8 Representing Basic Euclidean Transforms

The various blades introduced in the last section are very good representations for important Euclidean objects. The ability to transform such objects in a concise and simple manner is another powerful feature of the conformal model. Such blades can be transformed using CGA versors. A versor is the geometric product of one or more non-null vectors (vectors whose square is non-zero). Versors provide the following benefits for transforming CGA blades:

- CGA versors can be used to represent all Euclidean transformations in a generalized, compact, and computationally efficient manner.
- CGA versors can be applied to all CGA multivectors including blades and other versors. Thus new versors can be easily composed of applying versors to each other using the exact same transformation formula.
- CGA versors are invertible. A versor representing a Euclidean transform can be inverted to represent the inverse Euclidean transform. Thus versors provide the same functionality an orthogonal matrix provides but with added features and flexibility.
- A versor can be directly applied to all multivectors including points, lines, spheres, planes, circles, point pairs, directions, and tangents. An orthogonal matrix, on the other hand, can only be applied to vectors. The use of orthogonal matrices to transform objects is thus very limited compared to using versors.
- A versor $V$ can be applied to any multivector $A$ using the following relation:

$$B = VAV^{-1} \tag{3.92}$$

The following subsections introduce several important versors that can represent useful transformations on objects in Euclidean space.

### 3.8.1 Representing Translation

A translation versor can be constructed by using a direction vector $t_E = t_1 e_1 + t_2 e_2 + t_3 e_3$ as follows:

$$V_t = 1 - \tfrac{1}{2} t_E \wedge e_\infty \tag{3.93}$$

### 3.8.2 Representing Rotation around the Origin

A rotation around the origin in 3D Euclidean space can be represented in two ways. The first is to select an axis of rotation in the form of a direction vector $r_E$. The second way is to select a direction bivector $R_E = r_E \rfloor I_3$ to represent the plane in which the rotation happens. In either case, the construction of the rotation versor is:

$$V_R = \cos(\phi/2) - \sin(\phi/2)R \qquad (3.94)$$

Where $R = R_E / \|R_E\|$ and $\phi$ is the angle of rotation in radians.

### 3.8.3 Representing Reflection

Two types of reflections are considered here. The first type is reflection in a plane passing through the origin and having a normal vector $n_E$, the reflection versor associated with that plane is simply $V_n = n_E$. The second type is reflection in the origin point itself. The versor of such reflection has the form $V_o = e_o \wedge e_\infty$.

### 3.8.4 Representing Uniform Scaling

Strictly speaking, scaling is not a Euclidean transform since it does not preserve distances between transformed objects. Uniform scaling can nonetheless be used in many applications, thus its versor is introduced here.

A uniform scaling versor around the origin by a factor of $s$ can be constructed as follows:

$$V_s = \cosh(\gamma) + \sinh(\gamma)\, e_o \wedge e_\infty \qquad (3.95)$$

Where $\gamma = \frac{1}{2}\ln(s) \Leftrightarrow s = e^{2\gamma}, s > 0$.

## 3.9 Representing General Euclidean Transforms

From the previous discussions the modeling capabilities of CGA blades were illustrated. For applications such as computer graphics for example such capabilities are insufficient. Free form objects other than points, spheres, planes, circles, and lines are very important. There is no CGA blade that can represent other simple surfaces such as a torus, a cone, or a cylinder. This section opens the door for another powerful modeling aspect of CGA; the twist representation of 3D Euclidean objects.

With the exception of translations, all the transformations of the previous section were related to the origin. For example, if a rotation around an axis not passing through the origin is required none of the previous versors can perform it on its own. Luckily, a composition of such versors can be used to perform vary complex Euclidean transforms.

### 3.9.1 General Rotation

A general rotation is a rotation around an axis (line) not passing through the origin. Assuming a line passes through point $p_E$ and having a direction vector $v_E$, the general rotation versor $V_{GR}$ can be constructed by multiplying three versors. The first versor is a translation versor $V_{-p}$ that translates the object in the direction of $-p_E$. This makes the problem much simpler by transforming the rotation axis to the origin. The second versor is the rotation versor $V_v$ around the axis direction $v_E$. The third versor is a translation versor $V_p$ that restores the object to its original position in space. Hence the whole operation can be performed as follows:

$$\begin{aligned} B &= V_{GR} A V_{GR}^{-1} \\ &= V_p V_v V_{-p} A V_{-p}^{-1} V_v^{-1} V_p^{-1} \\ &= V_p V_v V_p^{-1} A V_p V_v^{-1} V_p^{-1} \end{aligned} \tag{3.96}$$

Where it is clear that $V_{-p} = V_p^{-1}$.

### 3.9.2 Twist Motion

A twist motion or twist transform is a composition of a general rotation followed by a translation along the rotation axis. Thus a twist versor $V_{TW}$ can be constructed by the geometric product of a general rotation versor $V_{GR}$ and a translation versor along the same axis direction $V_v$. The full operation is as follows:

$$\begin{aligned} B &= V_{TW} A V_{TW}^{-1} \\ &= V_v V_{GR} A V_{GR}^{-1} V_v^{-1} \\ &= V_v V_p V_v V_p^{-1} A V_p V_v^{-1} V_p^{-1} V_v^{-1} \end{aligned} \tag{3.97}$$

Twist versors are one of the most general continuous Euclidean transforms. They will be used in this work for object modeling discussed in chapter 6.

### 3.9.3 General Reflection and Uniform Scaling

As in the case of general rotation, general reflection can be done by composition of translation versors and reflection versors. Assuming a general reflection in a plane with a normal $n_E$ and passing through the point $p_E$ , a general reflection versor can be constructed as the product $V_{GRP} = V_p V_n V_p^{-1}$ .

For a general reflection in a point $p_E$ , the general reflection versor is $V_{GRO} = V_p V_o V_p^{-1}$ . Finally, a general uniform scaling versor in a point $p_E$ is constructed as $V_{GS} = V_p V_s V_p^{-1}$

# Chapter 4 : GMac System Architecture

This chapter provides a description for the proposed GA-based code generator. The function of the code generator, called GMac, is to convert a higher-level description of GA-based algorithms into a number of optimized low-level expressions acting on the coordinates' level of multivectors. In addition, the chapter provides a full description of GMacDSL as a high-level language for encoding geometric algorithms using geometric algebra. Finally, the chapter provides a comparison of the architecture and characteristics of GMac with two similar GA-based systems.

The chapter begins by defining the requirements of the code generator in section 4.1. Section 4.2 illustrates the basic operation principle upon which GMac is designed. Section 4.3 represents the general layout of GMac. Section 4.4 illustrates the inputs enabling GMac to perform code generation and the components of the core GMac engine. Section 4.5 presents the steps in which GMac works to generate its code. Section 4.6 presents some implementation details of GMac. Section 4.7 describes the syntax used inside GMacDSL code files. Section 4.8 details the syntax of GMacDSL macro statements. Section 4.9 describes the syntax of GMac binding points inside the target language. Section 4.10 gives an illustrative example of the inner workings of GMac and a discussion of the most important features of GMacDSL. A comparison between GMac and Gaalop is provided in section 4.11. Finally, a comparison between GMac and Gaigen 2 is given in section 4.12.

## 4.1 System Requirements

The proposed system is designed to fulfill some requirements when solving geometric problems using GA. These requirements are:

1. ***High-Level User Interface***: Geometric algebra provides a high-level mathematical system for modeling and solving geometric problems. The proposed system must take full benefit of the expressive power of geometric algebra. On the other hand, the system must be capable of generating optimized code in the target language based on the high level GA operations. Such a requirement is the main motivation behind the system to solve the problems preventing the wide adoption of geometric algebra in practice.

2. ***Full Code Separation***: Full Separation of GA algorithm description and source code is an important requirement. In modern software systems no single developer is capable of maintaining the full system alone. That is because most modern software systems are very large and complex. They usually require expertise from diverse fields of programming. To address this problem the proposed system must separate the code containing GA operations from the code requiring the geometric algorithm. Usually the developers working on both sides come from different backgrounds and require a good level of code separation to complete and maintain the system efficiently.
3. ***Ease of Code Generation***: Single-click, single-pass code generation is an advantageous requirement for the system. A good code generator must have a simple interface that takes the required inputs. It then generates the output code using a single click on the keyboard or mouse. In addition, it should perform code generation in the least time possible. The proposed system should be capable of performing code generation in a single pass on the input code using a very simple user interface.
4. ***Optimal Code Reuse***: Optimal code-reuse in both the target language and the GA algorithms is also an important requirement. On the target language side, the code generator should not generate unnecessary code. The code generator should not define additional classes to be added to the original system, for example, without a good reason for doing so. It should use the variables and classes already present in the target language code directly without introducing artificial variables and classes. Such unnecessary code additions would affect performance of the final software system. In addition it will decrease source code readability which can severely affect code maintenance. On the geometric algorithm side, the designer should be capable of reusing previously defined GA code inside new algorithms without problems. These two requirements of code reuse on both sides are primary concerns for the proposed system to be useful and effective.

5. ***Simplicity of Integration***: Simplicity of integration with the target language strengthens the applicability of the system. The integration between geometric algorithms and target language code must be simple to define and easy to maintain. This requirement is a complement to the requirement of full code separation.
6. ***Minimal System Components***: The number and complexity of components of the system must be kept minimal. If the system contains too many components with complex integration procedures between them it may cause more problems than it solves. The proposed system should have the smallest and simplest possible components for ease of use.

## 4.2 GMac Basic Operation Principle

Having a GA-based algebraic algorithm, the main function of the code generator is to transform the algorithmic steps into an optimized series of scalar assignment expressions that ultimately computes the outputs from the inputs in the least number of computational steps possible. To illustrate such process, assume the 3D Euclidean geometric algebra based on the orthonormal basis vectors $\langle e_1, e_2, e_3 \rangle_{\perp 1}^{\mathbb{R}^3}$. Assume that the inputs are two vectors $u, v$ defined as:

$$\begin{aligned} u &= \sin(x) e_1 + \cos(x) e_2 + e_3 \\ v &= \cos(x) e_1 + \sin(x) e_2 + e_3 \end{aligned} \qquad (4.1)$$

Where $x$ is a real scalar parameter. The required output is the vector $w$ defined as:

$$\begin{aligned} w &= (u \wedge v) \rfloor I_3^{-1}, \\ I_3 &= e_1 \wedge e_2 \wedge e_3 \end{aligned} \qquad (4.2)$$

The output vector is simply the normal to both input vectors and is also a function of the real parameter $x$.

First, it is typically required to describe the algorithm in the DSL of the code generator. Next, the DSL code is parsed and a kind of in-memory representation for the inputs, output, constants, and algorithmic steps is created. Mathematically, the proposed code generator re-formulates the problem as follows:

***Step 1: Reformulation***

$$
\begin{aligned}
&u = u_1 E_1 + u_2 E_2 + u_4 E_4, \\
&u_1 = \sin(x), \quad u_2 = \cos(x), \quad u_4 = 1 \\
\\
&v = v_1 E_1 + v_2 E_2 + v_4 E_4, \\
&v_1 = \cos(x), \quad v_2 = \sin(x), \quad v_4 = 1 \\
\\
&k = I_3 = r E_7, \\
&r = 1 \\
\\
&w = (u \wedge v) \rfloor k^{-1} = w_0 E_0 + w_1 E_1 + \ldots + w_7 E_7 \\
&w_0 = ?, \quad w_1 = ?, \quad \cdots, \quad w_7 = ?
\end{aligned}
\tag{4.3}
$$

Initially, the output of any GA-based algorithm is a general multivector with unspecified coefficients. The following processing steps aim at systematically deducing the computational relations between the coefficients of the input multivectors and the output multivector as follows:

***Step 2: Evaluation***

$$
\begin{aligned}
u \wedge v = \\
& u_1 v_1 e_1 \wedge e_1 + u_1 v_2 e_1 \wedge e_2 + u_1 v_4 e_1 \wedge e_3 + \\
& u_2 v_1 e_2 \wedge e_1 + u_2 v_2 e_2 \wedge e_2 + u_2 v_4 e_2 \wedge e_3 + \\
& u_4 v_1 e_3 \wedge e_1 + u_4 v_2 e_3 \wedge e_2 + u_4 v_4 e_3 \wedge e_3 \\
& = \\
& [u_1 v_2 - u_2 v_1] e_1 \wedge e_2 + \\
& [u_1 v_4 - u_4 v_1] e_1 \wedge e_3 + \\
& [u_2 v_4 - u_4 v_2] e_2 \wedge e_3,
\end{aligned}
\tag{4.4}
$$

***Step 3: Reformulation***

$$
\begin{aligned}
t_1 &= u \wedge v \\
&= a\, e_1 \wedge e_2 + b\, e_1 \wedge e_3 + c\, e_2 \wedge e_3 \\
&= a\, E_3 + b\, E_5 + c\, E_6 \\
a &= [u_1 v_2 - u_2 v_1], \\
b &= [u_1 v_4 - u_4 v_1], \\
c &= [u_2 v_4 - u_4 v_2],
\end{aligned}
\tag{4.5}
$$

***Step 4: Evaluation***

$$
\begin{aligned}
k^{-1} &= I_3^{-1} = r\, e_3 \wedge e_2 \wedge e_1 \\
&= -r\, E_7
\end{aligned}
\tag{4.6}
$$

***Step 5: Reformulation***

$$
\begin{aligned}
t_2 &= k^{-1} = d\, e_1 \wedge e_2 \wedge e_3 \\
&= d\, E_7, \\
d &= -r
\end{aligned}
\tag{4.7}
$$

***Step 6: Evaluation***

$$
\begin{aligned}
t_1 \rfloor t_2 = ad\, & [(e_1 \wedge e_2) \rfloor (e_1 \wedge e_2 \wedge e_3)] \\
& bd [(e_1 \wedge e_3) \rfloor (e_1 \wedge e_2 \wedge e_3)] \\
& cd [(e_2 \wedge e_3) \rfloor (e_1 \wedge e_2 \wedge e_3)] \\
& = -ad\, e_3 + bd\, e_2 - cd\, e_1
\end{aligned}
\tag{4.8}
$$

***Step 7: Reformulation***

$$
\begin{aligned}
w &= t_1 \rfloor t_2 \\
&= w_1 E_1 + w_2 E_2 + w_4 E_4, \\
w_1 &= -cd, \quad w_2 = bd, \quad w_4 = -ad
\end{aligned}
\tag{4.9}
$$

The previous sequence of steps continues until all output multivector coefficients are fully defined. Thus, the code generator deduces the following input-output computational sequence of scalar assignments:

***Step 8: Scalar Expressions Deduction***

$$
\begin{aligned}
u_1 &= \sin(x), \quad u_2 = \cos(x), \quad u_3 = 1, \\
v_1 &= \cos(x), \quad v_2 = \sin(x), \quad v_3 = 1, \\
r &= 1, \\
a &= [u_1 v_2 - u_2 v_1], \\
b &= [u_1 v_4 - u_4 v_1], \\
c &= [u_2 v_4 - u_4 v_2], \\
d &= -r, \\
w_1 &= -cd, \\
w_2 &= bd, \\
w_4 &= -ad, \\
w_i &= 0 \quad ; i = 0, 3, 5, 6, 7
\end{aligned}
\tag{4.10}
$$

The final step before actual code generation is to reduce the scalar expressions list to a more computationally efficient one as follows:

***Step 9: Expression List Optimization***

$$
\begin{aligned}
u_1 &= \sin(x), \\
u_2 &= \cos(x), \\
w_1 &= u_2 - u_1, \\
w_2 &= -w_1, \\
w_4 &= u_1^2 - u_2^2, \\
w_i &= 0 \quad ; i = 0, 3, 5, 6, 7
\end{aligned}
\tag{4.11}
$$

This final list is then simply converted to whatever target language the code generator is capable of producing code for.

In summary, the main function of the GA-based code generator is to transform the GA-based high-level algorithm given in (4.2) into a much more computationally efficient series of low-level assignment expressions given in (4.11). The conversion process is a systematic series of reformulation steps followed by GA-based evaluation steps that continue until all input to output relations are specified. Finally an optimization step is used for reducing the scalar expressions to a more efficient list of scalar expressions.

## 4.3 General Layout

The proposed system is called GMac. Its name stands for ***G***eometric ***Mac***ro and it is intended to fulfill all of the previous requirements as close as possible. The general layout of GMac is shown in Figure 4-1. The system operates on two sets of inputs. The first set is a high-level description of geometric algorithms used for solving a certain geometric problem; like ray tracing for example. The description is simply some code written in a special Domain Specific Language (DSL) called the GMacDSL. The GMacDSL uses geometric algebra-based operations to describe geometric algorithms.

The second set of inputs is a number of source code files written in a target language; like C# or Java. These source files contain normal code in the target language but with special code parts used as placeholders for GMac to process in a special way. These place-holders are called GMac binding points. Each binding point will contain the final code generated from GMac using one of the geometric macros defined using GMacDSL code.

As can be seen in Figure 4-1, GMac operates in three stages: parsing, compilation and code generation. During compilation and code generation stages GMac heavily interacts with a back-end symbolic processing engine; like Mathematica or Maple. After the third stage, GMac produces the exact same input source files in the target language but with one difference: all GMac binding points are replaced by optimized low-level code in the target language.

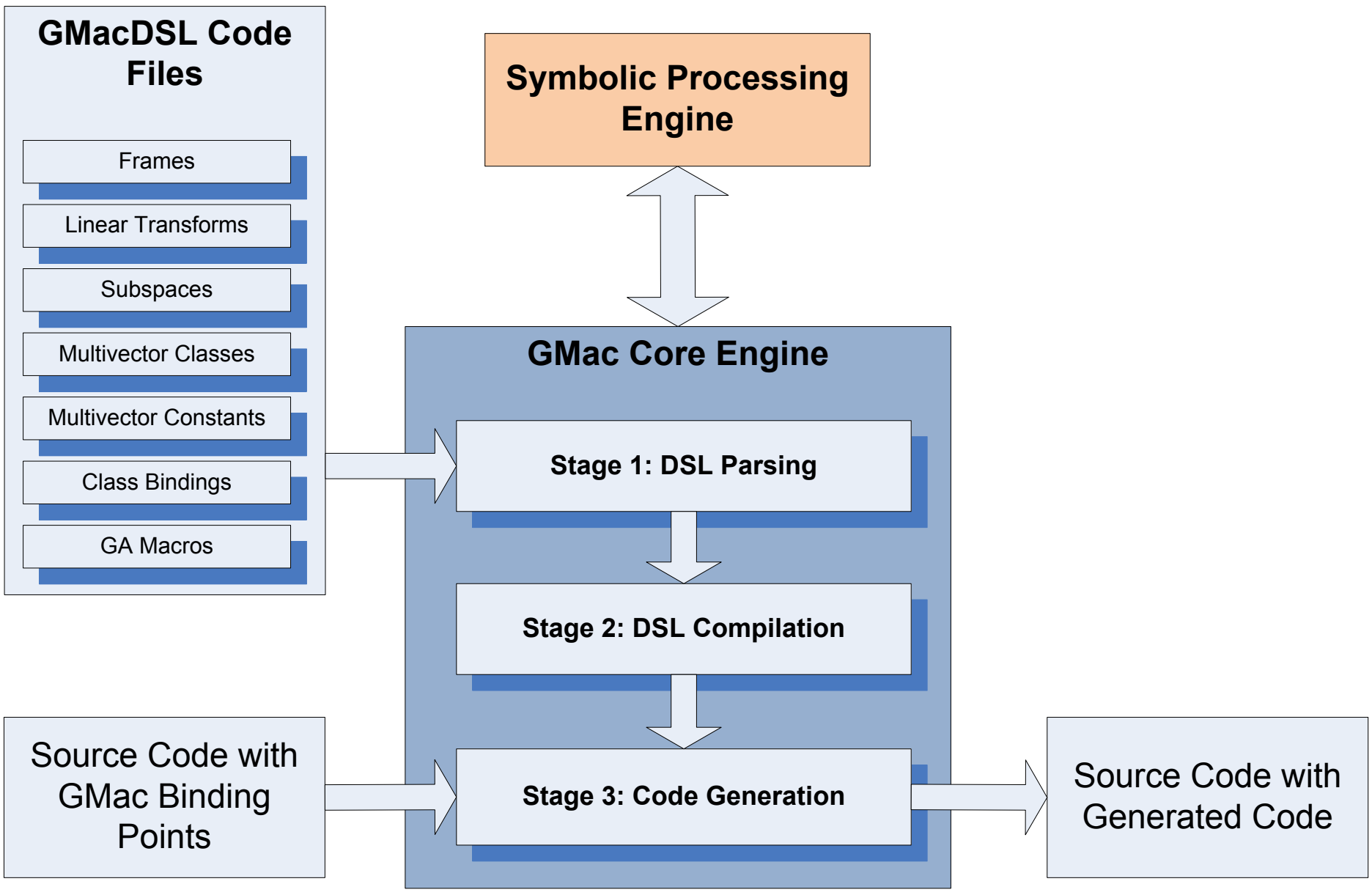

**Figure 4-1: GMac system layout**

## 4.4 GMacDSL Input Code Files

In order to describe geometric algorithms using GMacDSL several code files must be written to encode the necessary information required for code generation. The various DSL code files and their intended functions are described in the following subsections.

### 4.4.1 Frames

Any single geometric problem may require one or more linear spaces to fully describe its solution. In ray tracing, for example, 3D Euclidean space is required. In addition, the 5D Conformal Model is also necessary for implementing Euclidean transforms (represented as CGA versors) on GA objects (represented as CGA blades). Finally 2D Euclidean space is needed for some operations related to the viewing surface (the camera) and 2D texture coordinates. Mathematically, any single space of dimension n can be fully represented using a set of n-linearly independent vectors. Such a set is traditionally called a basis for the linear space. In GMacDSL such a set is called a frame of basis vectors, or a frame for short. Each n-dimensional frame with n basis vectors defines a corresponding $2^n$ dimensional geometric algebra with $2^n$ basis-blades. The basis blades include the n basis vectors and the outer products of all of their combinations. Figure 4-2 illustrates the relation between frames, transforms, subspaces, multivector classes, and constant multivectors in GMacDSL. Any single space may have several frames for its representation. Each frame may be suitable for some operations more than others. For example, some operations are more efficient when applied on multivectors represented using an orthonormal frame. On the other hand, some other operations may be more efficient when applied on multivectors represented using another related frame with a different signature for its basis vectors. Frames can be defined using a signature matrix called the Inner-Product Matrix (IPM) that defines the bilinear form of the GA. Frames can also be defined based on other frames using invertible linear transforms. Finally a frame can be defined as a subspace of another larger frame.

### 4.4.2 Linear Transforms on Multivectors

Linear transforms are essential tools used to transform multivectors between different frames as shown in Figure 4-2. A linear transform can be used within a GA macro to change the representation of a multivector from some frame to another in order to perform operations on the new representation as described in the previous subsection. A linear transform is represented in GMacDSL as a constant outermorphism [11] between two frames of the same dimension. A linear transform may or may not be invertible. If the linear transform is invertible, its inverse can be defined as a separate linear transform.

### 4.4.3 Subspaces

Any single frame of dimension n defines a geometric algebra space with dimension $2^n$. The basis of the frame consists of n LID vectors where the basis of the GA space consists of $2^n$ basis-blades. The $2^n$ basis blades contain the n vectors in addition to blades constructed from the outer product of all combinations of the basis vectors. Thus it is possible to take any subset of the basis-blades to construct a subspace of the GA space constructed from the frame. In Figure 4-3 a 3D GA frame is shown with four different subspaces defined. In (a) the subspace of vectors consists of basis blades $e_1$, $e_2$ and $e_3$. In

(b) the subspace of 2-blades consists of basis blades $e_1 \wedge e_2$, $e_2 \wedge e_3$ and $e_1 \wedge e_3$. In (c) the subspace of even versors (2-Versors) consists of basis blades $1$, $e_1 \wedge e_2$, $e_2 \wedge e_3$ and $e_1 \wedge e_3$. Finally (d) is the subspace formed by the geometric algebra defined on the two basis vectors $e_1$ and $e_2$ only. That subspace consists of basis blades $1$, $e_1$, $e_2$ and $e_1 \wedge e_2$. Subspaces are useful for defining multivector classes and performing some special projection operations in GA macros as will be seen in latter sections.

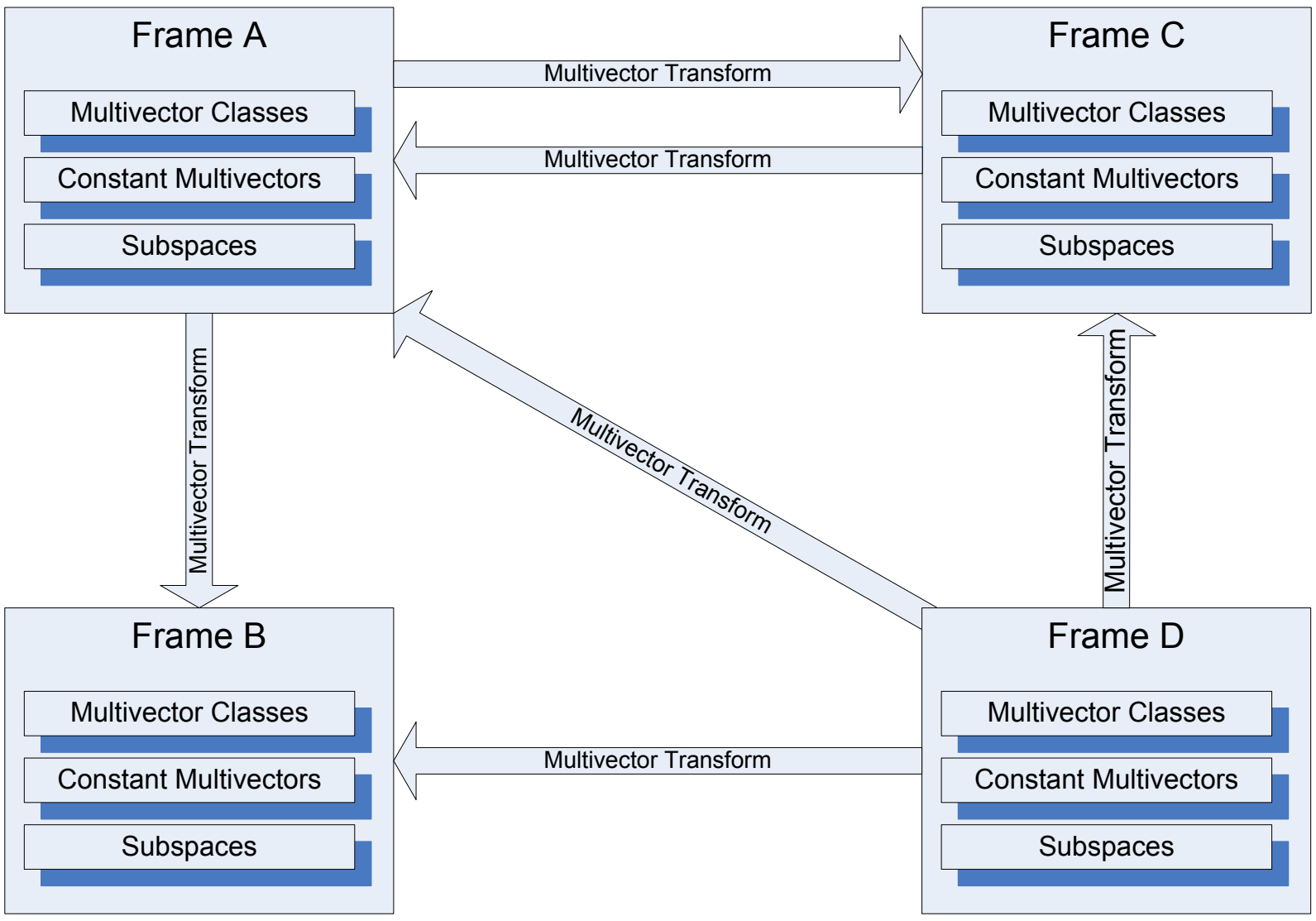

**Figure 4-2: GMacDSL Frames and Linear Transforms**

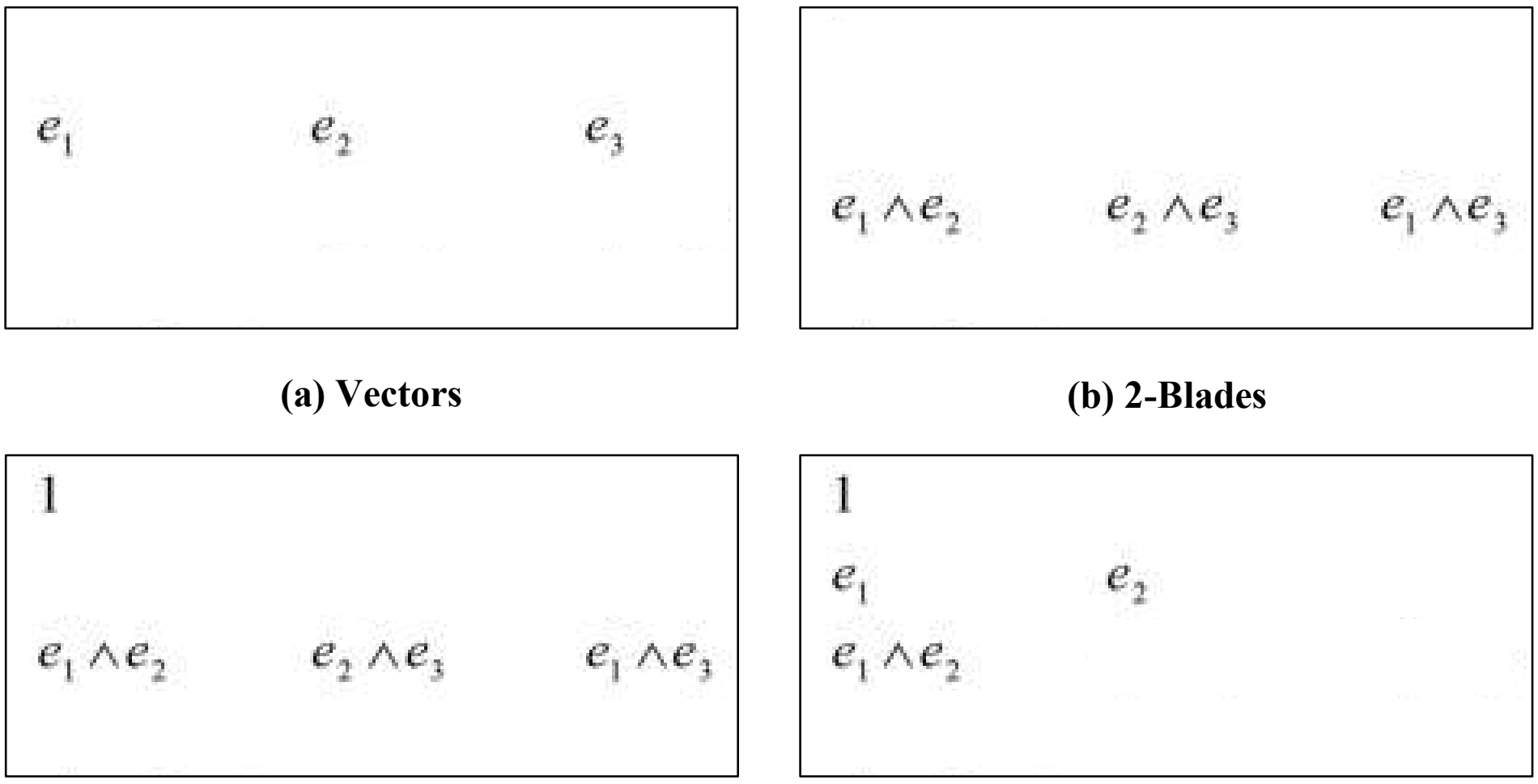

**(a) Vectors** **(b) 2-Blades**

**(c) Even versors** **(d) GA Span of $e_1$ and $e_2$**

**Figure 4-3: Examples of Subspaces in GMacDSL for a 3D Frame**

### 4.4.4 Multivector Classes

In GMacDSL a multivector class is the weighted sum of a number of basis-blades with possibly some of which having constant coefficients (weights) while the others having symbolic variable ones. In the conformal model for example, points, lines, planes, and spheres are all represented using different sets of basis-blades. In addition some basis-blades have constant coefficients like the case of normalized points. Each of these entities can be represented with a separate multivector class in GMacDSL.

### 4.4.5 Multivector Constants

In GA some constants play an important role in certain operations. An example is the pseudo-scalar used in defining the dual operation on multivectors. Another example is any sum of basis blades with constant coefficients defined on a single frame. In GMacDSL any number of constant multivectors can be defined per frame. Such constant multivectors can be used inside geometric macros like any other input, temporary or output multivectors.

### 4.4.6 Class Bindings

Using ray tracing as an example, the class representing a ray in C# code usually defines three double variables for holding the Euclidean coordinates of the origin of the ray; say ox, oy and oz. In addition it contains three double variables for holding the coordinates of the ray direction vector; say dx, dy and dz. Such a C# class can be bound in a fixed manner to two multivectors in 3D Euclidean space. One multivector holds the origin of the ray while the other holds its direction. Such a fixed binding with the ray class can be used inside all GMac binding points without redefining the bindings for each ray object over and over again.

Actually, class bindings are capable of more than that simple example. As shown in Figure 4-4, a class binding is a set of symbolic scalar functions, called binding functions, capable of operating on zero or more input variables and producing a single scalar output. The inputs of a binding function are class member variables. The output of a binding function is a coefficient associated with a basis blade in a multivector class.

### 4.4.7 GA Macros

Defining GA macros is the main target of the system. A single GA macro may take one or more input multivectors. In addition it may produce one or more output multivectors. Any single input\output multivector is always defined on a certain frame and restricted to be of a certain multivector class. The macro can perform GA operations on input multivectors and constant multivectors to produce output multivectors or temporary multivectors to operate on later. The GA macro is intended to be a high level description of the steps required to solve a sub-problem of the geometric problem using GA operations. In GMacDSL any GA macro can call another separate GA macro.

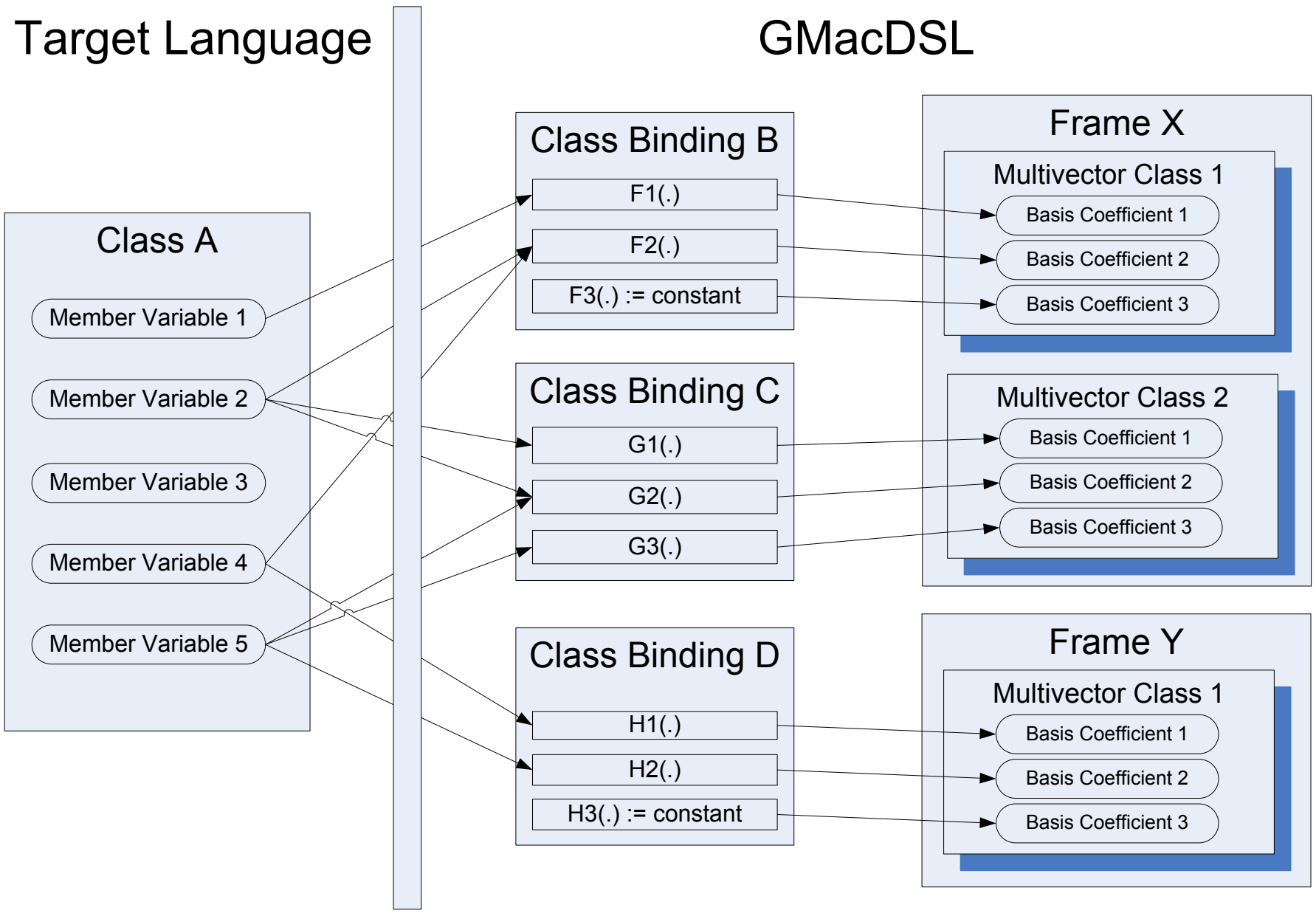

**Figure 4-4: Class Bindings in GMacDSL**

### 4.4.8 Language Binding

By itself the GMacDSL is not useful unless a specification of how to use its macros for code generation inside the target language is defined. The specification is defined in one or more binding points in the target language source code files. Each binding point defines three elements. The first is which GA macro to be used in generating the code of that binding point. The second is a mapping between input\output multivectors coefficients and the target language local and object variables. That mapping will be used to generate the final code in the target language source files. The third element is the maximum and minimum values for some or all of the bound variables. Such variable limits can be essential in producing optimized code using GMac and the symbolic processing engine in stage 3. As in class bindings, the target language variables are bound to GMacDSL multivectors through symbolic scalar functions as shown in Figure 4-5.

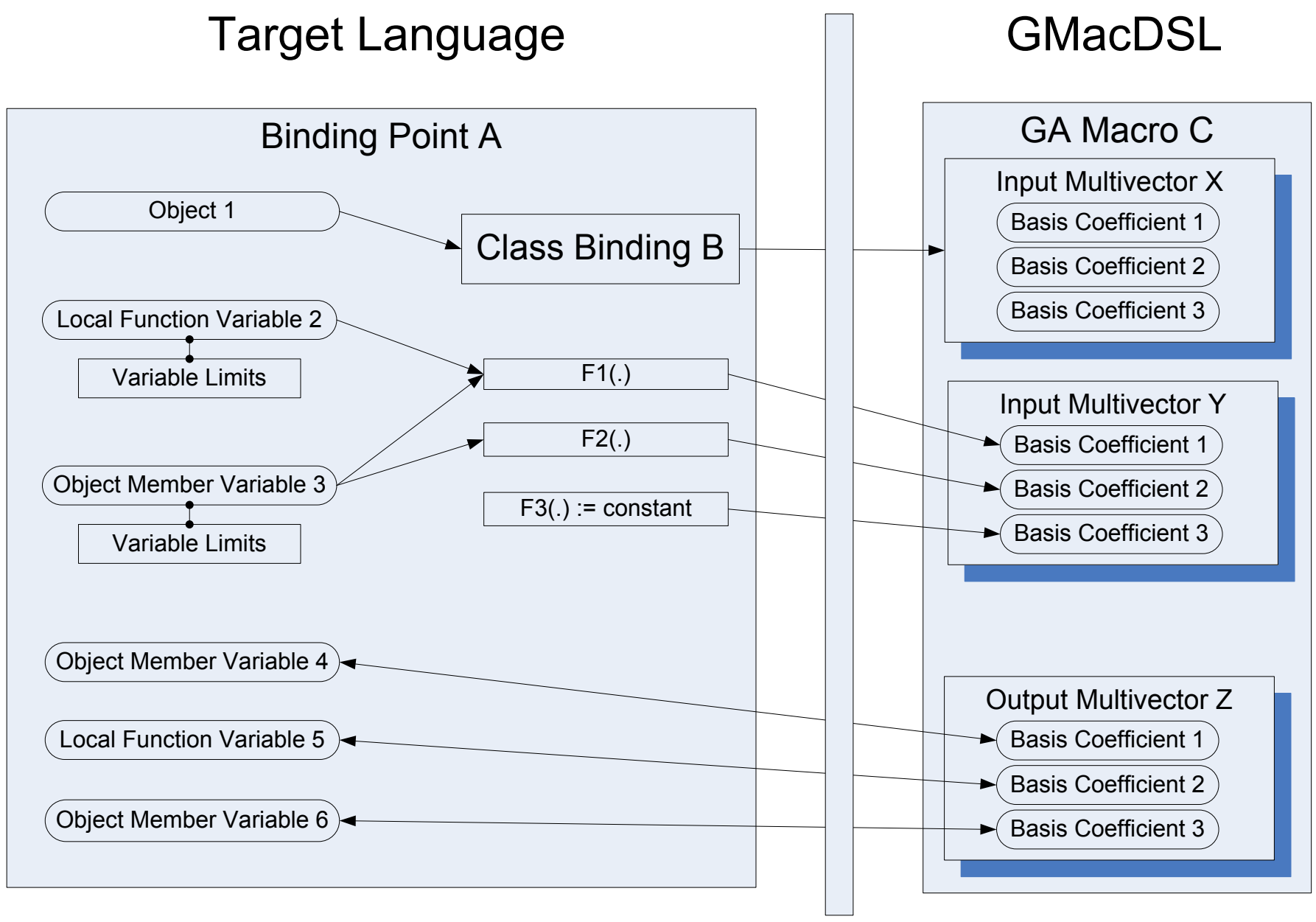

**Figure 4-5: Binding Points in GMacDSL**

### 4.4.9 GMac Core Engine

As shown in Figure 4-6, the GMac core engine consists of four sets of classes. The first set contains the in-memory DSL representation classes. These classes are direct images of the GMacDSL code files placed in main memory. These classes are typically initialized during the parsing stage and used during all three stages.

The second set contains the symbolic representations necessary to interact with the external symbolic processing engine. These classes are typically initialized during the compilation stage and used during the compilation and code generation stages. Such classes are responsible for generating lists of scalar expression sequences from GA macro operations on multivectors.

The third set contains classes for expression optimization. As will be seen later, GMac produces a list of expressions, called expression sequences, based on the specified GA macro for each binding point. The classes in the third set are used to hold such sequences and to perform necessary optimizations on them. The optimized sequences are then used for final code generation.

The final set contains some helping classes. For example the GMac project class is used for file input/output operations on GMacDSL code files and target language source files. The GMac parser class is used for string parsing operations of both types of code. Finally, the Language Binding class is a representation of a single GMac Binding Point inside the target language source files.

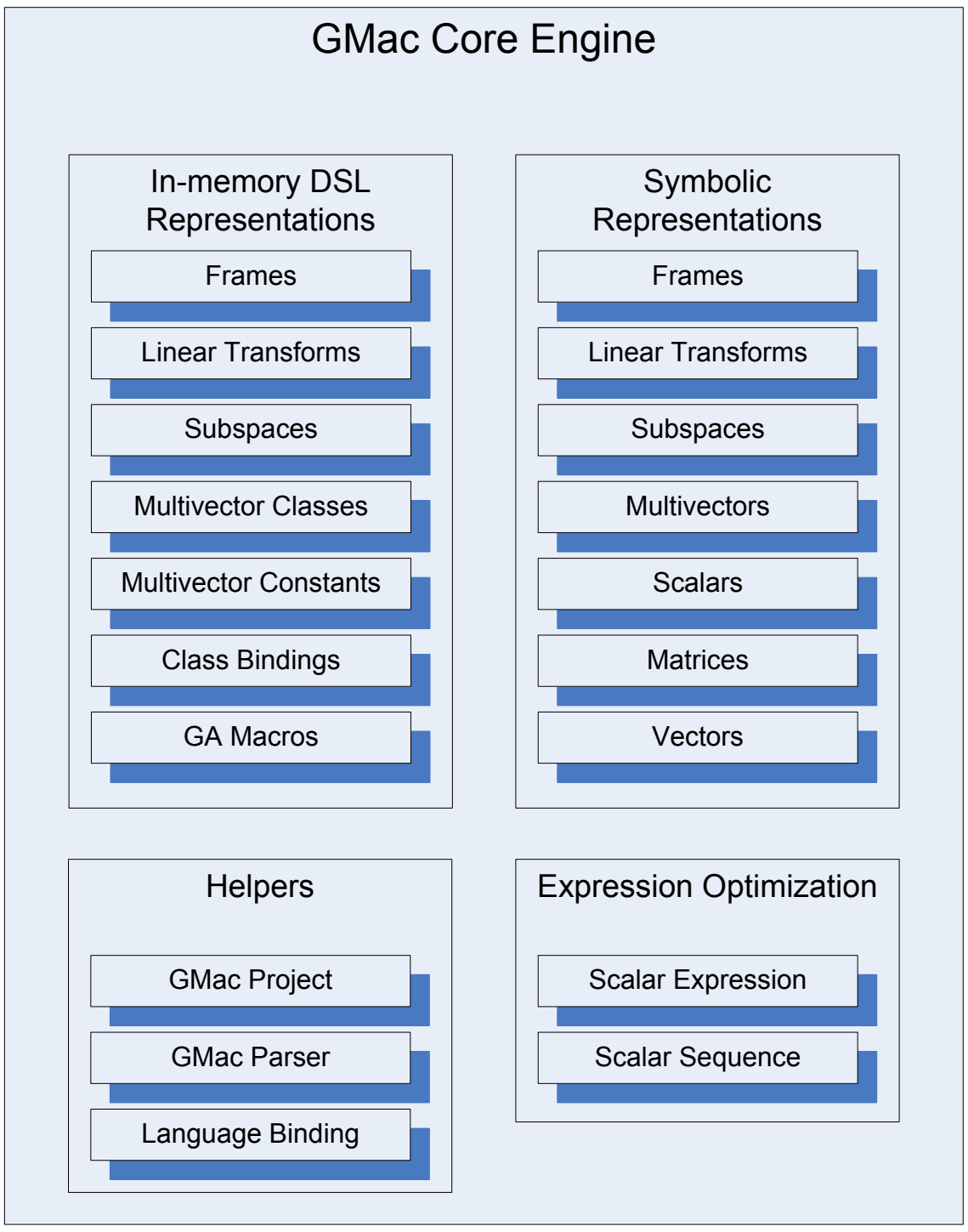

**Figure 4-6: GMac Core Engine Main Components**

## 4.5 System Operation

GMac operates through three stages. These stages are described in the following subsections in order.

### 4.5.1 Stage 1: DSL Parsing

The first stage is only executed once per project. In this stage, GMac reads the GMacDSL code files. It then constructs an in-memory image of the DSL code to be used during the second and third stages as shown in Figure 4-7. The in-memory representations are capable of writing the same GMacDSL code back to files. This can be a useful during the debugging of GMac Core Engine and for extending GMac in the future with a suitable Application Programming Interface (API).

### 4.5.2 Stage 2: DSL Compilation

In the compilation stage several parts of the GMacDSL in-memory representations are compiled. The compilation process creates the actual symbolic representation of frames, transforms, subspaces, and multivector classes and constants. In addition, macros are transformed to linear lists of GA operations on multivectors in a format suitable for processing by the core GMac engine during stage three. Stage two is also performed once per project as shown in Figure 4-7.

### 4.5.3 Stage 3: Code Generation

The third and final stage is performed once per binding point. In that stage the macro to be used for code generation is identified. In addition, the target language variables are bound with the input\output multivectors coefficients for the generated GA macro. The maximum and minimum values for the coefficients can also be defined inside the binding point. Next the compiled macro is symbolically evaluated on the input multivectors using GA operations. This is where the most intense interaction with the symbolic engine takes place. After this phase a list of symbolic expressions is obtained from the symbolic engine. That list is then optimized within the GMac core engine. The final phase is to substitute the target language variables into the optimized list of expressions. The resulting code is inserted in place of the binding point in the target language source code file. The entire process is illustrated in Figure 4-8.

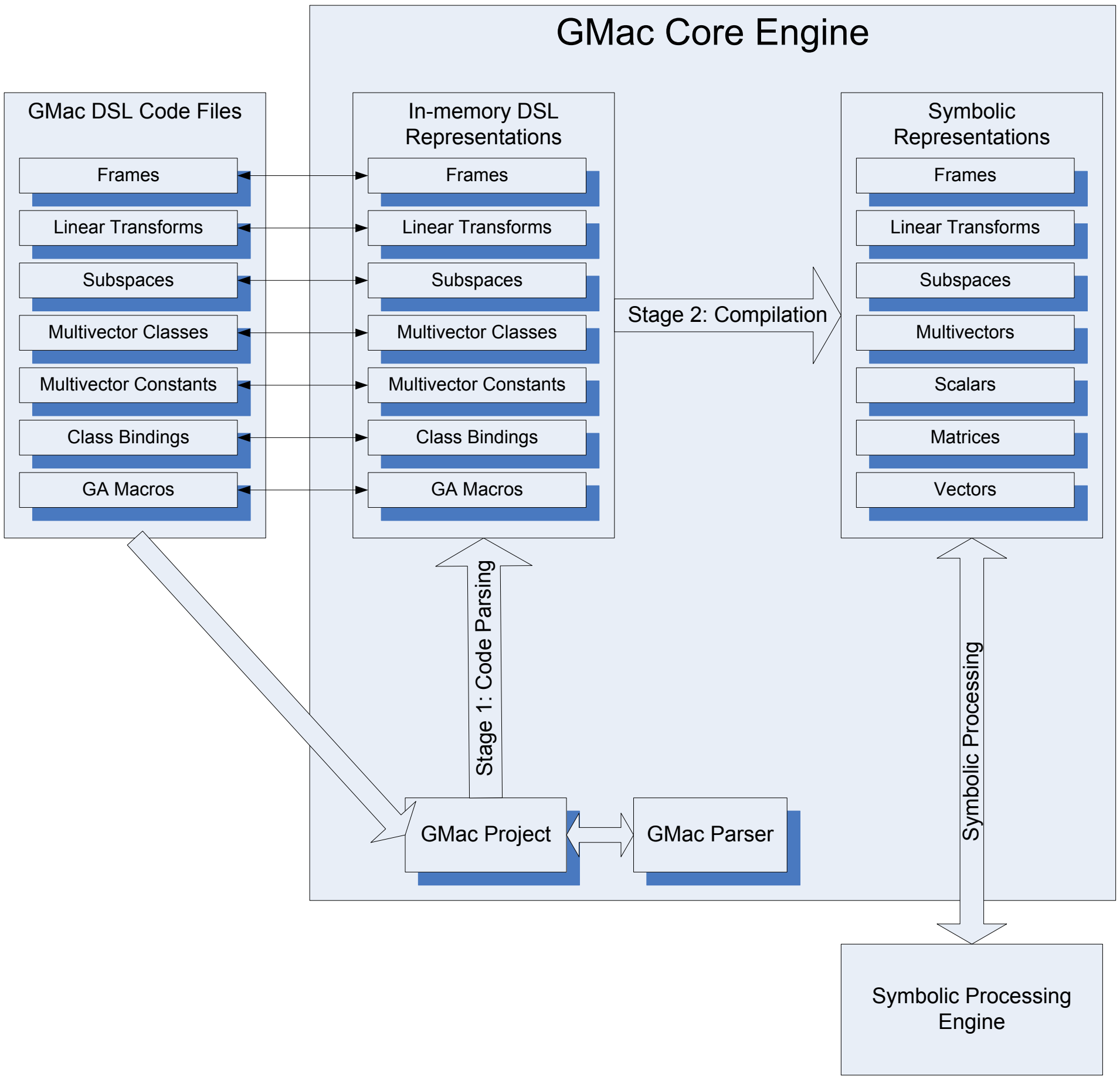

**Figure 4-7: GMac parsing and compilation stages**

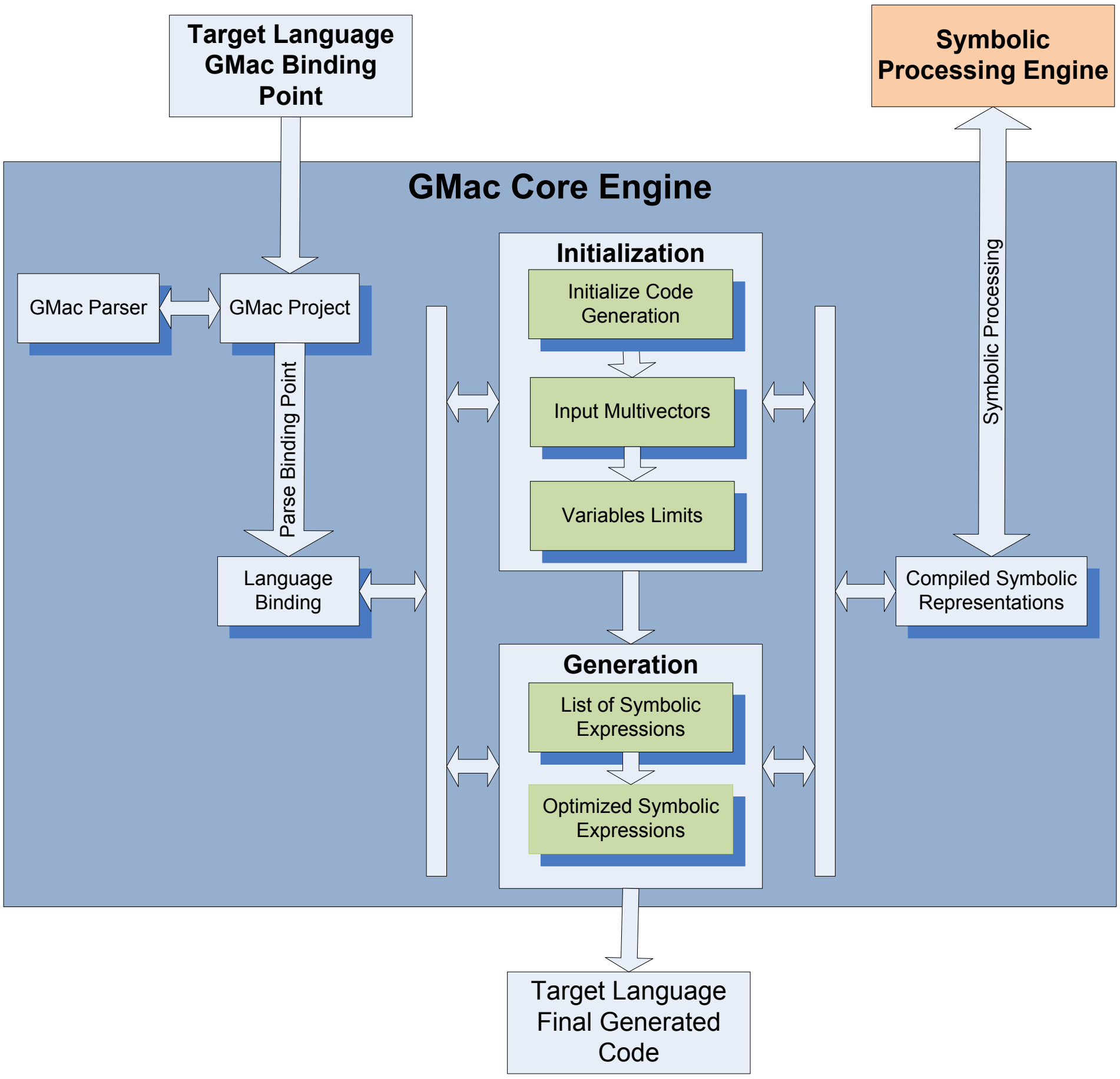

Figure 4-8: GMac Code Generation Stage

## 4.6 GMac Implementation

The current version of GMac is implemented on the Microsoft .NET framework 3.5 using Visual Studio 2008 IDE and C#. Wolfram Mathematica 6.0 is used as the external symbolic processing engine. The current GMac implementation can handle C# source files as the target language for code generation. The following subsections introduce some implementation details.

### 4.6.1 Implementation Choices

To implement GMac several choices had to be made regarding three elements. The first is choosing a software development platform to use in developing GMac. There are two competing technologies today that are used in the development of most complex software systems; the Java and the .NET platforms. The choice between using the Java platform and the .NET platform is in favor for .NET for implementing a system like GMac. GMac is mainly a desktop application. The .NET platform is better suited for desktop applications than the Java platform. In addition, applications created in .NET are not restricted to the Windows operating system. The Mono project is an implementation of most .NET framework features on the MacOS and Linux operating systems. Hence the .NET framework provides excellent software development experience without sacrificing application portability.

The second element is the language to implement GMac with. C# is a multi-paradigm programming language that encompasses functional, imperative, generic, object-oriented (class-based), and component-oriented programming disciplines. C# is intended to be a simple, modern, general-purpose, object-oriented programming language. The language include strong type checking, array bounds checking, detection of attempts to use uninitialized variables, source code portability, and automatic garbage collection. Such features ensure software robustness, durability and programmer productivity. In addition the language is designed such that applications written in C# are intended to be economical with regard to memory and processing power requirements. C# contains most features provided by Java in addition to being able to create faster applications more suitable for engineering and scientific fields. Compared to C++, C# is much more programmer friendly and type-safe. In addition, if the use of C++ code is required for some reason, it can always be integrated with existing C# code through the .NET framework Common Language Infrastructure (CLI) [34].

The third element is the selection of a good mathematical symbolic processing engine. Many good symbolic processing software exist in the scientific\engineering community. Names like Maple, MathCAD, Axiom, Euler, Maxima, and Reduce are well known as very good Computer Algebra Systems (CAS). The requirements for the GMac code generator on the symbolic engine are not hard to fulfill. The first requirement is simple and efficient communication with the GMac engine. The second requirement is the support of good simplification capabilities for real scalar symbolic expressions. Most good computer algebra systems can fulfill these requirements. The choice to use Mathematica does not exclude other CAS packages, especially the free ones like Reduce and Euler. Mathematica [35] was chosen for its powerful computational engine and simplicity of integration with any .NET application. Mathematica provides a rich ground for expanding GMac capabilities in its future implementations.

### 4.6.2 Symbolic GA Representation

The GMac core engine relies on a symbolic representation of multivectors. A multivector $A$ on a GA space can be represented as a linear combination of the basis blades of such space: $A = \sum_{k=0}^{2^n-1} a_k E_k$ . Inside the GMac core engine, each basis blade $E_k$ is encoded as an integer ranging from 0 to $2^n$ - 1 for a base vector space of dimension n. The integers are encoded as described in chapter 19 of [11]. A symbolic multivector is then defined as an associative map (a `SortedDictionary` object in C#). The keys to the associative map are the integers encoding basis blades having non-zero symbolic coefficients. The value associated with each key is a Mathematica scalar expression that is the coefficient of such key. As an example, the multivector $A = 5 + \sin(x) e_1 - y^2 z \; e_1 \wedge e_3$ is represented as the following associative map inside GMac core engine (assuming a 3D Euclidean GA space):

| Key | Value |
|---|---|
| 0 | "5" |
| 1 | "Sin[$x$]" |
| 5 | "$y^2 z$ " |

All bilinear GA products (geometric, outer, contraction …etc.) can thus be defined on such representation if a product table is constructed between basis blades. That is because all bilinear products satisfy the relation:

$$\begin{aligned} A \otimes B &= \left( \sum_{r=0}^{2^n-1} a_r E_r \right) \otimes \left( \sum_{r=0}^{2^n-1} b_r E_r \right) \\ &= \left( \sum_{m=0}^{2^n-1} \sum_{k=0}^{2^n-1} a_m b_k \left( E_m \otimes E_k \right) \right) \\ \forall A &= \sum_{r=0}^{2^n-1} a_r E_r, \quad B = \sum_{r=0}^{2^n-1} b_r E_r \end{aligned} \tag{4.12}$$

Thus, if all values of $E_m \otimes E_k$ are pre-computed, the symbolic evaluation of any bilinear product $\otimes$ on symbolic multivectors $A, B$ is simple. The same argument is valid for all linear unary operations on symbolic multivectors, like involutions and conjugates, and constant outermorphisms [11]. All other GA operations in GMacDSL can be defined using such products and operations. Thus, all GMacDSL macro statements can be symbolically evaluated by either constructing symbolic multivectors or performing GA operations on symbolic multivectors.

Such approach is the same one described in [11] and [20] with one major difference: GMac uses symbolic scalar coefficients for the purpose of code generation. Whereas the approach of [11] and [20] uses real double coefficients as a direct implementation for GA multivectors and operations.

### 4.6.3 Engine Optimizations

The main processing bottleneck in the system is the intense interaction between GMac and Mathematica during symbolic computations. The main purpose of the symbolic engine is to simplify a single input expression to produce a single output expression. Many inputs to the symbolic engine from GMac are repeated many times. In order to reduce that redundant interaction, a symbolic cache is used inside GMac. The symbolic cache can be used to store the results of previously simplified symbolic expressions. The symbolic cache is implemented as an associative mapping (a `SortedDictionary` object in the .NET framework). When a new expression is needed to be simplified, the symbolic cache is searched first. If the expression is found its associated output is obtained from the symbolic cache. If it is a new expression it is simplified by Mathematica instead and the result is stored in the symbolic cache for possible later use. That approach resulted in more than 100% speedup in symbolic processing operations.

### 4.6.4 Fulfilling the Requirements

The implementation of GMac is intended to fulfill all the requirements of section 4.1. For the first requirement of a high-level user interface, GMacDSL macros are defined using high-level GA operations on multivectors. In addition, the optimization processes on the generated expression sequences produce optimized low-level code in the target language.

GMac handles two separate sets of code. The first is the GMacDSL code used to define GA macros. The second is the target language source code with its GMac binding points. Each set of code can be developed, optimized and maintained separately from the other set. Thus the second requirement of full code separation is obtained.

The whole code generation process using GMac has a very simple interface as shown in Figure 4-9. The user just selects the paths containing the two sets of code and presses a button to begin generation. In addition, all source files are processed in a single pass. Thus, GMac is a single-click, single-pass code generator.

The generated code from GMac is a list of assignment statements in the target language. No external classes are generated from GMac. This approach is suitable for a wide spectrum of applications where a problem is divided into a large number of small code parts. Each code part is intended to perform a single task or algorithm in an optimized manner. GMac can produce different code based on the same GA macro given different inputs in any two binding points. In addition, any GA macro in GMacDSL can use another GA macro internally. Thus optimal code reuse on both code sides is obtained.

The integration of GMacDSL code and the target language source code is obtained through GMac binding points. These binding points are small portions of the target language code used to link it with a GA macro for code generation. In addition, the binding points are not removed from the target language code after generation. Instead the binding point code is commented out. This enables the possibility of re-generating the whole project after modifying a single binding. This form of integration is simple, maintainable and efficient thus fulfilling the requirement of simplicity of integration.

Finally, GMac is fully implemented using only two components. The first is the .NET framework and the second is the Mathematica symbolic engine. No external systems or libraries are required for normal GMac operation. Thus the requirement of minimal system components is obtained.

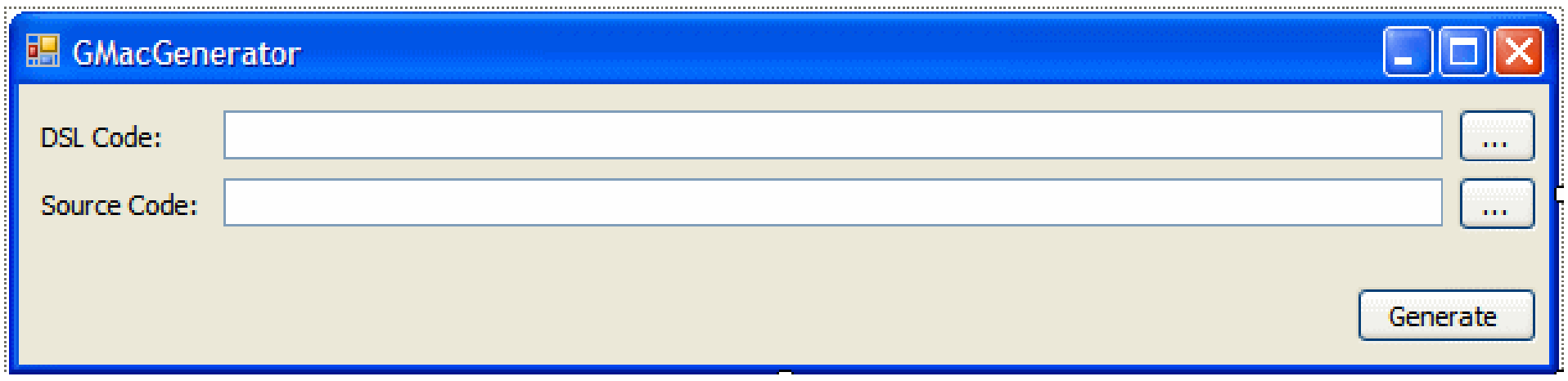

**Figure 4-9: GMac user interface**

## 4.7 GMacDSL Code Files

In the current GMac implementation there are seven separate files per GMac project. The code inside each of the files is described in what follows.

### 4.7.1 The Frames File

The frames file is the first file to be parsed by GMac in any GMac project. This file contains all definitions for frames used in the GA problem. A frame is defined using the following syntax:

```
define frame <frame name> as
    basis: {<list of basis names>}
    <frame definition>
end frame
```

Where:

***<frame name>*** is an identifier for that frame.

***<list of basis names>*** is a comma separated list of identifiers for the names of basis vectors for that frame.

***<frame definition>*** can take several forms shown in Table 4-1

**Table 4-1: Methods of defining frames in GMacDSL**

| `<frame definition>` | Meaning |
|---|---|
| `Euclidean` | A frame with all basis vectors having a signature of 1.0 |
| `subspace of` ***<source frame>*** | A subspace of another frame |
| `orthogonalize` ***<source frame>*** | Any orthogonal frame equivalent to another frame |
| `IPM =` ***<Mathematica matrix expression>*** | A frame with a given inner product matrix |
| `transform` ***<source frame>*** `by BCM =` ***<Mathematica matrix expression>*** | A frame obtained by transforming another frame using the given Basis Change Matrix (BCM) |

In Table 4-1 ***<Mathematica matrix expression>*** is any valid Mathematica expression that produces a square matrix with constant real scalar elements.

### 4.7.2 The Transforms File

The next file to be parsed by GMac is the transforms file. This file contains all fixed linear transforms between the previously defined frames. The syntax for defining a linear transform is as follows.

```
define transform <transform name> : <source frame> -> <destination
frame> as

      <transform definition>

end transform
```

Where

***<transform name>*** is an identifier for that transform

***<source frame>*** is the name of the frame defining the domain of the transform

***<destination frame>*** is the name of the frame defining the co-domain of the transform

***<transform definition>*** can take several forms shown in Table 4-2.

### 4.7.3 The Subspaces File

The subspaces file contains any number of subspace definitions related to the problem. The syntax of defining a single subspace is as follows:

```
define subspace <frame name>.<subspace name> as
      <subspace definition>
end subspace
```

Where

***<frame name>*** is the name of the frame the subspace is part of

***<subspace name>*** is an identifier for that subspace

***<subspace definition>*** can take several forms shown in Table 4-3.

**Table 4-2: Methods of defining transforms in GMacDSL**

| <transform definition> | Meaning |
|---|---|
| identity | An identity transform between two frames |
| alias of ***<transform name>*** | The exact same transform but with a new name |
| inverse of ***<transform name>*** | The inverse of a given transform |
| transpose of ***<transform name>*** | The transpose of a given transform |
| inverse transpose of ***<transform name>*** | The inverse transpose of a given transform |
| outermorphism using ***<source frame>***.BCM | An outermorphism defined using a Basis Change Matrix of a certain frame |
| outermorphism using ***<Mathematica matrix expression>*** | An outermorphism defined using a vector linear transform matrix |

**Table 4-3: Methods of defining subspaces in GMacDSL**

| <subspace definition> | Meaning |
|---|---|
| basis {***<list of basis blades>***} | A subspace defined using some basis blades |
| ga_span {***<list of basis vectors>***} | A subspace defined using the geometric algebra spanned by some basis vectors |

### 4.7.4 The Multivectors File

This file contains all definitions for multivector classes. Any single multivector class can be defined using the following syntax:

```
define multivector class <frame name>.<multivector class name> as
        <class definition>
end multivector class
```

Where

***<frame name>*** is the name of the frame the multivector class is part of

***<multivector class name>*** is an identifier for that multivector class

***<class definition>*** is semicolon separated list containing one or more subspace names and may optionally contain constant basis coefficients on the form ***<basis name>*** : ***<scalar value>*** where ***<scalar value>*** is any valid Mathematica real scalar expression.

### 4.7.5 The Constants File

All multivector constant are defined in the constants file. Any constant can be defined using the following syntax:

```
define constant <frame name>.<constant name> as
        multivector {<list of basis coefficients values>}
end constant
```

Where

***<frame name>*** is the name of the frame the multivector constant is part of

***<constant name>*** is an identifier for that multivector constant

***<list of basis coefficients values>*** is semicolon separated list containing one or more constant basis coefficients on the form ***<basis name>*** : ***<scalar value>*** where ***<scalar value>*** is any valid Mathematica real scalar expression.

### 4.7.6 The Bindings File

All C# fixed class bindings are defined in this file. The syntax for defining a binding is as follows:

```
define binding <binding name> as
        use frame <frame name>
        bind { <list of basis coefficients bindings> }
        min { <list of variables lower limits> }
        max { <list of variables upper limits> }
end binding
```

Where

***<binding name>*** is an identifier for that binding

***<frame name>*** is the name of the frame the bound multivector is part of

***<list of basis coefficients bindings>*** is semicolon separated list containing one or more scalar functions basis coefficients on the form ***<basis name>*** : ***<scalar function>*** where ***<scalar function>*** is any valid Mathematica real scalar function.

***<list of variables lower limits>*** is a semicolon separated list containing one or more scalar values on the form ***<C# variable>*** : ***<scalar value>*** where ***<scalar value>*** is any valid Mathematica real scalar constant value.

***<list of variables upper limits>*** is a semicolon separated list containing one or more scalar values on the form ***<C# variable>*** : ***<scalar value>*** where ***<scalar value>*** is any valid Mathematica real scalar constant value.

The parts used to define C# variables limits are optional for any binding.

### 4.7.7 The Macros File

The final file to be read is the file containing all GA macros definitions. The general syntax for defining GA macros is:

```
define macro <macro name> as
        inputs: {<list of input multivectors>}
        outputs: {<list of output multivectors>}
```

```
        performs:
                &lt;list of macro statements&gt;
end macro
```

Where

***<macro name>*** is an identifier for that macro

***<list of input multivectors>*** is semicolon separated list containing one or more multivector names on the form ***<multivector name>*** : ***<multivector class>*** where ***<multivector class>*** is the name of previously defined multivector class.

***<list of output multivectors>*** is semicolon separated list containing one or more multivector names on the form ***<multivector name>*** : ***<multivector class>*** where ***<multivector class>*** is the name of previously defined multivector class.

***<list of macro statements>*** is semicolon separated list containing one or more legal macro statements as described in the next section.

## 4.8 GMac GA Macros Statements

In the current implementation of GMacDSL, there are seven types of statements. The following subsections describe each of these types.

### 4.8.1 The multivector Statement

This statement constructs a multivector represented in a certain frame by directly assigning values to its basis coefficients. The values come from real scalar functions operating on the coefficients of other multivectors. This statement takes the form:

```
<multivector name> = <frame name>.multivector { <list of basis coefficients> }
```

Where

***<multivector name>*** is the name of the multivector holding the result of the statement.

***<frame name>*** is the name of the frame the resulting multivector belongs to.

***<list of basis coefficients>*** is semicolon separated list containing one or more scalar functions basis coefficients on the form ***<basis name>*** : ***<scalar function>*** where ***<scalar function>*** is any valid Mathematica real scalar function.

## 4.8.2 The Binary Operator Statement

This type of statement applies a binary operation on two multivectors in some frame to produce a third multivector belonging to the same frame. This statement takes the form:

***<multivector name>*** = ***<left multivector> <binary operator> <right multivector>***

Where

***<left multivector>*** and ***<right multivector>*** are the names of the operand multivectors

***<binary operator>*** is the applied binary operator as described in Table 4-4.

**Table 4-4: Binary operators on multivectors in GMacDSL**

| ***<binary operator>*** | Meaning |
|---|---|
| `gp` | Geometric Product |
| `op` | Outer Product |
| `sp` | Scalar Product |
| `lcp` | Left-Contraction Product |
| `rcp` | Right Contraction Product |
| `fdp` | Fat-Dot Product |
| `hip` | Hestenes Inner Product |
| `cp` | Commutator Product |
| `acp` | Anti-Commutator Product |
| `+` | Addition of Multivectors |
| `-` | Subtraction of Multivectors |

### 4.8.3 The Linear Transform Statement

This statement is used to apply a given linear transform on a multivector to produce another multivector. The frames of both multivectors must match the source and destination frames of the applied transform. This statement takes the form:

```
<multivector name> = <transform name> [ <multivector operand> ]
```

Where

***<transform name>*** is the name of the linear transform to be applied

***<multivector operand>*** is the name of the multivector to be transformed

### 4.8.4 The Unary Operator Statement

This type of statements applies special GA operators on multivectors to produce other multivectors in the same frame. There are two forms of unary operators. The first takes the form:

```
<multivector name> = <unary operator>( <multivector operand> )
```

And the second takes the form:

```
<multivector name> = <unary operator>( <multivector operand>,
<second operand> )
```

Where <unary operator> can take any of the values shown in Table 4-5.

In addition there are two other unary operators. The copy operator that takes the form:

```
<multivector name> = <multivector operand>
```

And the negation operator that takes the form:

```
<multivector name> = - <multivector operand>
```

### 4.8.5 The call macro Statement

This statement is used to call a GA macro from within another GA macro. The input\output multivectors of the called macro are substituted with multivectors from the calling macro. The statement takes the form:

```
call <macro name> { <list of multivector bindings> }
```

Where

***<macro name>*** is the called macro

***<list of multivector bindings>*** is a semicolon-separated list having items of the form ***<called macro multivector>*** : ***<calling macro multivector>***

**Table 4-5: Unary operators on multivectors in GMacDSL**

| ***<unary operator>*** | ***<second operand>*** | Meaning |
|---|---|---|
| `grade_inv` | | Grade Involution |
| `cliff_conj` | | Clifford Conjugation |
| `reverse` | | Reverse |
| `scale` | `<scalar function>` | Multiply by real scalar |
| `div_by_scalar` | `<scalar function>` | Divide by real scalar |
| `norm` | | Calculate norm |
| `norm2` | | Calculate squared norm |
| `quasi_norm` | | Calculate quasi-norm |
| `quasi_norm2` | | Calculate squared quasi-norm |
| `diff` | `<multivector coefficient>` | Parametric differentiation |
| `cast_to_grades` | `<multivector name>` | Cast to grades |
| `cast_to_subspace` | `<subspace name>` | Cast to subspace |
| `cast_to_class` | `<multivector class>` | Cast to multivector class |

### 4.8.6 The output Statement

The output statement instructs GMac to output the exact same text in the final generated code with possible substitution for multivector coefficients with their bound target language variables. The syntax is as follows:

output {***<target language code>***}

For example, to instruct GMac to generate a comment at a certain point in the final code the following statement can be used:

```
output {//This line is generated by GMac};
```

The output statement is useful for inserting target language related code from within GMacDSL to perform tasks not typically attainable by GMacDSL itself. The most important examples include conditional statements and loops. Since GMacDSL does not contain if statements or for loops, the output statement can be used to generate such statements. The main benefit here is convenience for the software designer to make the geometric algorithm more complete from within GMacDSL code. The following example inserts an if statement to test some multivector coefficient in the final generated code:

```
output { if (<mv1.e1^e2> != 0.0) };
```

The expression <mv1.e1^e2> in the previous statement will be replaced by a C# variable bound to the real coefficient of e1^e2 for the multivector mv1. The final generated code might look like the following in C#:

```
If (var0013 != 0.0)
```

Where var0013 is the C# variable bound with mv1.e1^e2.

### 4.8.7 The join on\off Statements

These two statements simply control how the final assignment statements in the target language will be generated. They take the simple form:

```
join on
```

And

```
join off
```

Without using these two directives, GMac outputs a list of assignment statements computationally equivalent to the GA operations in the macro. If the user requires outputting the minimal number of output expressions as direct functions of the input expressions without any temporary variables, the join on\off directives should be used around the macro statements. It is sometimes, however, more computationally efficient to use temporary variables than to use a full input-output expression. This decision is left for the user as GMac cannot predict the best way to use, or not use, the join directives.

## 4.9 GMac Binding Points

As stated earlier, GMac binding points are special portions of the code in the target language source files. These are the only means of integration with GMacDSL code files. In what follows is a description of the syntax of a binding point in a C# code file.

A binding point in C# takes the form:

```
#region GMac : <macro name>
<variable bindings>
<variable limits>
#endregion
```

Where

***<macro name>*** is the name of the GA macro to be generated inside the binding point.

***<variable bindings>*** is a semicolon-separated list of bindings between C# variables and the called macro multivector coefficients.

***<variable limits>*** is a semicolon-separated list of lower and upper limits for bound C# variables.

Any single binding statement can take one of two forms. The first is:

```
GMac.Bind("<multivector>.<basis blade>", "<scalar function>")
```

The second form is:

```
GMac.Bind("<multivector>", "<class binding name>", “<C# object name>”)
```

The first form is used to directly bind a multivector coefficient to a scalar function operating on C# variables. The second form uses any of the class bindings defined in GMacDSL to associate a multivector with the members of a certain C# object.

To define lower and upper limits for C# variables inside the binding point, the following forms are used:

```
GMac.AssumeMin("<C# variable name>", "<scalar value>")
```

```
GMac.AssumeMax("<C# variable name>", "<scalar value>")
```

## 4.10 Example and Discussion

This section provides a simple example to the inner operation of GMac when generating code for a simple GA expression. In addition a brief discussion of the strong and weak points of GMacDSL is provided.

### 4.10.1 An Example

The geometric problem in this example is simply to find a vector that is normal to two linear independent vectors in 3D Euclidean space. In traditional linear algebra this problem is solved by the cross product of the two vectors. The cross product can only be used in 3D Euclidean space and cannot be generalized to higher dimensions. Assuming the two input vectors are $v_1 = x_1 e_1 + x_2 e_2 + x_3 e_3, v_2 = y_1 e_1 + y_2 e_2 + y_3 e_3$ the equivalent GA expression to solve this problem is:

$$w = (u \wedge v) \rfloor I_3^{-1} \qquad (4.13)$$

Where $I_3^{-1} = e_3 \wedge e_2 \wedge e_1$ is the inverse of the GA space pseudo-scalar $I_3$.

The GMac code that can express the previous GA expression is:

```
define macro GetNormalToVectors as
        inputs: {
                u as e3d.Multivector;
                v as e3d.Multivector
        }
        outputs: {
                w as e3d.Multivector
        }
        performs:
                t1 = u op v;
                w = t1 lcp e3d.Ii;
end macro
```

In the main application code, the operation is to be defined inside a function taking two objects of type Vector3D and returning an object of the same type.

Assuming a Vector3D C# class contains three double members x, y, z to hold the coordinates of the vector, the function is initially defined as follows:

```
public static Vector3D GetNormalToVectors(Vector3D u, Vector3D v)
{
    double wx, wy, wz;

    #region GMac : GetNormalToVectors
    //GMac.Bind("u.e1", "<u.x>");
    //GMac.Bind("u.e2", "<u.y>");
    //GMac.Bind("u.e3", "<u.z>");
    //GMac.Bind("v.e1", "<v.x>");
    //GMac.Bind("v.e2", "<v.y>");
    //GMac.Bind("v.e3", "<v.z>");
    //GMac.Bind("w.e1", "<wx>");
    //GMac.Bind("w.e2", "<wy>");
    //GMac.Bind("w.e3", "<wz>");
    #endregion

    return new Vector3D(wx, wy, wz);
}
```

When GMac reaches the `GetNormalToVectors` function code, the `GetNormalToVectors` macro is flagged as the macro to be generated. The first step GMac performs is to prepare a table with input and output multivectors in memory. The data in the table are defined using the binding point data as shown in Table 4-6. After symbolically evaluating the macro, the values of the multivectors are as shown in Table 4-7.

**Table 4-6: Input\output multivectors for the initial step in the GMac code generation example**

| Multivector Name | Type | Calculated | Multivector Value | |
|---|---|---|---|---|
| | | | Basis Blade | Real Coefficient |
| u | input | yes | $e_1$ | u.x |
| | | | $e_2$ | u.y |
| | | | $e_3$ | u.z |
| v | input | yes | $e_1$ | v.x |
| | | | $e_2$ | v.y |
| | | | $e_3$ | v.z |
| w | output | Not yet | $e_1$ | wx |
| | | | $e_2$ | wy |
| | | | $e_3$ | wz |

The data in Table 4-7 results in GMac generating a list of un-optimized C# assignment statements on the form:

```
double var1 = -(u.y * v.x) + u.x * v.y;
double var2 = -(u.z * v.x) + u.x * v.z;
double var3 = -(u.z * v.y) + u.y * v.z;
wx = var3;
wy = -var2;
wz = var1;
```

After optimizing the expressions, GMac generates the final statements as:

```
wx = -(u.z * v.y) + u.y * v.z;
wy = u.z * v.x - u.x * v.z;
wz = -(u.y * v.x) + u.x * v.y;
```

**Table 4-7: Final values for multivectors in the GMac code generation example**

<table>
<tr><th rowspan="2">Multivector Name</th><th rowspan="2">Type</th><th rowspan="2">Calculated</th><th colspan="2">Multivector Value</th></tr>
<tr><th>Basis Blade</th><th>Real Coefficient</th></tr>
<tr><td rowspan="3">u</td><td rowspan="3">input</td><td rowspan="3">yes</td><td>$e_1$</td><td>u.x</td></tr>
<tr><td>$e_2$</td><td>u.y</td></tr>
<tr><td>$e_3$</td><td>u.z</td></tr>
<tr><td rowspan="3">v</td><td rowspan="3">input</td><td rowspan="3">yes</td><td>$e_1$</td><td>v.x</td></tr>
<tr><td>$e_2$</td><td>v.y</td></tr>
<tr><td>$e_3$</td><td>v.z</td></tr>
<tr><td rowspan="3">t1</td><td rowspan="3">temporary</td><td rowspan="3">yes</td><td>$e_1 \wedge e_2$</td><td>var1 = -u.y * v.x + u.x * v.y</td></tr>
<tr><td>$e_1 \wedge e_3$</td><td>var2 = -u.z * v.x + u.x * v.z</td></tr>
<tr><td>$e_2 \wedge e_3$</td><td>var3 = -u.z * v.y + u.y * v.z</td></tr>
<tr><td rowspan="3">w</td><td rowspan="3">output</td><td rowspan="3">yes</td><td>$e_1$</td><td>wx = var3</td></tr>
<tr><td>$e_2$</td><td>wy = -var2</td></tr>
<tr><td>$e_3$</td><td>wz = var1</td></tr>
</table>

From the previous example, several points are notable. In the geometric macro code the variable `e3d.Ii` is a constant multivector defined in the constants input DSL file. The macro uses the `e3d.Multivector` multivector class defined on the `e3d` frame. Such class can represent any general multivector defined on the frame. The un-optimized C# code initially generated by GMac directly corresponds to the low-level relations between the symbolic values of the coordinates for the participating multivectors. The internal representation of symbolic multivectors enable using the data in binding points to select the minimal number of symbolic coefficients necessary for the computation. After generating a suitable list of assignment statements, an optimization process is made to further reduce the number of computations required by the final generated code.

### 4.10.2 GMacDSL Discussion

GMacDSL is a simple domain specific language designed for coding geometric algebra expressions. Several important characteristics present in GMacDSL serves such goal. The use of Mathematica expressions inside GMacDSL to define real valued scalar coefficients of multivectors is one of the main features in GMacDSL. On the other hand, GMacDSL provides a simple but rich structure for adding GA capabilities to any modern programming language. Such feature is lacking in all modern languages since geometric programming [36] is still an emerging field in its beginnings. Geometric programming features are not yet accepted as part of general purpose programming languages' features list. GMacDSL can reduce the gap between modern general purpose languages, like C# and Java, and the geometric programming demands of modern applications like game engines, visual simulation systems, and other computer graphics applications. The use of GA as a base for GMacDSL is the main advantage of such simple yet useful DSL.

Other features of GMacDSL facilitate its use to describe many GA-based expressions. The multivector macro statement can be easily used to construct multivectors in any frame using calculated coefficients based on other multivectors. Such statement can provide a good method for implementing linear transforms between different GA spaces. Such feature is application dependent and should be left for the designer of the application.

Some features should be added to GMacDSL to make it easier to use. One feature that should be added is the ability to encode several GA operations in one GMacDSL line. A GA expression like $(a-b)\wedge(c \rfloor I^{-1})$ must be first decomposed into 5 GMacDSL steps before coding: a subtraction, an inverse, a contraction product, and an outer product. Such feature will make GMacDSL more capable of directly handling GA expressions compared to its current implementation. A second feature to be added is the addition of traditional control-of-flow constructs; like if else blocks, and for while loops. The addition of functional programming features like the ones present in F# [37] and OCaml [38] will be a great addition to GMacDSL. Such functional programming features are already present in PLaSM [36]; a geometric programming language for Computer Aided Design (CAD) applications.

## 4.11 Comparing GMac and Gaalop

As described in Chapter 2, Gaalop operates according to the same basic idea as GMac does. The idea is to use a symbolic processing engine or Computer Algebra System (CAS) to deduce optimized expressions based on GA multivector operations. The expressions are then rewritten to some target language as the final implementation. Despite the similarity, GMac and Gaalop are very different when it comes to architecture and operation.

Gaalop requires the presents of Maple as a symbolic processing engine. In addition it relies completely on the CLIFFORD package for Maple which is a package designed for Clifford algebra computations [39], [40]. Thus the development of Gaalop is firmly tied to such package. It will inherit any bugs or disadvantages in the external package. In addition, it will suffer from all of its limitations.

GMac, on the other hand, uses Mathematica as its symbolic engine. GMac does not rely on any external Mathematica package for its normal operation. GMac’s development is free from such concerns that Gaalop may suffer from. In addition, GMac’s core engine defines its own set of GA operations on symbolic multivectors. Such structure can easily be extended at any time regardless of any concerns related to the status of Mathematica as a CAS software.

Another very important difference between GMac and Gaalop is the GMacDSL. Gaalop does not contain a corresponding DSL for describing GA operations and algorithms. Instead it relies on the syntax given by Maple and the CLIFFORD package. The GMacDSL is very flexible and is almost independent of Mathematica. It only uses Mathematica syntax inside scalar functions and scalar values for some multivector operations as described in previous sections.

As described in [28], Gaalop is limited to implementing 5D CGA. GMac, on the other hand, is not limited to any specific GA. In addition GMac can easily handle multiple GAs in a single problem along with any interactions between them.

Table 4-8 compares GMac and Gaalop main features using the requirements of section 4.1. Clearly GMac is much more suitable for generating optimized software implementations based on GA than Gaalop.

## 4.12 Comparing GMac and Gaigen 2

Gaigen 2 is currently the most successful solution to the GA implementation problem discussed in chapter 1. As apparent from this chapter, GMac provides many architectural advantages over Gaigen 2. In the following subsections the main architectural differences between the two systems are described.

### 4.12.1 Comparing Requirements Fulfillment

Using the requirements of section 4.1, the features of GMac compared to Gaigen 2 are apparent. Both GMac and Gaigen 2 are capable of directly using the high level expressiveness power of GA expressions in their DSLs. Nonetheless, the most important feature present in GMac but not in Gaigen 2 is single pass code generation. As described in chapter 2, Gaigen 2 requires several iterations of profiling to perform optimal code generation. On the other hand, GMac only processes all of its inputs once.

Another point of strength of GMac compared to Gaigen 2 is its full separation of GA code from target language code. GMac never generates unnecessary code for classes requiring another stage of integration with main application code like Gaigen 2 does. The generated code from GMac is highly localized to the binding points of the target language source code. Gaigen 2 also suffers from being bulky and consisting of many unrelated components as described in detail in [20]. GMac on the other hand only relies on the .NET framework and Mathematica. For such reasons GMac is much better than Gaigen 2 in its simplicity of integration and optimality of code reuse. Table 4-8 summarizes the main features of GMac compared to Gaigen 2.

**Table 4-8: Comparing requirements fulfillment of GMac, Gaalop, and Gaigen 2**

| Requirement | GMac | Gaalop | Gaigen 2 |
|---|---|---|---|
| High-level User Interface | Yes | Lower | Yes |
| Full Code Separation | Yes | No | Lower |
| Ease of Code Generation | Yes | Yes | No |
| Optimal Code Reuse | Yes | Lower | Lower |
| Simplicity of Integration | Yes | No | Lower |
| Minimal System Components | Yes | Yes | No |

### 4.12.2 Comparing DSLs

Both GMac and Gaigen 2 utilize a high-level DSL based on geometric algebra. GMacDSL is much more versatile then Gaigen 2 DSL as described in [20]. The main strength of GMacDSL comes from the definition of transforms between frames and the use of scalar functions and values that can directly use the Mathematica syntax for scalar expressions. The symbolic optimization capabilities of Gaigen 2 are very limited compared to GMac or even Gaalop as they rely on very strong CAS software like Mathematica and Maple.

One technical advantage of Gaigen 2 over GMac is the ability of Gaigen 2 DSL to directly handle more useful control-of-flow statements. Such statements include if-else, for and while loops, and return statements. This technical advantage is not currently implemented in GMac although it can be added in future implementations. Such feature can be nonetheless compensated using the output statement of GMac as discussed in a previous section.

### 4.12.3 Comparing Outputs

As will be illustrated in chapter 7, the performance of GMac generated code is much better than Gaigen 2. The main reason is that the code generated by GMac is minimal compared to Gaigen 2. As an example, the following GA expression is used in [11] to perform the well known hidden-surface removal operation in computer graphics on a triangle:

$$\left\langle (v_3 - v_1) \wedge (v_2 - v_1) \right\rangle_{e_1 \wedge e_2} \tag{4.14}$$

}

Where $v_1, v_2, v_3$ are the three position vectors representing triangle vertices projected on the viewing plane. When this simple relation is used by GMac, the generated code is as follows:

```
double var0001 = -v1.m_c[0] + v2.m_c[0];
double var0002 = -v1.m_c[1] + v2.m_c[1];
double var0003 = -v1.m_c[0] + v3.m_c[0];
double var0004 = -v1.m_c[1] + v3.m_c[1];
B_e1e2 = -(var0002*var0003) + var0001*var0004;
```

While in Gaigen 2 the generated code is as follows:

```
B = (v2 - v1) ^ (v3 - v1);
B_e1e2 = B.e1e2();

inline vector subtract(const vector& x, const vector& y) {
       return vector(vector_e1_e2, (x.m_c[0] + (-1.0f * y.m_c[0])), ((-1.0f * y.m_c[1]) + x.m_c[1]));
}

inline vector operator-(const vector& arg1, const vector& arg2) {
       return ::e2ga::subtract(arg1, arg2);
}

inline bivector op(const vector& x, const vector& y) {
       return bivector(bivector_e1e2, ((x.m_c[0] * y.m_c[1]) + (-1.0f * x.m_c[1] * y.m_c[0])));
}

inline bivector operator^(const vector& arg1, const vector& arg2)
{
       return ::e2ga::op(arg1, arg2);
}

inline Float e1e2() const {
```

```
        return m_c[0];
}
```

It is not surprising that GMac code executes 7 times faster than Gaigen 2 code for this particular example. The `inline` directive used extensively by Gaigen 2 is useless in modern C++ compilers. Most modern C++ compilers simply ignore such directive and selectively inline functions as they see fit. That is because most of the time the decision made by the C++ compiler is much better than that made by the programmer. Thus the code generated by Gaigen 2 may or may not contain several levels of function invocations resulting in reduced performance compared to GMac generated code.

In addition, Gaigen 2 generates and uses many classes external to the original problem. Objects are created in the heap in main memory from such classes at runtime. This creation process has two disadvantages. First, the creation of objects on the heap requires significant processing, for heap allocation, compared to creating objects on the stack. Second, memory requirements of the final software are increased by such objects. GMac does not create any classes in the final generated code. Only temporary variables are created during runtime. Such variables are double numbers typically created on the stack not the heap. Thus no extra processing is required for heap allocation and no additional memory is needed beyond the basic requirements of the main application.

# Chapter 5 : Base Ray Tracer Software Architecture

This chapter provides a brief introduction to Ray Tracing. As will be illustrated in this chapter, ray tracing is a very geometry demanding application. Ray tracing requires good geometric models for all of its components. A ray tracer is a rather complex system requiring a high level programming approach. At the same time it is a very demanding application in its memory and processing needs. It provides a very good testing ground for using geometric algebra and GMac and illustrating their strengths and possible weaknesses. In addition, this chapter provides a description for the software architecture of the base ray tracer used in this work. The implementation is mostly based on [41] along with some additions and enhancements. The architecture is based on OOP concepts. The base ray tracer can be easily extended with additional geometric primitives, acceleration techniques, textures, tone-mapping techniques, ray-object intersection algorithms, and much more. Such extensibility is essential to using the base ray tracer as a testbed for new ray tracing techniques including GA enhancements discussed in later chapters.

The material presented in this chapter is mainly based on [41]. Section 5.1 is an introduction. Section 5.2 provides a brief description for the first stage of ray tracing; the scene modeling stage. Section 5.3 provides a brief description for the second stage; the scene rendering stage. Section 5.4 is a brief discussion related to the current research activities in ray tracing. Finally, section 5.5 gives a general overview of main ray tracer components.

## 5.1 Introduction to Ray Tracing

One of the best resources to begin studying ray tracing is [41]. This book describes ray tracing as being a branch of computer graphics. One of the main concerns of Computer graphics is to simulate the distribution of light in a virtual 3D environment. There are a few algorithms proving to be suitable for such simulation. Such algorithms can be loosely classified into two classes. The first class is called projective algorithms (also called object-space algorithms). This class projects each geometric primitive onto the image plane, with local shading taking care of the appearance of objects. Such algorithms are still widely used because they are amenable to pipeline processing and therefore to hardware implementation as evidenced by all modern graphics cards.

The other main set of algorithms is called Image-space algorithms. These algorithms compute the color of each pixel by trying to deduce where the light came from for that pixel. Here, the basic operation is to determine the nearest object along a line of sight. Following light back along a line has given this basic operation and the associated image-synthesis algorithm their name: ray tracing. This chapter is a brief introduction to the basic ray tracing algorithm.

Generally speaking, there are two main stages for ray tracing. The first stage is called the scene modeling stage, or modeling stage for short (Figure 5-1). This stage is concerned with defining the environment and its constituting objects to be rendered as a synthetic image. The second stage is called the scene rendering stage, or the rendering stage for short. In the rendering stage, the basic ray tracing algorithm (Figure 5-2) is executed for each pixel of the final image to deduce its color.

## 5.2 The Scene Modeling Stage

As shown in Figure 5-1, the scene modeling stage is mainly used to define three sets of objects in the simulated environment. The first set contains the physical objects that are part of the environment. The second contains the viewing system that receives the light from the environment. Finally the third set contains the light sources that illuminate the environment. The following is a brief description for each set.

### 5.2.1 Modeling the Objects

The rendered scene consists mainly of a set of objects. Each object must have some sort of mathematical model that is used to describe its characteristics relevant to ray tracing. Most mathematical object representation models rely on 2D surface representations in 3D space. In addition, 1D curve and 3D volumetric representations are also common in practice. The main goal of the mathematical representation of the object is to provide shape, location and orientation information of the object in space. Such information is required during the calculations to determine hit points with rays and shading information for shading calculations. In addition, any object must define its material characteristics. Properties of materials include texture, reflectivity parameters, and transparency properties. Such information is required extensively for the shading calculations.

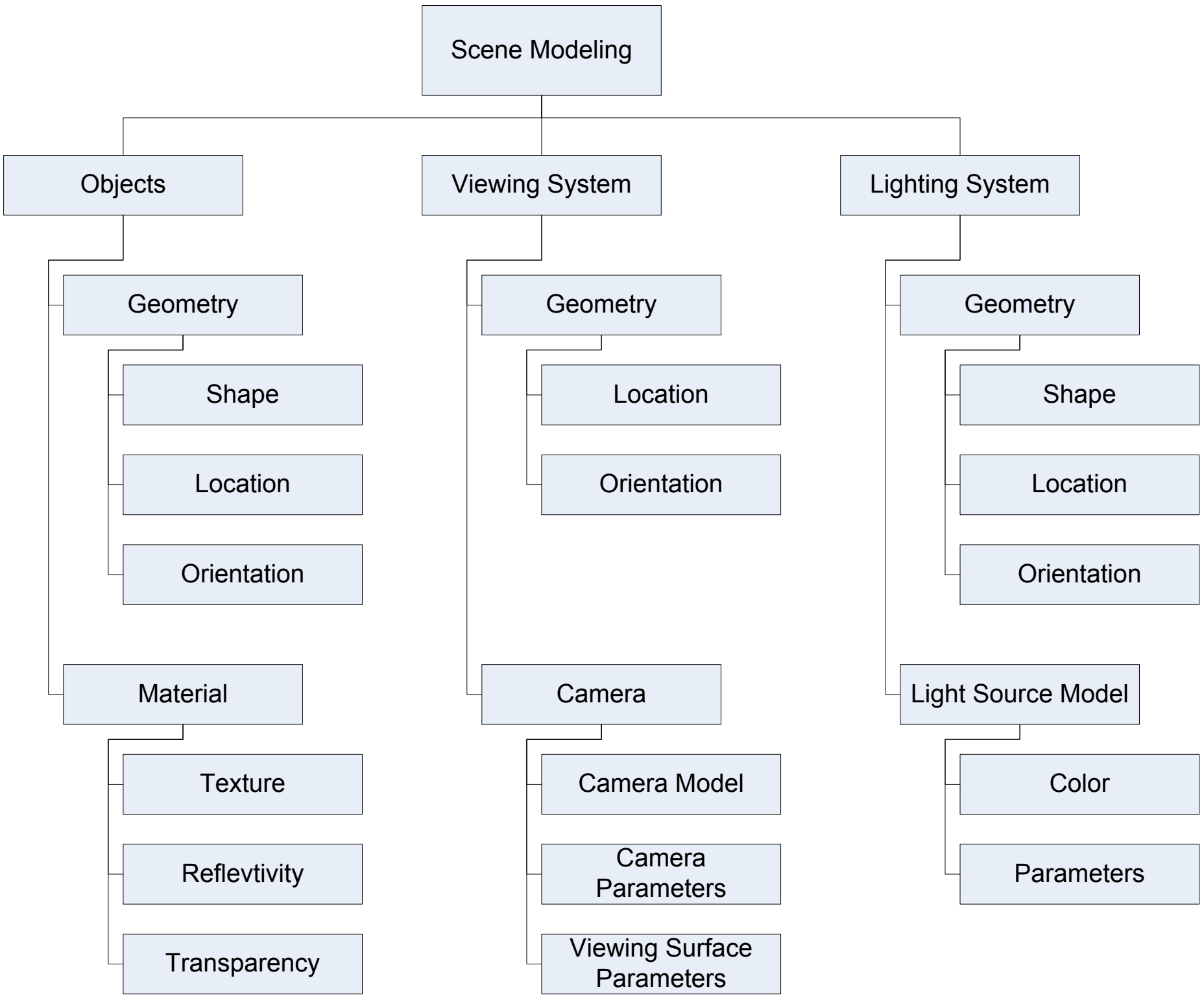

**Figure 5-1: The Scene Modeling Stage**

In order to reduce the number of ray intersection tests an acceleration scheme is usually applied. The main purpose of such scheme is to restrict the number of objects tested by any single ray to a small portion of the total scene objects. Many good acceleration schemes exist in current ray tracing literature. Some of the most effective schemes include regular grids[41], kd-trees [42] and bounding interval hierarchies (BIH) [43], [44], [45].

### 5.2.2 Modeling the Viewing System

The viewing system is where the scene is rendered as a final image. Its modeling consists of selecting a camera model, defining its parameters and defining the surface on which the image is created (called the viewing surface). A camera is located in the same space where the objects are defined with a location and orientation of its own. The pinhole camera model and the thin-lens camera model are among the most used camera models. Camera parameters vary from one model to the other. The viewing surface is usually taken to be a planar rectangular region in 3D space with a local 2D coordinate system that can be directly mapped to the pixels of the final image. Other camera models use non-planar viewing surfaces like the models for angular fisheye projection, spherical, and cylindrical panoramic projections. Such models are often called non-linear projection models [41].

### 5.2.3 Modeling the Lighting System

The lighting system consists of all light sources used to illuminate the objects in the scene. The light coming from light sources is scattered in different directions with different intensities at object surfaces. Such differences are what give objects their perceived colors and rich visual properties. The scene complexity increases rapidly with the increase of the number of light sources. Among the most used light source models are point lights, directional lights, and area lights [41]. Area lights are more complex models for light sources than point and directional lights. Their presence in a scene gives a more realistic simulation of soft-shadows on the expense of more processing requirements.

## 5.3 The Scene Rendering Stage

The main task of the scene rendering stage is to deduce the color of each pixel in the final rendered image. For each pixel the basic ray tracing algorithm shown in Figure 5-2 is performed. The recursive nature of the algorithm is very clear from the figure. In what follows a brief description of key steps in the algorithm is provided.

### 5.3.1 Sampling Pixels

Using a single ray per pixel is usually not enough to compensate for aliasing artifacts in the final image. Often more than one ray is required to produce images with low level of noise. The obvious problem with this technique is the rapid increase in the required processing resources. Nevertheless, it may be the only solution to handle severe noise patterns in many ray traced images. Many good sampling techniques are present to reduce noise in the final image with the least possible number of samples per pixel. Among these are random sampling, multi-jittered sampling, and Hammersley sampling [41].

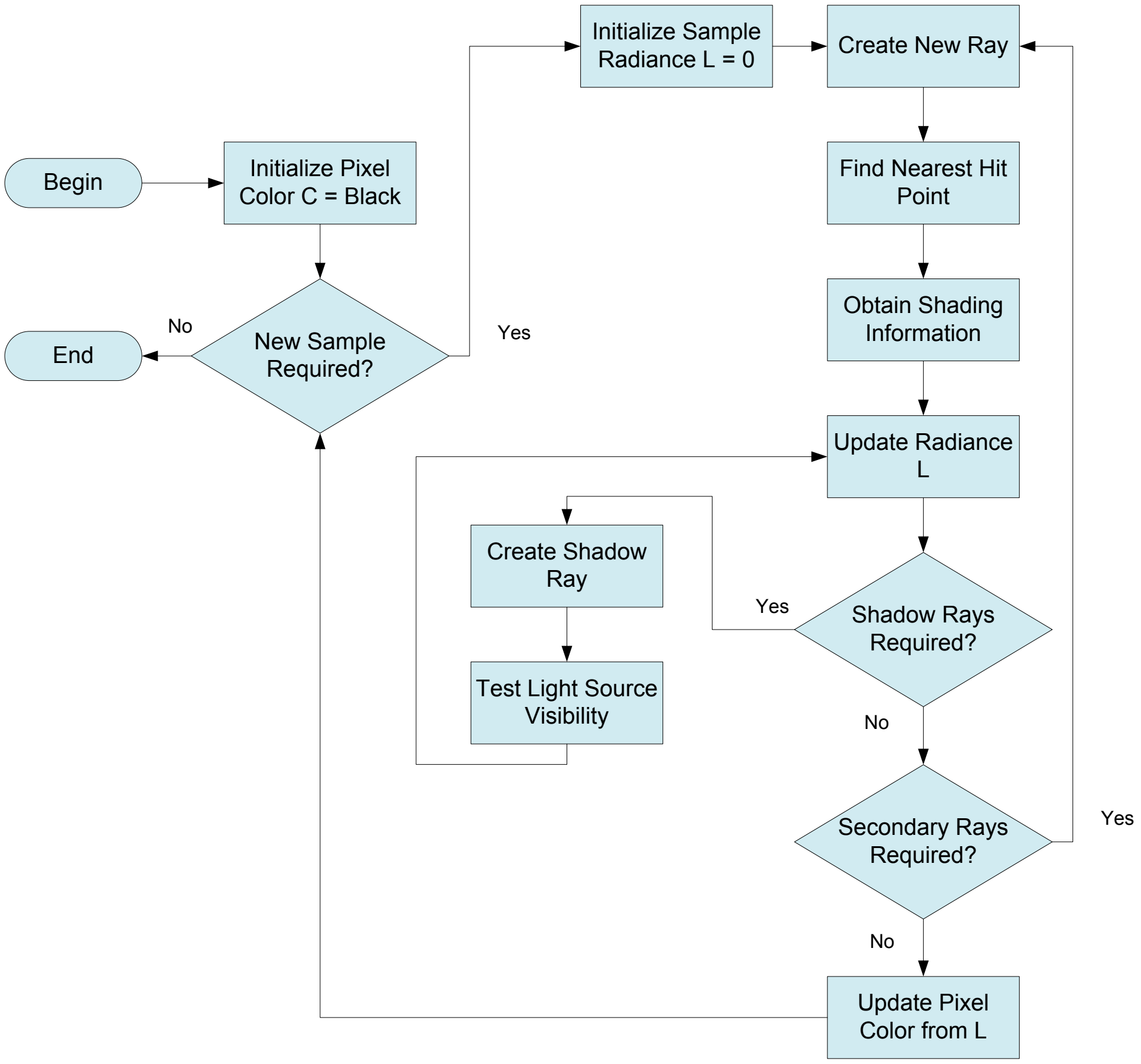

**Figure 5-2: Basic Ray Tracing Algorithm**

### 5.3.2 Ray Casting

The first step in the main body of the algorithm is to create what is called a primary ray. Such ray is created using the camera parameters and the mapped location of the pixel sample in 3D space. The next step is to use the acceleration structure already built in the modeling stage to find the nearest ray-object intersection point in 3D space. Next, the hit point is returned along with some important shading information related to the hit point. Such information include texture and color information, normal vector to object surface at the hit point, and material parameters. Such information is used to update the simulated radiance [41] for this particular pixel sample. This process is usually called ray casting and it is the main process in the ray tracing algorithm. Such process typically consumes 90% or more of the processing required for rendering the final image.

### 5.3.3 Spawning New Rays

In order to produce a more accurate simulation to light interaction with the environment more rays might be required. The rays originate at the hit point and travel in several directions. Two classes of rays are needed to complete the simulation. The first class consists of shadow rays. These rays are used to test the visibility of the hit point relative to each light source in the scene. If the point is visible then the radiance of the pixel sample is affected by light source. The other class of rays is called secondary rays. These are used to simulate light reflections and refractions through reflective and transparent material. The secondary rays are the source of recursion in the ray tracing algorithm as each secondary ray begins a new ray casting loop all over again. There must be defined a stop condition for generating secondary rays to limit the possible exponential growth of rays used to simulate the scene.

### 8.3.4 Deducing Pixel Color

After calculating the radiances of all pixel samples they are averaged giving an average value of radiance at that pixel location in 3D space. The problem with such process is that color representations in modern computers operate on limited intervals for color values. For example the RGB model can only hold red, green and blue values in the range 0.0 to 1.0. The calculated radiance, on the other hand, may be any real number not limited to such values. This problem requires some sort of mapping between the calculated radiances and the color representations. This mapping is often called tone mapping. Many techniques are used for tone mapping. A good discussion of such techniques is presented in [46].

## 5.4 Current Research in Ray Tracing

As evident from the previous discussion, ray tracing is a very active field of research in computer graphics. Ray tracing can produce high quality images with a good approximation to real world scenes using an algorithm well-suited to parallel processing techniques. The main problem with ray tracing is its processing requirements. Such requirements prevented ray tracing from being applied in real-time applications like computer games and real-time simulations. The main trend in current ray tracing research is thus to improve its performance. In addition to the performance problem, more object modeling techniques are investigated in ray tracing. On one hand, some of such techniques can improve ray-object intersection time or memory requirements for modeling the environment. On the other hand, some techniques are intended to increase the reality of rendered images without substantially increasing processing and memory requirements.

In order to appreciate the current research in ray tracing, several web sites provide good sample of such research. These web sites include [47], [48], and [49]. Most research in these sites focus on bringing ray tracing to become a real-time application [50], [51]. A typical ray tracer spends most of its time inside procedures for ray-object intersection. In order to improve the speed of any ray tracer such intersection tests must be made more efficient. According to [52], there are three different strategies to consider. The first is reducing the average cost of intersecting a ray with the objects of the environment [53], [54], [55]. The second is to reduce the total number of rays intersected with the environment objects [56], [57]. The third is by replacing individual rays with a more general entity [58]. All current attempts to obtain real-time ray tracing use one or more of such strategies simultaneously [59], [60], [61]. In addition, many papers focus on using the parallel nature of ray tracing to test several rays simultaneously using parallel processing hardware architecture [59], [62], [63].

Any geometric object is ultimately modeled using a standard set of primitive objects. Such primitives typically include planes, spheres, triangles, axis-aligned boxes, cones, and cylinders. Other modeling techniques include surfaces described by implicit equations [64], [65], [66] or parametric representations [67], [68]. Some applications are not satisfied by such primitives for some reasons. For example, an emerging new primitive for ray tracing is the point primitive [69]. Most 3D scanning hardware produce a stream of 3D points as output [70].

Many scientific visualization applications require the rendering of 3D volumetric data. Ray tracing can be used to perform such task giving excellent results but with increased processing demands [71], [72]. Ray tracing volume data is also important for simulating many interesting phenomena like participating media (fog and smoke for example) [73]. In addition other naturally occurring material can only be simulated using ray tracing volume data like hair, fur, and other 3D textures; called hyper-textures [74], [75], [76].

## 5.5 Main Ray Tracer Components

As evident from the previous sections, the base ray tracing algorithm is conceptually simple. Unfortunately the complexity of a ray tracer as a software system is increased by some factors. The first factor is the amount of information required to model the environment. Many objects are required to interact in a unified and well orchestrated framework to accurately describe the simulated environment. The other factor is the large amount of optimizations required to increase the speed of a typical ray tracer. Optimizations like acceleration schemes, helping data structures for efficient ray-object intersection, and low-level code optimizations are common in a typical implementation. Thus a ray tracer must be written using a well structured programming language in accordance to the Object Oriented Programming (OOP) discipline. The following sections introduce the OOP architecture and design of the base ray tracer used in this work. This base ray tracer is enhanced later by GA and GMac to illustrate their power in such type of application.

The base ray tracer is written in C# using the Visual Studio 2008 IDE. C# was selected because of its excellent OOP capabilities and integration with other languages like C++ through the .NET framework CLI technology. The ray tracer is mainly based on the C++ implementation of [41]. The implementation is enhanced and ported to C# manually. The resulting system is a good testbed that is capable of handling many ray tracing tasks. In addition its good OOP design enables its extension to be used for testing any ray tracing related techniques with minimum integration effort and maximum code reuse.

**Table 5-1: Namespaces of base ray tracer**

| Namespace | Main Purpose |
|---|---|
| `Acceleration` | Contains classes used for all ray-object intersection acceleration schemes like regular grids, kd-trees, BIHs and Octrees. |
| `BRDFs` | Contains classes used for BRDFs and BTDFs as described in [41]. |
| `Factories` | Contains factory classes for creating and initializing main ray tracer components. |
| `Geometry` | The largest namespace. It contains all classes related to objects and their geometric description used to define the environment. |
| `Lights` | Contains classes related to the light sources, their models and parameters as described in [41]. |
| `Materials` | Contains classes related to materials used to model surface properties of objects as described in [41]. |

**Table 5-1 (continued): Namespaces of base ray tracer**

| | |
|---|---|
| `Sampling` | Is used mainly to implement many sampling techniques to reduce noise and aliasing artifacts in final image as described in [41]. |
| `Scenes` | All scenes are encoded manually inside the classes of this namespace. |
| `Textures` | Contains all classes related to2D and 3D textures used to model surface properties of objects as described in [41]. |
| `Tracing` | Contains classes to implement several versions of the base ray tracing algorithm |
| `Util` | Contains helper classes and data structures used to exchange information throughout the system. |
| `Viewing` | Contains classes related to camera models and viewing surface parameters. |

The ray tracer is composed of 12 namespaces. A namespace in C# is a logical grouping of several related classes similar to Java packages. The 12 namespaces are shown in Table 5-1. The following subsections introduce some implementation details for those namespaces.

### 5.5.1 The Util Namespace

This namespace contains the main utility classes and information exchange data structures on which the system is built. The main classes of this namespace are shown in Table 5-2.

Most of the classes in the `Util` namespace are direct ports of the implementation in [41]. The only exceptions are the `ShadingRecord` and `TextureCoordinates` classes. The `ShadingRecord` class contains information related to the current hit point. Such information include the hit point coordinates, a reference to an object of type `TextureCoordinates`, the ray parameter value at the hit point, the normal vector at the hit point, the material object containing material information for the hit point, and the depth of the ray in the recursion tree of the ray tracing algorithm.

**Table 5-2: Main classes in the Util namespace**

| Class | Purpose |
|---|---|
| `AffineTransform` | Represent matrix-based affine transforms for translation, rotation, and scaling of vectors, points and normal vectors in 3D space |
| `BoundingBox` | Represent an axis-aligned bounding box for efficient ray-object intersection calculations. |
| `Normal3D` | Represents a normal vector in 3D space. |
| `Point2D` | Represents a point in 2D space. |
| `Point3D` | Represents a point in 3D space. |
| `RandomGenerator` | A random number generator for the system. |
| `Ray` | Represents a single ray with defined origin point and direction vector in 3D space. |
| `RGBColor` | Represents a color in the RGB model. |
| `ShadingRecord` | Holds the shading information for the current hit point. |
| `TextureCoordinates` | Holds local 2D and 3D texture coordinates of the current hit point. |
| `Utility` | A static class that contains mathematical functions for many low-level ray tracing tasks. |
| `Vector3D` | Represents a directional vector in 3D space. |

### 5.5.2 The Geometry Namespace

The `Geometry` namespace is the most important namespace in the system. It contains the `GeometricObject` abstract class that is the main class for representing objects of the environment. The base implementation contains the geometric objects of types shown in Table 5-3.

Universal objects include spheres, planes, cylinders, disks, triangles and axis-aligned boxes. All universal objects can have any location or orientation in 3D space. All the classes implementing such objects are ported from [41] except for the universal cylinder implemented based on [77].

Generic objects are geometric objects with special fixed position and orientation in space. In order to create an object of such type with general position and orientation, instancing must be used as described in [41]. Such objects include full generic cones, spheres, cylinders, disks, planes, rectangles, tori, and partial objects of all of them.

**Table 5-3: Object types in the Geometry namespace**

| Object Type | Purpose |
|---|---|
| Generic | Geometric objects with fixed position and orientation. |
| Compound | A collection of geometric objects. |
| Universal | General purpose primitives with universal location and orientation. |
| Instance | A transformed instance of an object. Uses ray-transformation (Instancing) for hit-tests. |
| Triangles | Planar triangle objects. |
| Binary | Objects with surface defined using binary texture maps. |

In addition to the geometric primitives described above, several classes for ray-object intersection calculations are defined. Such classes are called solvers. Solvers provide a clean separation of the necessary geometric calculations needed for ray-object intersections. Such separation is not present in [41]. Such separation is a very good feature since the intersection algorithms can be maintained, optimized, changed, or replaced at any time without affecting the structure of the ray tracer. In addition, a single solver class can be used for several primitives without any change. For example a universal sphere and a generic sphere can use the same ray-sphere intersection algorithm thus avoiding code duplication in both primitives.

A geometric instance object contains a reference to a geometric object. It also contains an invertible affine transform applied to that object. In order to intersect the ray with the transformed object, the ray itself is transformed by the inverse of the affine transform of the instance object. The transformed ray is then intersected with the untransformed geometric object. The point of intersection and the normal are transformed by the affine transform to obtain the final hit point and normal vector. The whole process is described in detail in [41]. Generic, compound and instance objects are direct ports of the implementation of [41] with minor enhancements. The whole instancing process is illustrated in Figure 5-3.

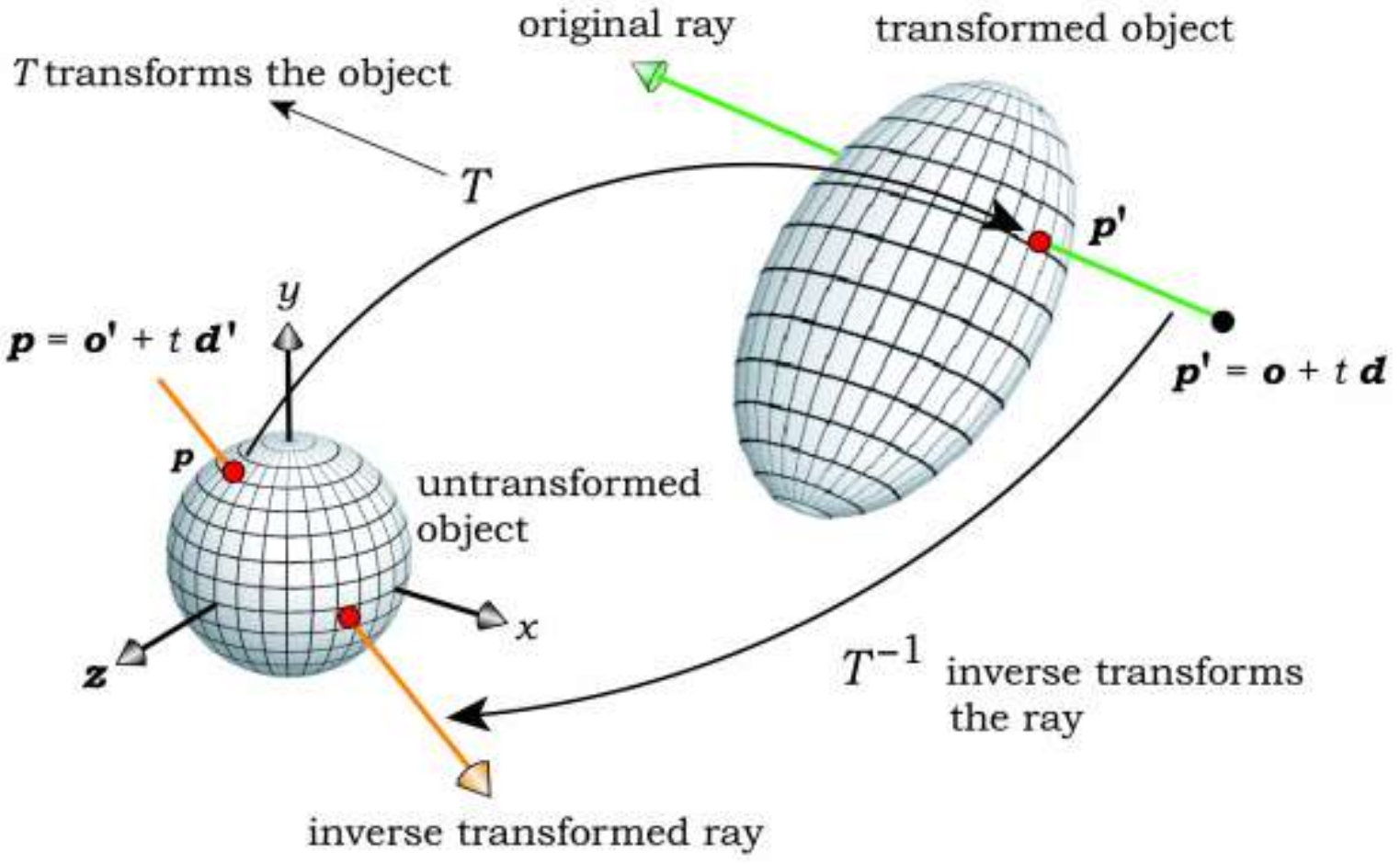

**Figure 5-3: Instancing Process [41]**

Binary objects are objects whose surface is definable by a base object plus a binary texture. For example a sphere with a mapped binary image texture having the earth map as an image with 1's for land pixels and 0's for sea pixels. Binary objects are not implemented in [41].

In addition to the `GeomatricObject` class, this namespace contains the `ObjectNature` base class. This class holds some important object characteristics like 2D texture mapping method, material information, and whether the object uses a 2D or 3D texture. This class can be inherited to implement other useful constructs like triangular meshes and point clouds. This class in not present in the implementation of [41].

The base ray tracer is also capable of constructing triangular meshes from PLY and STL files commonly found on the internet. The system can read 3D models stored in such files to construct vertex and normal information for the meshes and create triangles to be rendered in the second ray tracing stage.

### 5.5.3 The Acceleration Namespace

The only acceleration scheme implemented in [41] is the regular grid. Regular grids are not suitable for a wide spectrum of scenes. They can require a lot of memory storage and may not be suitable for scenes with large variations in object sizes. Other acceleration schemes exist that can perform better than regular grids. The `Acceleration` namespace enables the simple integration of such schemes. This namespace currently contains implementations for BIHs as described in [45] and regular grids as ported from [41]. BIHs are much better than kd-trees in their construction times while being much cheaper in their memory requirements than regular grids. All acceleration scheme classes inherit the `AccelerationScheme` abstract class that in turn inherits the `GeometricObject` class.

### 5.5.4 The Viewing Namespace

The `Viewing` namespace contains classes related to the viewing system. All camera model classes inherit the `Camera` abstract class. The base implementation includes classes for simulating the orthographic projection camera, the pinhole camera, and the thin-lens camera as implemented in [41].

In addition to defining camera models, this namespace contains the `ViewPlane` class. This class is used to define the characteristics of the final image like width, height, tone-mapping method, and output file name. The `ViewPlane` class is an enhancement to the corresponding implementation in [41]. The `ViewPlane` class is also capable of saving the final image to disk after applying tone mapping to the simulated radiances. The only tone mapping technique currently implemented is the basic trimming technique described in [41].

### 5.5.5 The Lights Namespace

All classes in the `Lights` namespace implement the `ILight` interface. Classes are provided for simulating ambient light, point lights, directional lights, environment light, and area lights. These are direct imports of the implementation of [41].

### 5.5.6 The Textures Namespace

The Textures namespace contains a redesigned, more flexible and maintainable version of the texture implementation presented in [41]. A texture is a mapping between 2D or 3D space and a set of values it can take. Four types of textures for binary, color, scalar, and vector textures can be defined. Binary textures take the values 0 or 1 according to some function defined in 2D or 3D space. Color textures (the most used type of textures) can take RGB color values. Scalar textures can take any real scalar value. Finally vector textures take 3D vectors as values.

Any type of texture can be defined on 3D space or within a 2D local texture space. Textures defined on 3D space implement the `IUVWMapped` interface while those defined on 2D texture coordinates implement the `IUVMapped` interface. Many types of mapping functions can be implemented as detailed in [41]. Such mappings include 2D checker textures, 2D image-based textures, 2D periodic textures, 2D and 3D noise based textures, and single color textures.

### 5.5.7 The Materials and BRDFs Namespaces

The `Materials` and `BRDFs` namespaces contain classes used for shading calculations as described in detail in [41]. The current implementation is a direct port of [41] with minor enhancements to the classes interfaces. A material class is a collection of BRDF and BTDF classes used to model the physical interactions between surfaces and light.

The classes are capable of handling many useful materials. Such materials include matte materials, Phong materials, reflective and glossy-reflective materials, emissive materials, simple transparent materials, and complex transparent dielectric materials. A material class implements the `IMaterial` interface. A BRDF class implements the `IBRDF` interface. Finally, a BTDF class implements the `IBTDF` interface.

### 5.5.8 The Sampling and Tracing Namespaces

In order to reduce aliasing artifacts and noise in the final generated image, a number of rays are used per pixel to deduce its final color. Many good sampling techniques exist to reduce such artifacts with minimum number of samples per pixel. The `Sampling` namespace contains classes for such techniques. The implementations of such classes is based on [41] but with some enhancements to their interface for ease of use. A sampling technique class implements the `ISamplingAlgorithm` interface. All sampling functionality is provided through the sealed class `Sampler`. Each instance of this class contains a reference to a single sampling technique selected in its initialization function.

The `Tracing` namespace contains several implementations for the basic ray tracing algorithm described in previous sections. Among these are implementations of the original Whitted ray tracer [41], [78] and another for a full path tracer [41]. The full path tracer can be used for simulating global illumination effects to produce more realistic results than the Whitted tracer. The main disadvantage of the full path tracer is its enormous processing requirements compared to the Whitted tracer. All tracer classes are derived from the `Tracer` abstract class.

### 5.5.9 The Scenes and Factories Namespaces

There are two methods for storing scene descriptions in ray tracers. The first is through external Scene Description Language (SDL) code files. The second is through scene description functions written in the language used to implement the ray tracer itself. The first approach is suitable for commercial ray tracers as it requires the programming of a special purpose DSL in the ray tracer itself. The other approach is much simpler and more suitable for research or educational purposes as it enables full access to all ray tracer capabilities without much effort. In addition, such approach frees the designer from writing a DSL interpreter to focus instead on ray tracing activities. This approach is used by [41] in the form of a single scene description function for each scene in the book. The base ray tracer of this work, however, uses a whole C# class for each scene. The class is inherited from the `SceneDescription` abstract class defined in the `Scenes` namespace. This approach has several advantages over the approach of [41]:

- To enable the definition of related scenes using a hierarchy of classes with maximum code reuse.
- To be used to generate animations using a sequence of ray traced images described as a series of calls to scene description classes. No discussion of such technique exists in [41].

Another namespace not present in [41] is the `Factories` namespace. This namespace contains C# static classes with methods to create and initialize important system components. Such components include camera models, geometric object definitions, BRDFs, BTDFs, light sources, and texture classes. It is a good OOP design practice to use such factory classes instead of the traditional constructor functions of each class.

# Chapter 6 : Geometric Algebra based Ray Tracer Enhancements

This chapter provides an illustration of some significant enhancements to the base ray tracer of the previous chapter. The enhancements are made to both the modeling and rendering stages. The main objective is to illustrate that GA-based code generation can greatly enhance the design and performance of such software when properly implemented. The use of a code generator like GMac to implement such enhancements has significant impact on the ease of design and maintainability of the proposed enhancements. The results of testing the performance of such enhancements will be illustrated in the next chapter. In addition, the application of geometric algebra to the problem of modeling 3D curves and surfaces is presented. The methods are based on using CGA versors to model shape in Euclidean space.

Section 6.1 provides a brief introduction to shape modeling techniques and a discussion of some of the shape modeling capabilities of the conformal geometric algebra. Sections 6.2 and 6.3 illustrate the application of GA to geometrically generate parametric representations for 3D Euclidean curves and surfaces. Section 6.4 discusses implementation details of geometrically generate parametric representations for 3D curves and surfaces through the Twister library. Section 6.5 offers a description of a generalized pinhole camera model made possible by the Twister library. Section 6.6 illustrates some enhancements to ray-object intersection algorithms.

## 6.1 Introduction to Shape Representation

Shape can be defined as the total of all information that is invariant under translations, rotations, and isotropic rescaling. Thus two objects can be said to have the same shape if they are similar in the sense of Euclidean geometry [79]. One of the main problems in computer graphics is constructing shapes of geometric objects using suitable mathematical representations. This field of research is called Computer Aided Geometric Design (CAGD) [80], [81]. In CAGD applications many mathematical representations for shape are used. Such representations include boundary representations (B-Rep) [82], Constructive Solid Geometry (CSG) [83], Euler operators [84], generative modeling [84], [85], parametric representations [86], [87], implicit representations [88], decomposition schemes [82], point-based methods [70], and volume modeling [82]. Each one of such representations has its own advantages, disadvantages and suitable fields of application. The powerful mathematical modeling characteristics of geometric algebra enable its effective contribution to such representations. In this chapter, such contribution will be illustrated through focusing on the generative modeling of parametric curves and surfaces. The generative modeling is built using geometric operations on points in 3D Euclidean space. The mathematical treatment will be performed through the 5D Conformal Geometric Algebra (CGA) that effectively represents 3D Euclidean space.

Parametric representations of shape were chosen over other common methods like implicit representations for several reasons:

- A parametric representation can be efficiently used to generate any number of points belonging to the represented shape in contrast to implicit representations where such process is hard to accomplish.
- Most parametric representations can be converted to efficient polygonal meshes using techniques such as [89] and [90].
- Many parametric representations exist for any single geometric shape [91]. Such different parameterizations are useful for holding more information about the shape like material and texture.
- As will be shown in this chapter, geometric algebra is a natural mathematical tool for generating parametric representations of surfaces and curves. The generation process is based on geometrically intuitive operations defined using GA operations on multivectors.

The approach of this work is to generate parametric representations using geometric operations on Euclidean points. Hence, this representation will be called Geometrically Generated Parametric Representation (GGPR) of curves and surfaces. The illustration given in this chapter is by no means exhaustive. GA can play an equally important role in the creation of other shape representations. Such role requires more mathematical investigations beyond the scope of this work.

CGA is one of the most applied geometric algebras in current literature. The main reason behind its wide application is its powerful representations of Euclidean space entities and transformations. For 3D Euclidean space the 5D CGA is used. Blades of the 5D CGA can effectively represent 3D Euclidean points, free direction vectors, normal vectors (through 3D Euclidean bivectors), planes, spheres, and lines. Through the meet and join operators of GA, the 5D CGA can also represent intersections and unions of such objects. A Circle can be represented as the blade resulting from the meet of a plane and a sphere or as the join of 3 points. A point-pair is the meet of a line and a sphere or the join of 2 points. Similar constructions result in blade representations for tangent vectors and tangent bivectors [11].

For CAGD applications, such unifying representations are certainly beneficial but not sufficient. For example there are no blades in 5D CGA that can represent simple geometric entities like a cylinder, torus or triangle. Unfortunately, most GA research in this point mainly focuses on CGA blades with very limited regard to the powerful CGA versors as geometric modeling operators. Versors are the other major type of multivectors in GA. They are typically used to perform orthogonal transformations on multivectors, including blades and versors. The geometric interpretation and use of 5D CGA versors in current GA literature is limited to using versors to define Euclidean motion operators [11]. Versors are seldom considered as powerful means for defining the geometric objects themselves.

A very distinctive exception is the so called Twist Representation of Shape presented in [10], [91], [92], and [93]. The twist representation relies on 5D CGA versors for representing 3D Euclidean curves and surfaces. The main motivation behind the twist representation was to solve pose estimation problems in computer vision as presented in [94] and [95]. The author of [91] points out that the twist representation can be successfully used in computer graphics applications. In [91] it is stated that this form of representation can be expected to have a great impact on both theory and practice in computer vision, computer graphics, and modeling of mechanisms. Unfortunately no other GA research is present to follow this important idea for computer graphics applications.

The terminology used in [91] is more related to group theory than to geometric algebra. The main principle of this chapter is nonetheless the same. The idea is to represent 3D curves and surfaces as the orbit of a point in 3D Euclidean space moving under the action of parameterized combinations of Euclidean versors (translation, rotation, and twist versors). The whole process is modeled within the 5D conformal geometric algebra. This chapter is a first attempt at illustrating the power of using 5D CGA versors and blades to represent shape in 3D Euclidean space for computer graphics applications. In addition, the treatment given in this chapter in GA terminology can be extended in future work to merge traditional continuous geometry with fractal geometry by allowing reflection versors in the model. Such unification can be used to study and model physical phenomena beyond Euclidean shapes. In addition, such extensions can be very useful in modeling naturally occurring phenomena related to fractal geometry in computer graphics using the same GA framework.

## 6.2 Geometrically Generated Parametric Curves

In the following subsections, details of the proposed GGPR for curves are presented. A mathematical definition is given followed by some example curves. Finally, a method for rendering such GGPR curves using ray tracing is described.

### 6.2.1 Mathematical Definition

A GGPR parametric curve is the image of a mapping $U$ between the parameter space $T = [t_1, t_2] \subset \mathbb{R}$ and 3D Euclidean space:

$$U : T \to E^3 \tag{6.1}$$

In this work, the mapping is to be defined as the orbit of a 3D Euclidean point $p_E = p_1 e_1 + p_2 e_2 + p_3 e_3$ moving under the influence of geometric operations in 3D Euclidean space having basis $e_1, e_2, e_3$. The geometric operations are applied through parameterized versors $V(t)$ of the impeding 5D CGA with basis $e_o, e_1, e_2, e_3, e_\infty$:

$$\begin{aligned} U(t) &= V(t) p V(t)^{-1}, \\ p &= e_o + p_E + \tfrac{1}{2} p_E^2 e_\infty \end{aligned} \tag{6.2}$$

The value of $U(t)$ is a normalized point in the 5D CGA. To extract the value of the equivalent position vector $u(t)$ in Euclidean space, the following relation is used:

$$\begin{aligned} U(t) &= e_o + u(t) + \tfrac{1}{2} u(t)^2 e_\infty, \\ u(t) &= u_1(t) e_1 + u_2(t) e_2 + u_3(t) e_3 \end{aligned} \tag{6.3}$$

Thus $u_1, u_2, u_3$ are the x, y, and z coordinates of the final transformed point for a given value of $t$ .

The allowed CGA versors are parameterized combinations of translation, rotation and twist versors as presented in chapter 3. The point $p$ will be called the seed point of the curve and is always constant with respect to $t$ . The values of the parameter interval limits $t_1, t_2$ are always finite real numbers usually in the range between 0 and 1.

One of the important quantities associated with 3D parametric curves is the tangent vector at a certain point $N(t)$. For differentiable parametric curves, the tangent vector can be obtained as the differentiation of $U$ with respect to $t$ :

$$N(t) = \frac{\partial U}{\partial t} = \lim_{h \to 0} \frac{U(t+h) - U(t)}{h} \tag{6.4}$$

As discussed in chapter 3, in 5D CGA $U(t)$ is a normalized Euclidean point. Thus the nominator difference is always a direction vector and the differentiation is well defined in 5D CGA.

### 6.2.2 Example Curves

A circle with radius $r$ and center $c = (c_1, c_2, c_3)$ in the x-y plane can be defined using the following geometric algorithm:

- Let $p = (c_1, c_2, c_3)$.
- Translate $p$ in the x-direction by r.
- Rotate $p$ in the x-y plane around the point $c$ by angle $2\pi t$ where $0 \le t < 1$.

All such steps can be directly defined using 5D CGA operations. This algorithm relies on a single constant versor; the translation versor of the second step. It also relies on a parameterized versor; the rotation versor of the third step. The composition of such two versors fully defines the GGPR of the circle. The net result is the traditional parametric equations of the circle parallel to the x-y plane:

$$\begin{aligned} u_1(t) &= c_1 + r\cos(2\pi t), \\ u_2(t) &= c_2 + r\sin(2\pi t), \\ u_3(t) &= c_3 \end{aligned} \tag{6.5}$$

A point of strength of such construction is immediately apparent. In order to define a circle in any general position it is enough to replace the unit vectors for the x,y, and z axis with any three orthogonal vectors. The algorithm remains the same in both its geometric and algebraic forms. Having a code generator like GMac can thus deduce the final parametric equations similar to (6.5) given the geometric algorithm. This is a huge simplification to the process of defining efficient computational representations for curves given a generation algorithm based on geometric operations.

A helix with constant radius $r$ and central axis being the z-axis can be geometrically defined as follows:

- Let $p = (0, 0, 0)$
- Translate $p$ in the x direction by $r$
- Rotate $p$ in the x-y plane by angle $2\pi kt$
- Translate $p$ in the z direction by $t$

Where $k$ is a real number representing a ratio that gives the pitch of the helix. This geometric algorithm relies on the composition of one constant versor (step 2) and two parameterized ones (steps 3 and 4). The final parametric equations are as follows:

$$\begin{aligned} u_1(t) &= r\cos(2\pi kt), \\ u_2(t) &= r\sin(2\pi kt), \\ u_3(t) &= t \end{aligned} \tag{6.6}$$

As in the circle case, any of the algorithm parameters can be changed without affecting the geometric or algebraic formulation of the algorithm. The radius and central axis can be given any valid values and the result can be directly obtained using a system like GMac.

The same geometric algorithm can be used to generate two different definitions for the curve. One definition may be having a constant value for the central axis while the other passes the central axis as a variable quantity. The code resulting from the first definition is more efficient while the later is more general. Thus it is up to the designer to select the most suitable generated definition according to the needs of the software system without making any significant changes to the GA algorithm itself.

### 6.2.3 Ray Tracing GGPR Curves

The probability of intersecting a ray with a curve in 3D space is typically zero. In order to use ray tracing to render a parametric curve, the curve must be given an artificial thickness. In addition, a method of finding the intersection point of a ray with the thickened curve must be found. The approach taken in this work is to approximate the curve with a number of line segments connecting the two endpoints of the curve and following its path as closely as possible. The line segments are then converted to cylinders of equal radius to give thickness to the curve. At each terminal point of every line segment a sphere with the same radius as the cylinders is defined. This construction ensures the correct rendering of the curve. It results in good lighting effects and can render parametric curves at any desired accuracy using simpler cylinder and sphere primitives. The linear approximation process is based on the procedure described in [90]. The procedure is called adaptive sampling with multiple random probing and is an extension of single random probing described in [96]. This procedure is easy to implement and has many degrees of freedom to help achieve a good sampling of the curve.

## 6.3 Geometrically Generated Parametric Surfaces

As in the case of GGPR curves, details of the proposed GGPR for 3D surfaces are presented in what follows. A mathematical definition is first given followed by some example surfaces. Finally, a method for rendering such GGPR surfaces using ray tracing is described.

### 6.3.1 Mathematical Definition

A GGPR parametric surface is the image of a mapping $S$ between the parameter space $T = [u_1, u_2] \times [v_1, v_2] \subset \mathbb{R}^2$ and 3D Euclidean space:

$$S : T \to E^3 \tag{6.7}$$

The mapping is to be defined as the orbit of a 3D Euclidean point $p_E = p_1 e_1 + p_2 e_2 + p_3 e_3$ moving under the influence of geometric operations in 3D Euclidean space with basis $e_1, e_2, e_3$. The geometric operations are applied through parameterized versors $V(u,v)$ of the impeding 5D CGA with basis $e_o, e_1, e_2, e_3, e_\infty$:

$$\begin{aligned} S(u,v) &= V(u,v)\, p\, V(u,v)^{-1}, \\ p &= e_o + p_E + \tfrac{1}{2} p_E^2 e_\infty \end{aligned} \tag{6.8}$$

The value of $S(u,v)$ is a normalized point in the 5D CGA. To extract the value of the equivalent position vector $s(u,v)$ in Euclidean space, the following relation is used:

$$\begin{aligned} S(u,v) &= e_o + s(u,v) + \tfrac{1}{2} s(u,v)^2 e_\infty, \\ s(u,v) &= s_1(u,v) e_1 + s_2(u,v) e_2 + s_3(u,v) e_3 \end{aligned} \tag{6.9}$$

Thus $s_1, s_2, s_3$ are the x, y, and z coordinates of the final transformed point for a given value of $u, v$ .

The allowed CGA versors are parameterized combinations of translation, rotation and twist versors as presented in chapter 3. The point $p$ will be called the seed point of the surface and is always constant with respect to $u, v$ . The values of the parameter interval limits $u_1, u_2, v_1, v_2$ are always finite real numbers and are usually in the range between 0 and 1.

One of the important quantities associated with 3D parametric surfaces is the normal vector at a certain point $N(u,v)$. An equivalent more suitable quantity in GA is the tangent bivector at such point $B(u,v)$. For differentiable parametric surfaces, the normal vector and tangent bivector can be obtained as follows:

$$\begin{aligned} B(u,v) &= \frac{\partial S}{\partial u} \wedge \frac{\partial S}{\partial v}, \\ N(u,v) &= B(u,v) \rfloor I_3^{-1}, \\ I_3 &= e_1 \wedge e_2 \wedge e_3 \end{aligned} \tag{6.10}$$

### 6.3.2 Example Surfaces

A sphere with radius $r$ and center $c = (c_1, c_2, c_3)$ can be defined using the following geometric algorithm:

- Let $p = (c_1, c_2, c_3)$.
- Translate $p$ in the z-direction by r.
- Rotate $p$ in the y-z plane around the point $c$ by angle $\pi u$ where $0 \le u < 1$.
- Rotate $p$ in the x-y plane around the point $c$ by angle $2\pi v$ where $0 \le v < 1$.

All such steps can be directly defined using 5D CGA operations. This algorithm relies on a single constant versor; the translation versor of the second step. It also relies on two parameterized versors; the rotation versors of the third and fourth steps. The composition of such three versors fully defines the GGPR of the sphere. The net result is the traditional parametric equations of the sphere:

$$\begin{aligned} s_1(u,v) &= c_1 + r\sin(\pi u)\sin(2\pi v) \\ s_2(u,v) &= c_2 + r\sin(\pi u)\cos(2\pi v) \\ s_3(u,v) &= c_3 + r\cos(\pi u) \end{aligned} \tag{6.11}$$

The same benefits obtained in the case of GGPR curves are present for GGPR surfaces. In addition, the mapping $s(u,v)$ can be directly used to map 2D texture and material information to the generated surface. This property is very useful for obtaining different texture mappings to the same geometric surface by using different generating algorithms. The following two definitions of a cylinder result in two different texture mapping for the same surface.

Assuming a circular cylinder with radius $r$ and unit height centered around the z axis. The first generating algorithm defines a parametric helix with constant radius $r$ and central axis being the z axis using the first parameter. The algorithm then rotates the helix around the z axis using the second parameter as follows:

- Let $p = (0,0,0)$.
- Translate $p$ in the x direction by $r$ .
- Rotate $p$ in the x-y plane by angle $2\pi ku$ .
- Translate $p$ in the z direction by $u$ where $0 \le u < 1$.
- Rotate $p$ in the x-y plane by angle $2\pi v$ where $0 \le v < 1$.

Where $k$ is a real number representing a ratio that gives the pitch of the helix.

The second algorithm creates a circle using rotation of a seed point using parameter $u$ . Then the algorithm translates the rotated point with a line segment constructed using the second parameter $v$ as follows:

- Let $p = (0,0,0)$.
- Translate $p$ in the x-direction by r.
- Rotate $p$ in the x-y plane around the origin by angle $2\pi u$ where $0 \le u < 1$.
- Translate the point $p$ using a translation versor in the direction of the y axis by $v$ where $0 \le v < 1$.

The results of the GGPR using the two algorithms are shown in Figure 6-1.

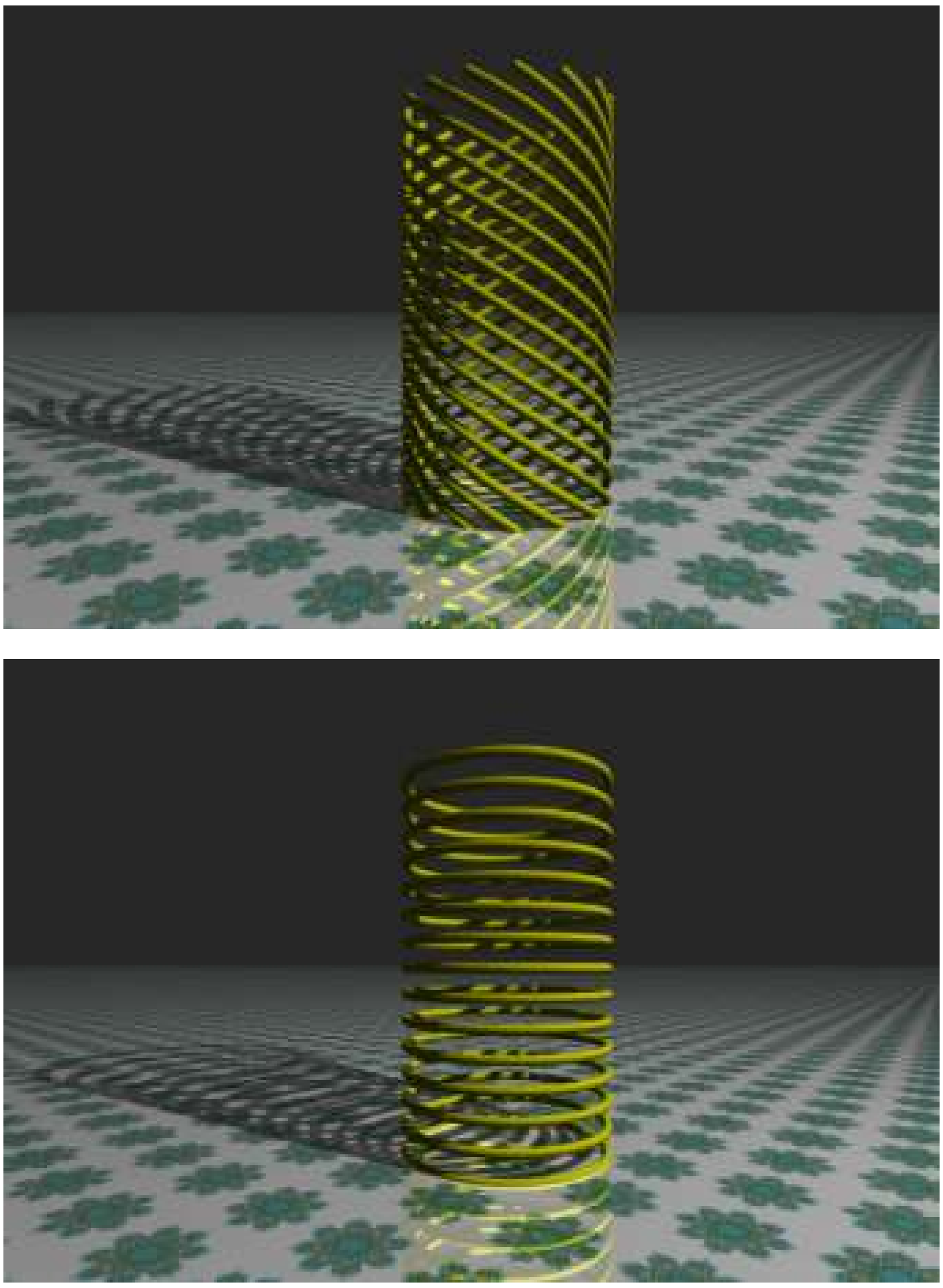

**Figure 6-1: Two Different GGPRs of a cylinder**

### 6.3.3 Ray Tracing GGPR Surfaces

Efficient direct ray tracing of general parametric surfaces is a difficult task. Many methods are present in current literature for performing such task [97], [98], [99]. The approach taken in this work is an indirect one. The GGPR surface is approximated by a set of connected triangles. The approximation process is based on the path-based adaptive polygonization algorithm described in [90]. The approximation algorithm combines adaptive curve sampling with simplicial subdivision. Complete edge sampling is used to avoid cracks. In addition area scanning is used to guide the subdivision process. This strategy allows better adaptation, trivially ensures global consistency, and produces meshes with an optimal number of polygons.

## 6.4 Enhancing Ray Tracer Object Modeling Capabilities

In the last two sections a method for geometrically generating parametric representations for curves and surfaces based on GA was presented. A small C# library, called the Twister library, was added to the base ray tracer of the last chapter to implement such GGPRs. This section discusses the Twister library and illustrates the various modeling enhancements it brought to the ray tracer.

### 6.4.1 Twister Library Architecture

The use of geometric algebra for solving a problem such as geometric modeling has a fascinating aspect. Geometric algebra enhances the accessibility of the designer to complex mathematical concepts and ideas. The main obstacle facing any good designer for such problems is the amount of mathematics the designer can handle and the available algorithms the designer can master. The more such tools are available to designers the more productive they become. Geometric algebra provides an excellent ground for increasing accessibility of designers to such concepts and algorithms. Although this is a rather subjective point of view, it is common among all researchers that have used GA for solving diverse problems. The design of the GGPR library was a proof for such argument.

The library defines some classes to represent simple 3D Euclidean objects like vectors, points, and bivectors to be transformed by twists. All such classes implement the `ITwistable` interface as shown in Figure 6-2. The library relies on the representation of twists as an abstract class, as shown in Figure 6-3, where a twist has the following characteristics:

- A twist is a software representation for a Euclidean transformation.
- Twists can be cascaded; a twist can be applied to the result of another twist.
- A twist can be directly applied to 3D points, vectors, and to other twists.

- A twist can be transformed by another twist including special functions for rotations around the three axes and translation along a vector.

Several types of twists were inherited from the abstract `Twist` class. The first type is the identity twist, represented by the `NullTwist` class, which is meant to represent a transform that produces an output as an exact copy of its input. The second type is a pure translation twist. A translation twist, the `TranslationTwist` class, has a translation vector $v_E$, a real translation coefficient $c_t$, and a real valued parameter mapping function $f_t(u,v)$ associated with it. When the translation twist is applied to a GA entity, the final translation vector $t_E$ used to construct the versor is created as follows:

$$t_E = c_t f_t(u,v) v_E \tag{6.12}$$

The third type of twist is a pure general rotation twist represented by the `RotationTwist` class. It consists of a rotation axis vector $r_E$, a rotation axis origin position vector $p_E$, a real rotation coefficient $c_r$ and a real valued parameter mapping function $f_r(u,v)$. The translation versor for the general rotation is constructed from the position vector $p_E$. The rotation angle and rotation bivector are constructed as follows:

$$\begin{aligned} \phi &= c_r f_r(u,v) \\ R_E &= p_E \rfloor I_3 \\ R &= R_E / \|R_E\| \end{aligned} \tag{6.13}$$

The fourth type is a full twist which is a cascade or composition of a pure translation and a pure general rotation where the translation axis and rotation axis are identical. Such type is represented by the `FullTwist` class.

The parameter mapping functions can be, but is not limited to, any of the following:

- $f(u,v):(u,v) \mapsto 0$
- $f(u,v):(u,v) \mapsto 1$
- $f(u,v):(u,v) \mapsto -1$

- $f(u,v):(u,v)\mapsto u$
- $f(u,v):(u,v)\mapsto v$

Twist based curves, with classes shown in Figure 6-4, use a parameter mapping function that is constant or only changes with $u$ . Twist based surfaces, with classes shown in Figure 6-5, can be constructed in two ways. The first is to use a single twist that is a function of both $u,v$ (the `SingleTwistSurfece` class). The second method is to use two curves depending on $u$ and connecting them with a line segment depending on $v$ (The `RolledTwistSurfece` class).

The code for applying all types of twists to points, vectors, and bivectors was generated using GMac. Although this code only represents about 10% of the architecture, it is the most important part of the code. The expressions of the code are too complex and low level to be written by hand, as seen in Figure 6-6, even when the equations are copied from a textbook. GMac and GA enable the rapid implementation of the Twister library with little work, few bugs, and immediate integration with the base ray tracer. This architecture for representing twists enabled the ray tracer of modeling many free form objects with very little effort. The next two subsections illustrate some of the capabilities of the library for modeling free form objects.

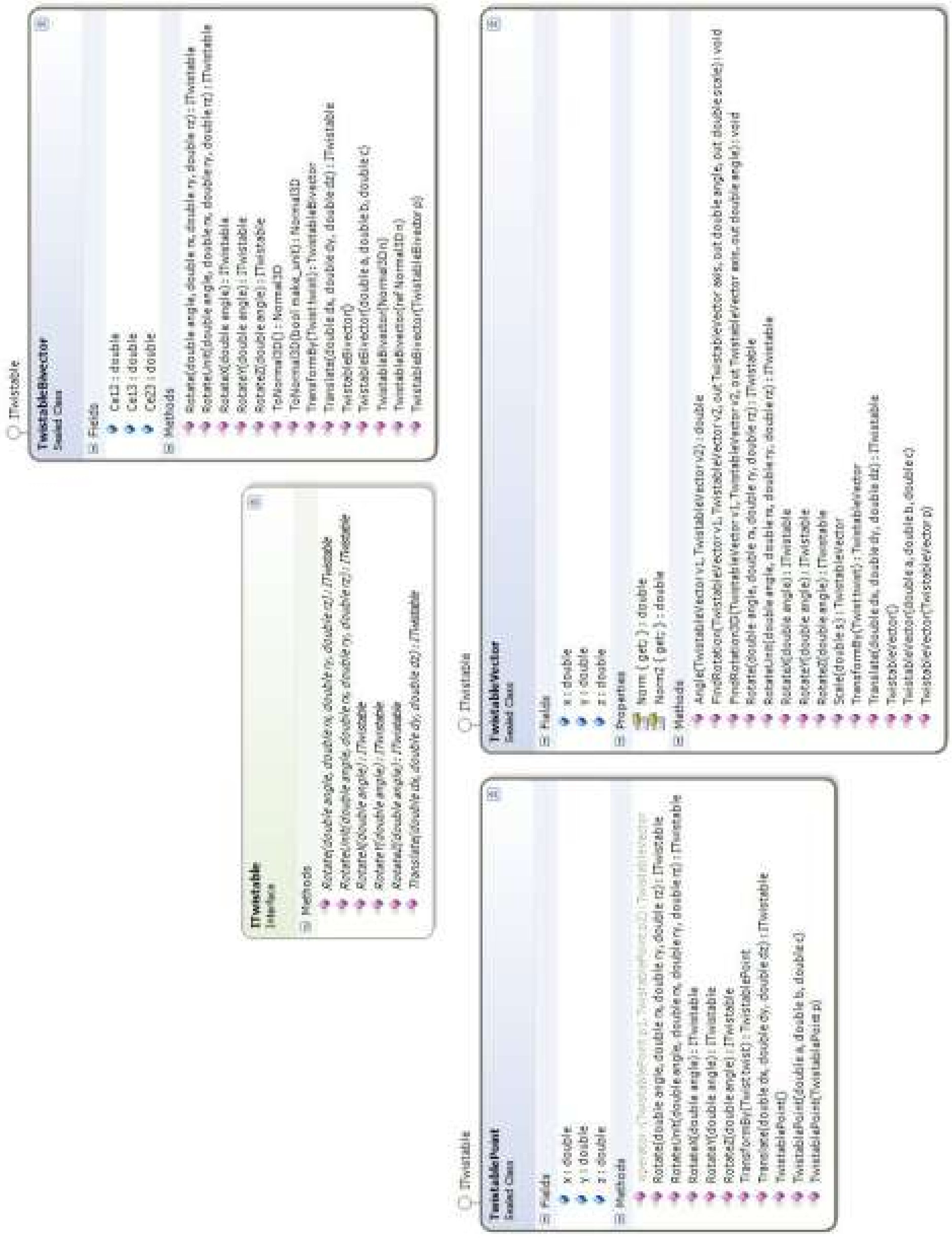

**Figure 6-2: ITwistable interface and classes**

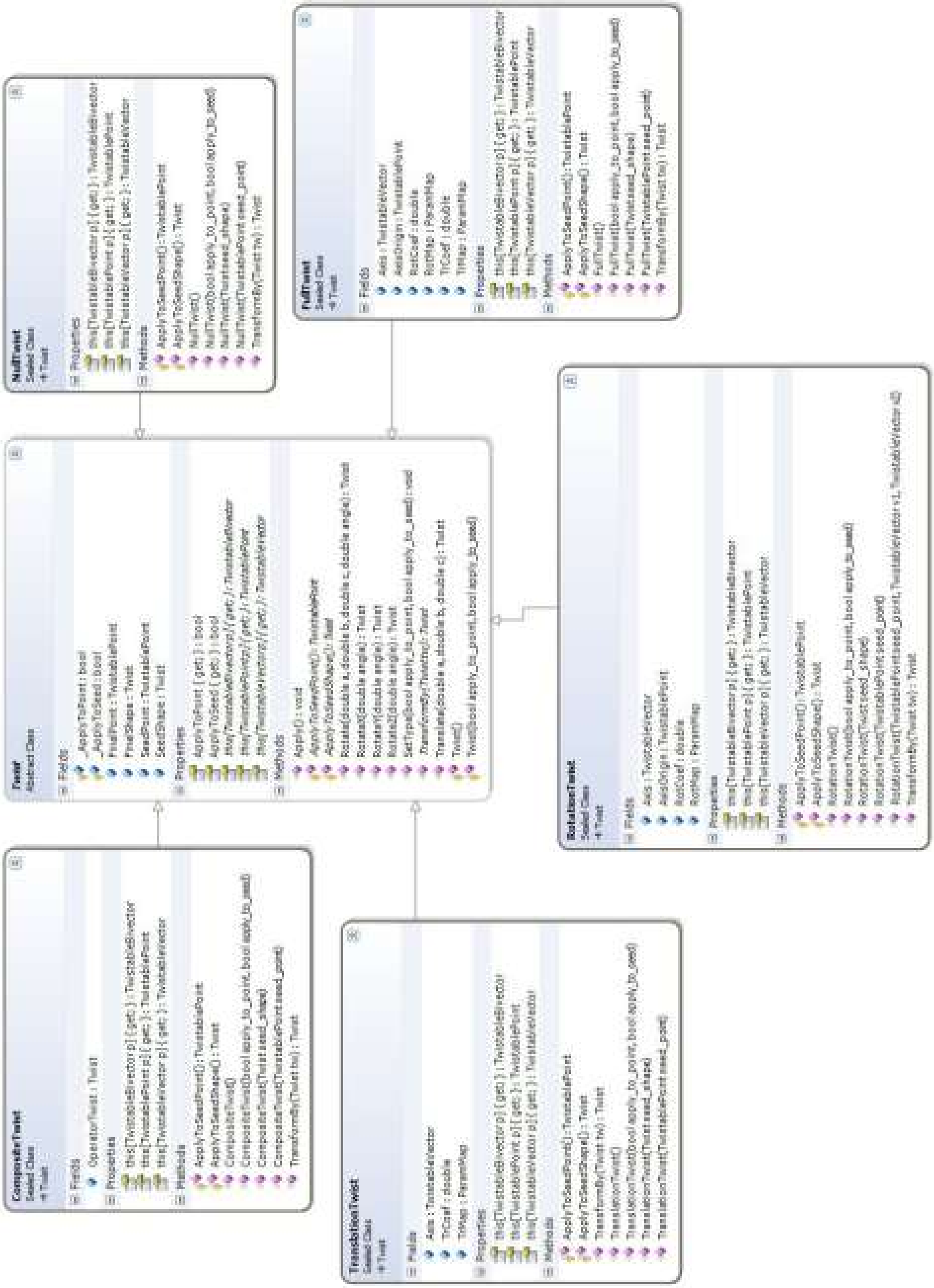

**Figure 6-3: The Twist base class and derived classes**

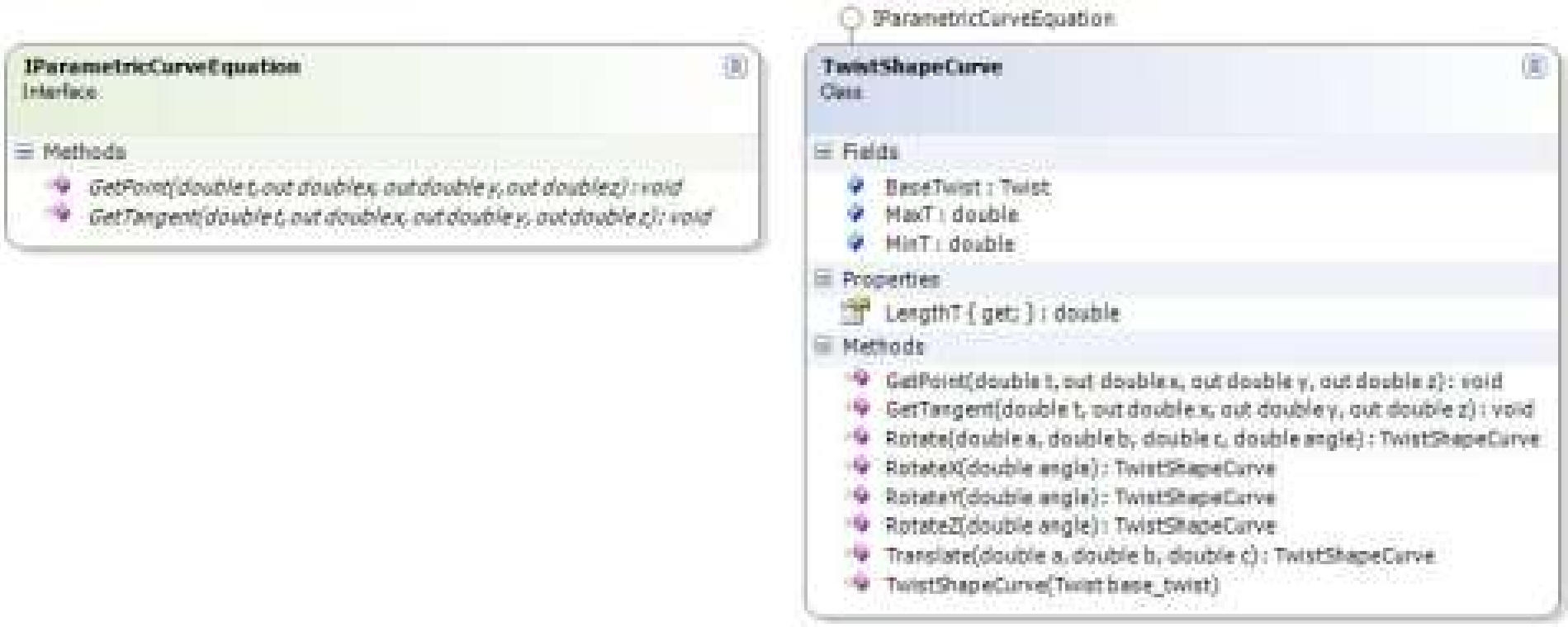

**Figure 6-4: Twist-based curve class**

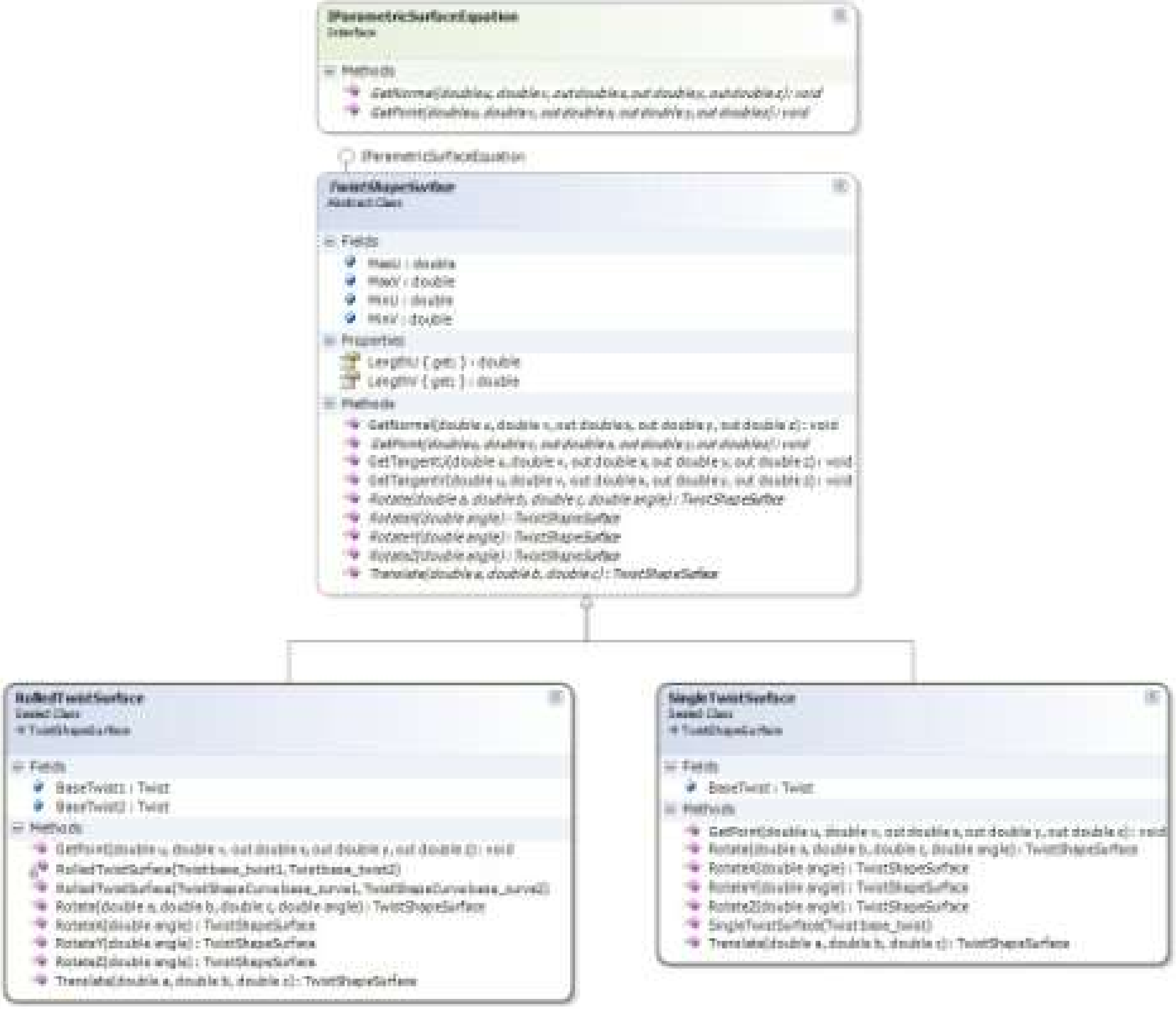

**Figure 6-5: Twist-based surface classes**

```
public override TwistablePoint this[TwistablePoint p]
{
    get
    {
        double x, y, z;
        double ct = TrMap() * TrCoef;
        double cr = RotMap() * RotCoef;

        #region GMac : ApplyTwistToPoint3D
        //GMac.Bind("rot_axis", "Vector3D", Axis);
        //GMac.Bind("tr_origin", "Vector3D", AxisOrigin);
        //GMac.Bind("angle.1", cr);
        //GMac.Bind("tr_param.1", ct);
        //GMac.Bind("in_mv", "Vector3D", p);
        //GMac.Bind("out_mv.e1", x);
        //GMac.Bind("out_mv.e2", y);
        //GMac.Bind("out_mv.e3", z);
        double var0085 = Math.Sqrt(Math.Pow(Axis.x, 2) + Math.Pow(Axis.y, 2)
            + Math.Pow(Axis.z, 2));
        double var0086 = -(Axis.z / var0085);
        double var0087 = Axis.y / var0085;
        double var0088 = -(Axis.x / var0085);
        double var0092 = var0086 * Math.Sin(cr / 2.0);
        double var0093 = var0087 * Math.Sin(cr / 2.0);
        double var0094 = var0088 * Math.Sin(cr / 2.0);
        double var0103 = p.x - AxisOrigin.x;
        double var0104 = p.y - AxisOrigin.y;
        double var0105 = p.z - AxisOrigin.z;
        double var0108 = -(var0092 * var0104) - var0093 * var0105
            + var0103 * Math.Cos(cr / 2.0);
        double var0109 = var0092 * var0103 - var0094 * var0105
            + var0104 * Math.Cos(cr / 2.0);
        double var0111 = var0093 * var0103 + var0094 * var0104
            + var0105 * Math.Cos(cr / 2.0);
        double var0114 = -(var0094 * var0103) + var0093 * var0104
            - var0092 * var0105;
        double var0127 = Math.Pow(var0092, 2) + Math.Pow(var0093, 2)
            + Math.Pow(var0094, 2) + Math.Pow(Math.Cos(cr / 2.0), 2);
        double var0131 = Math.Cos(cr / 2.0) / var0127;
        double var0132 = var0092 / var0127;
        double var0133 = var0093 / var0127;
        double var0134 = var0094 / var0127;
        double var0135 = var0092 * var0132 + var0093 * var0133 + var0094 * var0134
            + var0131 * Math.Cos(cr / 2.0);
        double var0136 = var0108 * var0131 - var0109 * var0132 - var0111 * var0133
            - var0114 * var0134;
        double var0137 = var0109 * var0131 + var0108 * var0132 + var0114 * var0133
            - var0111 * var0134;
        double var0139 = var0111 * var0131 - var0114 * var0132 + var0108 * var0133
            + var0109 * var0134;
        x = AxisOrigin.x * var0135 + Axis.x * ct * var0135 + var0136;
        y = AxisOrigin.y * var0135 + Axis.y * ct * var0135 + var0137;
        z = AxisOrigin.z * var0135 + Axis.z * ct * var0135 + var0139;
        #endregion

        return new TwistablePoint(x, y, z);
    }
}
```

**Figure 6-6: A GMac-generated function that applies a full twist to a 3D point**

### 6.4.2 Representation Capabilities for Curves

The modeling capabilities of the base ray tracer were greatly enhanced by the Twister library for curves. Initially, a parametric equation for the curve is required to perform the polygonal approximation algorithm used to render the curve. Obtaining the parametric equation is the problem that was solved by the Twister library. Many interesting curves can now be modeled without the designer referring to a book on curve equations to encode them by hand. The curves that can be generated by the Twister library include line segments, circles, several types of helixes, and all roulettes [100], [101] (trochoids, epitrochoids, and hypotrochoids). Any part or whole of such curves can be modeled and rendered with little effort. Some example curves are shown in the following figures.

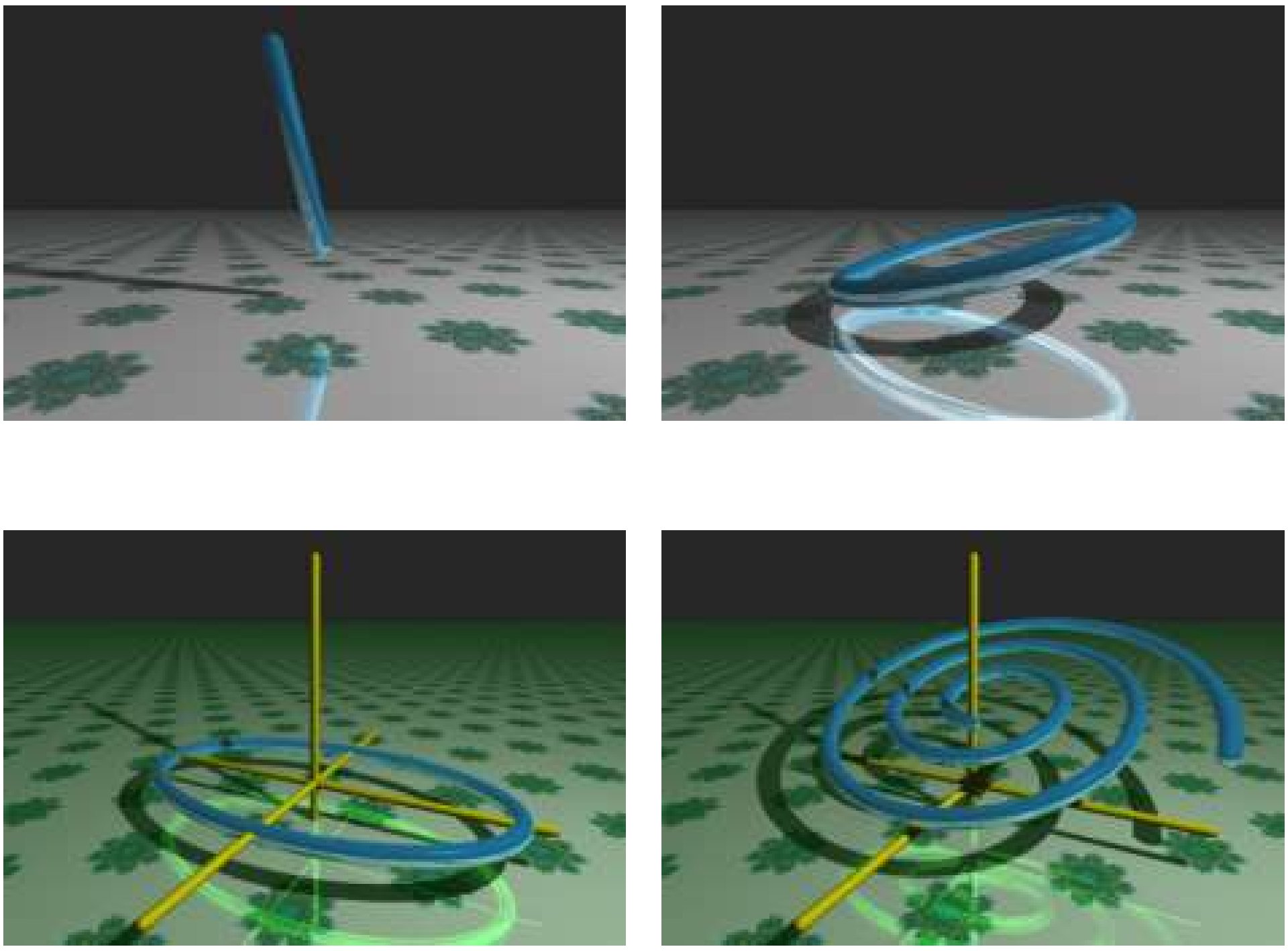

**Figure 6-7: Twister library curves**

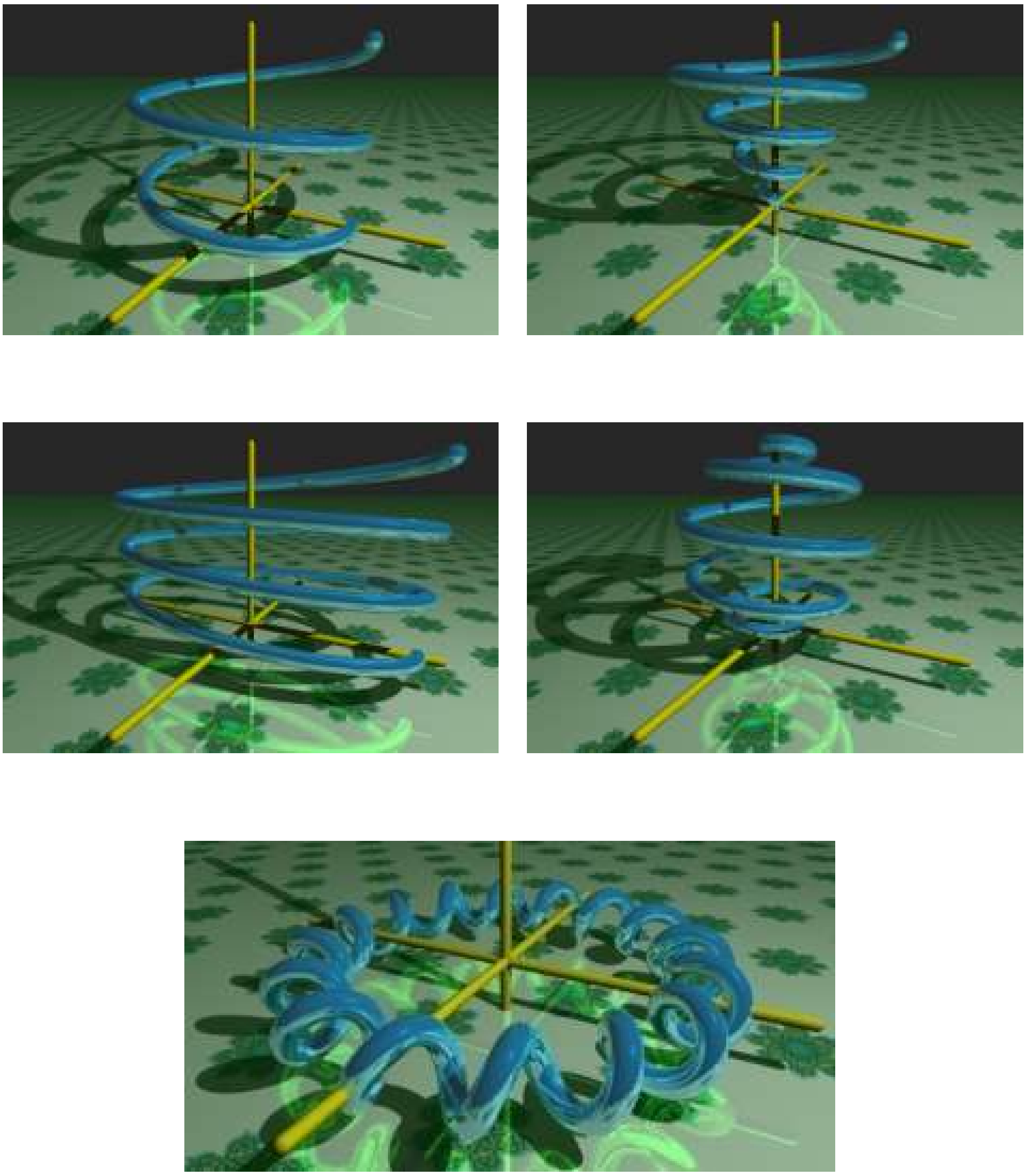

**Figure 6-8: Twister library curves (helixes)**

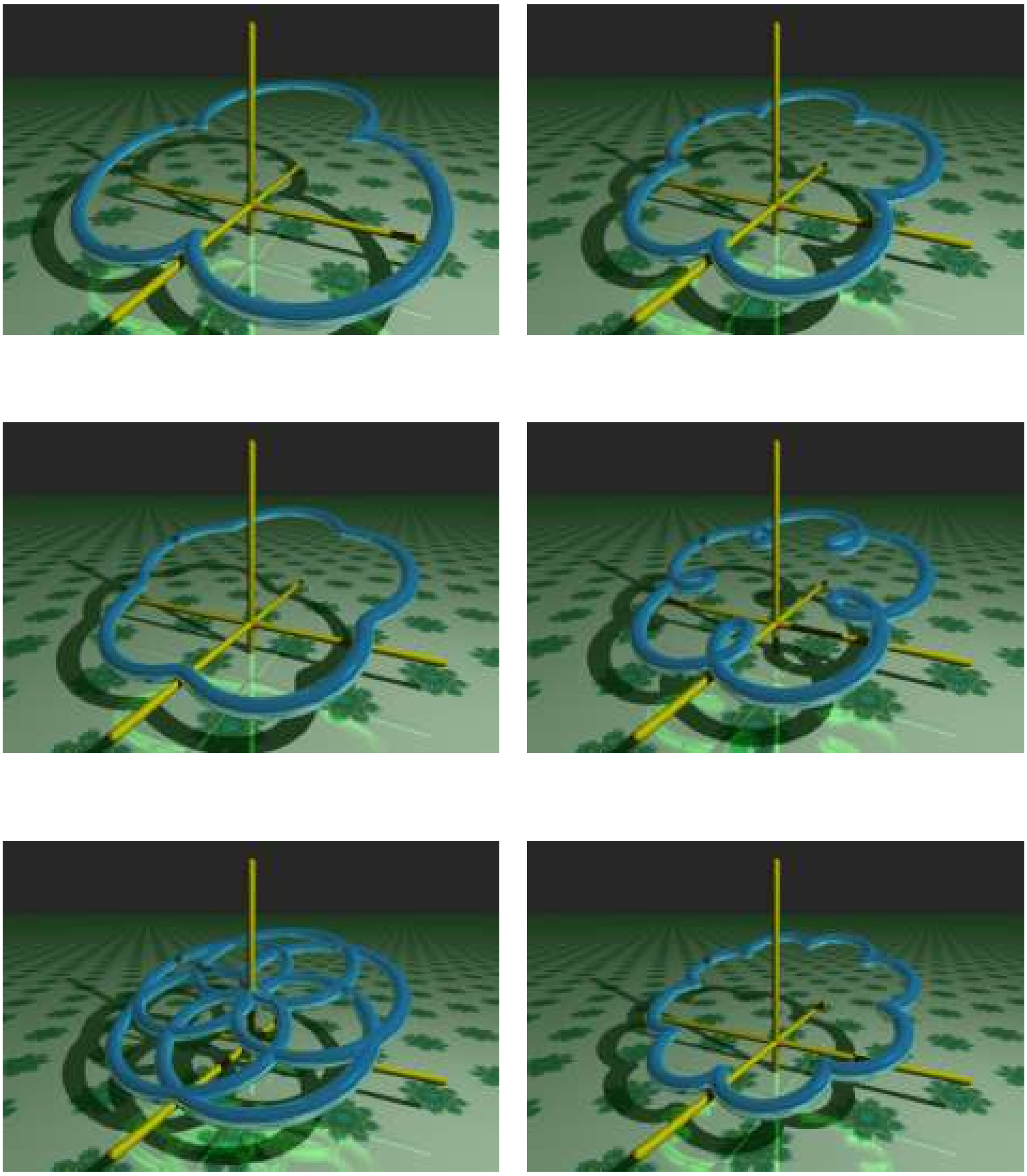

**Figure 6-9: Twister library curves (epitrochoids)**

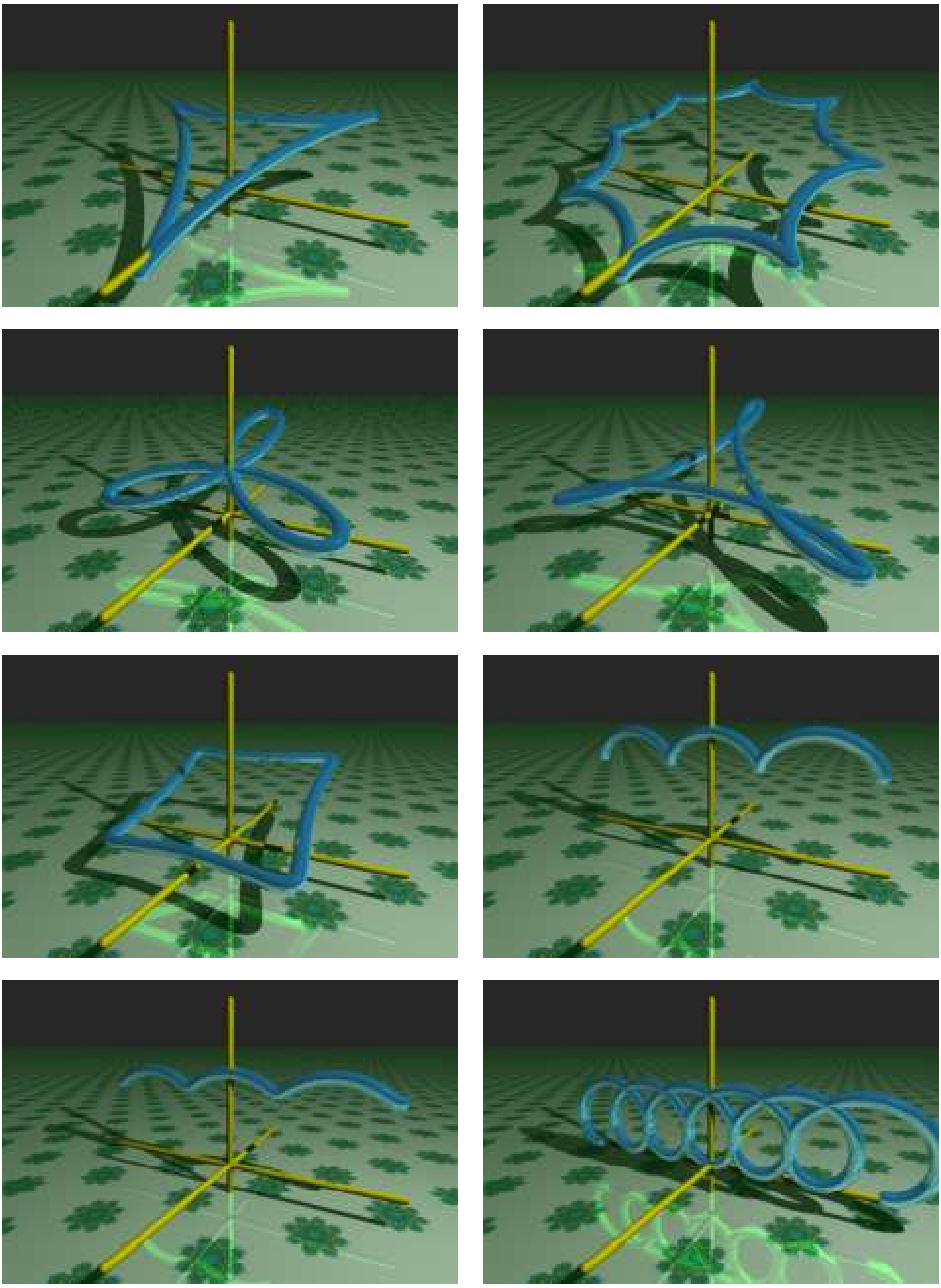

**Figure 6-10: Twister library curves (Hypotrochoids, roses, and trochoids)**

### 6.4.3 Representation Capabilities for Surfaces

As in the case of curves, The Twister library enabled the easy modeling of very complex 3D surfaces. The polygonal approximation algorithm of [96] was used to render the surfaces. The parametric equations were provided by the Twister library. Any part or whole of such surfaces can be modeled and rendered with little effort. Some example surfaces are shown in the following figures.

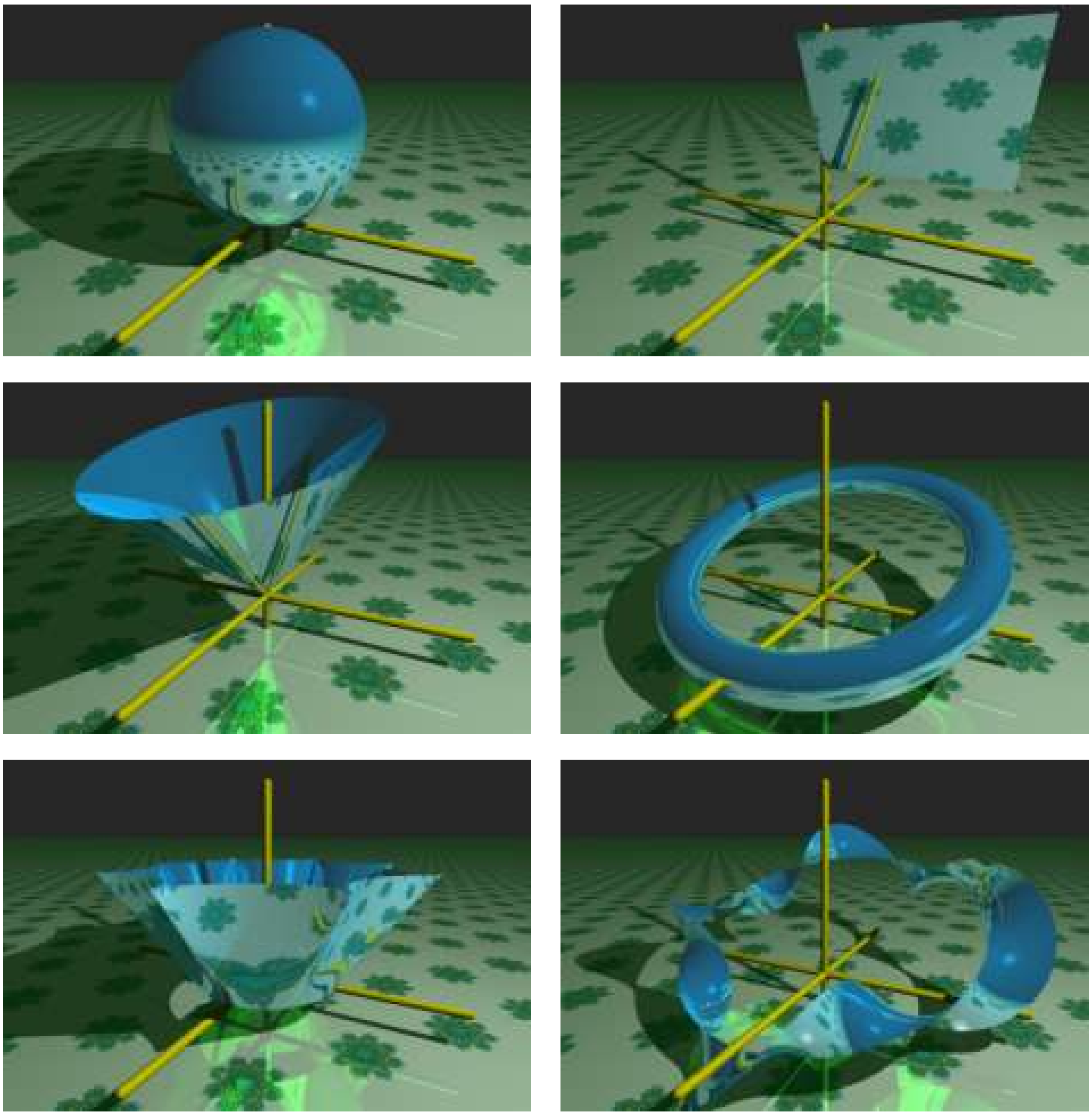

**Figure 6-11: Twister library surfaces**

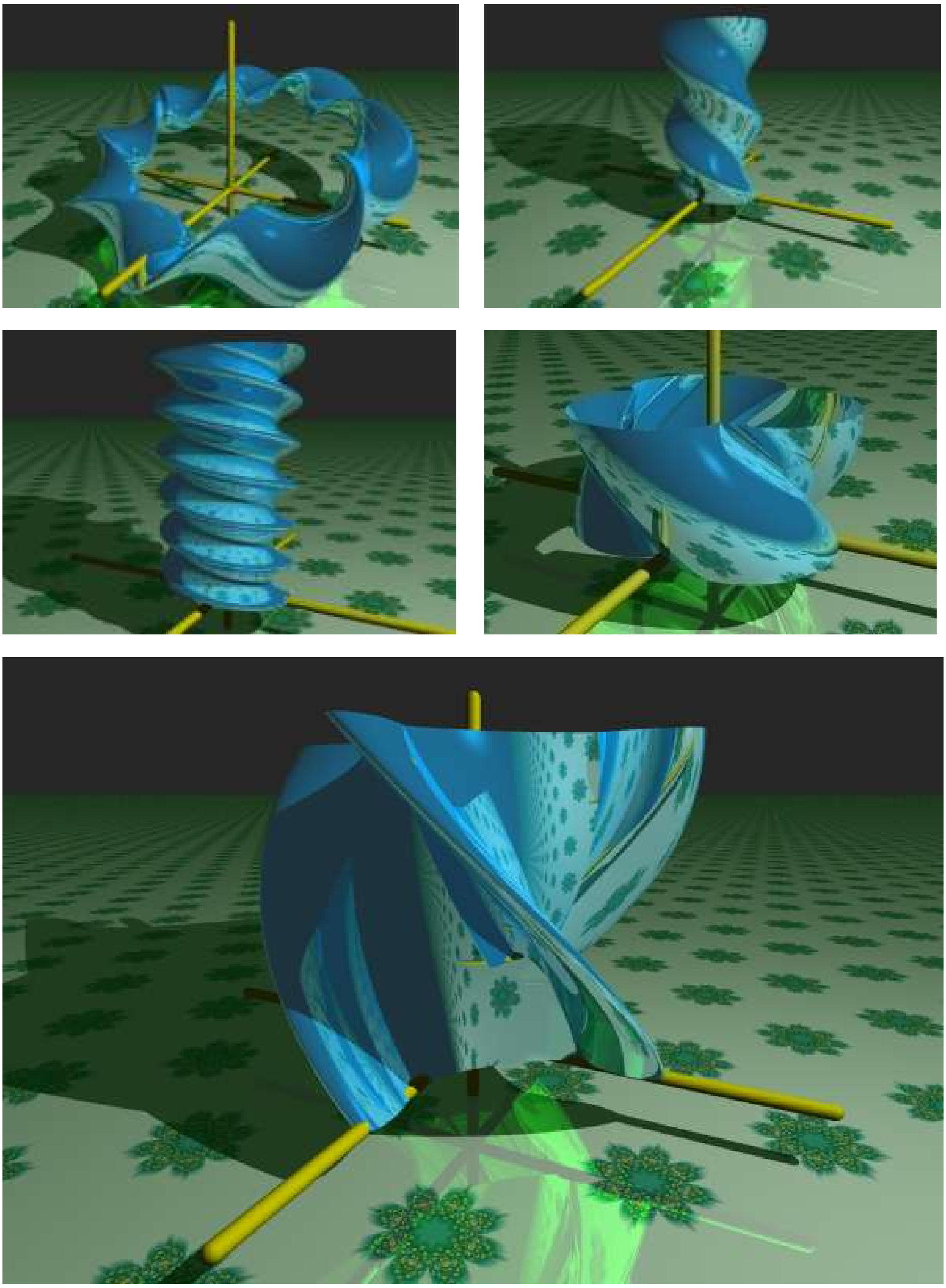

**Figure 6-12: Twister library surfaces (continued)**

## 6.5 Generalized Pinhole Camera Model

The pinhole camera model is one of the most used projection techniques in CG. One of the advantages of ray tracing compared to other rendering algorithms is its ability to simulate other more complex projection techniques with relative ease. Several other projection techniques are described in [41]. Such techniques include fisheye projection, spherical panoramic projection, and cylindrical panoramic projection. The treatment provided in [41] separately handles each technique and manually deduces its projection equations. GMac and the Twister library provide a more powerful alternative. This section describes the use of GMac and the Twister library to create a generalized projection model for ray tracing applications. The model is capable of simulating all projection techniques discussed in [41], in addition to many more, using the same unified model. No manual deduction of projection equations is required. Instead, the model is geometrically described and the projection equations are procedurally provided by the Twister library. This model shall be called the Generalized Pinhole Camera Model (GPCM).

### 6.5.1 GPCM Architecture

A GPCM consist of two main elements. The first element is the eye point $p_e$; a point representing the "pinhole" of the model. All primary rays traced through the model lie on lines passing through the eye point. The second element is a general 2D surface $S$ in 3D space. The surface is geometrically described like any other Twister library surface through a series of Euclidean versors acting on a seed point. A third auxiliary element is an arbitrary line segment with direction vector $v_k$ used for GPCM orientation in 3D space as will be illustrated in the following subsection. The whole model is illustrated in Figure 6-13.

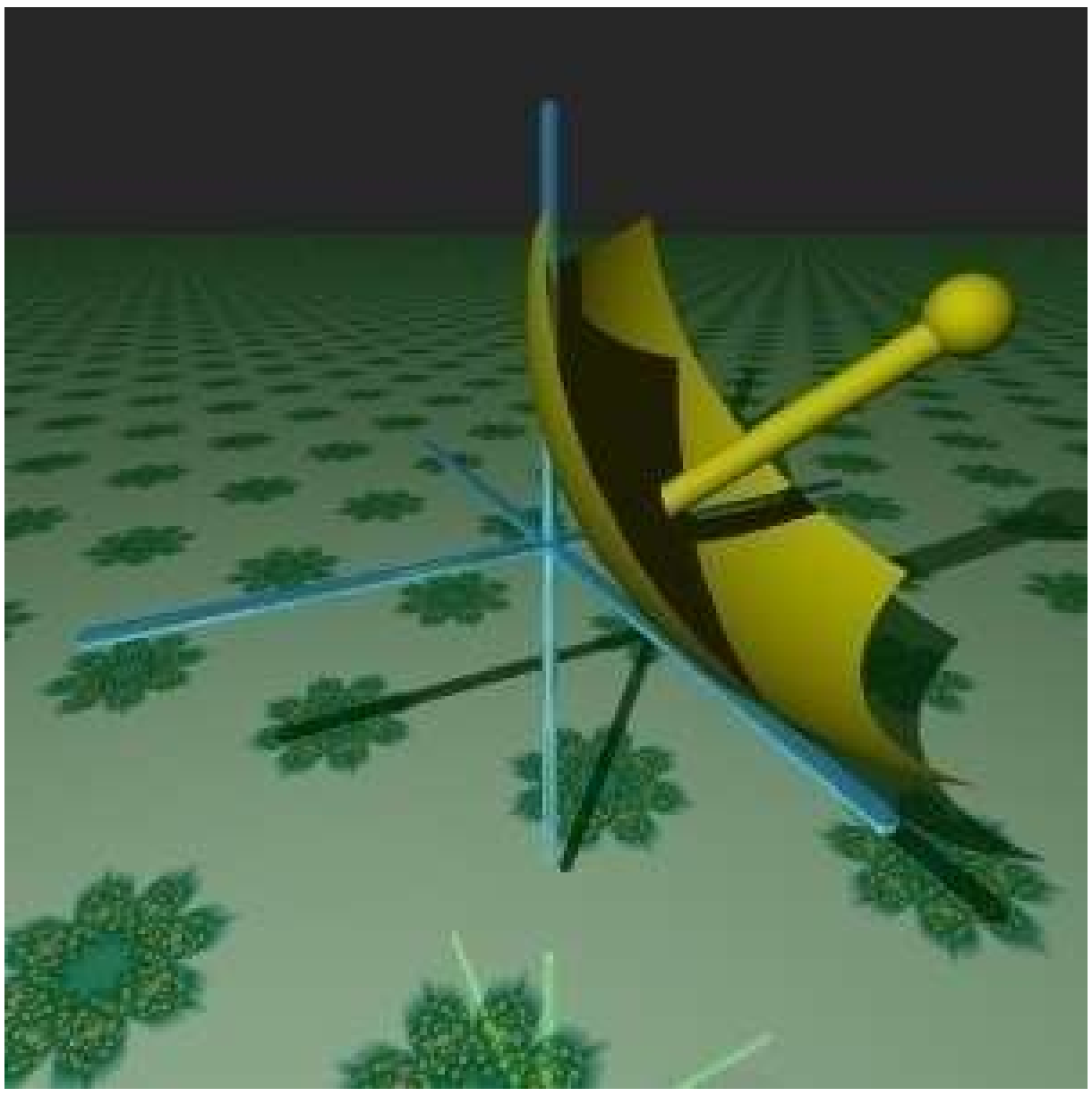

**Figure 6-13: Main elements of the GPCM**

### 6.5.2 Orienting the GPCM

Figure 6-14 illustrates one possible orientation procedure for the GPCM. In [41], the traditional pinhole camera model is oriented through the use of two points and a vector. The first point is the new position for the eye point $p'_e$. The second point is called the look-at point $p'_k$. The vector is called the up-direction $v_u$ and is used to simulate the up-down directions on earth. A look-at direction vector $v'_k = p'_k - p'_e$ is also used in the orientation process. The final projection equations in [41] are manually deduced depending on such elements using only vectors, inner products, and cross products. These same elements can be used for orienting the GPCM without any manual deduction of equations. Just simple geometric steps that are automatically executed by the Twister library and GMac generated code. This way, any other orientation procedure can be implemented with ease. In addition, any designer with good geometric intuition can implement, maintain, and enhance such procedures without the need for low-level algebraic manipulations on vectors.

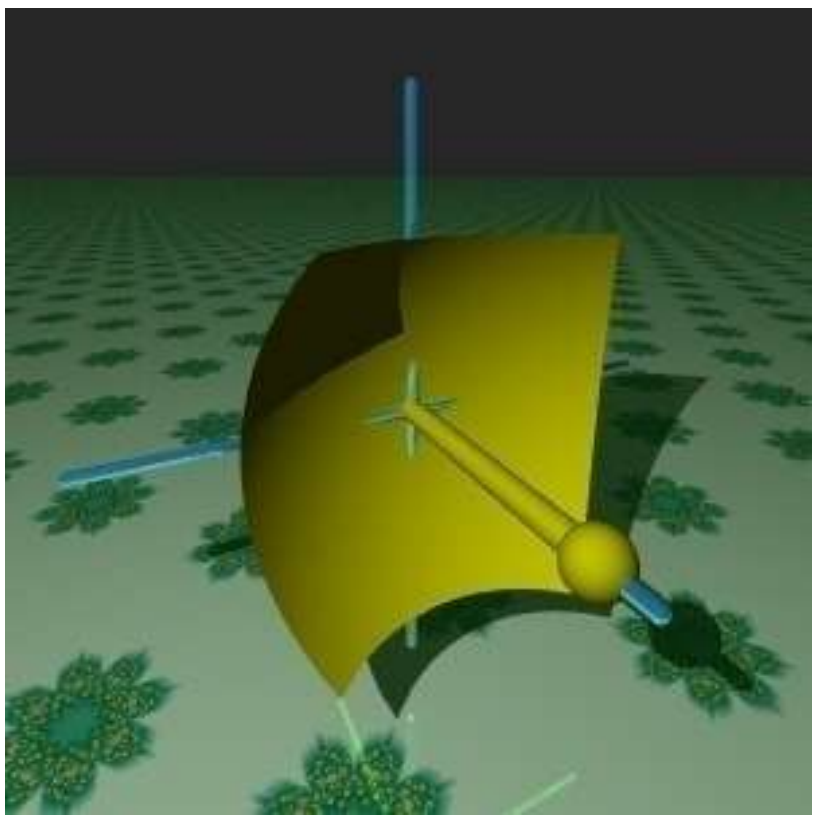

**(a) Initial standard position**

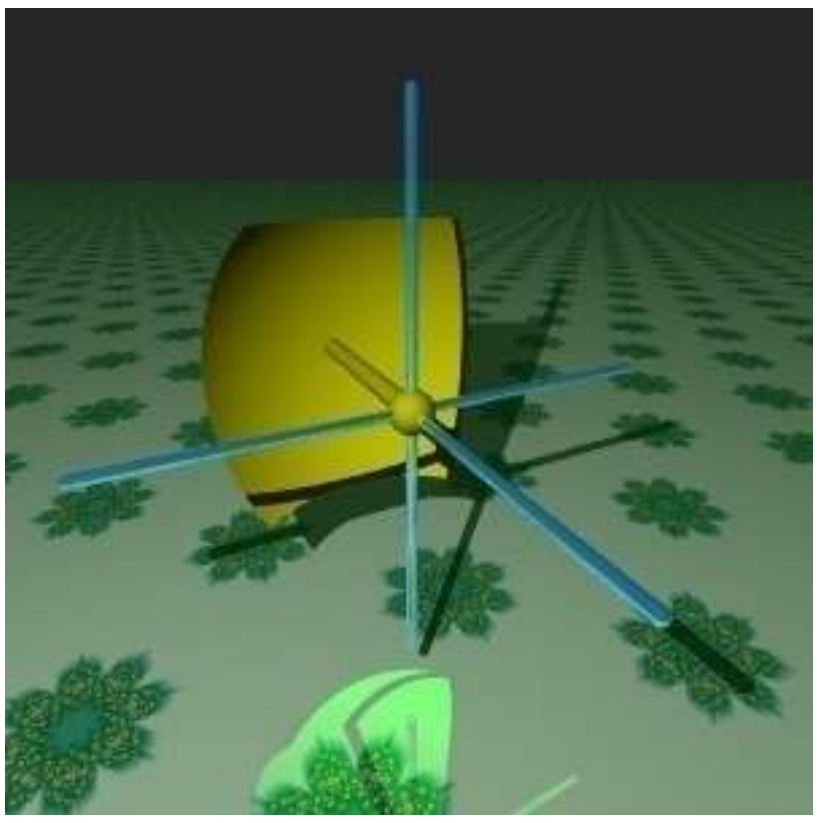

**(b) Translation of initial eye point to origin**

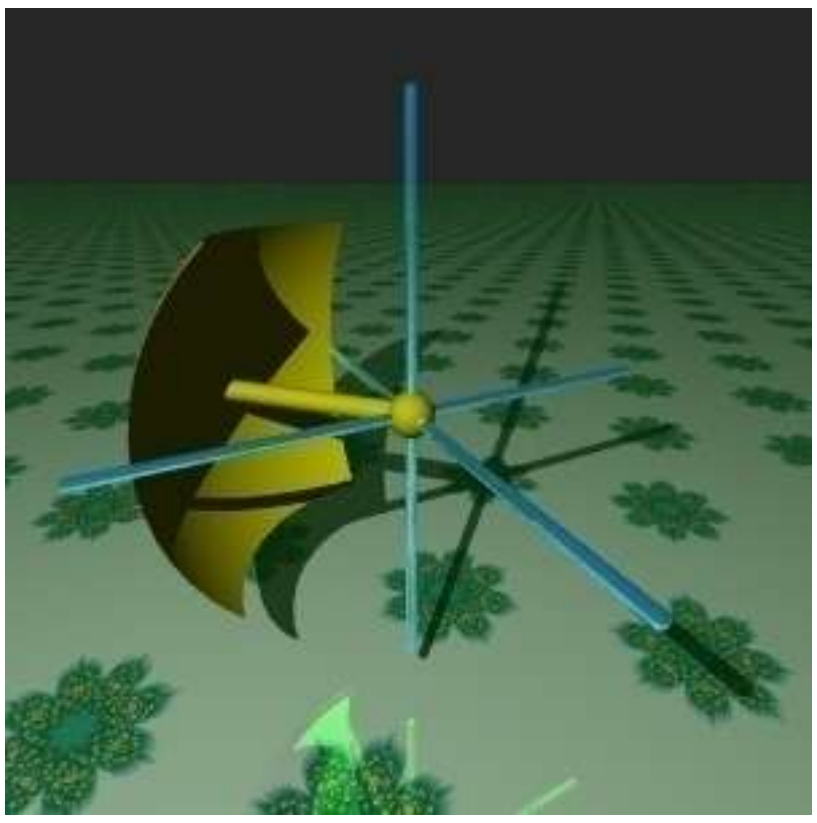

**(c) Rotation to Coincide with** $v_k''$

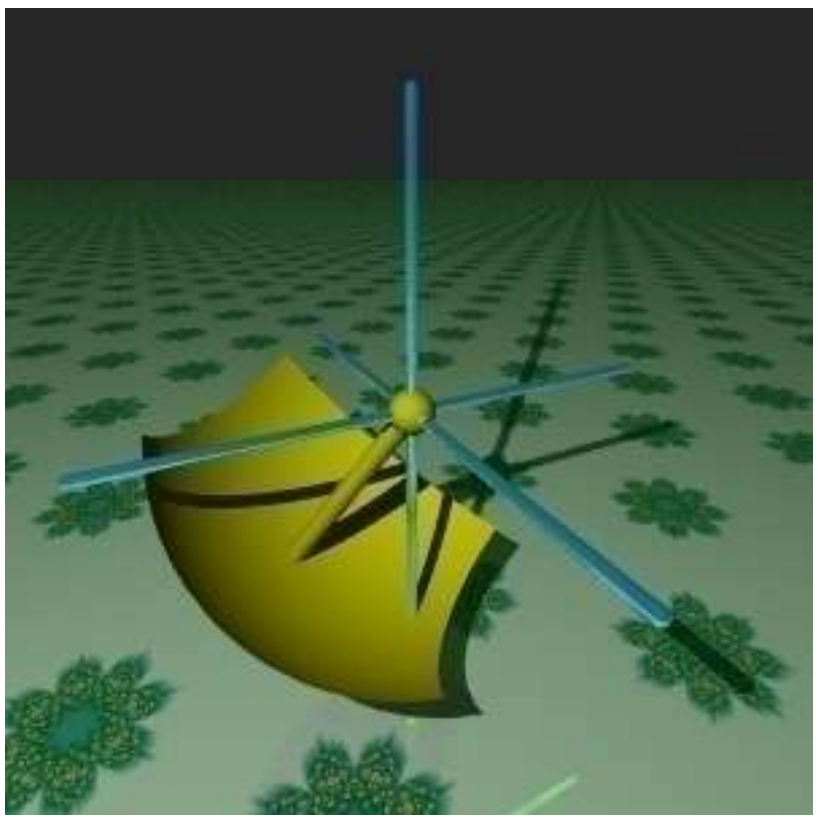

**(d) Rotation to Coincide with** $v_k'$

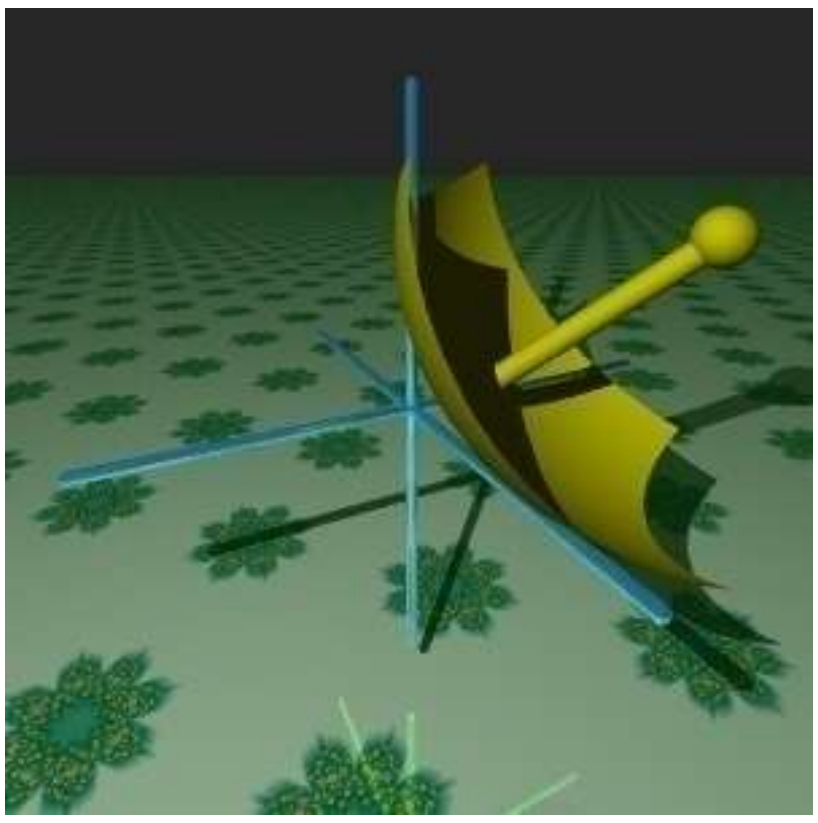

**(e) Translation to final eye point**

**Figure 6-14: Orientation procedure for the GPCM**

Assuming the up-direction is the y-axis for simplicity, the look-at direction $v'_k$ is projected on the xz-plane to a vector $v''_k$. The GPCM is initially constructed at a standard position and orientation as shown in Figure 6-14a. The eye GPCM is then translated so that the eye point is positioned at the origin. The third step is to rotate the GPCM so that the GPCM vector $v_k$ coincides with the projected look-at vector $v''_k$ as shown in Figure 6-14c. The fourth step is to rotate the GPCM so that the projected look-at vector $v''_k$ coincides with the final look-at direction $v'_k$ as shown in Figure 6-14d. The final step is to translate the GPCM so that the eye point is positioned at its final position at $p'_e$.

The Twister library is used for the construction of the projection surface $S$ and for all orientation steps of GPCM. GMac is used to find the suitable axis and angle of rotation that can rotate a vector to another vector as required in steps 3 and 4. The whole process is completely independent of $S$ and purely geometric in nature; in contrast with the treatment of [41].

### 6.5.3 Generation of Primary Rays

As discussed in chapter 5, a primary ray is a ray generated by the camera with origin point $p_r$ and direction vector $v_r$ to initiate a tracing cycle. Having an image plane with m by n pixels, there are two modes of parametric mappings that can be used in the GPCM library as shown in Figure 6-15. The projection surface parameters $u,v$ are always assumed to be in the range 0 to 1. For each image pixel, a sample is generated as discussed in chapter 5. The sample coordinates are converted to uv-coordinates using one of the selected mappings shown in Figure 6-15. Next, a corresponding point on the projection surface $p_s = S(u,v)$ in 3D space is procedurally calculated through the Twister library. Several options for constructing the new ray are thus available. First, a ray can have an origin point identical to the GPCM eye point $p_r = p'_e$ or to the surface point $p_r = p_s$. Second, the ray direction can be from the eye point to the surface point $v_r = p_s - p'_e$ or in the reverse direction $v_r = p'_e - p_s$. Thus, the scene can be rendered using a "cutting surface" when $p_r = p_s$ with rays going in either directions relative to the projection surface $S$. The polar uv-mapping is used for models like the fisheye camera model [41] while the rectangular uv-mapping is used for other types like spherical and cylindrical panoramic projections and traditional rectangular pinhole models.

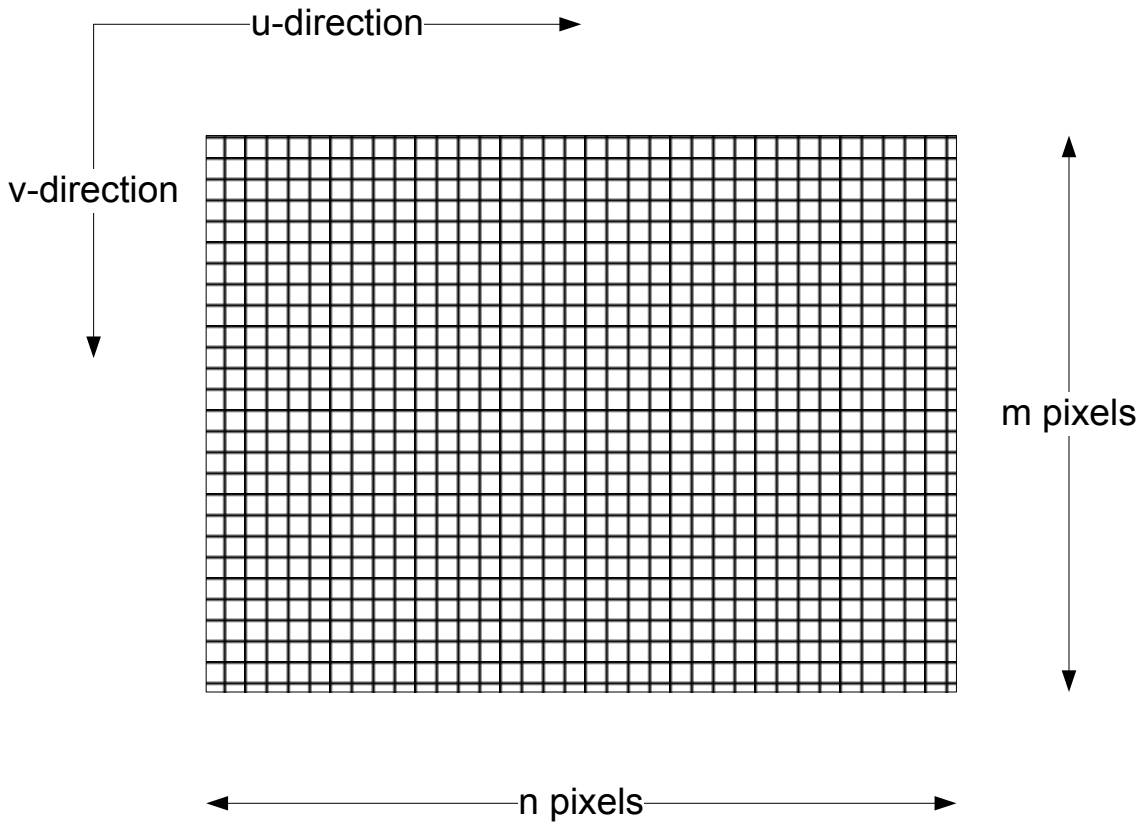

**(a) Rectangular uv-mapping**

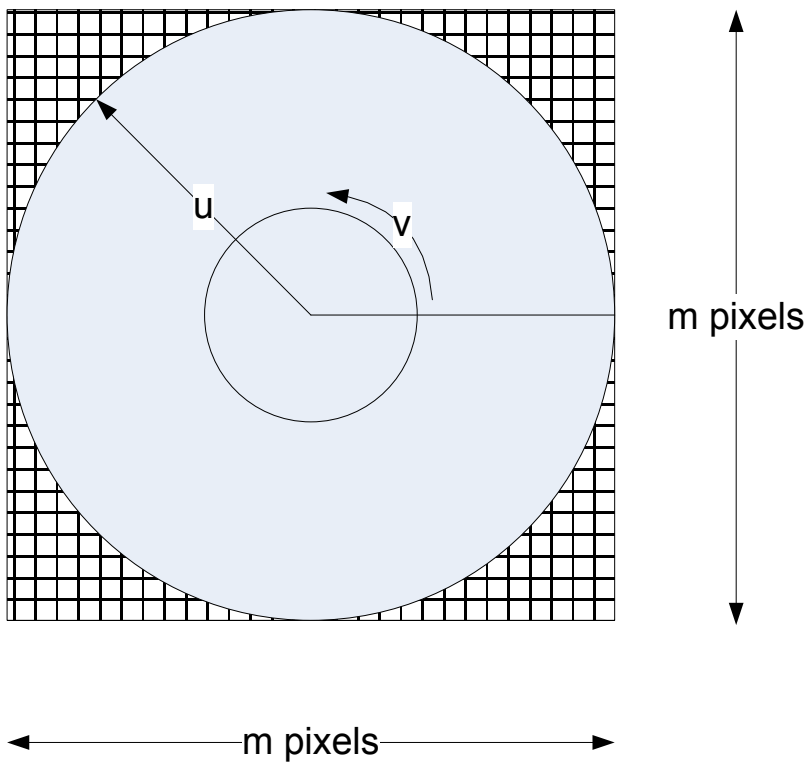

**(b) Polar uv-mapping**

**Figure 6-15: GPCM uv-mapping modes**

### 6.5.4 GPCM Modeling Capabilities

GPCM modeling capabilities are illustrated in Figure 6-16 to Figure 6-24. The classical pinhole camera model utilizes a rectangle as a projection surface as illustrated in Figure 6-16. Cylindrical panoramic projection is possible by using a cylinder as a projection surface. In addition, a part-cylinder can also be used for a partial field of view as illustrated in Figure 6-17 and Figure 6-18. Conic panoramic projection, not described in [41], is shown in Figure 6-19 and Figure 6-20. Spherical panoramic projection is illustrated in Figure 6-21 and Figure 6-22. Finally, fisheye projection is illustrated in Figure 6-23 and Figure 6-24. The GPCM is not limited to such models. Any surface that can be generated by the Twister library can be used in the GPCM. In addition, any camera orientation can be obtained using the Twister library capabilities.

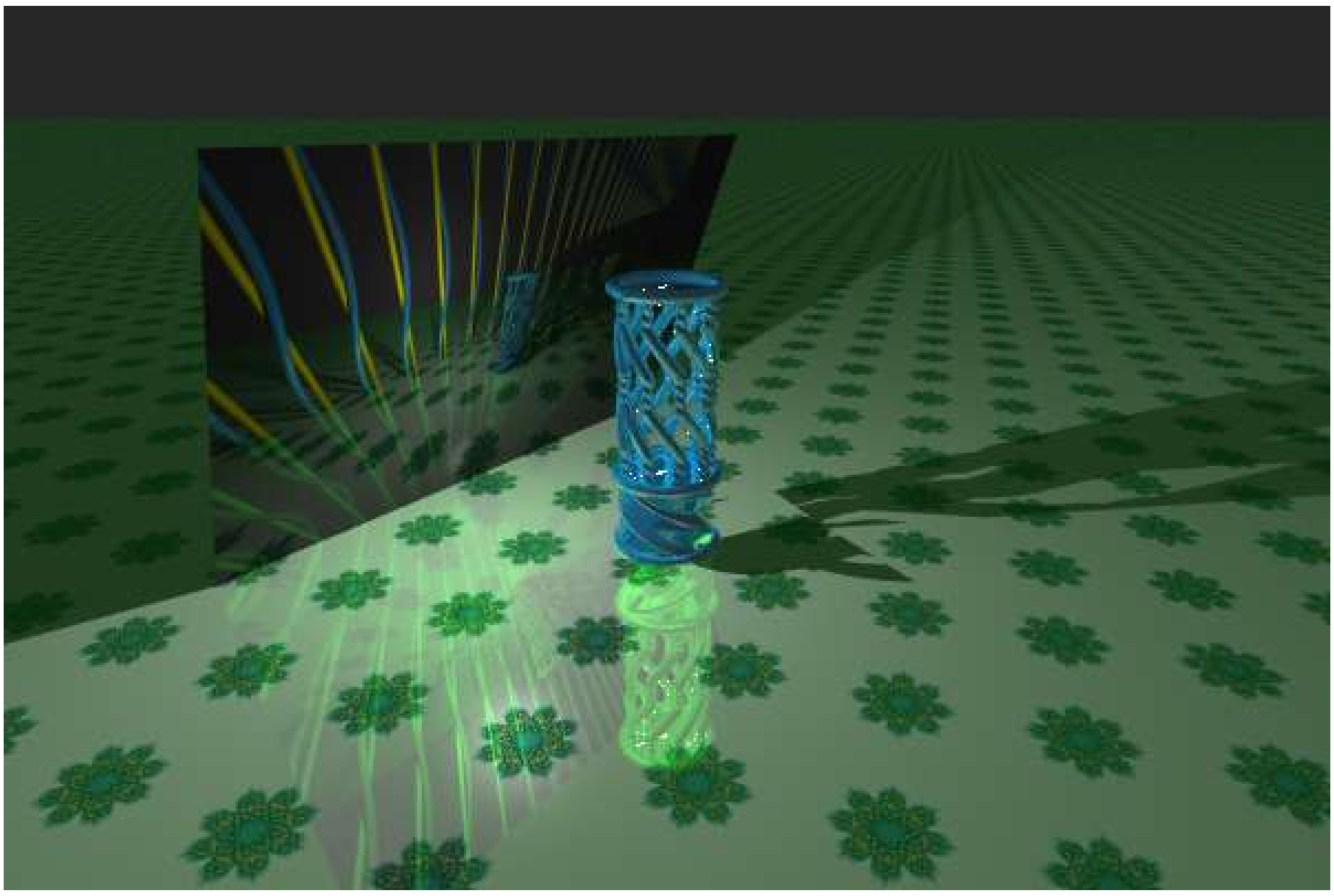

**(a) Rectangular GPCM layout and orientation**

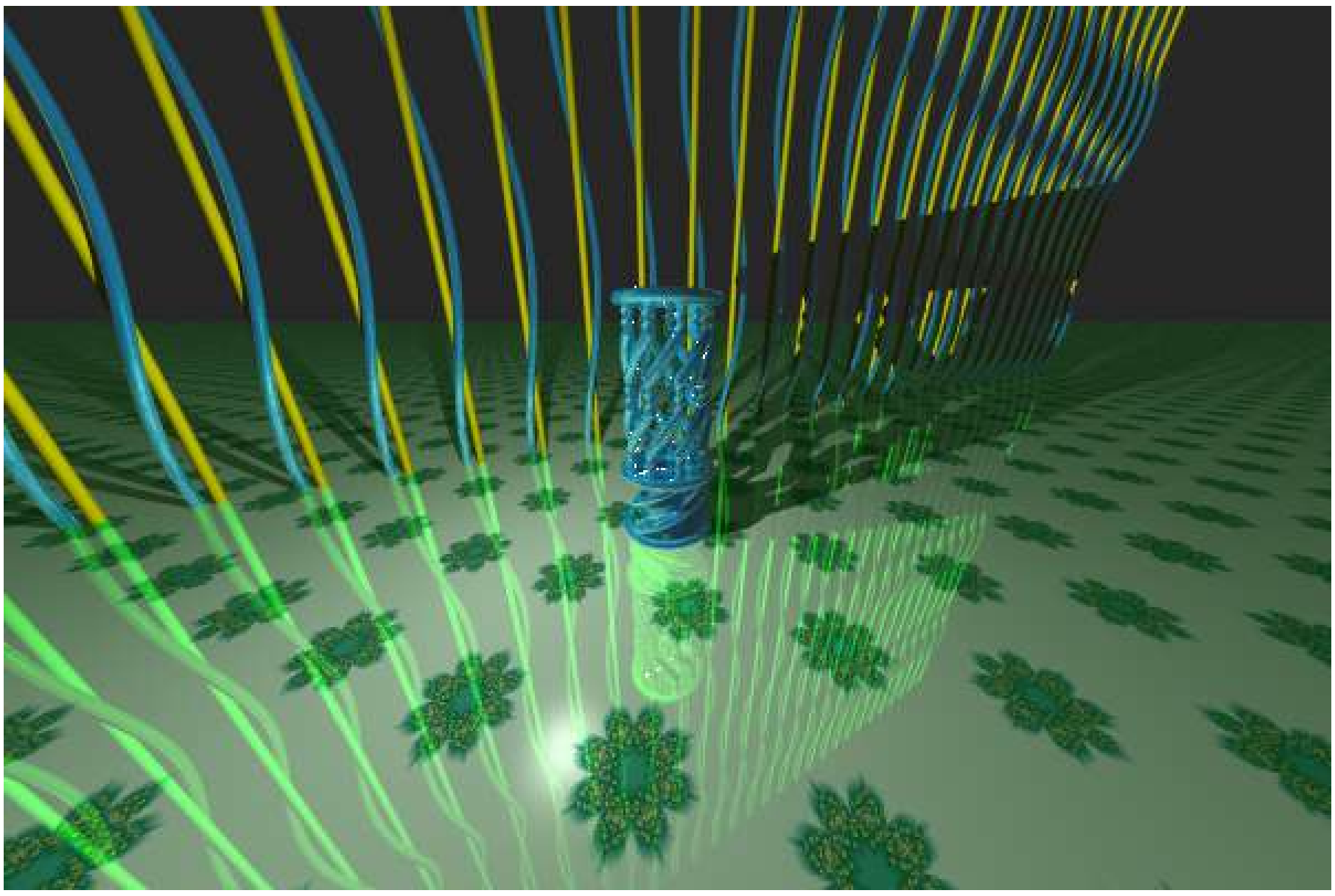

**(b) Rectangular GPCM Rendered image**

**Figure 6-16: Rectangular GPCM**

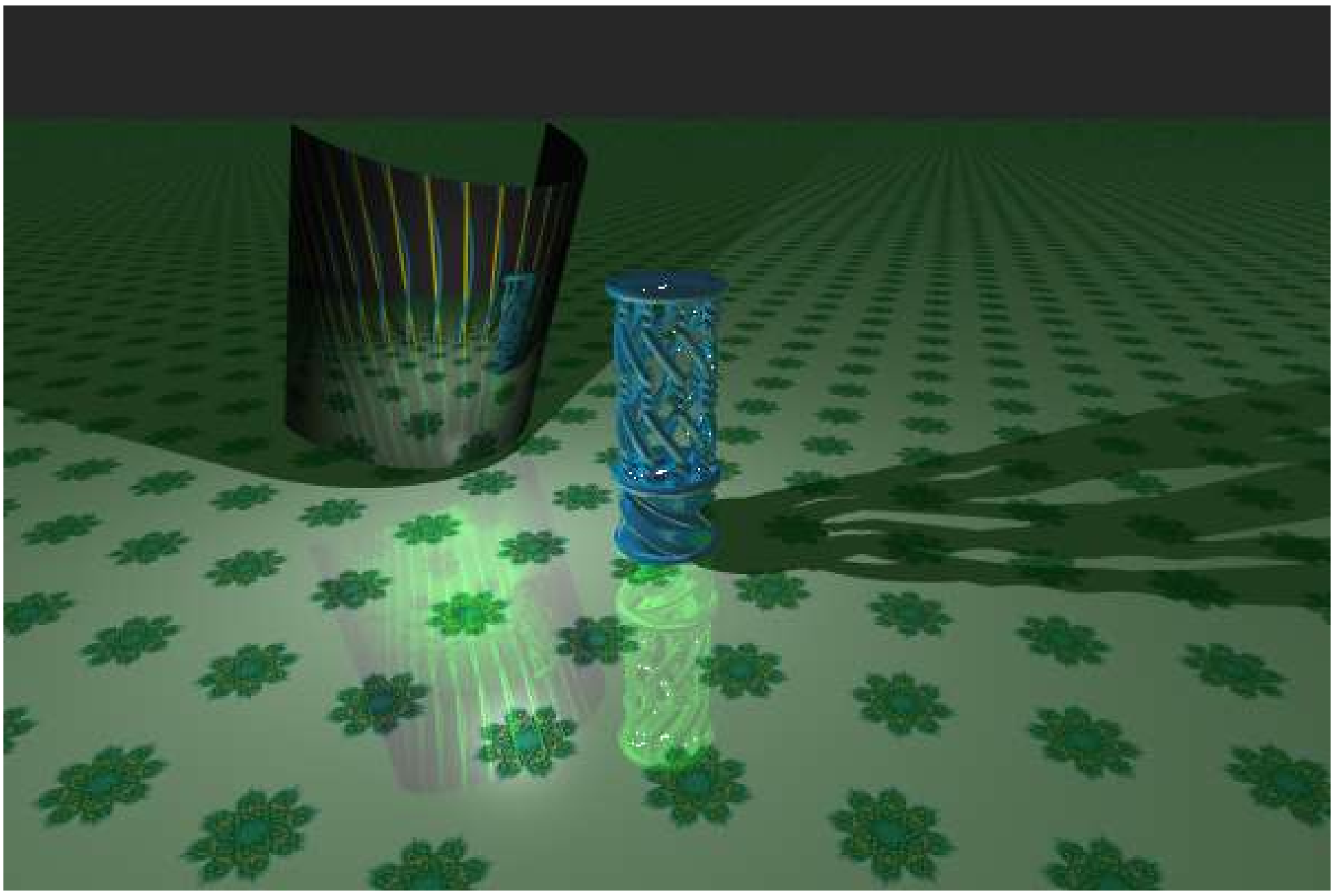

**(a) Cylindrical GPCM layout and orientation**

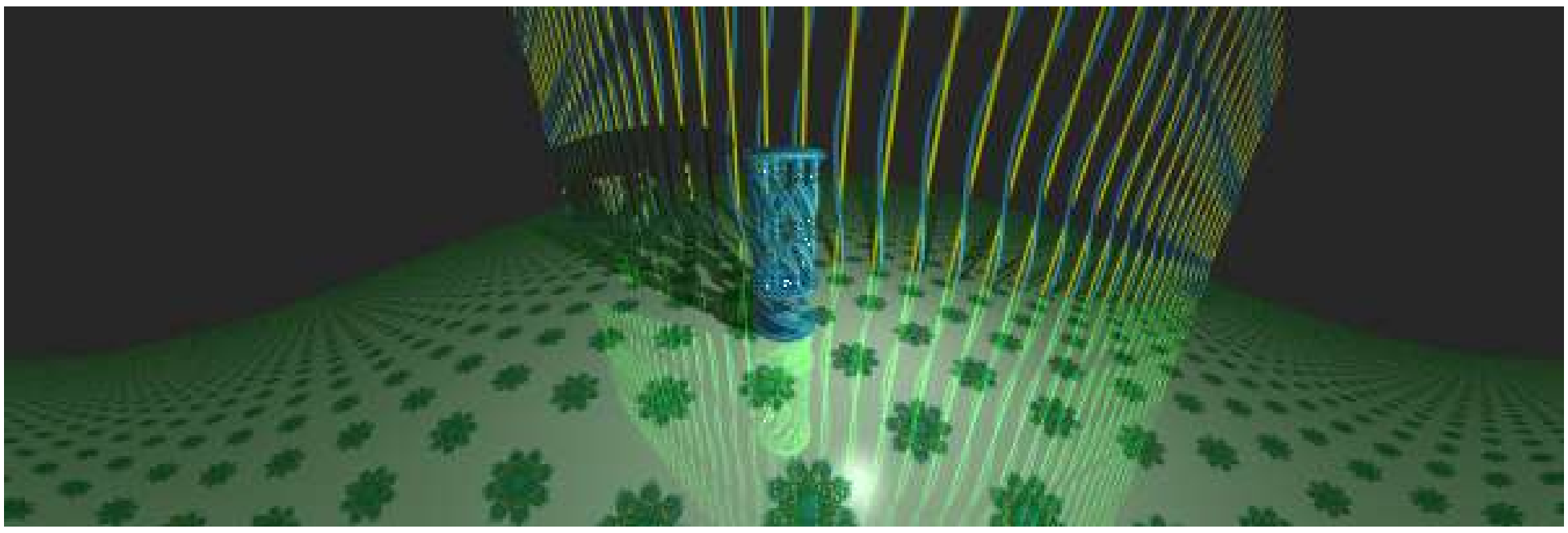

**(b) Cylindrical GPCM rendered image (360 degrees)**

**Figure 6-17: Cylindrical GPCM**

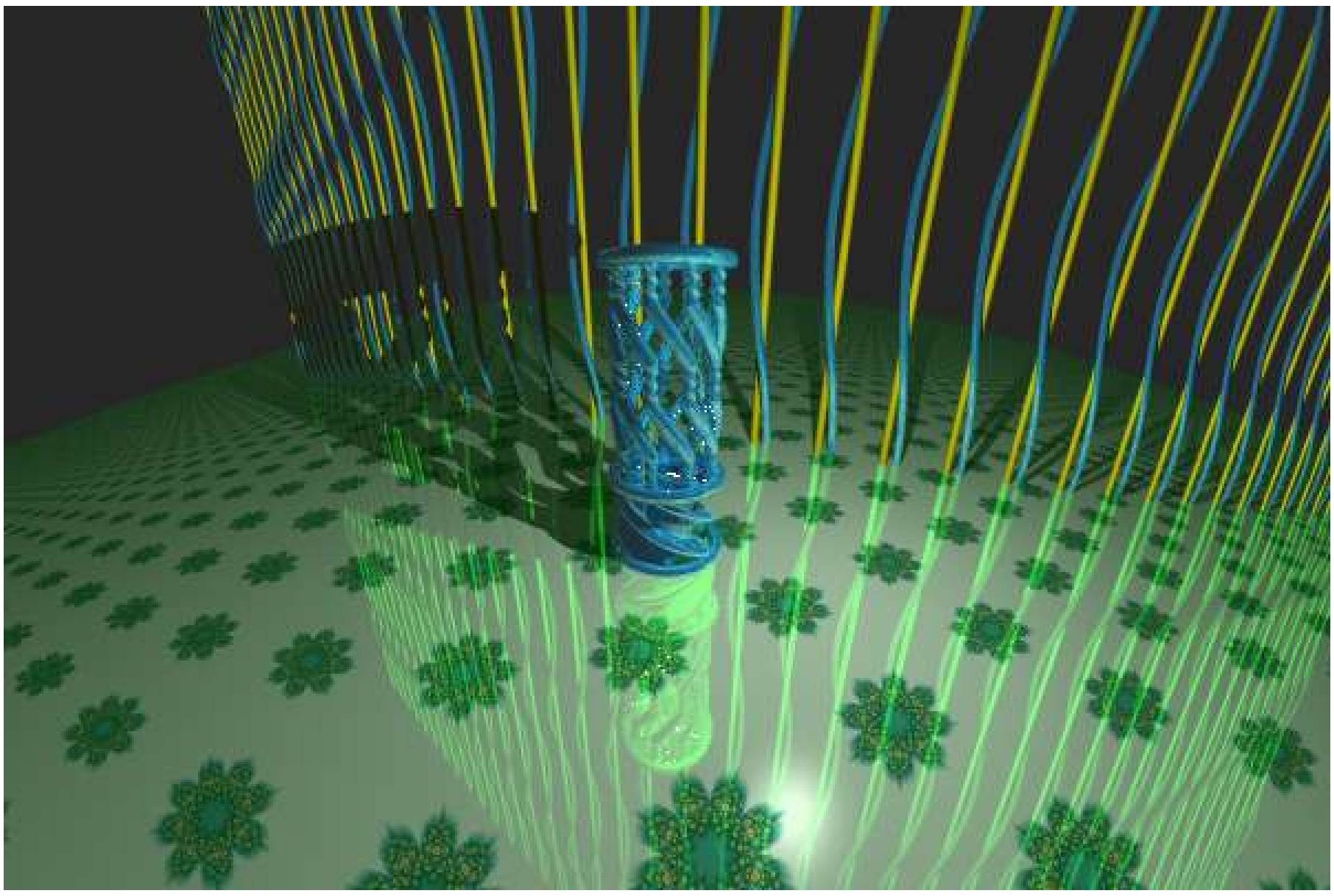

**(a) Cylindrical GPCM rendered image (180 degrees)**

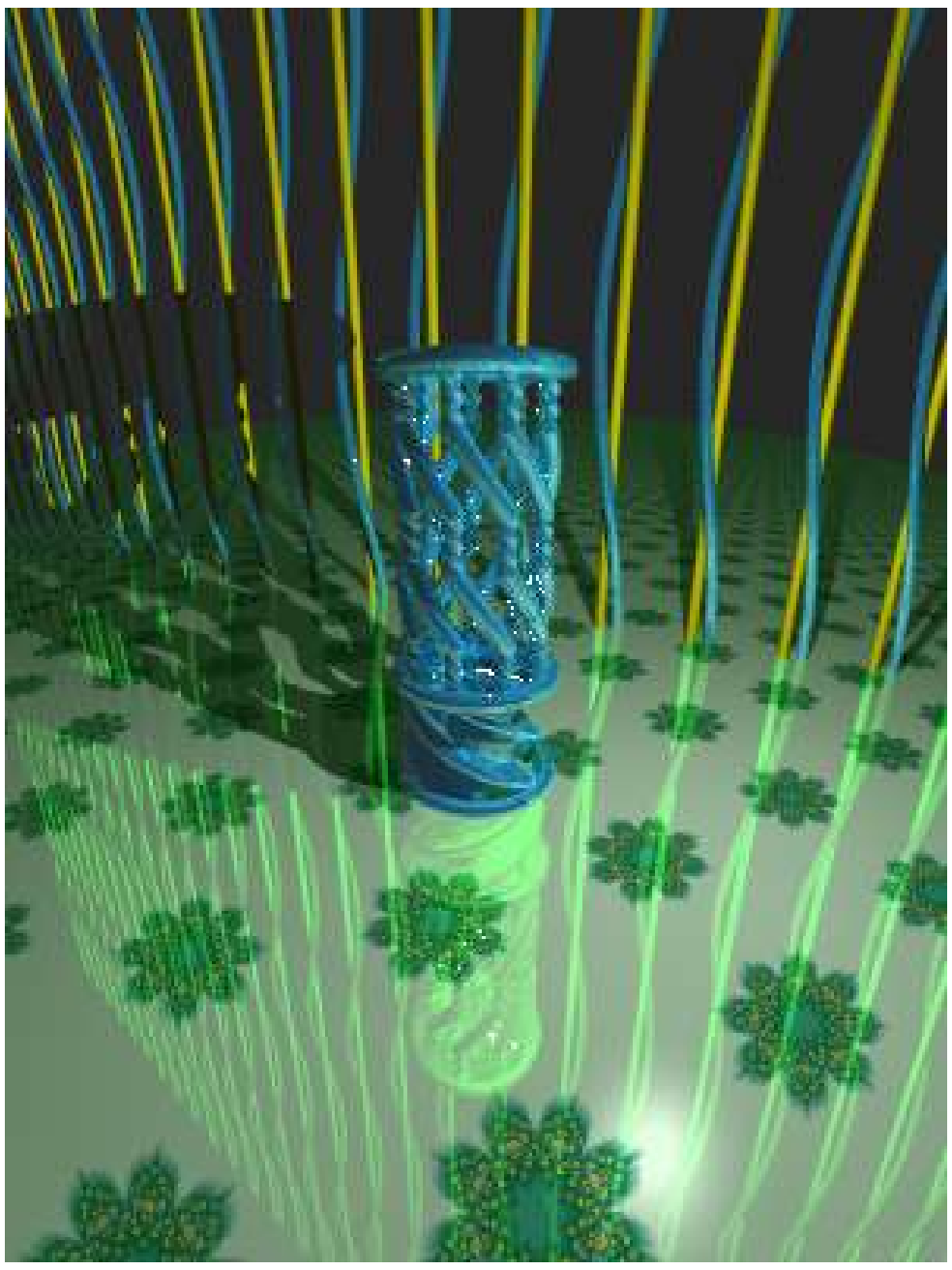

**(b) Cylindrical GPCM rendered image (90 degrees)**

**Figure 6-18: Cylindrical GPCM (continued)**

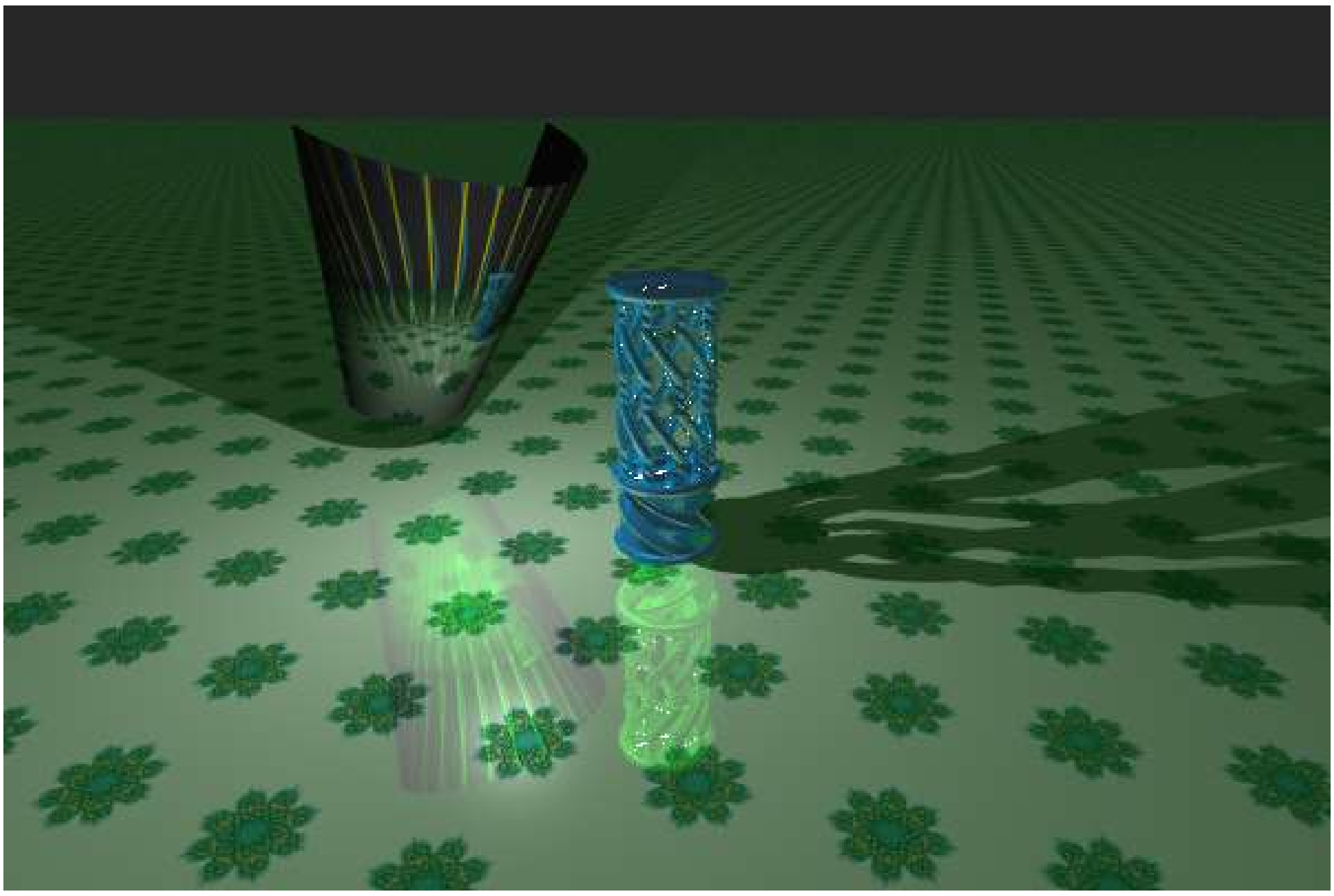

**(a) Conic GPCM layout and orientation**

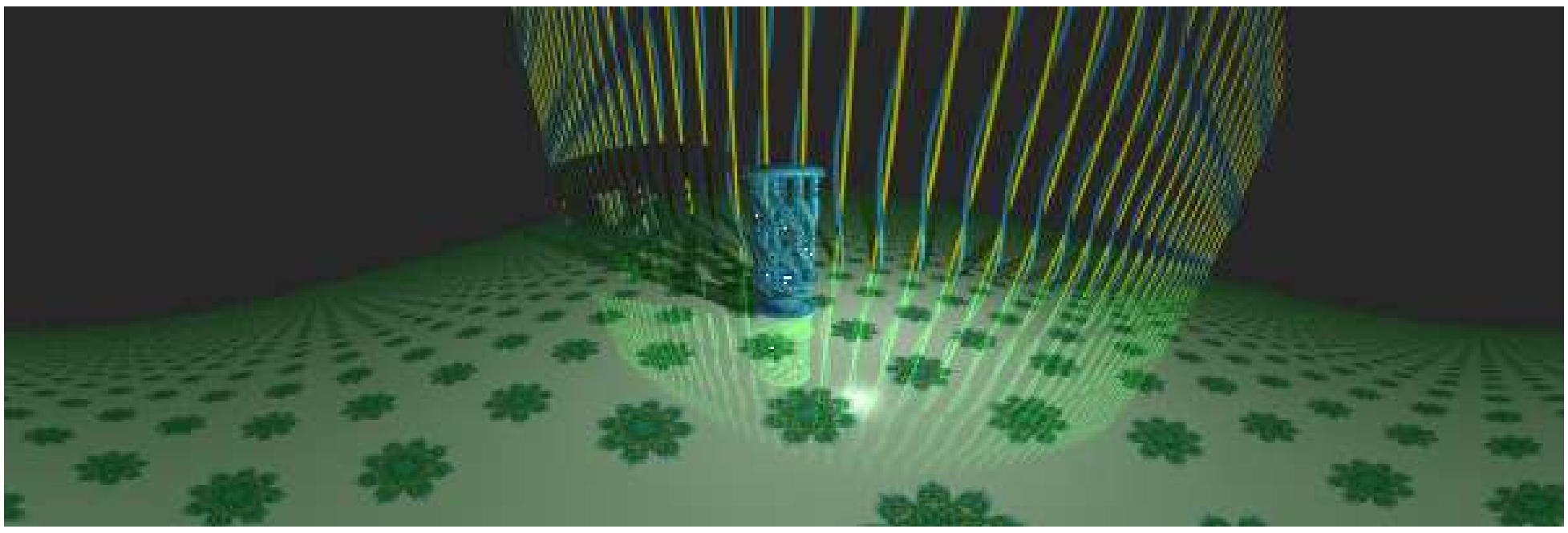

**(b) Conic GPCM rendered image (360 degrees)**

**Figure 6-19: Conic GPCM**

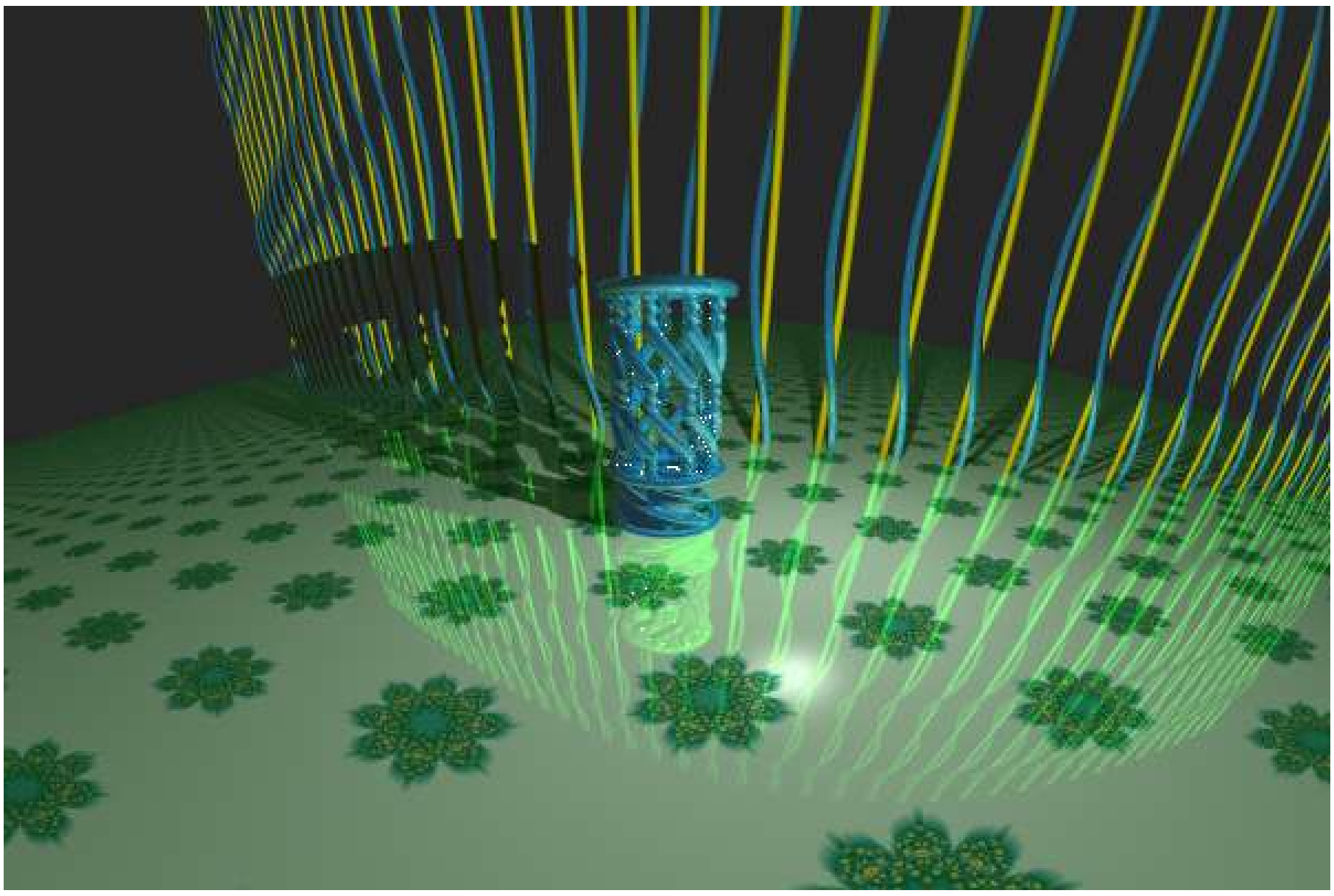

**(a) Conic GPCM rendered image (180 degrees)**

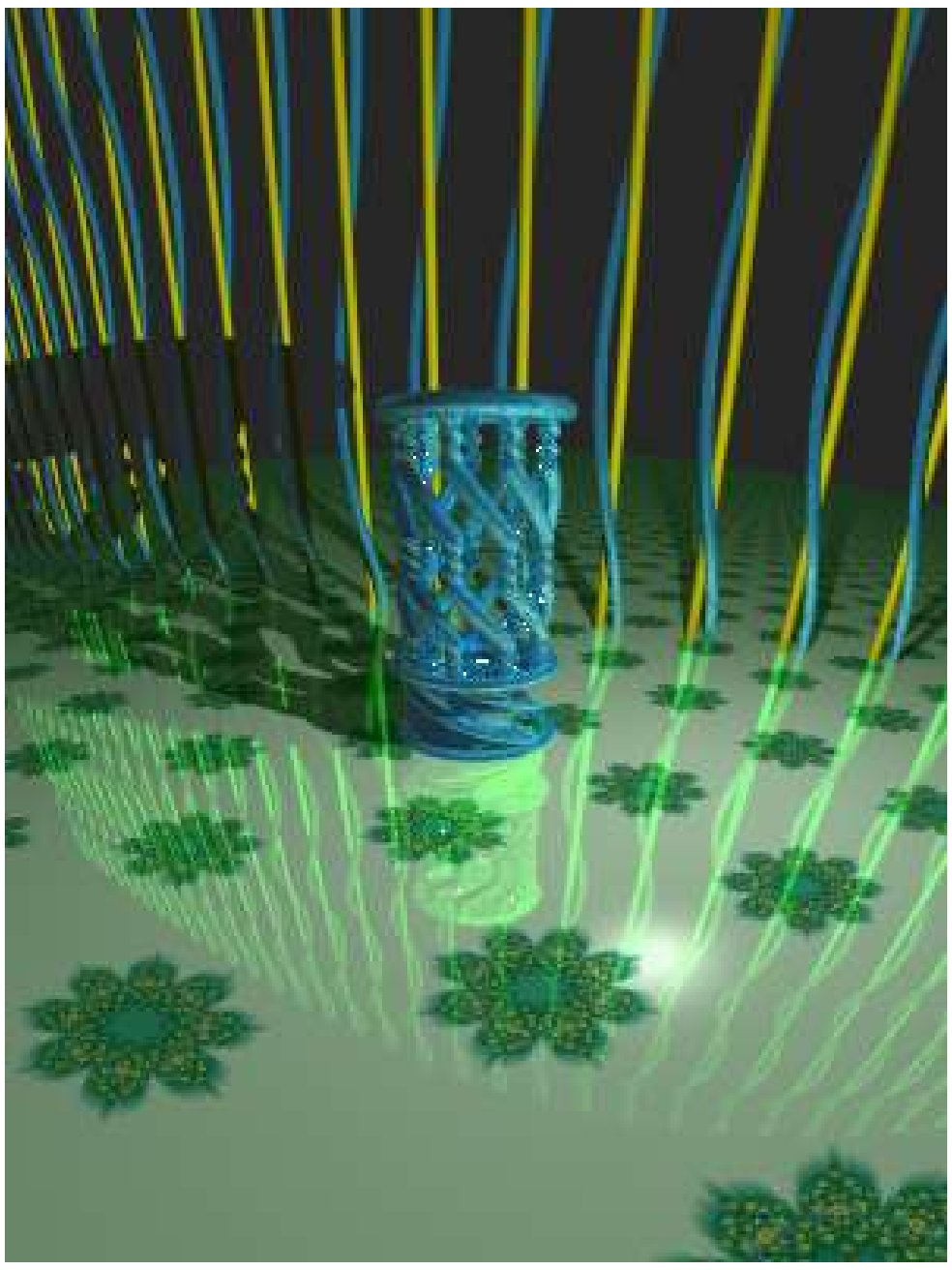

**(b) Conic GPCM rendered image (90 degrees)**

**Figure 6-20: Conic GPCM (continued)**

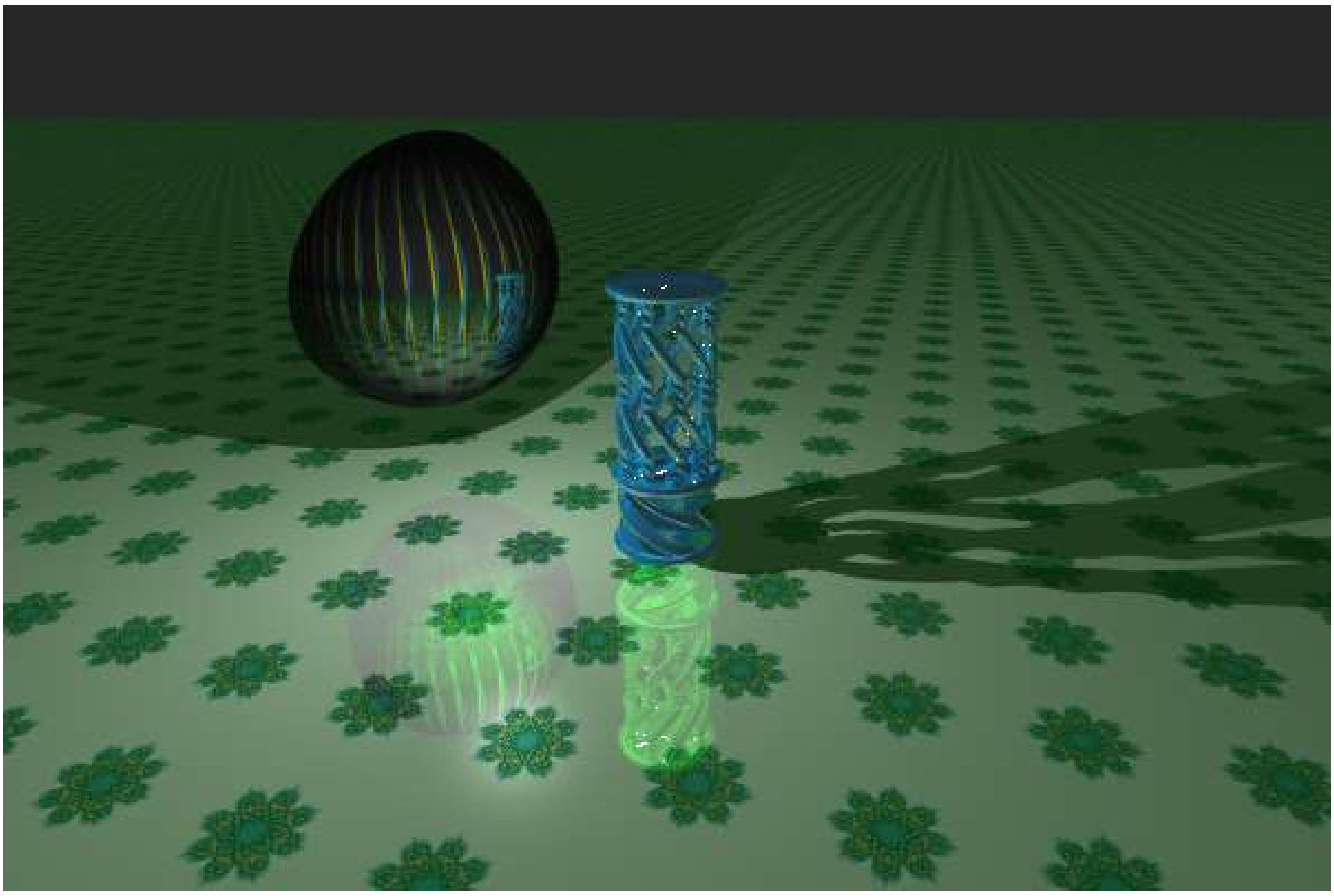

(a) Spherical GPCM layout and orientation

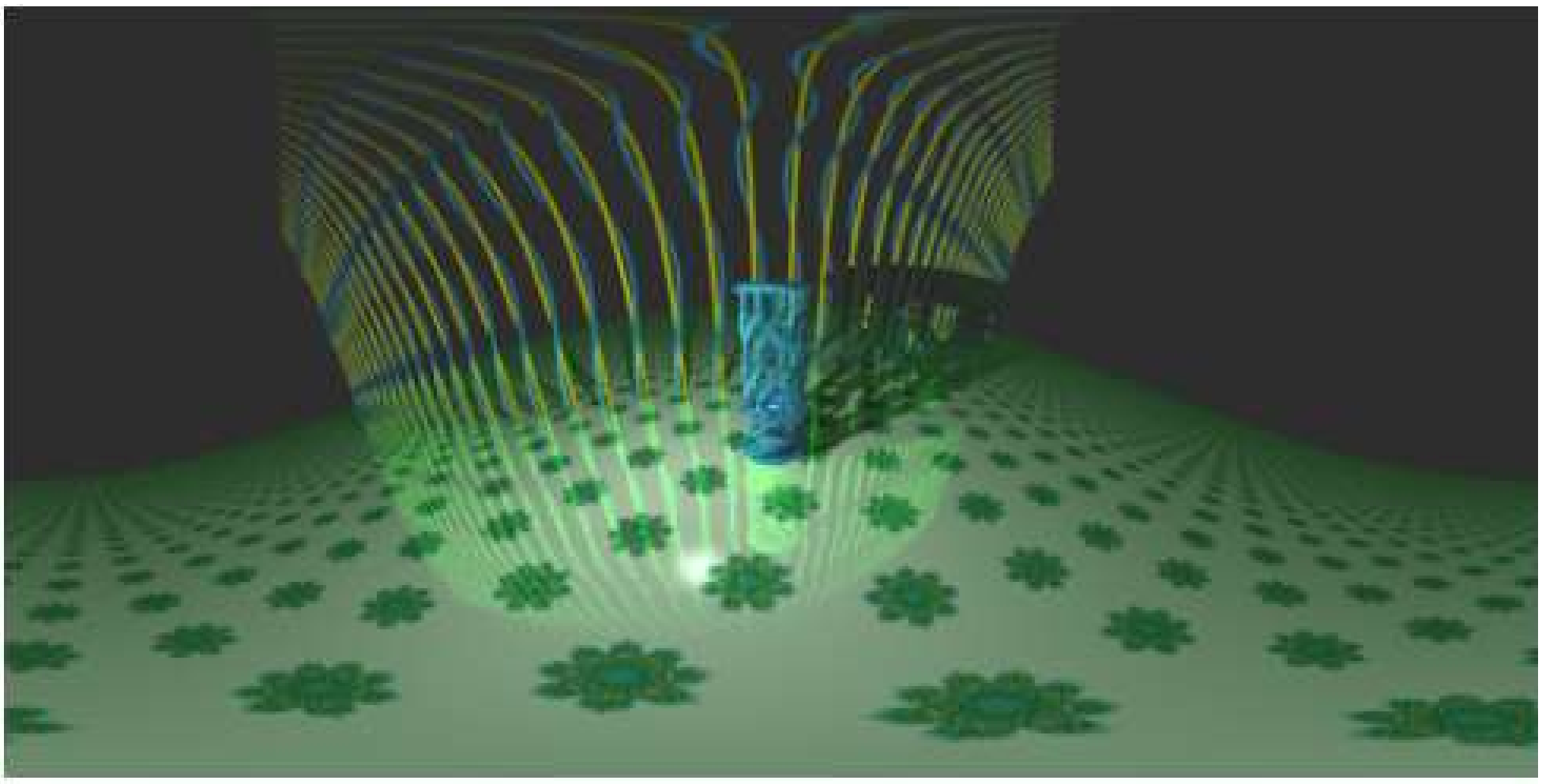

(b) Spherical GPCM rendered image (360 degrees)

Figure 6-21: Spherical GPCM

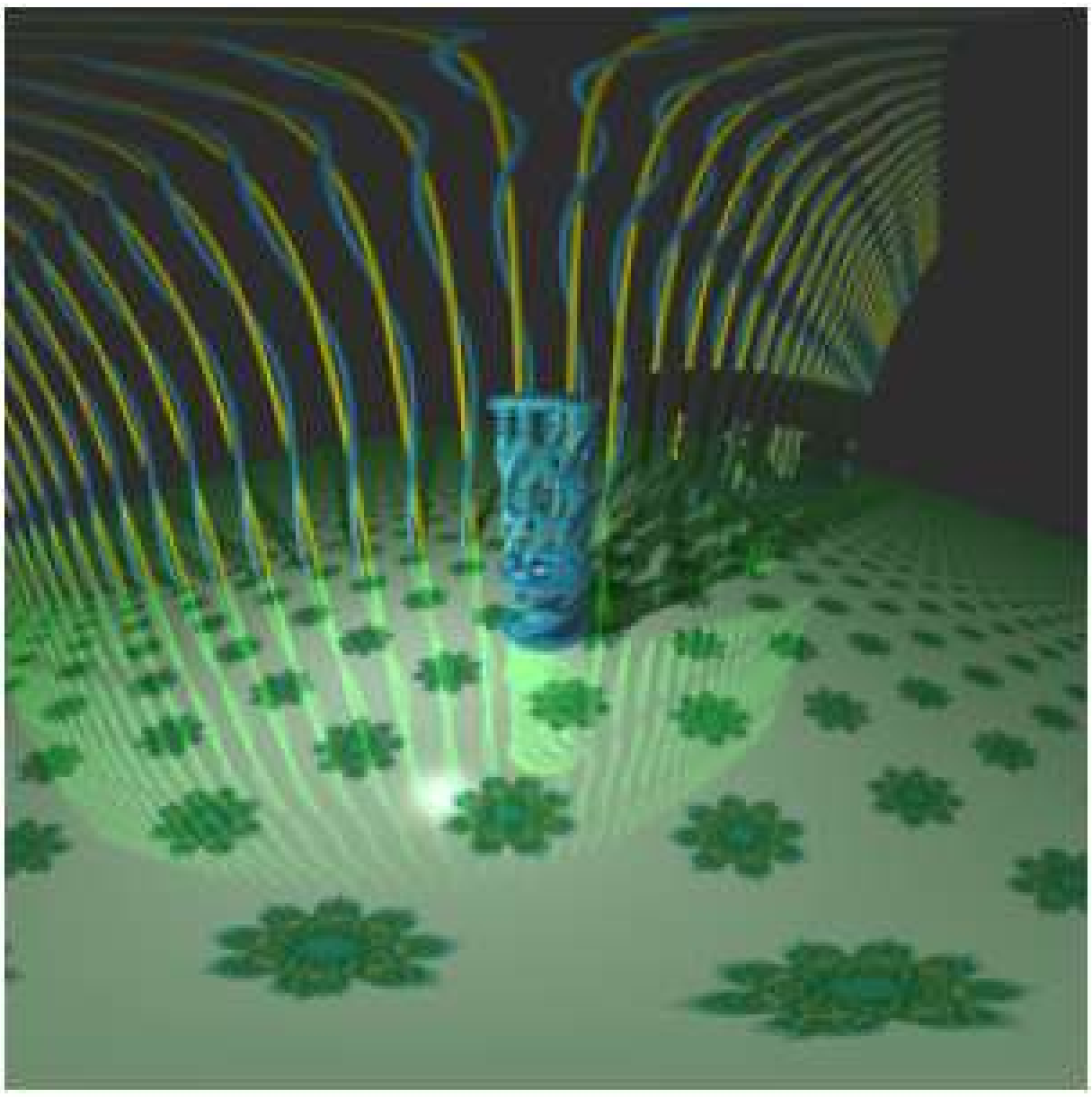

**(a) Spherical GPCM rendered image (180 degrees)**

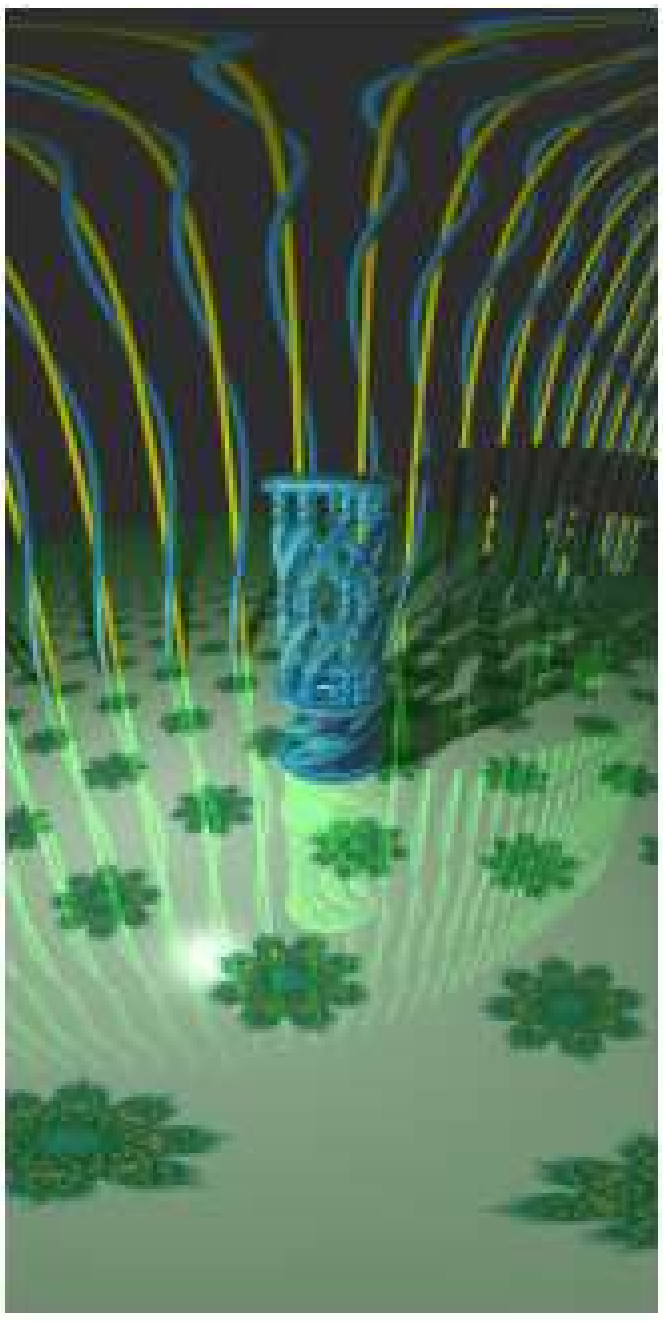

**(b) Spherical GPCM rendered image (90 degrees)**

**Figure 6-22: Spherical GPCM (continued)**

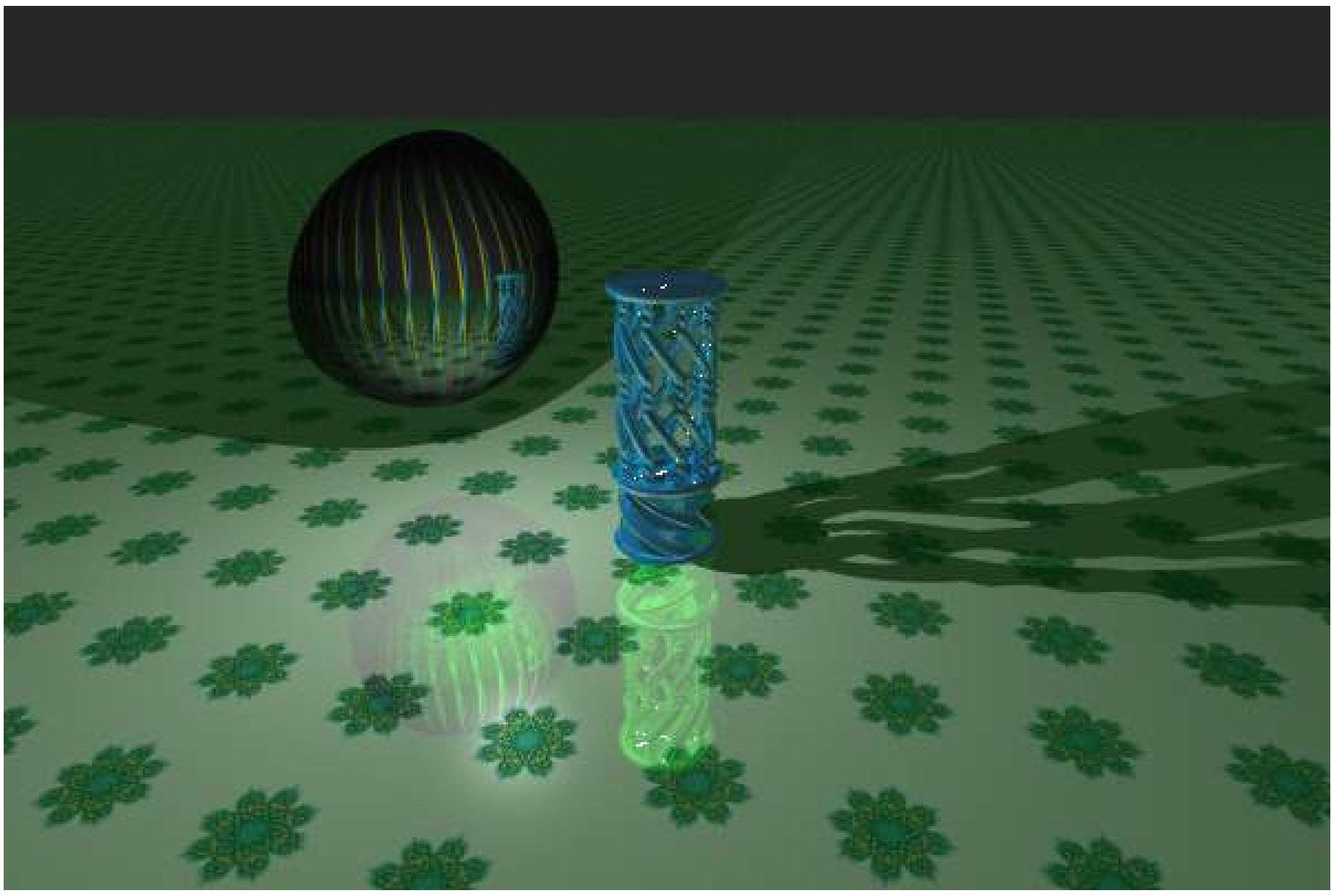

**(a) Fisheye GPCM layout and orientation**

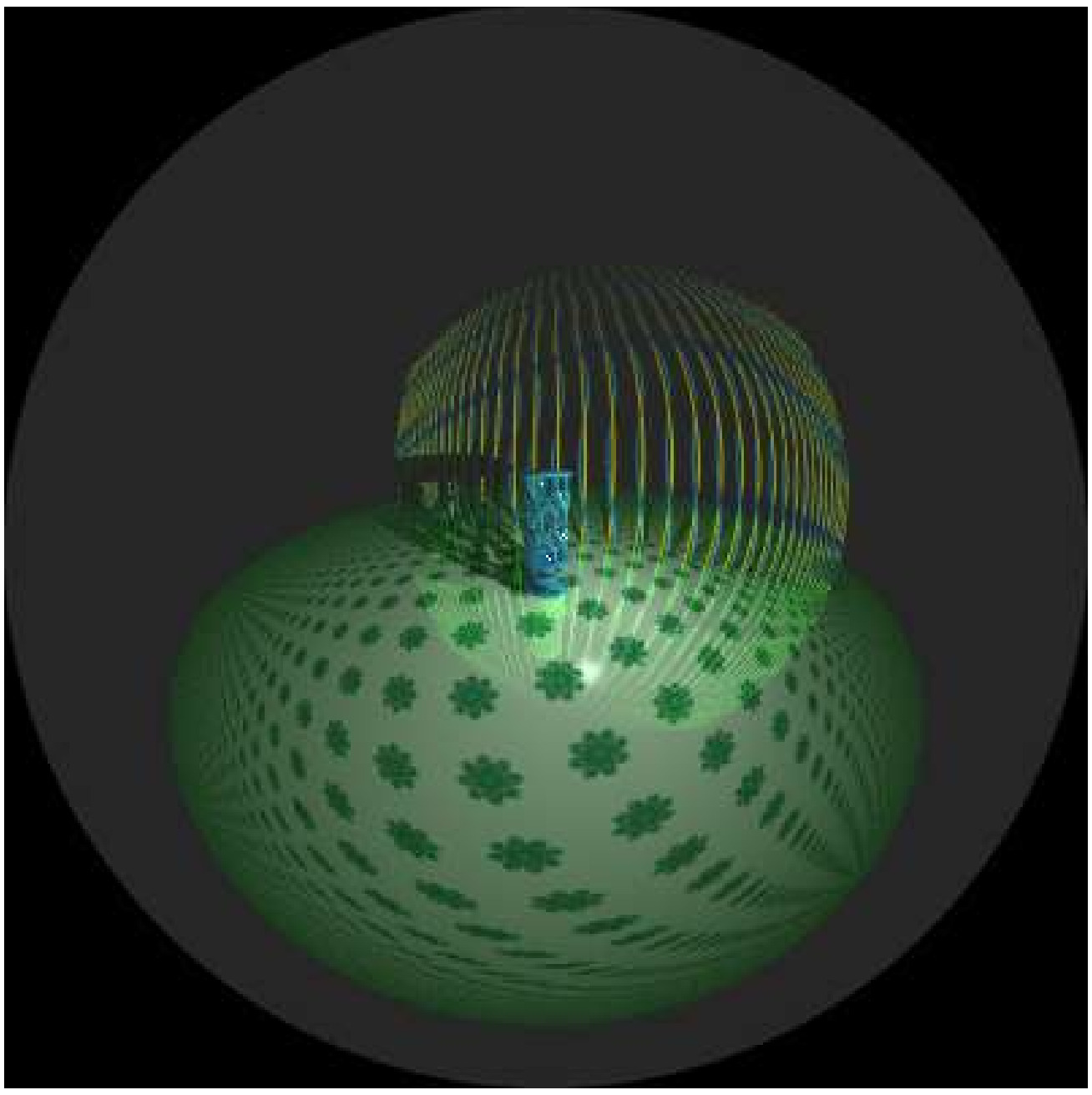

**(b) Fisheye GPCM rendered image (360 degrees)**

**Figure 6-23: Fisheye GPCM**

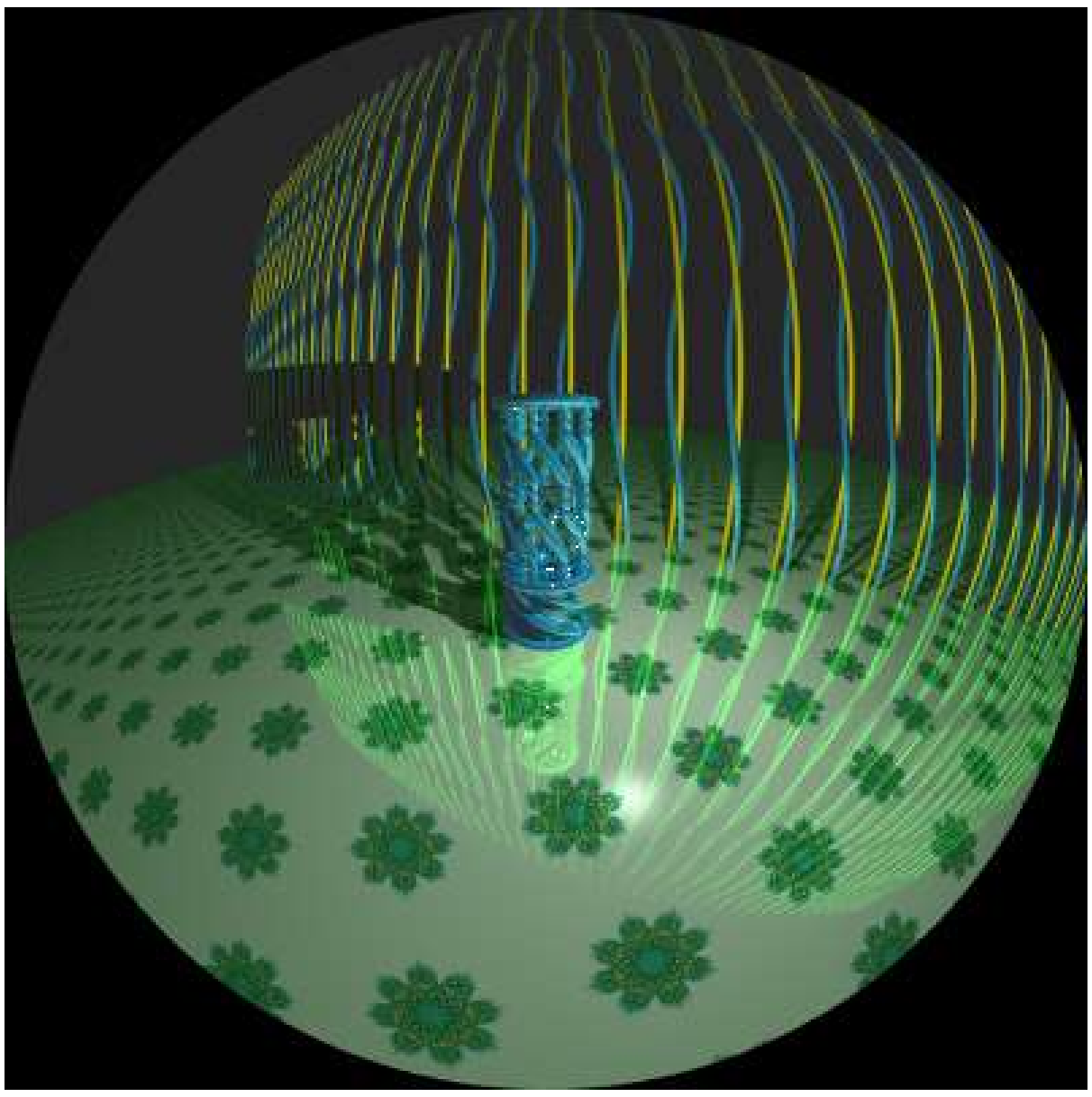

**(a) Fisheye GPCM rendered image (180 degrees)**

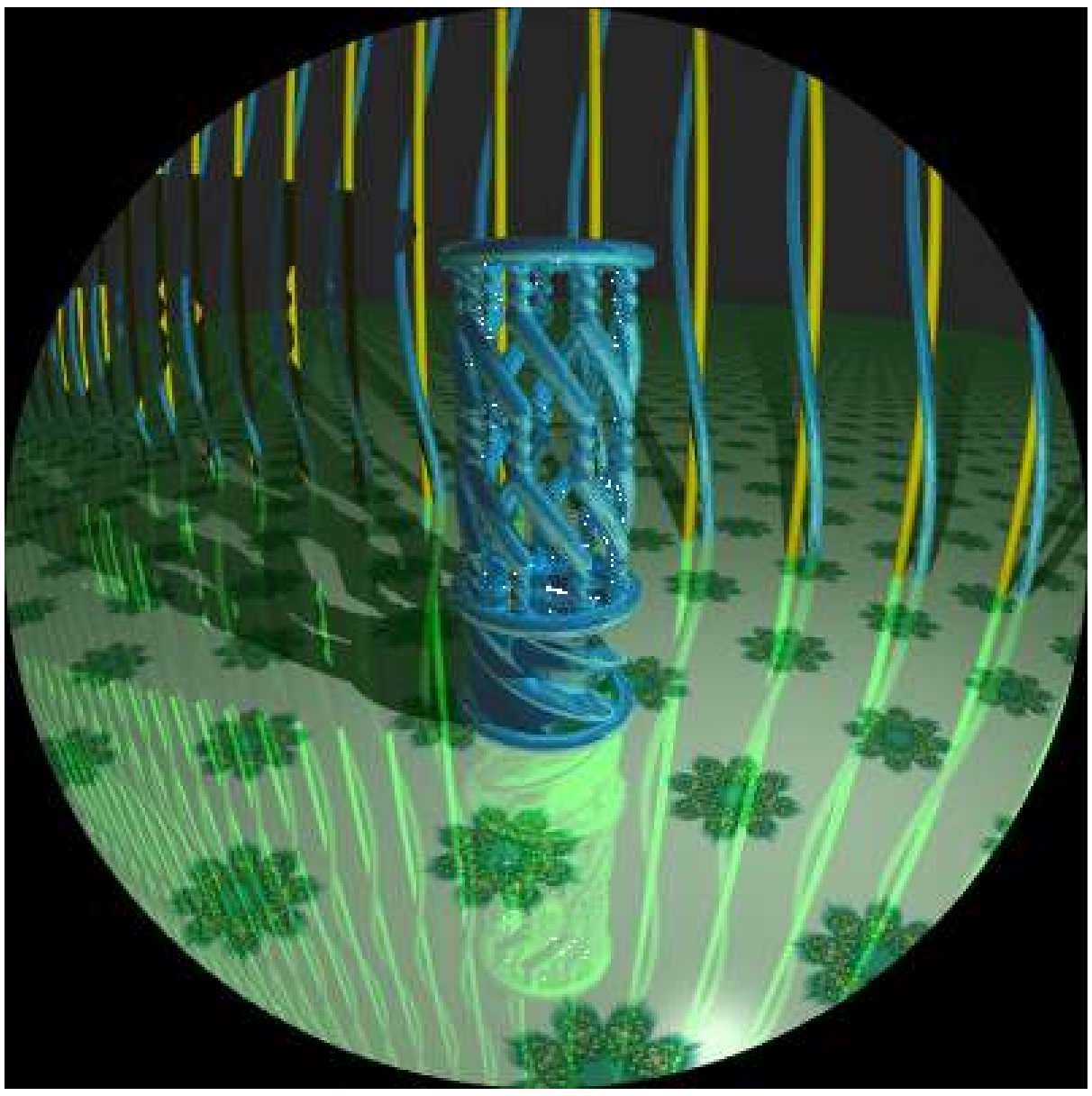

**(b) Fisheye GPCM rendered image (90 degrees)**

**Figure 6-24: Fisheye GPCM (continued)**

## 6.6 Enhancing Ray-Object Intersections

Any typical ray tracer spends most of its time making ray-object intersection tests. Geometric algebra can be used to design better algorithms with clearer geometric semantics and enhanced performance. The next three subsections illustrate some enhancements to ray-object intersection tests. Such enhancements are made possible by the compact modeling nature of GA combined with the automatic generation of efficient code provided by GMac.

### 6.6.1 TSR Instancing

One of the difficult problems in ray tracing is the deduction of a fast and accurate ray-object intersection algorithm for the selected mathematical representation of such object. Instancing [41] is one of the techniques used to generalize ray-object intersection algorithms to objects having any position, orientation, or scale in 3D space based on an object in a "standard" position and orientation in space. The technique was described in chapter 5 based on the discussion in [41]. An instance of a geometric object is the same object stored along with a 3D transform and its inverse transform. Instead of transforming the object itself, the inverse transform is applied to the ray to determine the hit point and normal using the transformed ray as shown in Figure 5-3. The traditional treatment of 3D transforms relies on the use of 4 by 4 matrices to encode the transform and its inverse. Such method is capable of representing all projective transforms; which typically include affine and thus Euclidean transforms. Such general capability is not always required in all modeling situations. If only Euclidean transforms are needed, such general approach is unnecessary. In such situation, geometric algebra and GMac can provide a more compact and efficient alternative. The alternative is compact in the sense that it only requires 12 coefficients Instead of storing 32 matrix coefficients as the traditional approach does. In addition, the alternative solution has less processing requirements than the traditional approach. The alternative approach can handle arbitrary translations, general rotations, and uniform scaling. Such subset can typically cover most of the modeling needs for describing many scene objects. The alternative approach is implemented through a class called the `TSRInstance` class in the base ray tracer of chapter 5. The approach relies on the fact that any composition of translations, rotations, and uniform scaling in 3D Euclidean space can be put on the standard form: TSR (Rotate then Scale then Translate the object) [11]. The `TSRInstance` class only stores the necessary coefficients for performing such standard composition. This approach will hence be called the TSR Instancing technique (TSRI).

Like the `Instance` class of chapter 5, the `TSRInstance` class stores a reference to the base geometric object. The 3D transform is stored as a set of 12 real coefficients. Three coefficients are used for the vector of the translation transform T. Four coefficients are used for the axis and angle of the rotation transform R. One coefficient is used for the scaling transform S. Finally, four other coefficients are used to store other (ray-independent) auxiliary values to speed up ray-instance hit calculations. The rotation and scaling transforms are always around the origin.

Initially, the `TSRInstance` takes a reference to the base object. In addition, a temporary set of two vectors $v_x, v_y$ (initially being the x and y axis unit vectors $e_1, e_2$) and a point $p_o$ (initially being the origin) is defined with each new instance. This temporary set is a “frame of reference” that gets transformed with the object. The user then applies any combination of translations, general rotations, and uniform scaling to the temporary set of two axis and point. All such transformations of this phase are applied using the Twister library. After all the required transforms are applied, a procedure is invoked to analyze the final position, orientation, and scale of the temporary set relative to its initial state. The analysis is based on optimized GA procedures applied using GMac. The final outputs of the analysis procedure are the 12 coefficients of the `TSRInstance`.

Assuming $v_T$ is the translation vector for T, $v_R, \phi_R$ are the rotation axis and rotation angle around the origin, $s$ is the scale factor for S, then the following relations are used to deduce these values from the final state of $v_x, v_y, p_o$. The value of $v_T = p_o$; the exact value of the position vector $p_o$. The scale factor $s = \|v_x\| = \|v_y\|$; the final length of any of the two temporary vectors. To find $v_R, \phi_R$ a process of three steps is required as shown in Figure 6-25. First, a

rotation axis and angle $v_{xR}, \phi_{xR}$ that rotates $e_1$ (the unit x-axis) into $v_x$ is found using the same procedure of the Twister library used for GPCM orientation procedure in section 6.5.2. This step is illustrated in Figure 6-25b. Next, the rotation defined by $v_{xR}, \phi_{xR}$ is applied to $e_2$, the unit y-axis, and the resulting vector is $v'_y$ as shown in Figure 6-25c. A second rotation $v_{yR}, \phi_{yR}$ is then deduced that rotates $v'_y$ into $v_y$ as in Figure 6-25d and Figure 6-25e. The final rotation is then the combination of these two rotations:

$$
\begin{aligned}
&R = \cos(\phi_R / 2) - \sin(\phi_R / 2) v_R^{\odot} = R_y R_x \\
&Where: \\
&R_x = \cos(\phi_{xR} / 2) - \sin(\phi_{xR} / 2) v_{xR}^{\odot}, \\
&R_y = \cos(\phi_{yR} / 2) - \sin(\phi_{yR} / 2) v_{yR}^{\odot}
\end{aligned}
\qquad (6.14)
$$

The final rotation axis $v_R$ is shown in Figure 6-25f and the deduced total rotation is shown in Figure 6-25g.

For each ray to be tested with the instance, the inverse of the TSR transform must be applied to the two components of the ray: its origin point $p_r$ and its direction vector $v_r$. This is the process requiring the most processing time in the ray-instance intersection tests. GMac is again used to generate optimized code for such critical process. The net performance results are discussed in the following chapter.

Several advantages are apparent from using this technique over the traditional one of [41]. The first advantage is that the TSRI technique is mainly based on pure geometric ideas. No matrices whatsoever were required during the design of the technique. GMac and the Twister library are used for all code related to the application and deduction of the TSR transform. The net result is an algorithm that requires less memory, less processing power, less debugging, and less design effort.

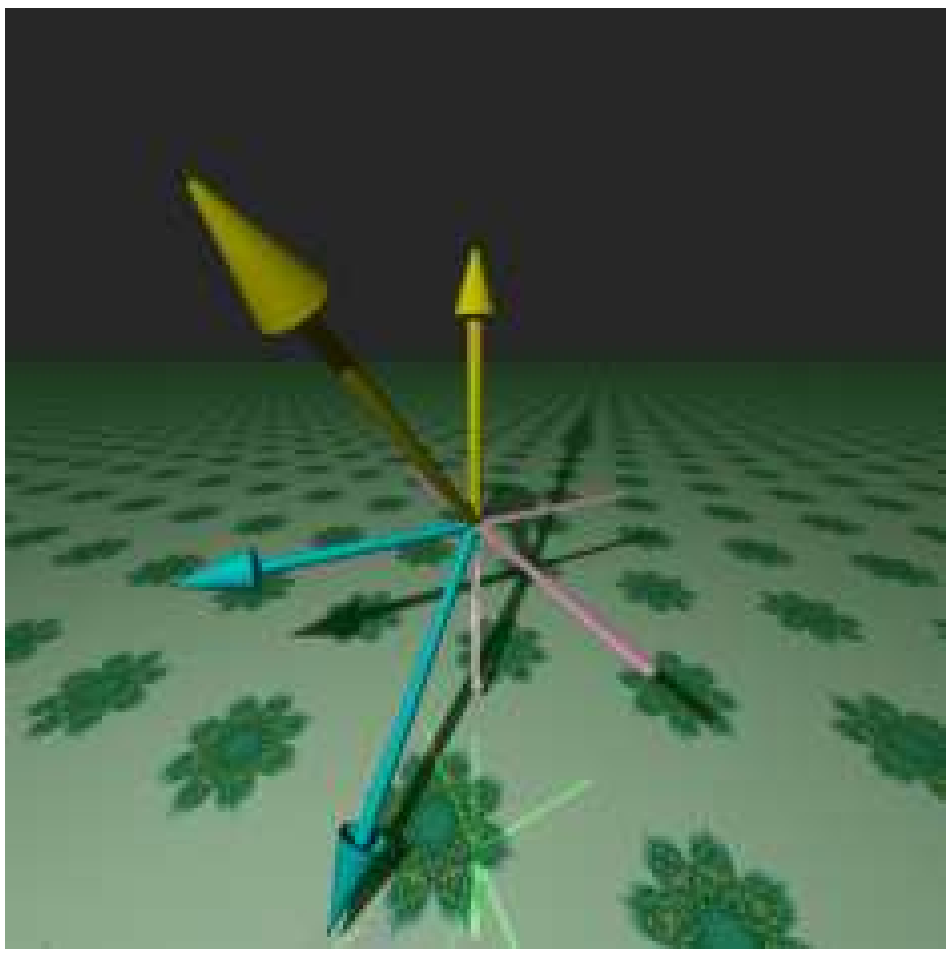

(a) Initial and final x and y axes

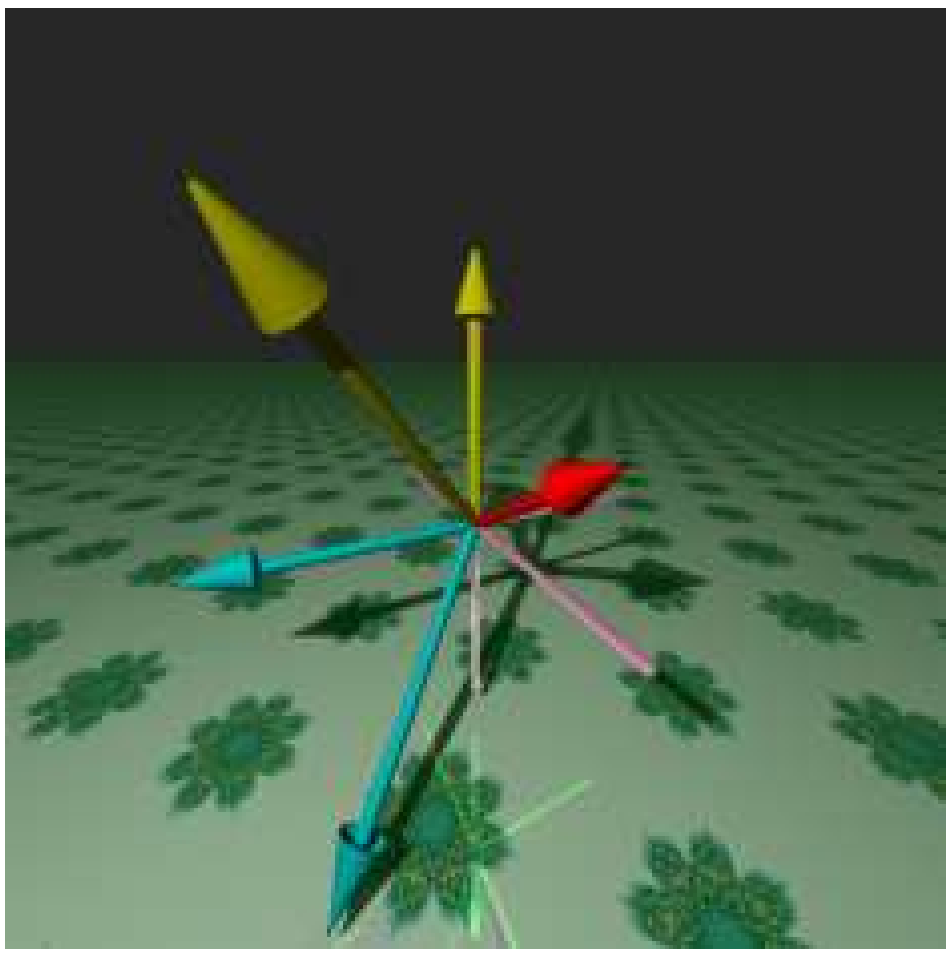

(b) First rotation taking initial x-axis to final x-axis

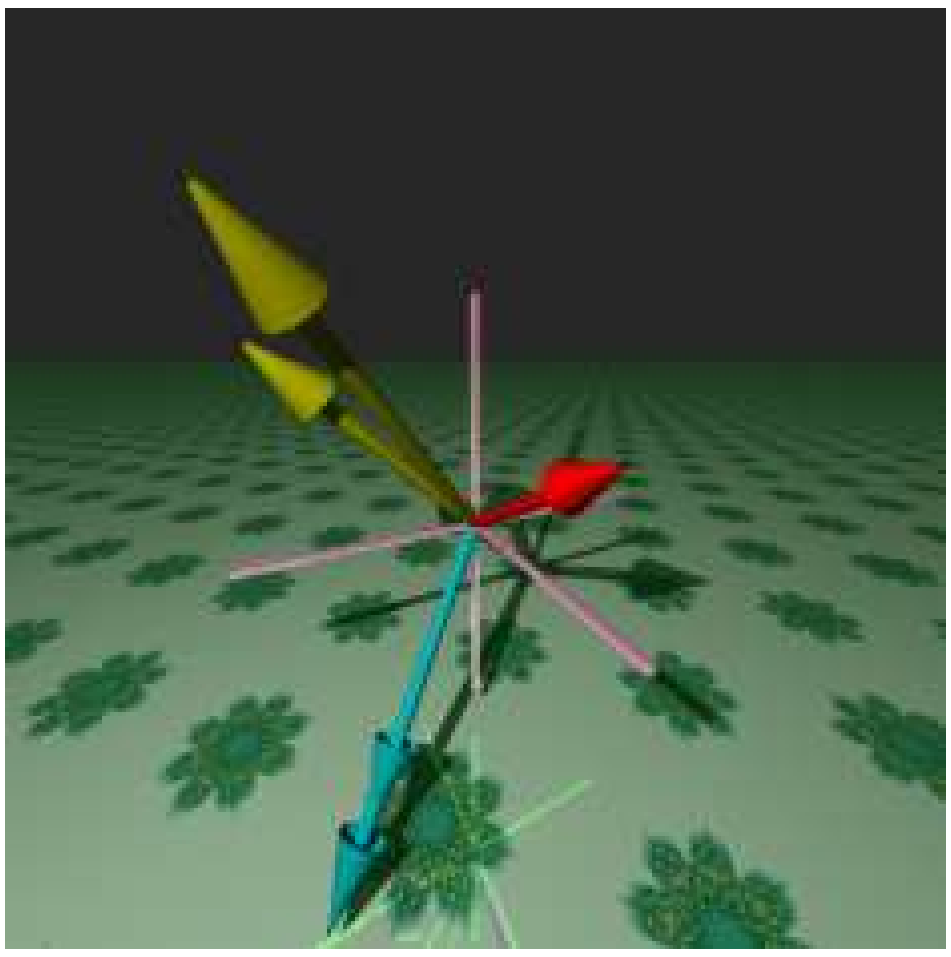

(c) First rotation applied to initial x and y axes

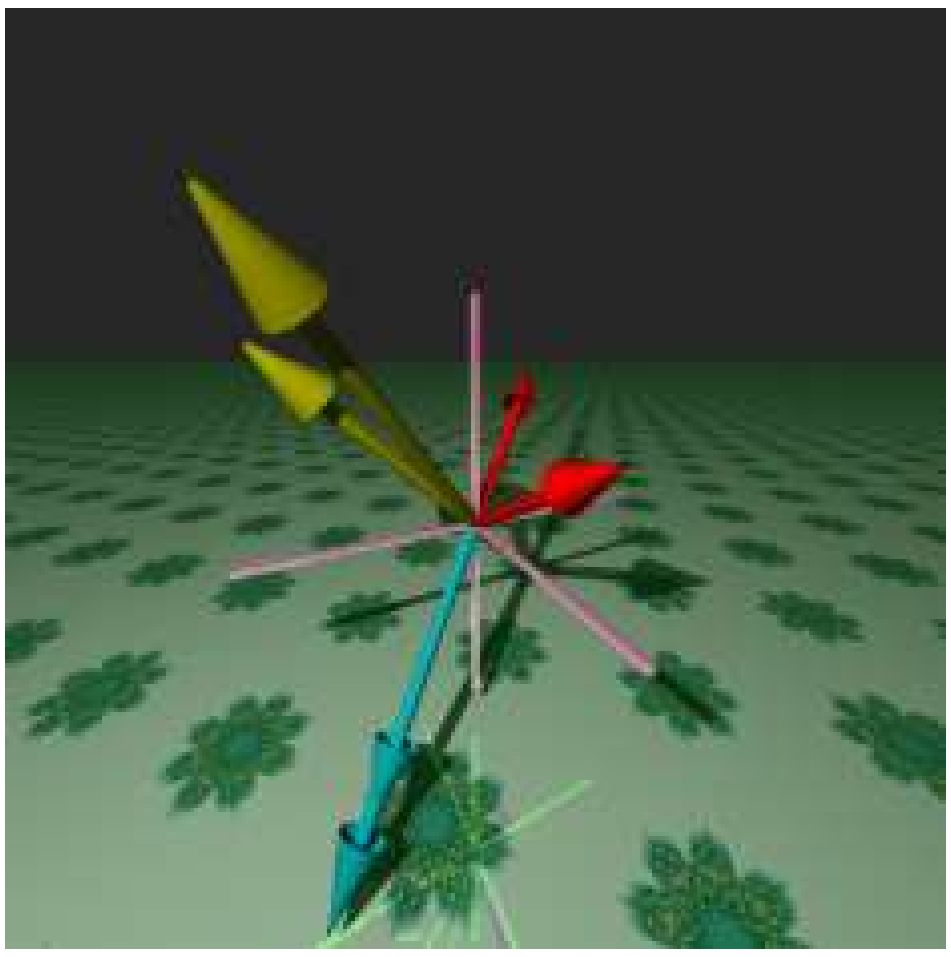

(d) Second rotation taking rotated y-axis to final y-axis

**Figure 6-25: Deduction of final rotation in TSRI transform**

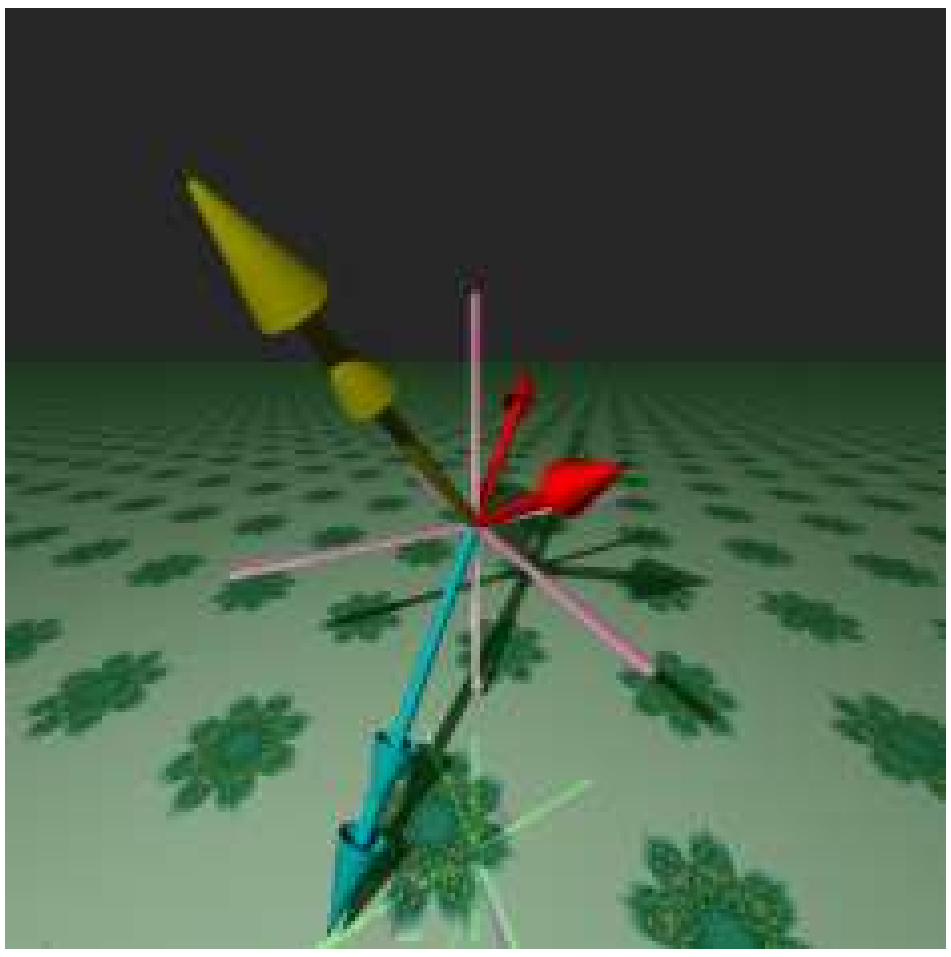

(e) Second rotation applied to rotated y-axis

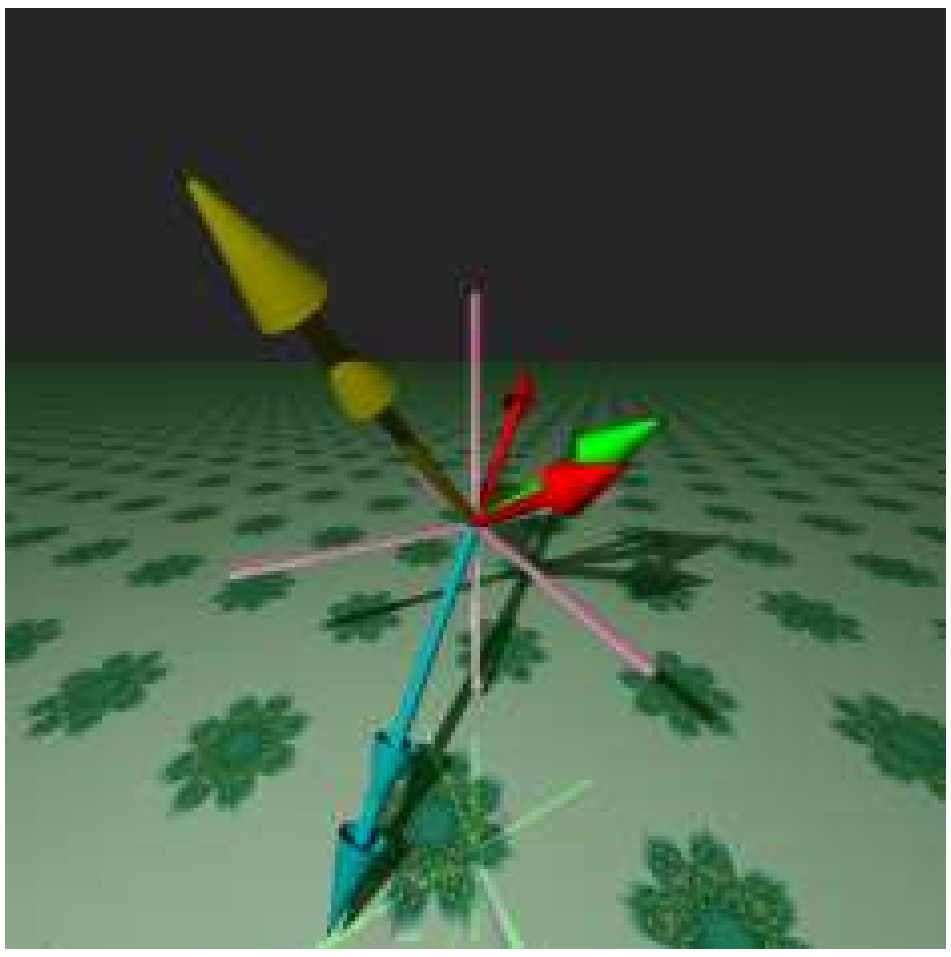

(f) Axis of total rotation (green)

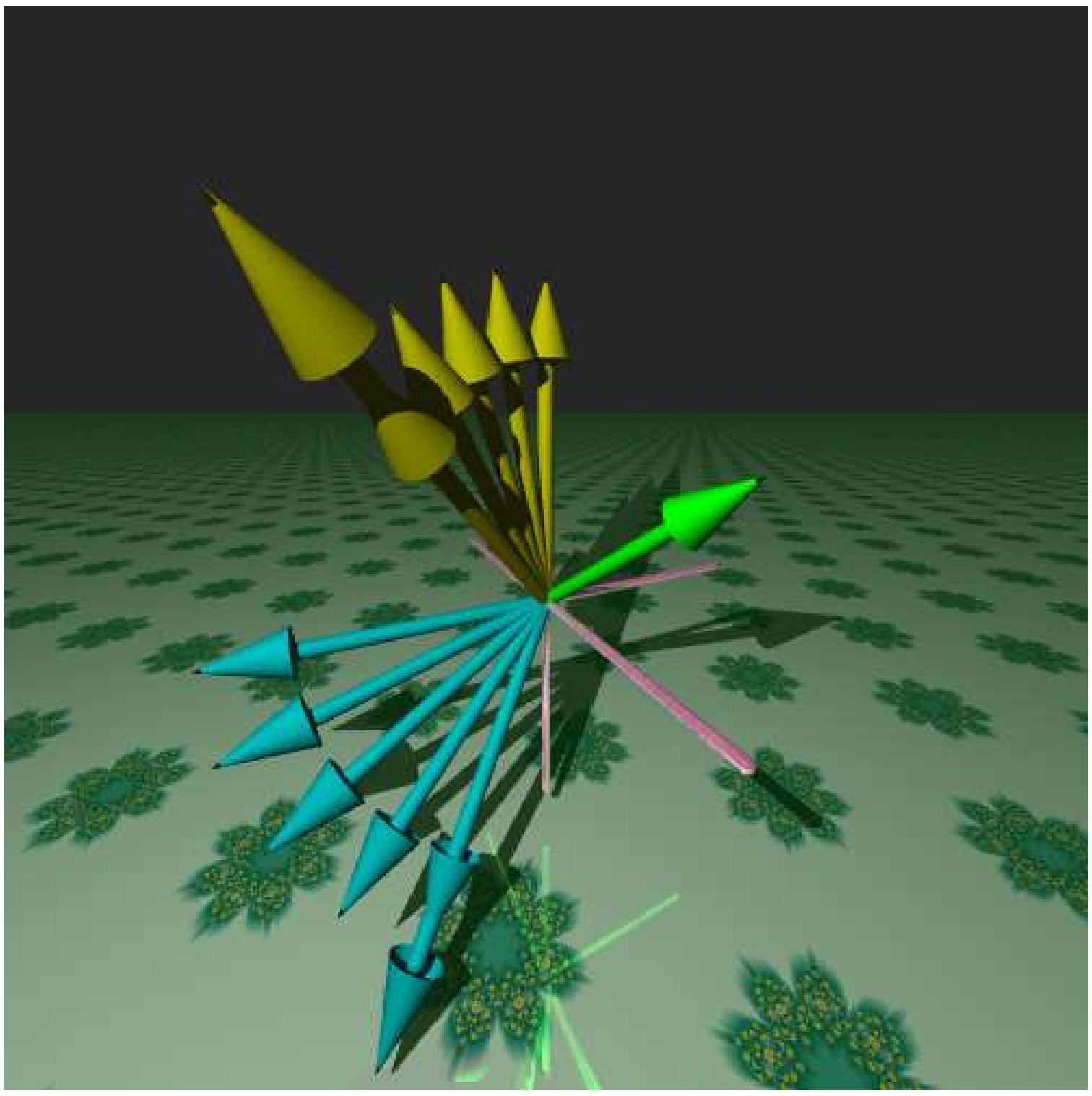

(g) Total rotation applied to initial x and y axes to obtain final x and y axes

**Figure 6-25 (continued): Deduction of final rotation in TSRI transform**

### 6.6.2 Rotationally-Symmetric TR Instancing

Many primitives commonly used in ray tracing are rotationally symmetric around an axis; like the y-axis for example. Such primitives include spheres, cylinders, tori, disks, and cones. When such primitives are used in a scene without the need for accurate 2D texture mappings, the full TSRI technique is not required for placing such primitives in arbitrary position and orientation in space. Another simpler and more efficient technique can be used, also based on GA and GMac. This technique will be called Rotationally-Symmetric TR Instancing (RSTRI).

As shown in Figure 6-26a, the y-axis rotationally-symmetric object (a torus) is given an arbitrary position $p_o$ and orientation $v_y$ in space. The same technique used in the TSRI is applied here with several major simplifications. First, there is no scaling assumed, the original object never changes size, thus no S transform is needed. Second, the rotation is simplified because the object is rotationally symmetric around the y-axis as shown in Figure 6-26c. This leads to the third simplification of using a temporary set (frame of reference) consisting of a single point $p_o$ and a single direction vector $v_y$. Such simplifications result in enhanced performance of ray-object intersection tests for rotationally-symmetric objects compared to TSRI or traditional instancing.

### 6.6.3 Ray-Triangle Intersection

As pointed out in [41], triangle meshes play a very important role in CG applications, including ray tracing. Whatever object modeling techniques are used (NURBS, subdivision surfaces, CSG …etc.), the objects are usually rendered using triangle approximations to the objects surfaces for performance reasons. Any speedup in the ray-triangle intersection algorithm is thus important as most models consist of thousands to millions of triangles.

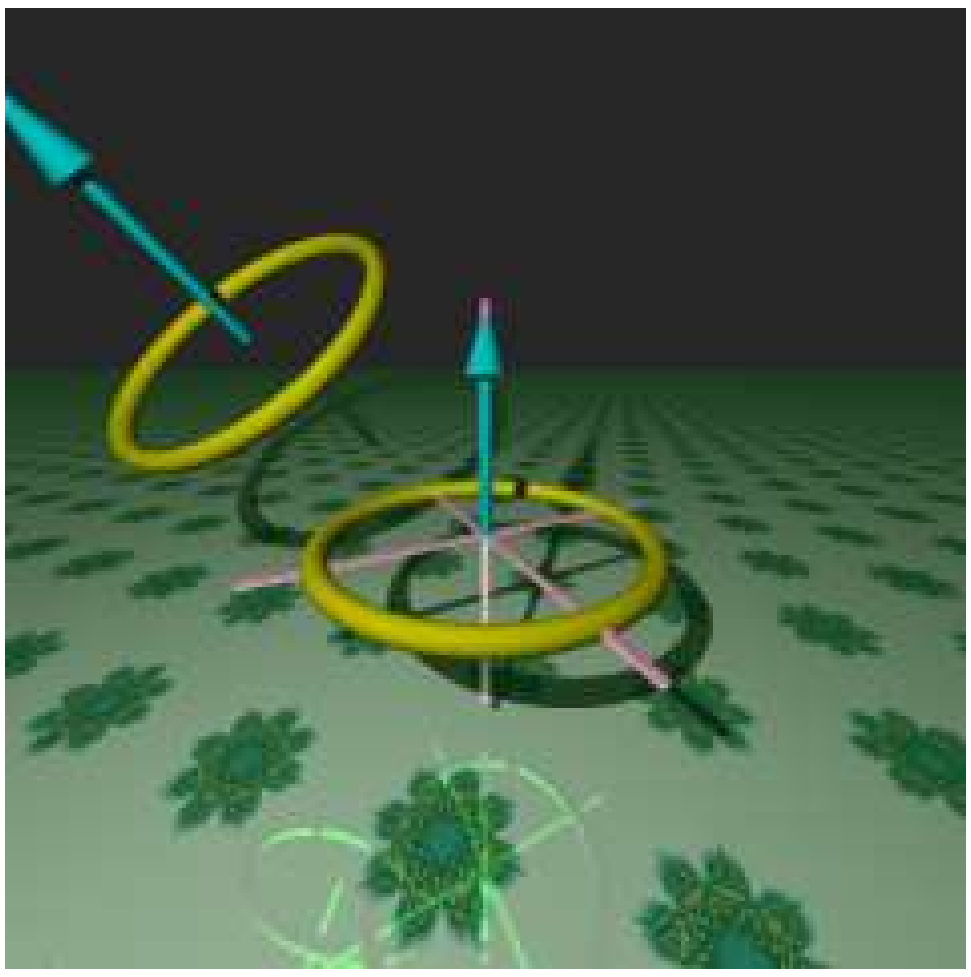

(a) Initial and final positions of rotationally-symmetric object

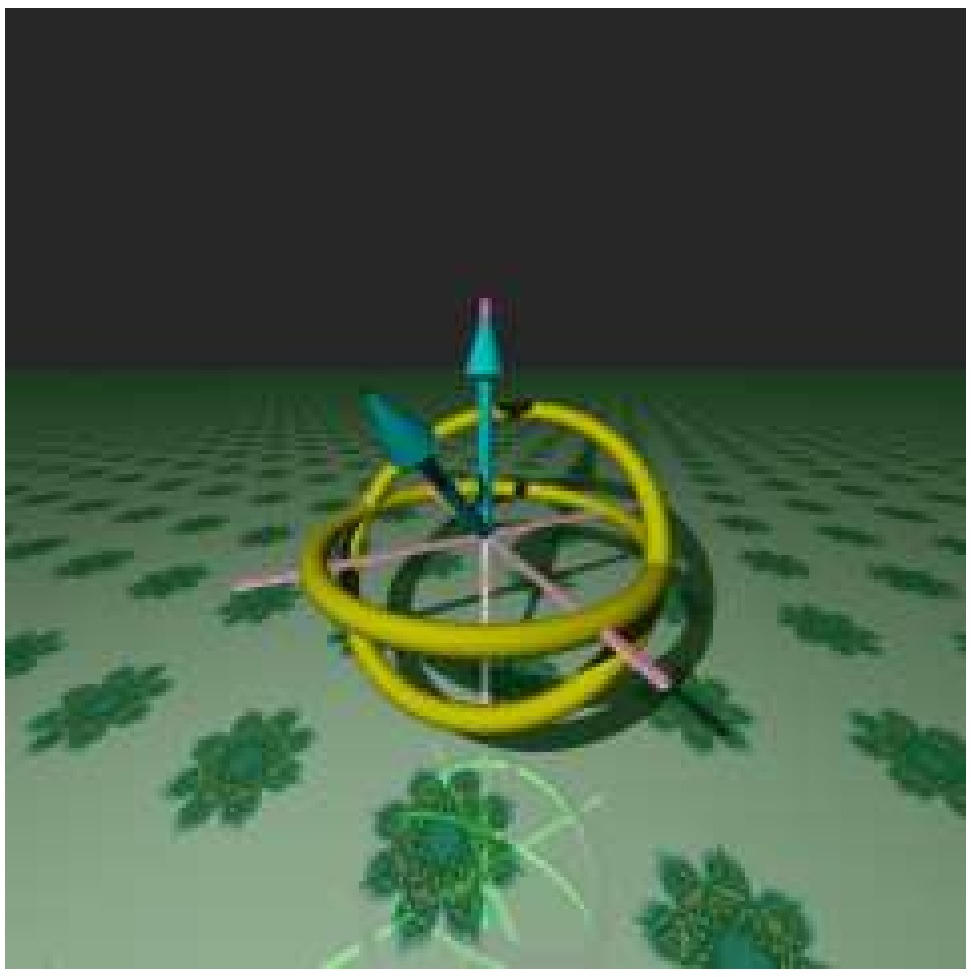

(b) Inverse translation applied to final object

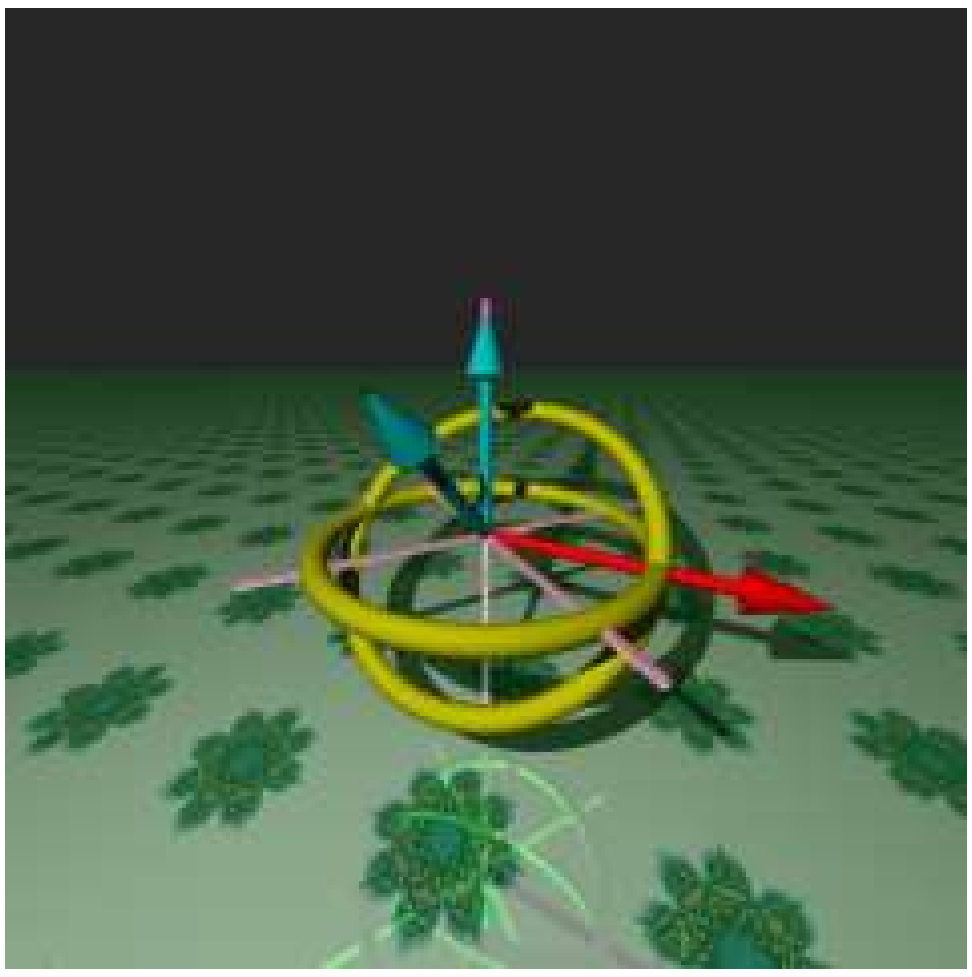

(c) Total rotation axis (red) for rotation transform

**Figure 6-26: Deduction of final translation and rotation in RSTRI transform**

The ray-triangle intersection algorithm implemented in the base ray tracer of chapter 5 is a direct import of the one in [41]. This test shall be called the algebraic hit test. The algebraic hit test first finds the intersection point of the ray with the plane containing the triangle. The algebraic test then tests if the hit point is inside or outside the triangle itself by calculating the barycentric coordinates $(\alpha_1, \alpha_2, \alpha_3)$ of the hit point. If all three coordinates lie in the range [0, 1], the hit point is inside the triangle. Barycentric coordinates are also used in rendering triangle meshes with smooth shading as discussed thoroughly in [41]. The algorithm of [41] is mainly based on manual algebraic manipulations of ray and triangle-related equations. The geometric significance is lost after one or two steps in algorithm deduction.

In [59], another type of ray-triangle intersection test is used. The algorithm is based on Plücker coordinates for lines and planes [11]. As a first step, the Plücker test determines if the ray passes between the three edges of the triangle; i.e. passes inside the triangle. If so, the algorithm finds the hit point with the plane of containing the triangle and possibly calculates the barycentric coordinates of the hit point if required.

Referring to Figure 6-27, if any of the three triangle edges, arranged in a cyclic order, is transformed with the ray so that the ray points to the up-direction in 3D space, the rotation induced by the edge around the ray is either clockwise or anti-clockwise. If all three rotations of the edges around the ray have the same handedness, the ray passes through the triangle; else it passes outside the triangle. Plücker coordinates provide an easy method to calculate the "signed" minimum distance between two skew lines in 3D space. The sign of the resulting distance is the same as the handedness just described.

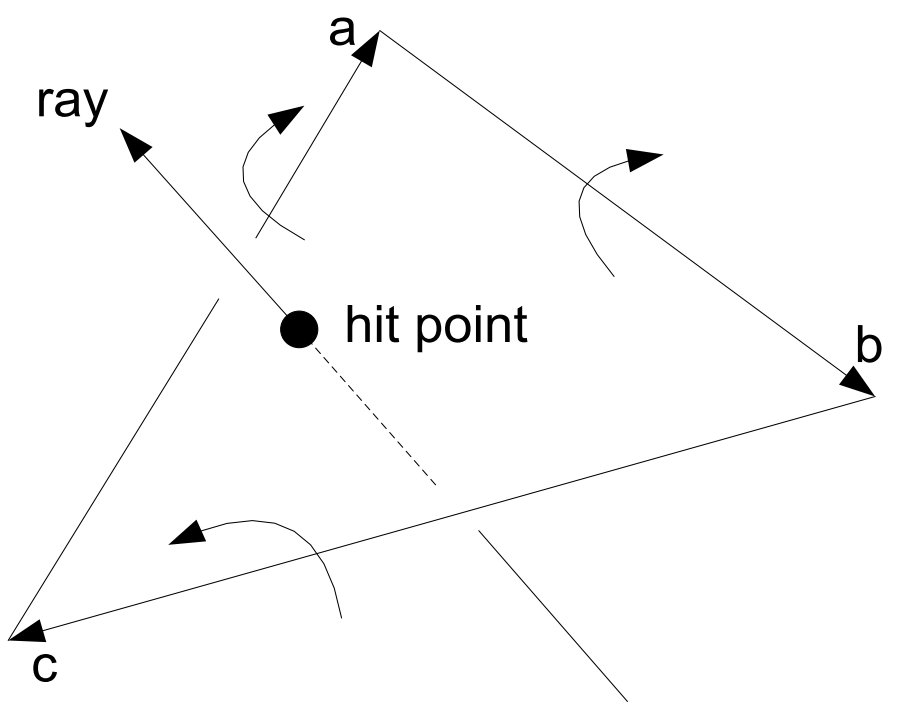

**Figure 6-27: Plücker ray-triangle intersection test**

The treatment of [59] uses no reference to GA or the natural interpretation of Plücker coordinates within the projective and conformal GA as discussed in [11]. The Plücker test was implemented in the base ray tracer of chapter 5 using GA concepts and GMac generated code. The performance results are shown in the next chapter.

Although the GA algorithm for the Plücker test can be written using CGA, the projective GA [11] having orthonormal basis vectors $e_1, e_2, e_3, e_o$ is better suited for such task. In this projective space, GA expressions for Plücker coordinates for lines and planes are simpler and more intuitive than their CGA counterparts. Assuming a triangle having vertices with position vectors $v_1, v_2, v_3$ and a ray with origin point having position vector $p_r$ and direction $v_r$ the following expressions are the Plücker test in the language of projective GA:

$$l_r = (p_r + e_o) \wedge v_r \tag{6.15}$$

$$v'_i = v_i + e_o \quad ; i = 1,2,3 \tag{6.16}$$

$$l_1 = v'_1 \wedge v'_2 \quad , l_2 = v'_2 \wedge v'_3 \quad , l_3 = v'_3 \wedge v'_1 \tag{6.17}$$

$$d_i = (l_r \wedge l_i)^* = (l_r \wedge l_i) \rfloor I^{-1} \quad ; i = 1,2,3 \tag{6.18}$$

$$\alpha_i = \frac{d_i}{d_1 + d_2 + d_3} \quad ; i = 1,2,3 \tag{6.19}$$

$$t = [v_1 \wedge v_2 \wedge (v_1 - p_r)][v_1 \wedge v_2 \wedge v_r]^{-1} \tag{6.20}$$

Where $l_r, l_i$ are the projective blades representing lines of the ray and the 3 edges respectively, $d_i$ are the signed minimum distances between $l_r$ and $l_i$, $(\alpha_1, \alpha_2, \alpha_3)$ are the barycentric coordinates of the hit point, $t$ is the ray parameter at the hit point, and $I = e_1 \wedge e_2 \wedge e_3 \wedge e_o$ is the space pseudo-scalar.

Two points make the Plücker algorithm of [59] much better than the algebraic one of [41]. The first point is that the Plücker test is a pure geometric algorithm; if viewed within the GA framework as discussed in [11]. The second point is illustrated in [59] as an intrinsic problem with the algebraic test. When testing a ray with an arbitrary triangle with the algebraic test, there is a high probability of the ray hitting the plane of the triangle while not hitting the triangle itself. This high probability results in unnecessary ray-plane intersection calculations. Thus slowing down the algebraic test of [41] relative to the Plücker test of [59].

# Chapter 7 : Experiments and Results

This chapter provides some comparisons to illustrate the effectiveness of using GA and GMac in several applications. Two separate sets of tests are required for such task as shown in Figure 7-1. The first set of tests aims at generating two software implementations using GMac and Gaigen 2 to compare their performance given the exact same GA algorithm. The second set of tests aims at comparing the performance of GA-based algorithms, implemented through GMac, compared to traditional geometric algorithms implemented manually. Thus the first set compares different implementations for the same algorithm while the second set typically compares different algorithms.

The chapter starts by section 7.1 that compares the performance of code generated by GMac and Gaigen 2 for several GA-based algorithms. Section 7.2 compares test results related to applying GA to ray tracing using GMac. Section 7.3 presents a summary of all results of the chapter.

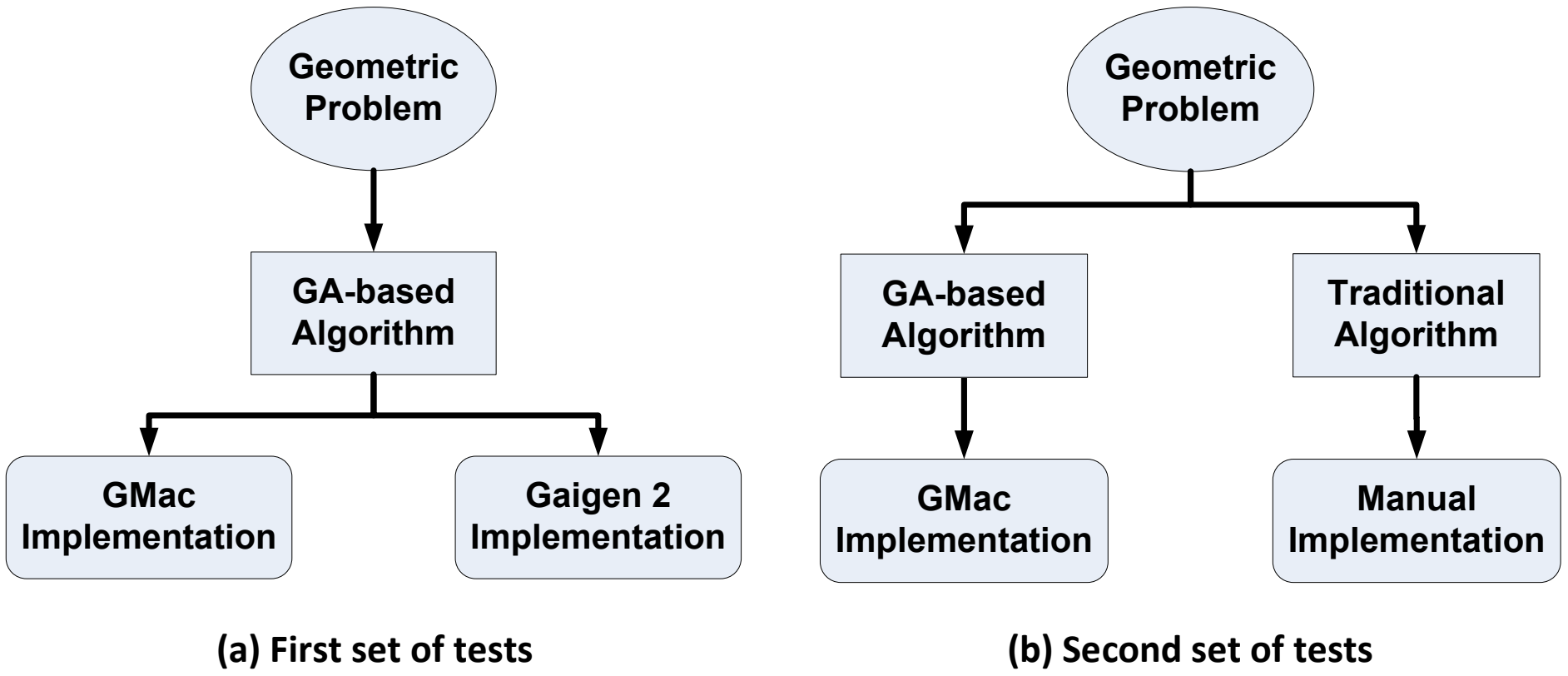

**Figure 7-1: Two sets of tests for testing GMac code performance**

## 7.1 GMac and Gaigen 2 Code Performance

In chapter 4, several architectural advantages of GMac over Gaigen 2 were presented. Such advantages result in superior code performance generated by GMac compared to Gaigen 2 generated code. In order to illustrate such performance gains, several GA applications were selected from [11] and [102]. The performance of GMac code was measured against the Gaigen 2 code used in [102]. In all cases, GMac code performed much better than Gaigen 2 code by a factor of 1.32 to 39.23 in terms of processing speed. In each application, the implementation using Gaigen 2 code was used. Only geometrically-critical sections were modified by generating code from GMac to replace Gaigen 2 code. Only such code sections were used for performance comparisons. No other part of the application implementation was modified or tested. The following subsections contain the details of the performance comparisons. Figure 7-2 contains a summary of the comparisons results. In each application, 25 trials were made to measure execution speed for GMac and Gaigen 2 generated code. An average execution time was calculated and the ratio between the two averages was taken as the speedup factor.

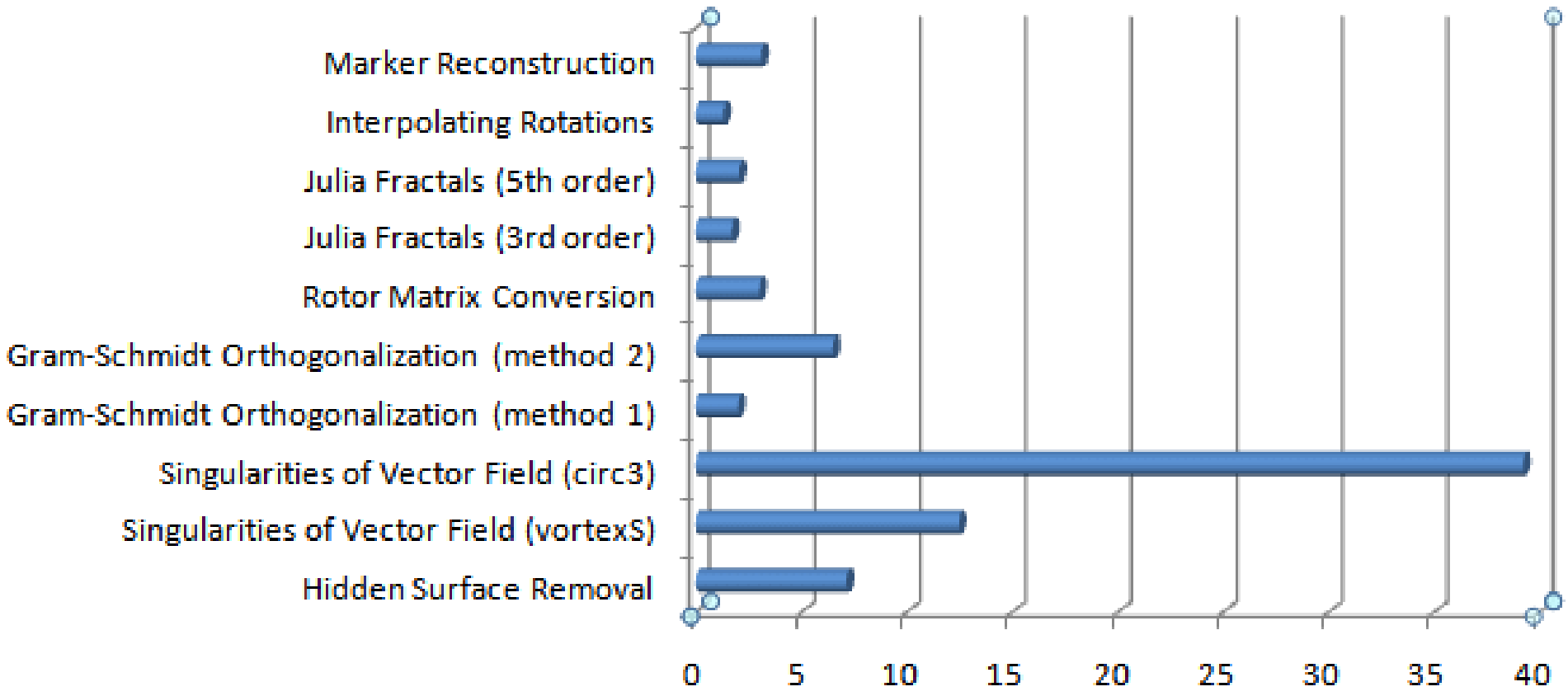

**Figure 7-2: Summary of GMac vs. Gaigen 2 code performance tests**

### 7.1.1 Hidden Surface Removal

In computer graphics, many rendering algorithms require that 3D models be built from a number of smaller polygons. Hidden surface removal is one of the standard problems in computer graphics [103], [82]. It involves determining whether a polygon (usually a triangle) should be rendered based on its orientation in space relative to the camera. Such query is used to reduce processing time during model rendering. In [11], the following GA expression is used to answer that query on a 2D triangle with vertices represented by 2D vectors $v_1, v_2, v_3$:

$$\langle (v_3 - v_1) \wedge (v_2 - v_1) \rangle_{e_1 \wedge e_2} \tag{7.1}$$

The 2D triangle is the projection of the 3D triangle on the viewing plane of the camera having basis vectors $e_1, e_2$ as described in detail in [11].

For this relation, GMac was able to generate code that executed faster than Gaigen 2 code by a factor of 7.15. The main reason behind the seven times faster execution is that GMac does not generate any additional function calls or class definitions like Gaigen 2 does. Instead, GMac only generates the necessary assignment expressions required for the problem as described in chapter 4.

### 7.1.2 Singularities in Vector Fields

As described in [11], a vector field is defined by a function that assigns a direction vector to every point in space. A singularity in a vector field occurs at any point where the function equals the zero vector. In vector field analysis, it is important to find the locations of such singularities. The algorithm implemented in [102] is described in detail in [11] and [104]. Only subroutines concerned with defining the vector field functions were re-generated by GMac.

A very notable disadvantage of using Gaigen 2 to define the vector field functions is apparent from the code in [102]. The performance of the Gaigen 2 code was very low that the author was forced to write the subroutines directly on the coordinates of the vector fields. The high-level geometric algebra implementation of Gaigen 2 was not fast enough to be usable in such application. The main reason behind such low performance is the complexity of the required GA expressions as seen in Table 7-1. This result indicates that Gaigen 2 scales poorly with increased complexity of GA expressions; contrary to GMac.

GMac, on the other hand, was capable of generating code that executed faster by factors of 12.48 and 39.23 for two of the vector field functions shown in Table 7-1. In the case of the circ3 function, the code generated by GMac was even faster than the manual code written by the author by a factor of 1.65.

**Table 7-1: Two Vector Field Functions**

| **Function** | **GA Expression** | **Speedup Factor** |
|---|---|---|
| vortexS | $v \leftarrow \omega(p \rfloor a)a^{-1} + \gamma[(p \wedge a)a^{-1}][a^{\odot}]$ | 12.48 |
| circ3 | $c_1 \leftarrow p \rfloor e_1,\quad c_2 \leftarrow p \rfloor e_2,\quad c_3 \leftarrow p \rfloor e_3,$ <br> $v \leftarrow 4(c_1^2 + c_2^2 - 1)c_1 e_1 + 4(c_1^2 + c_2^2 - 1)c_2 e_2 + 2c_3 e_3$ | 39.23 |

In Table 7-1, $p$ is the vector representing a point in 3D Euclidean space, $v$ is the output vector of the function, $e_1, e_2, e_3$ are the basis vectors of the space, $a$ is a constant vector, $\omega, \gamma$ are two real numbers.

### 7.1.3 Gram-Schmidt Orthogonalization

Another common geometric problem in computer graphics is the orthogonalization of three linearly independent vectors in 3D Euclidean space. There are two approaches for solving such problem using GA presented in [11]. The first is by using the following relations:

$$\begin{aligned} &w_1 \leftarrow v_1, \\ &w_2 \leftarrow v_1 \rfloor (v_1 \wedge v_2), \\ &w_3 \leftarrow (v_1 \wedge v_2) \rfloor (v_1 \wedge v_2 \wedge v_3) \end{aligned} \tag{7.2}$$

The other approach depends on the following relations:

$$\begin{aligned} &w_1 \leftarrow v_1, \\ &w_2 \leftarrow (v_1 \wedge w_1) w_1^{-1}, \\ &w_3 \leftarrow (v_3 \wedge w_1 \wedge w_2)(w_1 \wedge w_2)^{-1} \end{aligned} \tag{7.3}$$

Where $v_1, v_2, v_3$ are the three input LID vectors and $w_1, w_2, w_3$ are the three output orthogonal vectors.

GMac generated code that performed 2.01 times faster than Gaigen 2 code for the first approach. In addition, GMac code was faster than Gaigen 2 code by a factor of 6.49 for the second approach.

### 7.1.4 Rotor-Matrix Conversion

In traditional linear algebra, the usual tool for rotating vectors is a rotation matrix. A rotation matrix is a square matrix having orthonormal column vectors. A much better alternative provided by GA is the rotor multivector. A rotor is a special type of multivector capable of rotating any blade including vectors, planes …etc. Unfortunately, many computer graphics packages like OpenGL rely on rotation matrices. A method must be used to convert back and forth between a rotor multivector and a rotation matrix in order to link GA-based code with such libraries.

In [11] two methods are given for such conversion from rotor to matrix. The first method uses Gaigen 2 generated code to perform this operation. The second is a manual approach based on the symbolic simplification of the first method by hand. The first method deduces the columns of the rotation matrix $\mathbf{M}$ by applying the normalized rotor $R$ to the three basis vectors of 3D Euclidean space $e_1, e_2, e_3$ as follows:

$$\begin{aligned} v_1 &\leftarrow R\, e_1 \tilde{R}, \\ v_2 &\leftarrow R\, e_2 \tilde{R}, \\ v_3 &\leftarrow R\, e_3 \tilde{R}, \\ \mathbf{M} &\leftarrow [v_1 v_2 v_3] \end{aligned} \qquad (7.4)$$

The second method manually deduces the items of the rotation matrix using the coordinates of the rotor multivector and the basis vectors as follows:

$$\begin{aligned} v_1 &= R\, e_1 \tilde{R} \\ &= (w^2 + x^2 - y^2 - z^2)e_1 + (-2wz + 2xy)e_2 + (2wy + 2xz)e_3 \end{aligned} \qquad (7.5)$$

$$\Rightarrow v_1 = (1 - 2x^2 - 2z^2)e_1 + (-2wz + 2xy)e_2 + (2wy + 2xz)e_3 \qquad (7.6)$$

Where:

$$R = w + xe_2e_3 + ye_3e_1 + ze_1e_2; \qquad w^2 + x^2 + y^2 + z^2 = 1 \tag{7.7}$$

A similar relation can be used to find $v_2, v_3$ and then the three column vectors are combined to form the desired rotation matrix $\mathbf{M}$.

GMac was able to correctly deduce expressions equivalent to relation (7.5) but not relation (7.6). The reason behind that is the normalization constraint on the rotor $R$ given as $w^2 + x^2 + y^2 + z^2 = 1$. In the current implementation of GMac such constraints on multivector classes are not supported. Fortunately, Mathematica supports such constraints and adding such capability to GMac should not be difficult in future implementations.

The code generated by GMac performed faster than Gaigen 2 code by a factor of 3.02 while performing slower than the manual code by a factor of 0.4 due to the absence of the final simplification step of relation (7.6). The manual code performed faster than Gaigen 2 code by a factor of 7.55.

### 7.1.5 Julia Fractals

Geometric algebra subsumes complex numbers as explained in [11]. One of the interesting applications of complex numbers and complex analysis is fractal geometry [105]. An example application for generating the Julia fractals using geometric algebra instead of complex numbers is given in [11]. The implementation relies on recursive expressions of the form:

$$x \leftarrow xexex + c \tag{7.8}$$

$$x \leftarrow xexexexex + c \tag{7.9}$$

Where $x$ is a 2D variable vector representing a pixel in the image plane, $e$ is the first basis vector of the 2D plane, and $c$ is a constant 2D vector. The first relation generates Julia fractals of the 3rd degree while the second generates a 5th degree Julia fractal.

Using GMac, the performance was speed up by a factor of 1.72 for the first relation, and a factor of 2.05 for the second compared to Gaigen 2 code.

### 7.1.6 Interpolating Rotations

In geometric algebra, a rotor can be used to perform rotations of arbitrary blades. In 3D Euclidean space, any rotor $R$ can be expressed as an exponential of a bivector $B$ where the exponential function is defined for this particular case as follows:

$$R = e^{B} = \cos(\phi) + \frac{1}{\phi}\sin(\phi)B \qquad (7.10)$$
$$;\phi = \sqrt{-B^{2}}$$

An inverse logarithm function can thus be defined to obtain a bivector from the rotor as follows:

$$B = \frac{R_{\langle 2\rangle}}{\left\| R_{\langle 2\rangle} \right\|} \arctan\left( \frac{\left\| R_{\langle 2\rangle} \right\|}{\left\| R_{\langle 0\rangle} \right\|} \right) \qquad (7.11)$$

Using these two functions and assuming having two rotors $R_1, R_2$ any intermediate rotor can be found using the following relation:

$$R_{\alpha} = R_1 e^{\alpha \log(R_2 R_1^{-1})} \qquad (7.12)$$

This relation enables the interpolation of two rotations using a simple real parameter $0 \le \alpha \le 1$ as described in detail in [11].

An illustration of such interpolation is implemented in [102]. The code generated by GMac performed faster than Gaigen 2 code by a factor of 1.32.

### 7.1.7 Marker Reconstruction in Optical Motion Capture

One of the important problems in computer vision is to reconstruct the 3D motion of an object using a set of calibrated cameras [106]. One technique to solve such problem is by using physical markers on the object to facilitate the reconstruction process especially for non-rigid objects like the human body. This technique is used in [11] for marker reconstruction in an optical motion capture application. One of the repeatedly executed operations in such algorithm is to compute the closest points on two skew lines in 3D Euclidean space. The algorithm then computes a scalar value for each line representing the parameter value of that line corresponding to one of the two closest points.

Assuming the two lines having direction vectors $D_1, D_2$ and passing through the two points represented by vectors $P_1, P_2$ and having scalar parameters $t_1, t_2$ respectively; then the following GA algorithm can be used to find the values of $t_1, t_2$ corresponding to the two closest points:

$$\begin{aligned} D &\leftarrow P_2 - P_1, \\ I &\leftarrow (D_1 \wedge D_2)^{-1}, \\ t_1 &\leftarrow (D_2 \rfloor I) \rfloor D, \\ t_2 &\leftarrow (D_1 \rfloor I) \rfloor D \end{aligned} \tag{7.13}$$

Where the points themselves can be calculated using the parametric equations of the two lines:

$$\begin{aligned} p_1 &= P_1 + t_1 D_1, \\ p_2 &= P_2 + t_2 D_2 \end{aligned} \tag{7.14}$$

For this problem, GMac generated code performing faster by a factor of 3.1 compared to the Gaigen 2 code used in [102].

## 7.2 Ray Tracing Performance

This section illustrates some performance comparisons related to ray tracing. The first subsection illustrates the performance of TSRI and RSTRI techniques compared to traditional matrix-based instancing. The second subsection compares the performance of the algebraic ray-triangle intersection test of [41] to the Plücker test of [59] implemented through GA and GMac.

### 7.2.1 Instancing Performance

The four scenes, shown in Figure 7-3, were used to compare the performance of different instancing techniques. Cylinders were the main primitive to be used in all scenes. A cylinder in general position and orientation in space can be fully geometrically defined by three quantities: the center point of its base circle, its axis of symmetry (defining its orientation and height) and its radius. Five methods were used to render cylinders in each of the four scenes. The first method (U-Cyl) is a manual implementation based on [77] for describing any general circular cylinder in space. The second method (GAU-Cyl) is based on the impeding of RSTRI into a generic cylinder around the y-axis to get the most efficient ray-cylinder intersection code possible using the RSTRI technique of chapter 6. The third method (I-Cyl) is based on the instancing technique of [41] applied to a generic cylinder. The fourth method (TSRI-Cyl) is based on the TSRI technique of chapter 6 applied to a generic cylinder. The final method (RSTRI-Cyl) is based on the RSTRI technique of chapter 6 applied to a generic cylinder.

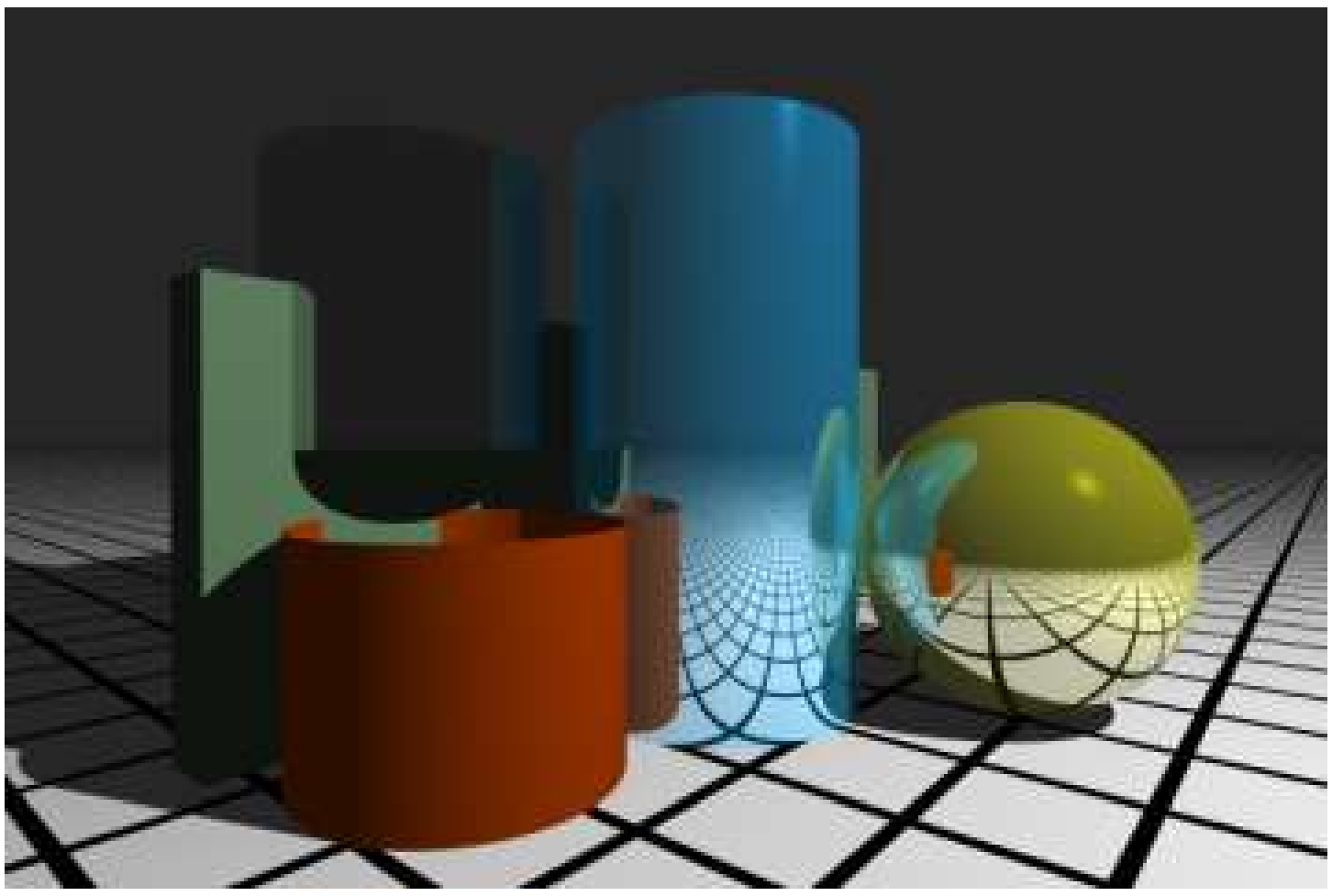

**(a) Reflective Blocks**

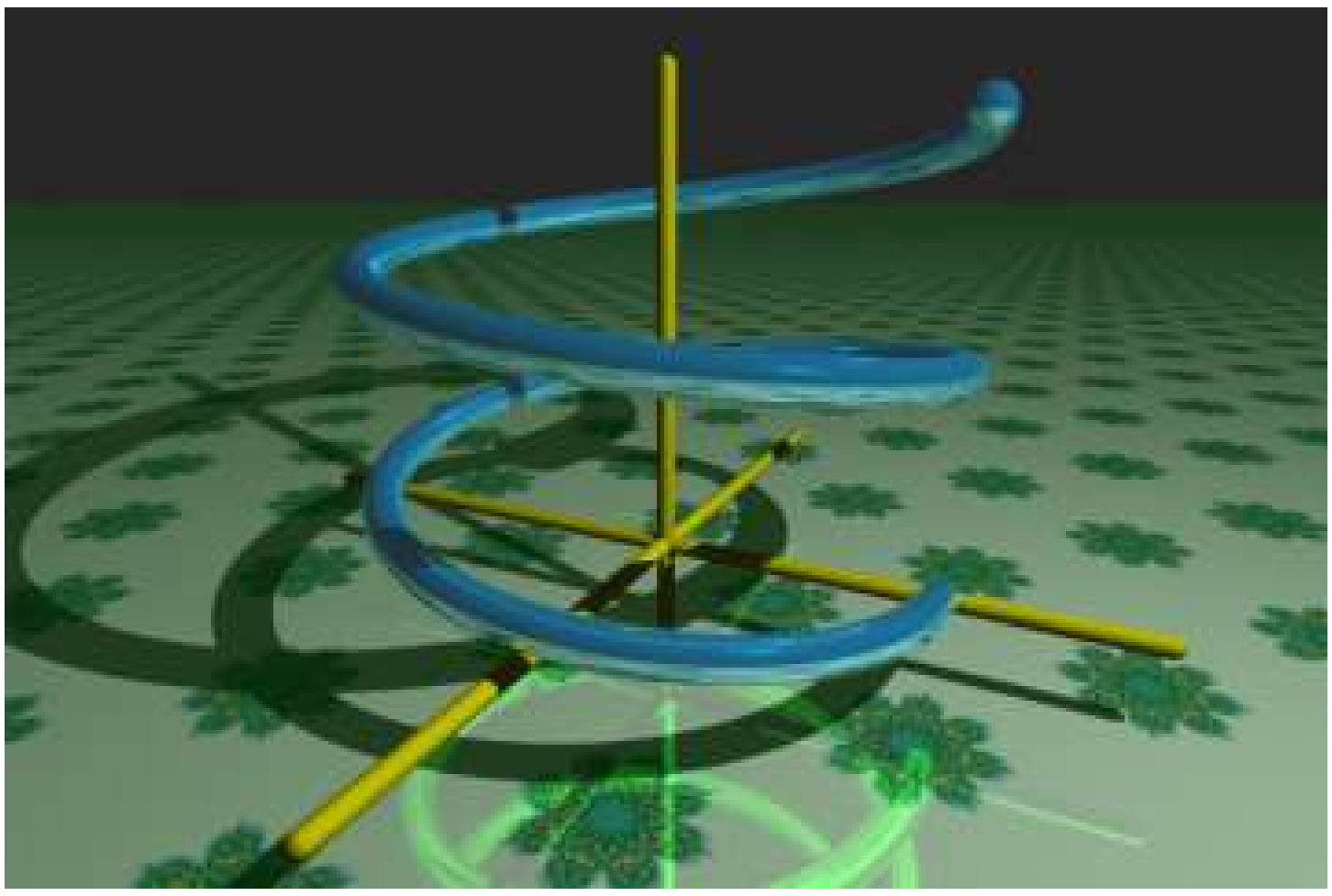

**(b) Helix**

**Figure 7-3: Test scenes for instancing performance**

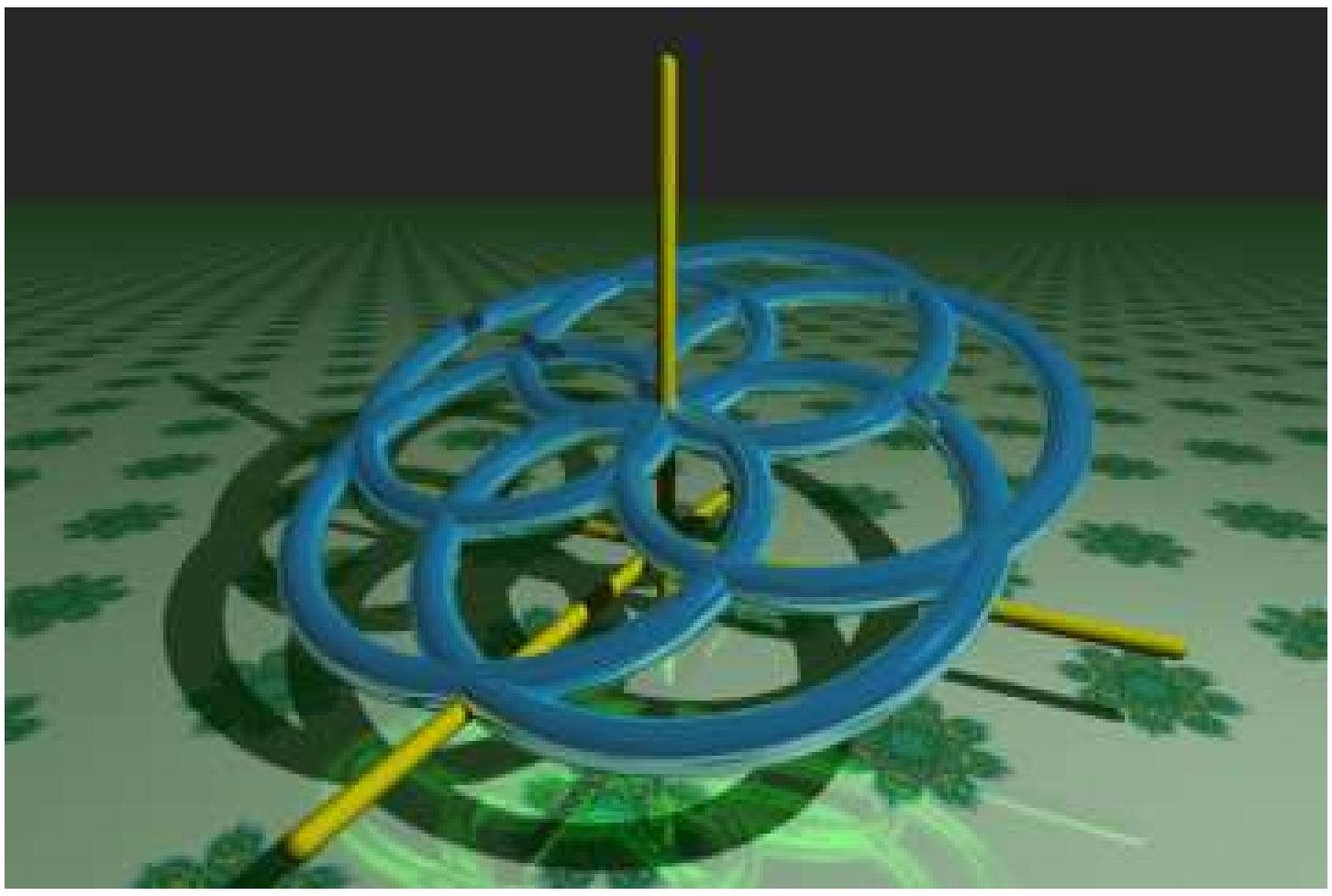

**(c) Epitrochoid**

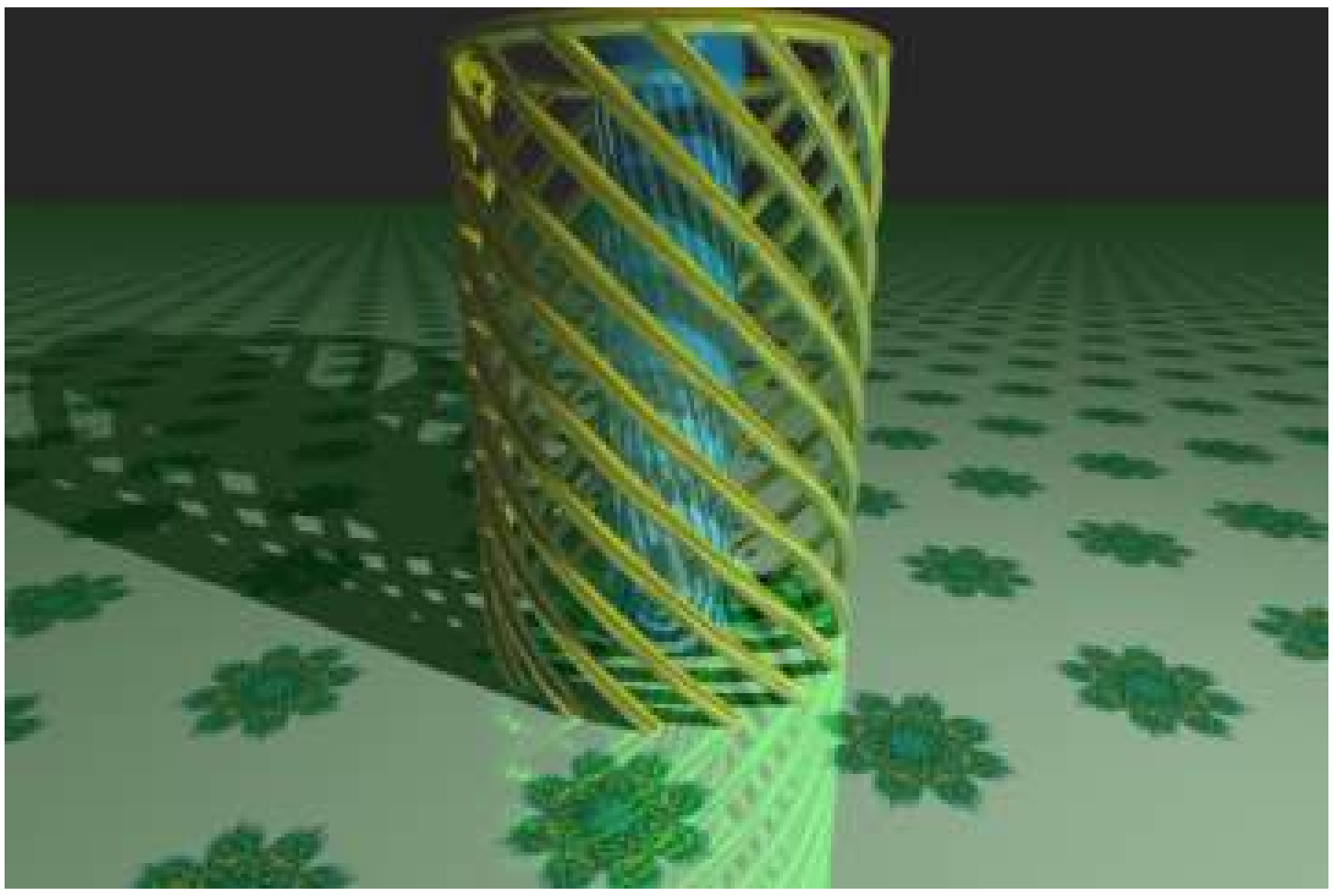

**(d) Twisted Column**

**Figure 7-3 (continued): Test scenes for instancing performance**

In summary, there are 20 tests based on 4 cylinder-based scenes using 2 types of universal cylinders and 3 types of instancing generic cylinders. The first test scene contains 3 large cylinders with different types of materials. The other three scenes are generated using the GGPR for curves and the Twister library described in chapter 6. Any typical GGPR curve contains a number of smaller cylinders to render its pricewise-connected line segments approximating the curve. The results of the 20 tests are shown in Table 7-2, Figure 7-4, Figure 7-5,

Figure 7-6, and Figure 7-7. The tests were performed on an Intel Pentium Dual CPU T2330 1.6 GHz machine with 1 GBytes of RAM. Each test was performed 5 times to measure an average for the total test time for each scene.

**Table 7-2: Performance comparison of 5 methods for rendering cylinders in general positions**

| | | **Blocks** | **Helix** | **Epitrochoid** | **Column** |
|---|---|---|---|---|---|
| **Total Cylinders** | | 3 | 333 | 772 | 1,193 |
| **Total Rays** | | 507,032 | 649,117 | 672,938 | 771,665 |
| **Total Test Invocations** | | 1,480,284 | 3,448,193 | 6,135,036 | 1,724,006 |
| **Total Test Time (msec.)** | U-Cyl | 625 | 727 | 2,363 | 720 |
| | GAU-Cyl | 112 | 201 | 354 | 151 |
| | I-Cyl | 876 | 1,934 | 3,364 | 999 |
| | TSR-Cyl | 781 | 1,703 | 2,949 | 901 |
| | RSTR-Cyl | 745 | 1,620 | 2,899 | 895 |
| **Total Instance\Cylinder Memory (Kbytes)** | U-Cyl | 5.765 | 54.633 | 126.656 | 195.727 |
| | GAU-Cyl | 6.047 | 85.852 | 199.031 | 307.570 |
| | I-Cyl | 6.703 | 176.906 | 410.125 | 633.781 |
| | TSR-Cyl | 6.586 | 163.898 | 379.968 | 587.180 |
| | RSTR-Cyl | 6.422 | 145.688 | 337.75 | 521.938 |

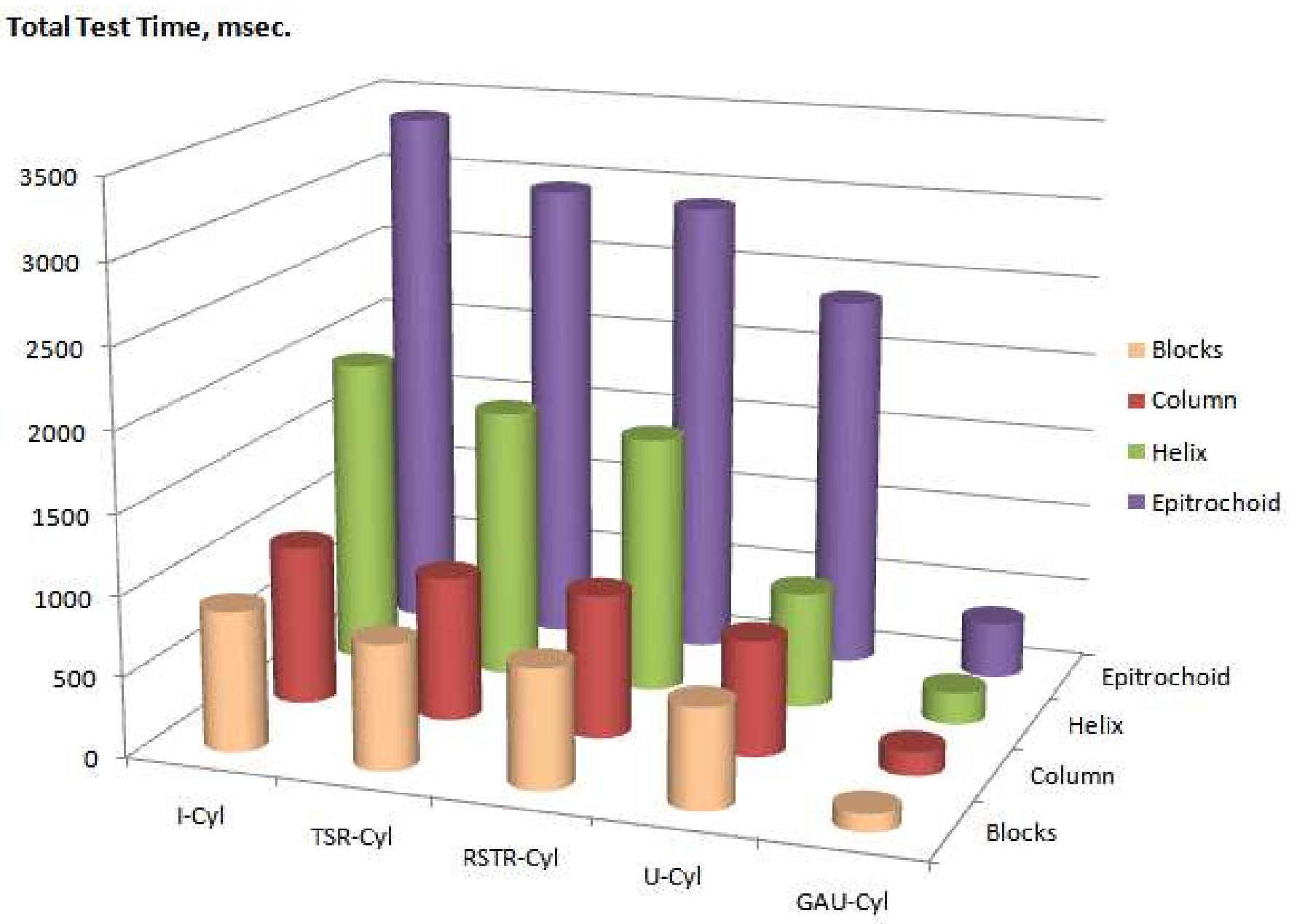

**Figure 7-4: Processing requirements in msec. for all five methods in four test scenes**

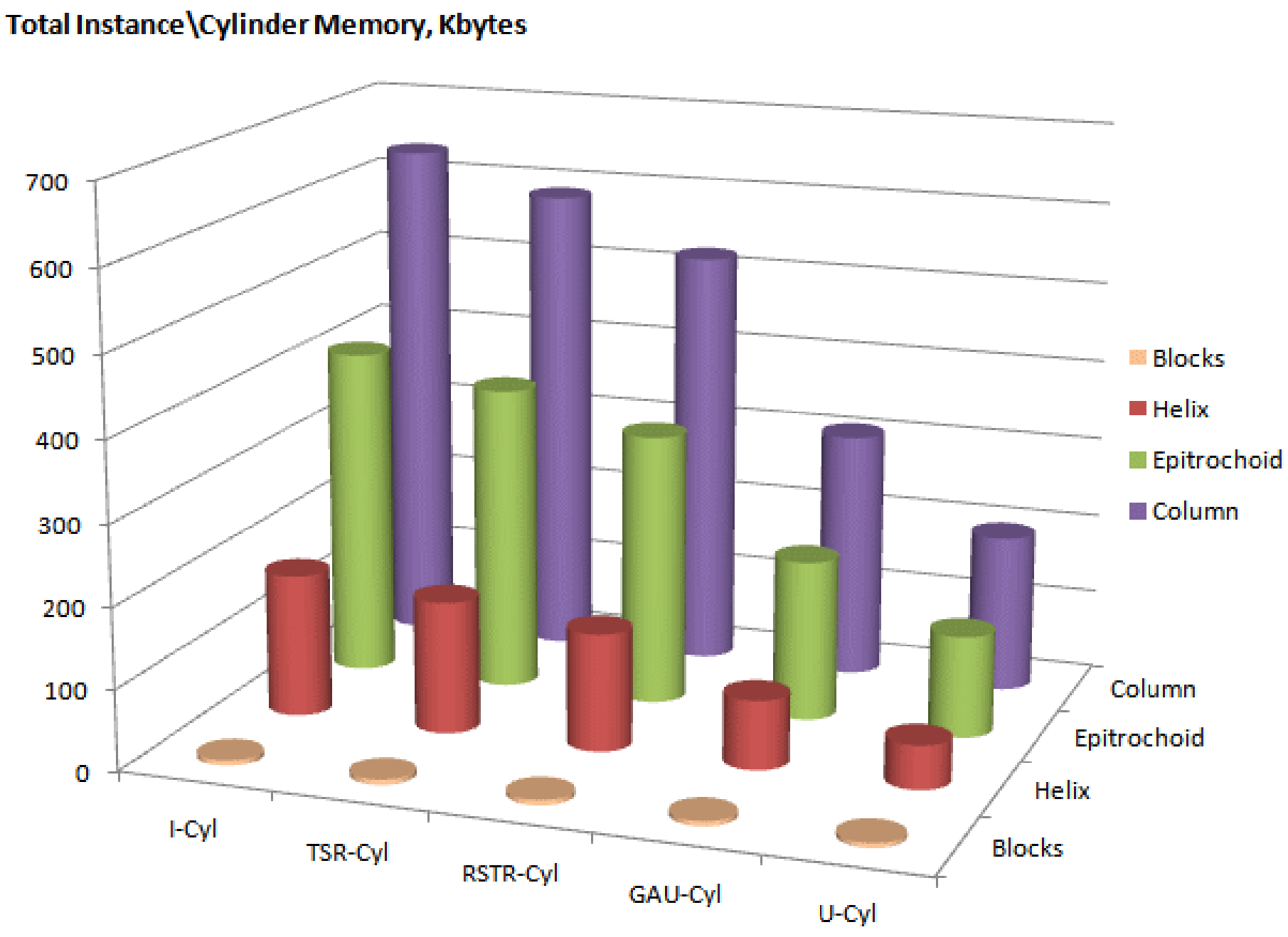

**Figure 7-5: Memory requirements in Kbytes for all five methods in four test scenes**

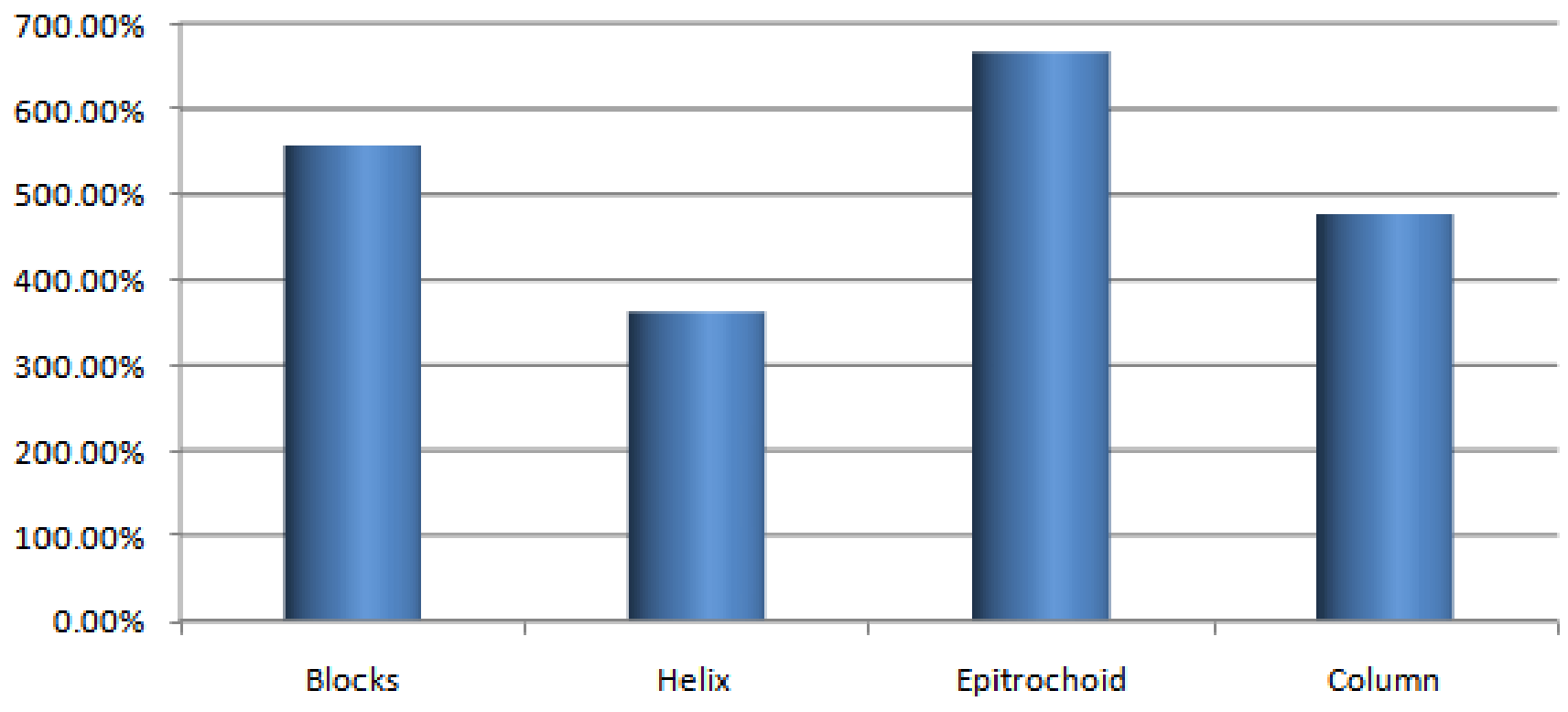

**Figure 7-6: Percentage of performance gain for using GAU-Cyl compared to U-Cyl**

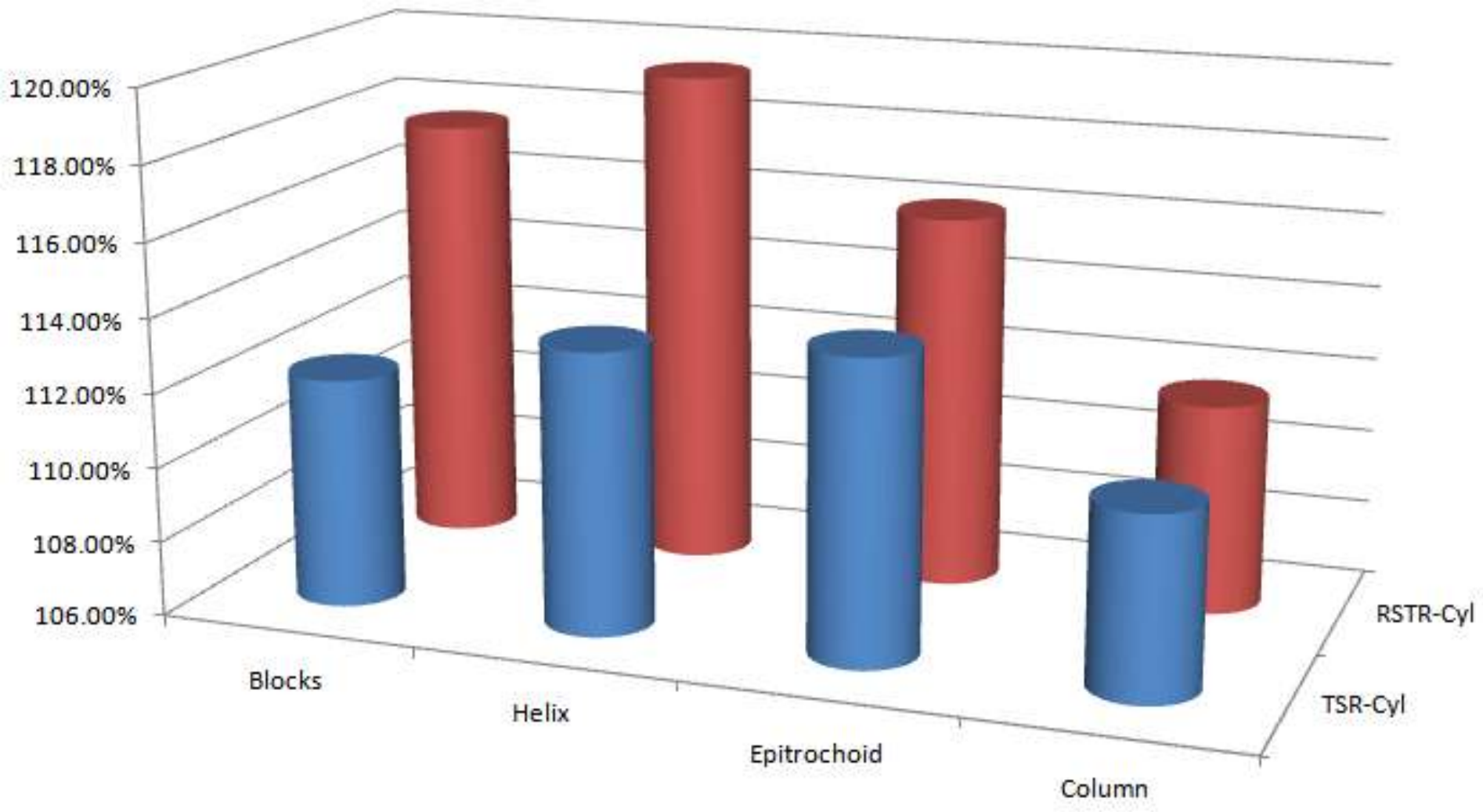

**Figure 7-7: Percentage of performance gain for using TSR-Cyl and RSTR-Cyl compared to I-Cyl**

The results illustrate the advantages and disadvantages of each method. The U-Cyl method is difficult to implement as it relies on manual deduction and coding requiring traditional vector algebra manipulations. The U-Cyl is specifically tailored to cylinders and cannot be directly generalized to any other primitive. The main advantage of the U-Cyl method is its low memory requirements compared to other methods.

The GAU-Cyl method is based on GA and implemented using GMac. The GAU-Cyl method is much easier to understand and implement than the U-Cyl method. The GAU-Cyl method has the best performance among all methods and comes second to U-Cyl in its memory requirements. The GAU-Cyl method can be extended to other rotationally-symmetric primitives with a small amount of manual coding. The main limitation of the GAU-Cyl method is its limitation to rotationally-symmetric primitives only. The GAU-Cyl method is not suitable for primitives like polygons and cubes for example.

The I-Cyl method is the most general among all methods. The I-Cyl method is based on matrices that can represent all projective transforms [106]. Thus, the I-Cyl method can represent all translated, rotated, sheared, and non-uniformly scaled primitives. In addition, the I-Cyl method can be used on “groups” of objects at once; like whole triangular meshes. The generality of the I-Cyl method comes with the price of being the most demanding in its memory and processing requirements. The I-Cyl method is not suitable for simple primitives like cylinders.

The TSR-Cyl method has most of the advantages of the I-Cyl method with a good reduction in memory and processing requirements. The TSR-Cyl method is much easier to design and implement since it relies on GA and GMac with no matrices involved. The TSR-Cyl can be applied to any translated, rotated, and uniformly scaled primitive or group of primitives.

For rotationally-symmetric objects, the RSTR-Cyl method is very similar to TSR-Cyl with reduced memory and processing requirements. The main limitation of RSTR-Cyl compared to TSR-Cyl is that it is limited to rotationally-symmetric objects only. The main advantage of RSTR-Cyl compared to GAU-Cyl is that it can be used directly without any manual modification to objects. The disadvantage being less efficient in processing and memory requirements compared to GAU-Cyl.

### 7.2.2 Ray-Triangle Intersections

In order to compare the performance of the traditional algebraic (A) and GA Plücker (P) ray-triangle intersections tests, five scenes based on triangle meshes were used as shown in Figure 7-8. The scenes contain several effects common to ray tracing. Such effects include transparent materials, shadows, reflective materials, and smooth triangle shading through normal interpolation [41]. The acceleration data structure used in all scenes is the BIH [45]. The results are summarized in Table 7-3 and Figure 7-9.

In Table 7-3 the rows having an "A" in the second column are for the algebraic test, the rows with a "P" are for the Plücker test, and the rows with the "%" are calculated as (1 - P/A) %. The tests were performed on an Intel Pentium Dual CPU T2330 1.6 GHz machine with 1 GBytes of RAM. Each test was performed 5 times to measure an average for the total test time for each scene.

Several observations can be noted from Table 7-3. First, the bunny and hand scenes have the largest number of rays. This is primarily due to using transparent material resulting in a large number of secondary rays. Second, in some scenes there is a small difference between the total numbers of triangle tests in both algorithms (total test invocations). This difference may come from the difference in round off errors in both algorithms. Such difference results in some small percentage of rays hitting triangles near their edges to pass or fail the tests at different rates in both algorithms. Third, in all scenes the GA Plücker test performed nearly as the algebraic test as shown in Figure 7-9. The main point here is that the Plücker test is generated automatically from its geometric description. On the other hand, the algebraic test is manually deduced and coded. The large reduction in design effort highly outweighs the small performance difference in some test scenes.

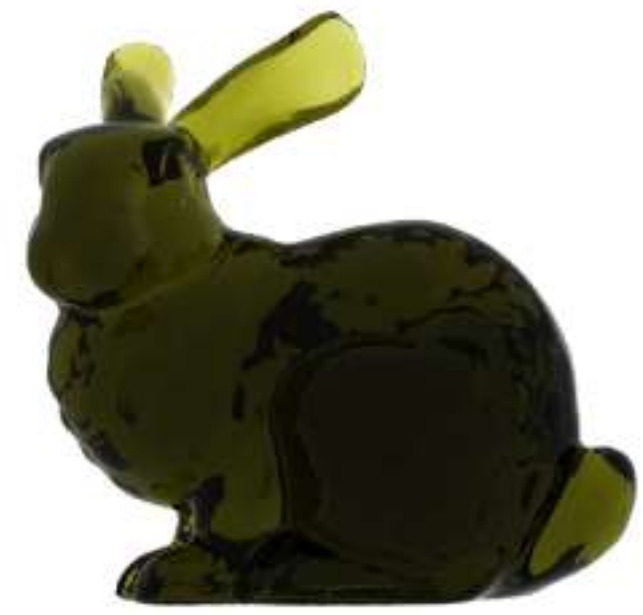

(a) Transparent bunny

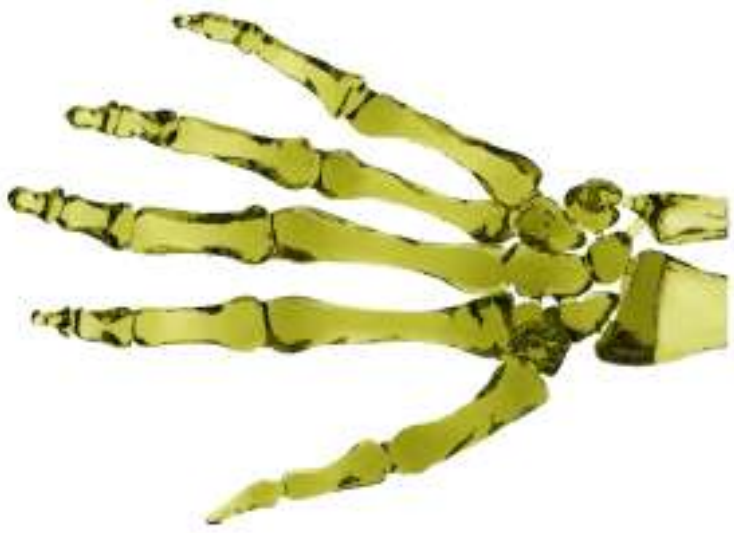

(b) Transparent hand

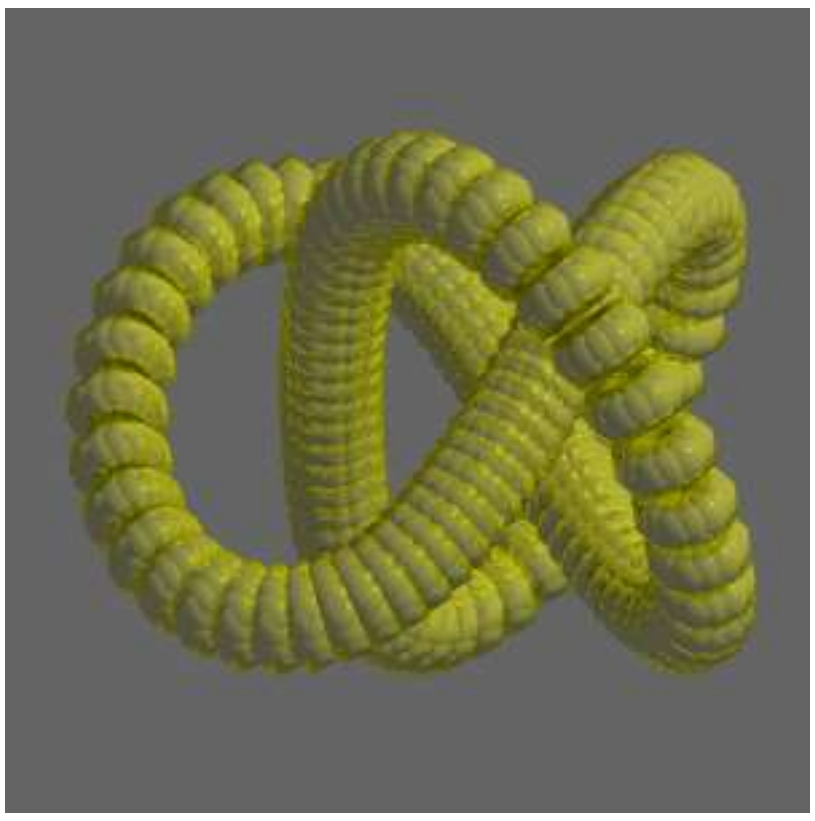

(c) Pyramob

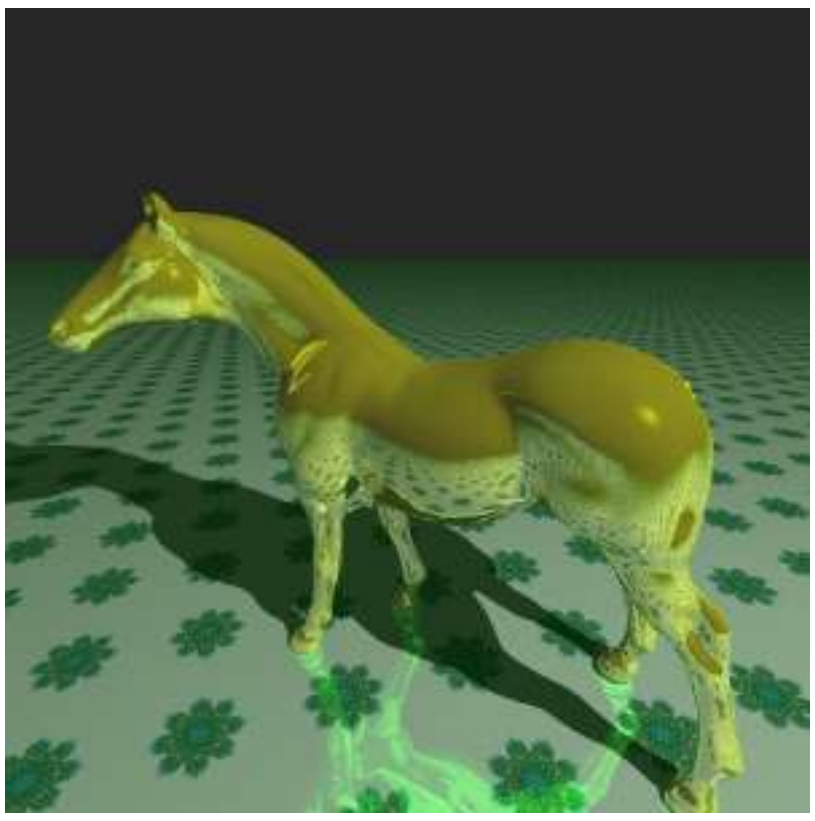

(d) Horse

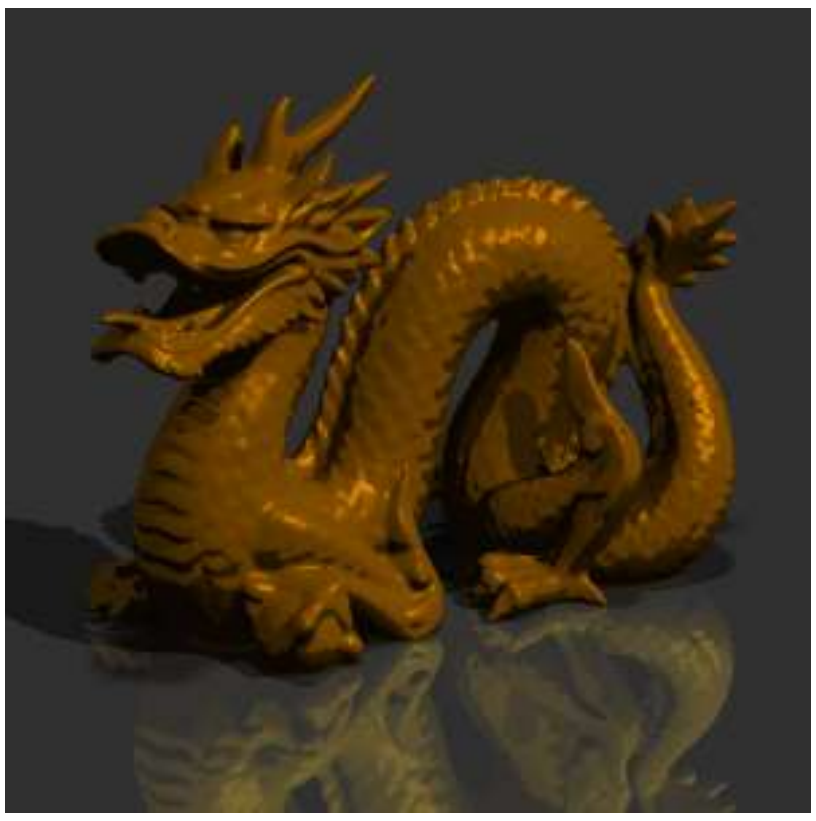

(e) Dragon

**Figure 7-8: Test scenes for ray-triangle intersection performance**

**Table 7-3: Performance comparison of algebraic and GA-based Plücker tests**

| | | Bunny | Hand | Pyramob | Horse | Dragon |
|---|---|---|---|---|---|---|
| **Triangles** | A | 69,451 | 654,666 | 31,460 | 96,966 | 871,414 |
| | P | | | | | |
| **Total Rays** | A | 464,588 | 589,078 | 173,355 | 136,779 | 21,730 |
| | P | | | | | |
| **Total Test Invocations** | A | 9,786,047 | 13,100,290 | 2,650,311 | 615,810 | 178,043 |
| | P | 9,786,020 | | | | 177,847 |
| **Total Test Time (msec.)** | A | 7,142 | 9,528 | 2,046 | 463 | 141 |
| | P | 7,056 | 9,284 | 2,071 | 472 | 131 |
| | % | 1.2% | 2.6% | -1.2% | -2% | 7.1% |
| **Total Triangles Memory (Mbytes)** | A | 9.013 | 84.910 | 4.226 | 10.357 | 113.017 |
| | P | | | | | |

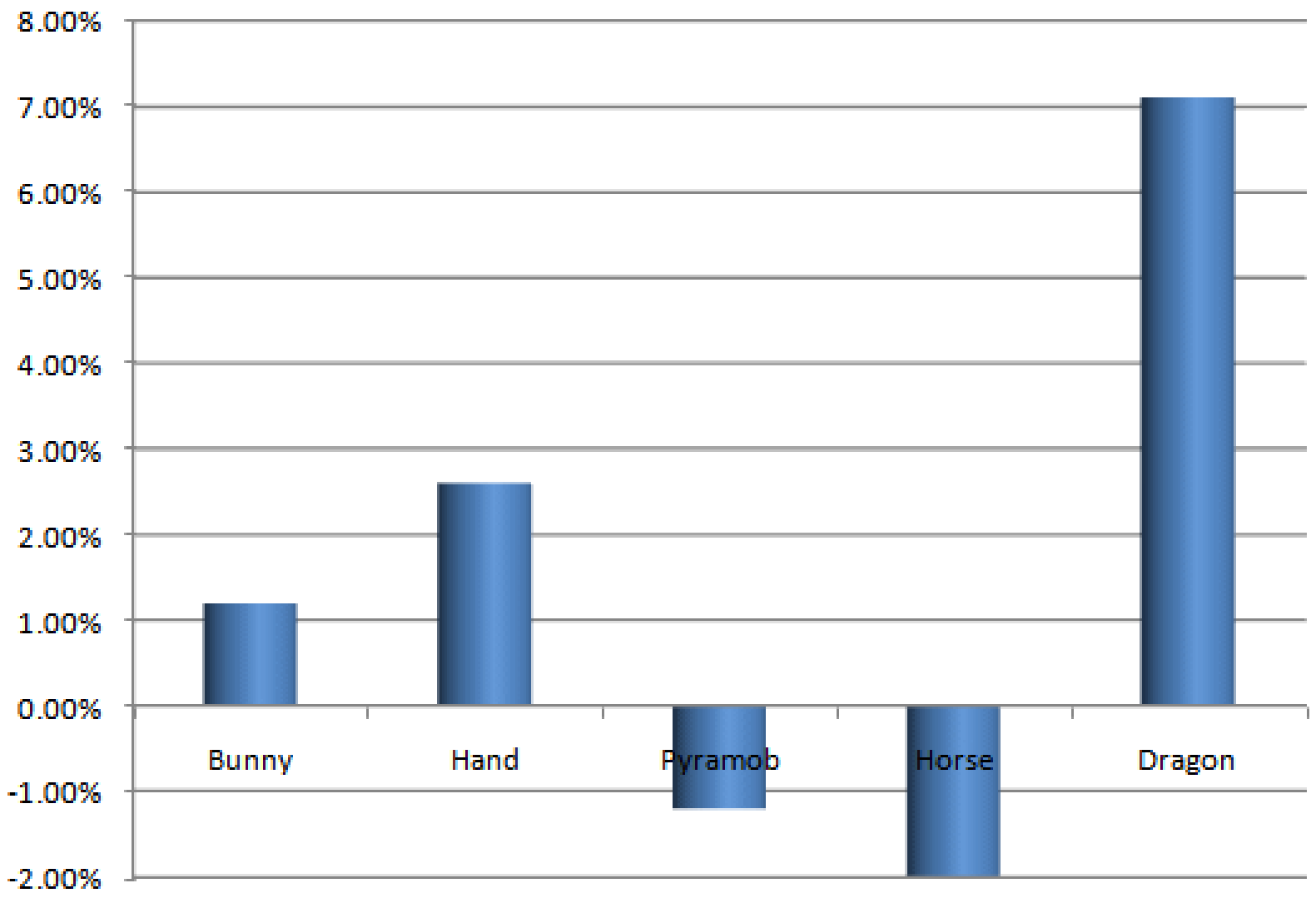

**Figure 7-9: Percentage of performance gain for the Plücker test compared to the algebraic test**

## 7.3 Summary of Results

This section presents a summary of all results presented in this chapter. Table 7-4 contains a summary of important features related to the performance of Gaigen 2 and GMac generated code. GMac code is clearly better than Gaigen 2 code in all aspects of processing and memory requirements.

Table 7-5 contains a summary of advantages and disadvantages for each of the five ray-cylinder intersection techniques presented earlier. From the comparisons the benefits of thinking in geometric algebra are apparent. Geometric algebra can provide much better alternatives suitable for different situations as in the case of TSRI and RSTRI techniques. In addition, implementing such alternatives using GMac results in efficient code. Such code possesses suitable memory requirements and good speed compared to traditional techniques.

Table 7-6 compares the traditional and GA-based ray-triangle intersection techniques. The memory and speed performance of the two techniques is very similar. The main advantage of the GA-based algorithms is its ease of design and implementation compared to the traditional test. Thus, geometric algebra and GMac can provide ease of design and implementation for geometric algorithms while retaining similar or better performance of traditional approaches.

**Table 7-4: Comparison of features of Gaigen 2 and GMac generated code**

| | Gaigen 2 | GMac |
|---|---|---|
| **Performance of Generated Code** | Low | High |
| **Scaling with More Complex GA Expressions** | Poor | Much Better |
| **Memory Requirements for Generated Code** | Higher | Low |

**Table 7-5: Comparison of five ray-cylinder intersection techniques**

| | U-Cyl | GAU-Cyl | I-Cyl | TSR-Cyl | RSTR-Cyl |
|---|---|---|---|---|---|
| **Generality** | Specific for cylinders in general position and orientation | Specific for cylinders in general position and orientation | Euclidean Transforms, Uniform and non Uniform Scaling of any Object | Euclidean Transforms and Uniform Scaling of any Object | Euclidean Transforms of Rotationally-Symmetric Objects only |
| **Traditional\GA based** | Traditional | GA | Traditional | GA | GA |
| **Speed** | Much slower than GAU-Cyl | Fastest | Slowest | Slower than U-Cyl | Slower than TSR-Cyl |
| **Memory** | Lowest | Higher than U-Cyl | Highest | Higher than RSTR-Cyl | Higher than GAU-Cyl |
| **Ease of Implementation** | Difficult (Manual Coding) | Easy (using GA and GMac) | Difficult (Manual Coding) | Easy (using GA and GMac) | Easy (using GA and GMac) |
| **Can be extended to other objects** | No | Require small amount of manual coding for rotationally-symmetric objects | Already suitable for any object | Already suitable for any object | Already suitable for any rotationally-symmetric object |

**Table 7-6: Comparison of two ray-triangle test techniques**

| | **Algebraic** | **GA Plücker** |
|---|---|---|
| **Algorithm Deduction** | Mainly Algebraic with manual optimizations | Mainly geometric with automatic optimizations |
| **Speed** | Similar | Similar |
| **Memory** | Similar | Similar |
| **Ease of Implementation** | Difficult (Manual Coding) | Easy (using GA and GMac) |

From the data in the three tables several conclusions can be made. From the first set of tests it is apparent that GMac is much easier to use compared to the only other good GA-based code generator; Gaigen 2. In addition, GMac provides much better performance for the generated code. The second set of tests illustrate that GA can be easily applied to geometrically demanding problems in computer science and engineering. The enhancements made to the base ray tracer of chapter 5 were tested. The results show that GA provides a much better alternative to traditional matrix and vector algebra methods. The performance of GMac code compares very well with hand coded procedures based on traditional mathematical methods. Hence, geometric algebra should not be ignored anymore as it provides both elegance and efficiency when suitably used and implemented. In addition, the possibility of adding efficient geometric computation abilities to modern general purpose languages through code generators like GMac is finally attainable.

# Chapter 8 : Conclusions and Future Work

The work presented in this thesis is very small compared to the doors open for research in geometric code generation and geometric algebra. This chapter presents the main conclusions of this work. In addition, the chapter points to some of the possible points upon which more research could be conducted.

Section 8.1 provides the main conclusions of this work. The following sections discuss some of the future developments that can be based on this work. Section 8.2 discusses possible enhancements to GMac and GMacDSL. Section 8.3 illustrates possible applications of GA in computer graphics. Section 8.4 focuses on some important mathematical aspects that could be very beneficial for the development and application of GA in practice. Section 8.5 discusses the importance of teaching GA in schools and universities.

## 8.1 Thesis Conclusions

This work presents a design for a code generator, called GMac, for automatic generation of software implementations for geometric algorithms. GMac is based on the universal algebraic system for geometric modeling called Geometric Algebra. GMac is capable of automatic software code generation for low-dimensional geometric problems. The operation of GMac is to transform a geometric algorithm written in a high-level geometric DSL into optimized low-level code. The DSL and transformation process are both based on the geometric algebra. GMac is compared against two similar GA-based code generators called Gaigen 2 and Gaalop. The architecture of GMac is illustrated to be superior to both systems architectures. In addition, a series of experiments aiming at comparing the speed of execution of generated code of GMac and Gaigen 2 illustrate the significant improvement of execution speed provided by GMac. Finally, GMac is used for enhancing some of the capabilities of a typical ray tracer. The enhancements are based on geometric algebra algorithms and the results are compared to traditional approaches typically used in ray tracing. The first enhancement is to create a small library, the Twister library, capable of procedurally generating and controlling free-form 3D curves and surfaces. The second enhancement is the creation of a generalized projection technique capable of procedurally creating and orienting a camera with any desired projection surface. The third enhancement is the use of a GA-based ray-triangle intersection algorithm for use in the rendering stage of ray tracing. The last enhancement is the proposition of two GA-based instancing techniques for modeling and rendering general objects in any location and orientation in space. This section presents the conclusions of this work.

### 8.1.1 Code Generation for Geometric Problems

From the illustrations and discussions provided in this work, it can be concluded that automatic code generation for low-dimensional geometric problems is a feasible and efficient geometric software development approach. The main reason behind such feasibility is the use of geometric algebra as the base algebraic system for modeling geometric concepts. The application of GA in such problems, through the use of the proposed code generator, has two main benefits. First, geometric ideas are more accessible through the high-level language of GA. Second, the resulting software is more efficient and maintainable. The use of a GA-based code generator like GMac made it possible to focus on the higher level, more difficult, and more abstract geometric problems. Such possibility is illustrated through the enhancements made to the ray tracer presented in this work.

From the performance comparisons of the previous chapter, it can be concluded that GMac is capable of generating code with enhanced performance compared to Gaigen 2. The use of geometric algebra for modeling geometric algorithms and GMac for their implementation is compared to traditional approaches. Results show that using GA and GMac, as geometric design and implementation tools, compares well with traditional approaches. GA provides a higher-level universal algebraic language compared to other algebraic systems. GMac provides a suitable tool for generation of efficient implementations based on geometric algebra. Together; a software designer can use GA and GMac to reach both goals of efficient code execution and ease of design for low-dimensional geometric problems.

### 8.1.2 Embedding GA in Computer Languages

Most current mainstream computer languages lack essential mathematical capabilities to be called general purpose languages. The need for geometric and algebraic computations is not restricted to scientific applications anymore. Many non-scientific applications require the use of intense geometric calculations. Game development software and educational software are major examples. It is about time that major computer languages be augmented with powerful geometric processing capabilities to meet such demand. A quick look at the OpenGL or DirectX APIs, for example, is enough to understand the major problems with current software languages and libraries used in computer graphics. Geometric algebra can act as an excellent base on which powerful addition to computer languages can be built.

### 8.1.3 Mutual Benefit of GA and Applications

As seen in the work of this thesis and of other research ([12], [14], [11], [10]), GA is a very dynamic field of research. GA can benefit from applications by adding more structure, relations, and geometric interpretations. Such additions are the natural byproducts of applying GA in diverse scientific fields.

On the other hand, as GA gains more structure, older GA applications will automatically benefit from its development. This mutual benefit effect between GA and applications has a very important consequence. GA can act as a common river of ideas that every field of application can add to and benefit from.

## 8.2 Future Development of GMac and GMacDSL

The current implementation of GMac and GMacDSL is just a prototype. Many enhancements can be added to GMac to greatly increase its power and applicability. The following subsections point to some of these enhancements.

### 8.2.1 Adding More Features

Many features should be added to GMac and GMacDSL. For example, more control of flow statements, like if and for loops, should be added to GMacDSL macros. GMac should be able to handle more target languages like Paython, Java, C++, VB, Matlab and more.

In its current implementation, GMac is not able to handle general blades or versors with unknown grades. Such capability can be added by implementing the GA multiplicative implementation, fully described in [20], in GMac.

The GMac interface requires more enhancements for code generation error reporting. In addition, GMacDSL should be redesigned to be more modular and perhaps to have functional language features like F# [37] and OCaml [38]. GMacDSL can also be integrated in the .NET framework to benefit from the excellent development environment of Visual Studio .NET.

The GMac core engine should be able to perform more optimizations before generating its final code. Current symbolic optimizations provided by Mathematica are good in most cases. Additional optimizations are nonetheless possible and should be investigated to output more efficient code.

### 8.2.2 Code Generation for GPUs

The expressions generated by GMac contain many independent computations. If a method is devised to identify and exploit such computations, parallel implementation is attainable. One active field of research is the use of Graphics Processing Units (GPUs), present in all modern PCs, for performing general purpose parallel computations. Such field is called General Purpose GPUs (GPGPU) [107], [108], [109]. GMac could be developed to output code suitable for such paradigm to automatically boost the performance of generated software.

### 8.2.3 Creating a DSL for Twists

The twist representation of shape presented and applied in chapter 6 can be very useful. The investigations of this thesis were limited to translations, rotations, and twists only. Other transformations, like reflections and scaling, can be added to the investigation and more complex shape can be represented.

The Twister library of chapter 6 is a good testbed for such idea. A better approach for practical application would be to create a specialized DSL, similar to GMacDSL, to handle 3D Euclidean transforms on points. Such DSL can produce excellent geometric representations with efficient generated code if properly merged with GMac.

## 8.3 Future Work in Computer Graphics

It is evident that computer graphics is one of the most geometrically demanding fields of application. Many techniques in computer graphics can benefit from the power of GA and GMac as illustrated in this work. More of the possible applications that should be investigated are pointed out in the following subsections.

### 8.3.1 Volume Data and Point Sampled Geometry

In this work, the twist representation of objects was used to create shape representations for curves and surfaces. Such approach can be easily extended to generating 3D volumetric free-form shapes using the similar treatments. Such generated 3D geometry can be used to simulate many naturally occurring phenomena like fog, hair, clouds, water, and many more using simpler techniques provided by GA. When such representations are combined with suitable rendering techniques like [73], a powerful addition to computer graphics is obtained. Such addition should be more accessible to wider spectrum of designers due to the nature and structure of geometric algebra.

In addition, many processing tasks related to point-sampled geometry [70] can be performed efficiently through GA. Attempts of such processing are already appearing [110]. Point sampled geometry techniques can equally benefit from and add to the development of geometric algebra.

### 8.3.2 Ray Tracing and Photon Mapping

The ray tracer of this work can be extended to handle many more techniques to enhance its performance and modeling capabilities. Examples of such techniques include [89], [111], [112], [113], [114], and [115]. Geometric algebra through GMac can be applied to such techniques to reduce programming effort while attaining high performance.

On the other hand, photon mapping [116] is much more suitable than ray tracing when a high level of reality is required in the rendered image. Photon mapping can benefit greatly from the capabilities of geometric algebra just like ray tracing did in this work. Applying GA to photon mapping is yet to be investigated.

### 8.3.3 Modeling Geometric Objects

As stated in chapter 6, geometric algebra can and should be used to represent shape in 3D Euclidean space through the many representations available to the CAGD community [80]. Several shape representation schemes can benefit from using GA. Such schemes include CSG [83], parametric representations [86], [87], point-based representations [70], and generative models [84], [85], [36]. Generative modeling would benefit the most from GA's capabilities. GA can be used to program the construction of shapes through geometric operations on simple primitives like points, circles and line segments. Such geometric constructions are not limited to Euclidean transforms. Other geometric operations can be used like the dual, projection, and subspace operations on blades and multivectors [117], [118].

## 8.4 Future Mathematical GA Developments

The development of geometric algebra in various application domains should be pursued. For such development to be attainable, more mathematical treatments must be investigated. The following subsections point to such mathematical treatments in several domains.

### 8.4.1 Interval Analysis

Interval analysis [119], [120] is a mathematical discipline where real number computations are represented by intervals rather than single real numbers. It has many applications in all fields of numeric computation especially in global optimization problems. The main reason behind such importance is the intrinsic uncertainty in many applications. In addition to errors in data, round off errors inside computers is another problem solvable by interval analysis methods.

The integration of interval analysis techniques with geometric algebra can lead to many geometric algorithms with very good results. For example, several problems in computer graphics could be solved through interval analysis and generative modeling as illustrated in [85], [121], and [122]. Such integration of interval analysis and geometric algebra can lead to automatic code generation of numerically optimized geometric algorithms in CG problems. Such problems include ray-parametric surface intersections and other global optimization related problems in computer graphics [121].

### 8.4.2 Wavelets and Fourier Transforms

In [123], a definition for the GA Fourier Transformation (GA FT) for GA of 3D Euclidean space is motivated and given. Known applications include uncertainty, linear shift invariant filters (smoothing, edge detection), signal analysis, image processing, fast multivector pattern matching, visual flow analysis, sampling, and multivector field analysis. In addition, GA FTs can be discrete; and fast GA FT algorithms are available. The paper then introduces several types of the so called Quaternion Fourier Transforms (QFTs). Their applications include partial differential systems, color image processing, filtering, multivector wave packet analysis in physics, and directional uncertainty with additional geometric insight. Next a local GA wavelet concept in 2D and 3D Euclidean space is introduced. The paper gives an example of a GA Gabor multivector wavelet.

In [10], the connection between the twist representation of shape and Fourier descriptors [124] is investigated. Such descriptors have been used extensively for object recognition and representation [125]. In [126], the creation of a new technology allowing real-time radiosity in videogames utilizing commodity graphics processing hardware is noted. The technology is based on geometric algebra wavelet technology.

All the previous examples point to the importance of integrating GA with wavelets and Fourier analysis techniques. Such integration may result in many good algorithms for shape representation, signal analysis, computer vision, pattern matching, and image processing applications.

### 8.4.3 Multiplicative Representation for Blades and Versors

All multivector representations in current GA software implementation packages rely on representing a multivector as a linear combination of basis blades. GMac is no different than other GA libraries in this aspect. Such method of representing multivectors is called the additive representation. Another very good representation is called the multiplicative representation; fully described in [20]. The multiplicative representation is based on matrix algebra to represent blades and versors of arbitrary geometric algebras. Such representation has an advantage over the traditional additive one. The advantage is that it is not restricted to low-dimensional GAs as the additive representation is. It was illustrated in [20] that the additive representation is less efficient than the multiplicative one for GAs of higher dimension. GMac should be augmented with capabilities for such advanced multivector representation scheme.

### 8.4.4 More Geometric Algebras

The conformal geometric algebra is one of the noted successes of GA in practical applications. There are, nonetheless, other important geometric algebras currently under investigation in GA literature. Such geometric algebras include the 6D GA that represent conic sections described in [10]. Another GA is the 8D geometric algebra representing the conformal conics, also described in [10]. A third geometric algebra is the one proposed in [127] for image processing applications. As more GAs appear in literature, the applicability of GA to many fields of application will certainly increase and considerably benefit.

## 8.5 The Future of Teaching Geometric Algebra

In 2002, the American Association of Physics Teachers awarded Prof. David Hestenes its Oersted Medal for notable contributions to the teaching of physics [31]. In September 2003, the Research Institute for Mathematical Sciences in Kyoto, Japan held an international symposium with an explicit focus on geometric algebra for teaching mathematics. One of the papers in this symposium [128] discussed the close dependence between pedagogic methods and lines of research. The paper illustrated the existence of continuous interchange between new advances in science and new methods of teaching such science. The present line of research could become the new subjects to teach. The paper gives some examples in differential calculus, algebra, and arithmetic. The paper refers mainly to the teaching of GA in high schools.

The previous two examples in mathematics and physics illustrate that GA is gaining attention among teachers in schools and universities. Unfortunately, no organized attempts are currently available for teaching GA in computer science and engineering. The work in this thesis can help in the development of teaching tools that both rely on and target GA for schools and universities. Such step is essential to the development of GA as a practical tool for applying geometry in systematic and successful ways.

# Appendix A: Mathematical Fundamentals

This appendix introduces some important mathematical background concepts for the development of geometric algebra. The main purpose here is to provide a relatively self-contained mathematical background suitable for the work provided in this thesis.

Geometric algebra is constructed on a vector space over some scalar field. These two concepts can be found in many mathematics books such as [129] and [130]. Section A.1 introduces fields. Section A.2 is about vector spaces and their properties. Section A.3 is about subspaces. Section A.4 is about bilinear and quadratic forms used to define inner products and orthogonality of vectors. Section A.5 is the algebraic definition of Clifford algebra.

## A.1 Fields

A field is a set of scalars $\mathcal{F}$ with two binary operations: addition '$+$' and multiplication '$\cdot$' satisfying the following:

- $\mathcal{F}$ is closed under addition:

$$\alpha + \beta \in \mathcal{F} \quad \forall \alpha, \beta \in \mathcal{F} \tag{A.1}$$

- Addition is commutative:

$$\alpha + \beta = \beta + \alpha \quad \forall \alpha, \beta \in \mathcal{F} \tag{A.2}$$

- Addition is associative:

$$\alpha + (\beta + \gamma) = (\alpha + \beta) + \gamma \quad \forall \alpha, \beta, \gamma \in \mathcal{F} \tag{A.3}$$

- Presence of a zero scalar:

$$\exists\, 0 \in \mathcal{F} : \alpha + 0 = \alpha \quad \forall \alpha \in \mathcal{F} \tag{A.4}$$

- Presence of a negative for each scalar:

$$\forall \alpha \in \mathcal{F} \quad \exists -\alpha \in \mathcal{F} : (-\alpha) + \alpha = 0 \tag{A.5}$$

- $\mathcal{F}$ is closed under multiplication:

$$\alpha\beta \in \mathcal{F} \quad \forall \alpha, \beta \in \mathcal{F} \tag{A.6}$$

- Multiplication is commutative:

$$\alpha\beta = \beta\alpha \quad \forall \alpha, \beta \in \mathcal{F} \tag{A.7}$$

- Multiplication is associative:

$$\alpha(\beta\gamma) = (\alpha\beta)\gamma \quad \forall \alpha, \beta, \gamma \in \mathcal{F} \tag{A.8}$$

- Multiplication is distributive over addition:

$$\alpha(\beta + \gamma) = \alpha\beta + \alpha\gamma \quad \forall \alpha, \beta, \gamma \in \mathcal{F} \tag{A.9}$$

- Presence of a unity scalar:

$$\exists 1 \in \mathcal{F} : \alpha 1 = \alpha \quad \forall \alpha \in \mathcal{F} \tag{A.10}$$

- Presence of an inverse for each nonzero scalar:

$$\forall \alpha \in \mathcal{F}, \alpha \neq 0 \quad \exists \alpha^{-1} \in \mathcal{F} : \alpha^{-1}\alpha = 1 \tag{A.11}$$

In applied mathematics, the two most important fields are the fields of real numbers $\mathbb{R}$ and complex numbers $\mathbb{C}$.

## A.2 Vector Spaces

An attempt to generalize ordinary real numbers to quantities having magnitude and direction resulted in the concept of vectors. Vectors have been a very important representation for many quantities in mathematics, physics, engineering, and computer science for the past 100 years. For mathematicians, vectors are a special instance of a more general concept called linear spaces. A linear space is a set that enables taking "linear combinations" of its elements. Many linear spaces are present in practice besides vectors. Examples include matrices, polynomials, real functions, and many other. Vectors are not sufficient to describe many important geometric concepts. Vectors can be, and should be, treated as representatives for 1D spaces. When vectors are used in practice they are usually abused to represent 2D, 3D and higher dimensional subspaces with operations not suited to its one-dimensional nature. This results in complex expressions and inefficient representations of geometric concepts. This is where GA becomes useful because it generalizes vectors (and linear spaces) to any dimension in a natural and intuitive way. This work is concerned mainly with finite-dimensional vector spaces defined on the field of real numbers. Hence, the terms “linear space” and “vector space” will be used in this work to refer to the same concept.

### A.2.1 Definition of Vector Space

A linear space (or vector space) over a field $\mathcal{F}$ is a set $\mathcal{V}$ along with the two operations of vector addition '$\oplus$' and scalar multiplication '$\otimes$' satisfying the following:

- $\mathcal{V}$ is closed under addition:

$$a \oplus b \in \mathcal{V} \quad \forall a, b \in \mathcal{V} \tag{A.12}$$

- Addition is commutative:

$$a \oplus b = b \oplus a \quad \forall a, b \in \mathcal{V} \tag{A.13}$$

- Addition is associative:

$$a \oplus (b \oplus c) = (a \oplus b) \oplus c \quad \forall a, b, c \in \mathcal{V} \tag{A.14}$$

- Presence of a zero vector:

$$\exists \mathbf{0} \in \mathcal{V} : a \oplus \mathbf{0} = a \quad \forall a \in \mathcal{V} \tag{A.15}$$

- Presence of a negative for each vector:

$$\forall a \in \mathcal{V} \quad \exists -a \in \mathcal{V} : (-a) + a = \mathbf{0} \tag{A.16}$$

- $\mathcal{V}$ is closed under scalar multiplication:

$$\lambda \otimes a \in \mathcal{V} \quad \forall \lambda \in \mathcal{F}, a \in \mathcal{V} \tag{A.17}$$

- Scalar multiplication is distributive:

$$(\lambda + \mu) \otimes a = (\lambda \otimes a) \oplus (\mu \otimes a) \quad \forall \lambda, \mu \in \mathcal{F}, a \in \mathcal{V} \tag{A.18}$$

$$\lambda \otimes (a \oplus b) = (\lambda \otimes a) \oplus (\lambda \otimes b) \quad \forall \lambda \in \mathcal{F}, a, b \in \mathcal{V} \tag{A.19}$$

- Scalar multiplication is associative with multiplication over $\mathcal{F}$ :

$$(\lambda \mu) \otimes a = \lambda \otimes (\mu \otimes a) \quad \forall \lambda, \mu \in \mathcal{F}, a \in \mathcal{V} \tag{A.20}$$

- Scalar multiplication with unity of $\mathcal{F}$ :

$$1 \otimes a = a \quad \forall a \in \mathcal{V} \tag{A.21}$$

The mathematical system $(\mathcal{F}, \mathcal{V}, \oplus, \otimes)$ is called a vector space (or linear space) and the members of $\mathcal{V}$ will be called vectors (or linear elements).

In this work, the symbol for addition of vectors '$a \oplus b$' will be replaced by '$a+b$' for simplicity. The difference between addition over a field and addition over a vector space will be apparent from context. Similarly, the symbol for scalar multiplication '$\lambda \otimes a$' will be replaced by '$\lambda a$'. In addition, the set $\mathcal{V}$ will be called a vector space where the context implies the full system $(\mathcal{F}, \mathcal{V}, \oplus, \otimes)$.

### A.2.2 Properties of Vector Spaces

- The zero vector is unique

$$\text{If } a+c=b+c=c \quad \forall c \in \mathcal{V} \quad \text{Then } a=b=\mathbf{0} \tag{A.22}$$

- For any vector, the negative vector is unique

$$\forall a \in \mathcal{V} \quad \text{If } a+b=a+c=\mathbf{0} \quad \text{Then } b=c=-a \tag{A.23}$$

- The scalar product with the scalar 0 is the zero vector

$$0\,a=\mathbf{0} \quad \forall a \in \mathcal{V} \tag{A.24}$$

- The scalar product with the scalar -1 is the negative vector

$$(-1)\,a=-a \quad \forall a \in \mathcal{V} \tag{A.25}$$

- The scalar product of any scalar with the zero vector is the zero vector

$$\lambda\,\mathbf{0}=\mathbf{0} \quad \forall \lambda \in \mathcal{F} \tag{A.26}$$

- A linear combination of a set of vectors $a, b, \ldots, x \in \mathcal{V}$ is any expression of the form:

$$\lambda a+\mu b+\cdots+\gamma x \in \mathcal{V} \quad ; \lambda, \mu, \ldots, \gamma \in \mathcal{F} \tag{A.27}$$

- A set of vectors $a, b, \ldots, x \in \mathcal{V}$ is said to be linearly independent (LID) iff the equation:

$$\begin{aligned} &\lambda a+\mu b+\cdots+\gamma x=\mathbf{0} \text{ implies that} \\ &\lambda=\mu=\cdots=\gamma=0 \quad ; \lambda, \mu, \ldots, \gamma \in \mathcal{F} \end{aligned} \tag{A.28}$$

Else, the set is called linearly dependent.

- A set of vectors $a, b, \ldots, x \in \mathcal{V}$ is said to span the vector space $\mathcal{V}$ if any vector in $\mathcal{V}$ can be expressed as a linear combination of the vectors of the set:

$$\mathcal{V} = span(a, b, \ldots, x) \Leftrightarrow w = \lambda a + \mu b + \cdots + \gamma x \\ ; \lambda, \mu, \ldots, \gamma \in \mathcal{F} \quad ; \forall w \in \mathcal{V} \tag{A.29}$$

$\mathcal{V}$ is thus said to be spanned by the set $a, b, \ldots, x \in \mathcal{V}$

- A set of vectors $e_1, e_2, \ldots, e_n \in \mathcal{V}$ is said to be a basis for the vector space $\mathcal{V}$ if the set is linearly independent and spans $\mathcal{V}$:

$$\langle e_1, e_2, \ldots, e_n \rangle^{\mathcal{V}} \Leftrightarrow \mathcal{V} = span(e_1, e_2, \ldots, e_n) \text{ and} \\ e_1, e_2, \ldots, e_n \text{ are linearly independent} \tag{A.30}$$

- Any basis for the vector space $\mathcal{V}$ contains the same number of vectors. The dimension of $\mathcal{V}$ is thus defined as the number of vectors in any basis for $\mathcal{V}$:

$$\dim(\mathcal{V}) = n \quad ; \langle e_1, e_2, \ldots, e_n \rangle^{\mathcal{V}} \tag{A.31}$$

- Coordinates of a vector:

For a vector space $\mathcal{V}$ having a basis $\mathbf{E} = \langle e_1, e_2, \ldots, e_n \rangle^{\mathcal{V}}$ the coordinates of a vector $x \in \mathcal{V}$ with respect to basis $\mathbf{E}$ are:

$$\text{Rep}(x)_{\mathbf{E}} = (x_1, x_2, \ldots, x_n)^T \text{ where } x = x_1 e_1 + x_2 e_2 + \cdots + x_n e_n \tag{A.32}$$

## A.3 Subspaces of Vector Spaces

A subspace $\mathcal{W}$ of a vector space $\mathcal{V}$ is a subset of $\mathcal{V}$ that is itself a vector space over the same field and operations of $\mathcal{V}$. Subspaces are very important geometric objects in practice. For example in 3D Euclidean space, any line or plane passing through the origin is a subspace.

- Any set $\mathcal{W} \subseteq \mathcal{V}$ is a subspace of $\mathcal{V}$ if it is closed under addition and scalar multiplication:

$$\mathcal{W} \leq \mathcal{V} \text{ iff } \mathcal{W} \subseteq \mathcal{V} \text{ and } \lambda a + b \in \mathcal{W} \quad \forall \lambda \in \mathcal{F}; a, b \in \mathcal{W} \tag{A.33}$$

- The trivial subspace:

$$\{\mathbf{0}\} \leq \mathcal{V} \text{ is called the trivial subspace of } \mathcal{V} \tag{A.34}$$

Any other subspace of $\mathcal{V}$ contains an infinite number of vectors.

- The intersection of two subspaces of a vector space $\mathcal{V}$ is a subspace of $\mathcal{V}$:

$$\mathcal{W} \cap \mathcal{U} \le \mathcal{V} \quad \forall \mathcal{W} \le \mathcal{V}, \mathcal{U} \le \mathcal{V} \tag{A.35}$$

- The sum of subspaces:

$$\begin{aligned} &\forall \mathcal{W} \le \mathcal{V}, \mathcal{U} \le \mathcal{V} \quad \mathcal{W} + \mathcal{U} \le \mathcal{V} \text{ where} \\ &\mathcal{W} + \mathcal{U} = \{x : x = w + u \quad ; w \in \mathcal{W}, u \in \mathcal{U}\} \end{aligned} \tag{A.36}$$

The sum of subspaces is a consistent way to express the union of the subspaces.

- The direct sum of subspaces:

Having a number of subspaces $\mathcal{W}_1 \le \mathcal{V}, \mathcal{W}_2 \le \mathcal{V}, \ldots, \mathcal{W}_k \le \mathcal{V}$ their sum $\mathcal{W}$ is called a direct sum if the intersection of each two is the trivial space:

$$\begin{aligned} &\mathcal{W} = \mathcal{W}_1 \oplus \mathcal{W}_2 \oplus \cdots \oplus \mathcal{W}_k \text{ iff} \\ &\mathcal{W} = \mathcal{W}_1 + \mathcal{W}_2 + \cdots + \mathcal{W}_k \text{ and} \\ &\mathcal{W}_i \cap \mathcal{W}_j = \{\mathbf{0}\} \quad \forall i = 1,2,\ldots,k ; j = 1,2,\ldots,k ; i \ne j \end{aligned} \tag{A.37}$$

In addition, the dimension of $\mathcal{W}$ is equal to the sum of the dimensions of the subspaces:

$$\mathcal{W} = \mathcal{W}_1 \oplus \mathcal{W}_2 \oplus \cdots \oplus \mathcal{W}_k \quad \Rightarrow \quad \dim(\mathcal{W}) = \sum_{i=1}^{k} \dim(\mathcal{W}_i) \tag{A.38}$$

- The orthogonal complement of a subspace:

Having a subspace $\mathcal{W} \le \mathcal{V}$ its orthogonal complement in $\mathcal{V}$ denoted by $\mathcal{W}^{\perp}$ is defined as the set of all vectors in $\mathcal{V}$ that are orthogonal to every vector in $\mathcal{W}$:

$$\mathcal{W}^{\perp} = \{x \in \mathcal{V} : y \perp x \quad \forall y \in \mathcal{W}\} \tag{A.39}$$

The orthogonal complement satisfies the following:

$$\mathcal{V} = \mathcal{W} \oplus \mathcal{W}^{\perp} \quad \forall \mathcal{W} \le \mathcal{V} \tag{A.40}$$

$$x \perp y \quad \forall x \in \mathcal{W}, y \in \mathcal{W}^{\perp} \quad , \mathcal{W} \le \mathcal{V} \tag{A.41}$$

$$(\mathcal{W}^{\perp})^{\perp} = \mathcal{W} \quad \forall \mathcal{W} \le \mathcal{V} \tag{A.42}$$

## A.4 Bilinear Forms and Quadratic Forms

Bilinear forms and quadratic forms are the mathematical tools for defining what is meant by orthogonality for a given vector space. In what follows, a mathematical description for their meaning and use is presented. Other useful definitions and properties can be found in [130].

### A.4.1 Bilinear Forms on Vector Spaces

A bilinear form $\mathrm{B}$ on a vector space $\mathcal{V}$ over a field $\mathcal{F}$ is a mapping $\mathrm{B}:\mathcal{V}\times\mathcal{V}\to\mathcal{F}$ that is linear in both arguments:

$$\begin{aligned}
&\mathrm{B} \text{ is a bilinear form on } (\mathcal{F},\mathcal{V},\oplus,\otimes) \text{ iff}\\
&\mathrm{B}:\mathcal{V}\times\mathcal{V}\to\mathcal{F} \text{ where}\\
&\mathrm{B}(\lambda\otimes u_1\oplus u_2,v)=\lambda B(u_1,v)+B(u_2,v) \text{ and}\\
&\mathrm{B}(u,\mu\otimes v_1\oplus v_2)=\mu B(u,v_1)+B(u,v_2)\\
&\forall u,u_1,u_2,v,v_1,v_2\in\mathcal{V} \quad ,\lambda,\mu\in\mathcal{F}
\end{aligned} \tag{A.43}$$

Sometimes the bilinear form is written as $\mathrm{B}(u,v)=\langle u,v\rangle$

A bilinear form is called symmetric if the order of its arguments is irrelevant to its value:

$$\begin{aligned}
&\text{A bilinear form } \mathrm{B} \text{ is symmetric iff}\\
&\mathrm{B}(u,v)=B(v,u) \quad \forall u,v\in\mathcal{V}
\end{aligned} \tag{A.44}$$

Having a basis for $\mathcal{V}$: $\mathbf{E}=\langle e_1,e_2,\ldots,e_n\rangle^{\mathcal{V}}$ then the matrix of the bilinear form $\mathrm{B}$ is defined as:

$$\begin{aligned}
&\mathbf{A}_{\mathrm{B}}=(a_{ij}) \quad where\ a_{ij}=\mathrm{B}(e_i,e_j)\\
&\forall i=1,2,\ldots,n \text{ and } j=1,2,\ldots,n
\end{aligned} \tag{A.45}$$

The value of the bilinear form can be calculated using its matrix:

$$\begin{aligned}
&\text{If } \mathbf{E}=\langle e_1,e_2,\ldots,e_n\rangle^{\mathcal{V}},u,v\in\mathcal{V} \text{ where}\\
&\mathrm{Rep}_{\mathbf{E}}(u)=(u_1,u_2,\ldots,u_n)^T=x\in\mathbb{R}^n\\
&,\mathrm{Rep}_{\mathbf{E}}(v)=(v_1,v_2,\ldots,v_n)^T=y\in\mathbb{R}^n\\
&\Rightarrow \mathrm{B}(u,v)=\langle u,v\rangle=x^T A_{\mathrm{B}} y
\end{aligned} \tag{A.46}$$

The matrix of the bilinear form is symmetric iff the bilinear form is symmetric:

$$\mathbf{A}_{\mathrm{B}} = \mathbf{A}_{\mathrm{B}}^{T} \Leftrightarrow \mathrm{B(u,v)=B(v,u)} \quad \forall \mathrm{u,v} \in \mathcal{V} \tag{A.47}$$

A bilinear form is called non-degenerate if its matrix is non-singular. In this work, the matrix of the bilinear form will be called the Inner-Product Matrix (IPM) for the vector space.

A bilinear form is called reflexive iff:

$$\mathrm{B}(u,v) = 0 \Leftrightarrow \mathrm{B}(v,u) = 0 \quad \forall u,v \in \mathcal{V} \tag{A.48}$$

A bilinear form is reflexive if and only if it is symmetric and alternating:

$$\begin{aligned} &\mathrm{B} \text{ is reflexive iff:} \\ &\mathrm{B}(u,v) = \mathrm{B}(v,u), \quad \mathrm{B}(u,u) = 0 \quad \forall u,v \in \mathcal{V} \end{aligned} \tag{A.49}$$

For reflexive bilinear forms, the concept of orthogonality can be defined as:

$$\begin{aligned} &\text{If } u,v \in \mathcal{V} \text{ then } u \perp v \text{ iff } \mathrm{B}(u,v) = \mathrm{B}(v,u) = 0 \\ &\text{where } \mathrm{B} \text{ is a reflexive bilinear form on } \mathcal{V} \end{aligned} \tag{A.50}$$

Thus, two vectors are orthogonal with respect to a reflexive bilinear form if the value of their bilinear form is zero.

Having a basis for $\mathcal{V}$: $\mathbf{E} = \langle e_1, e_2, \ldots, e_n \rangle^{\mathcal{V}}$ and a bilinear form $\mathrm{B}$ then $\mathrm{B}$ is said to have the signature $p,q,r$ iff:

$$\begin{aligned} &\mathrm{B}(e_j, e_j) > 0 \quad \forall j = i_1, i_2, \ldots, i_p \ ; e_j \in \mathbf{E} \\ &\mathrm{B}(e_k, e_k) < 0 \quad \forall k = i_{p+1}, i_{p+2}, \ldots, i_{p+q} \ ; e_k \in \mathbf{E} \\ &\mathrm{B}(e_m, e_m) = 0 \quad \forall m = i_{p+q+1}, i_{p+q+2}, \ldots, i_{p+q+r} \ ; e_m \in \mathbf{E} \\ &; p + q + r = n \end{aligned} \tag{A.51}$$

The vectors $e_j$ are said to have positive signature, the vectors $e_k$ are said to have negative signature and the vectors $e_m$ are said to be null vectors (or have zero signature). The signature of the bilinear form is independent from the selected basis $\mathbf{E}$. Hence, the signature of a particular bilinear form is a characteristic for that bilinear form.

The inner product between two vectors in 3D Euclidean space defines their relative relation in terms of direction in space. Two vectors are orthogonal if their inner product is zero. Hence, the inner product can be used to define orthogonality of vectors. Symmetric bilinear forms can represent inner products. For example, the usual 3D space $\mathbb{R}^3$ is said to have "orthonormal" basis $\langle e_1, e_2, e_2 \rangle_{\perp 1}^{\mathbb{R}^3}$ if their inner product is defined as:

$$\begin{aligned} &e_1 \cdot e_1 = e_2 \cdot e_2 = e_3 \cdot e_3 = 1 \\ &, e_1 \cdot e_2 = e_1 \cdot e_3 = e_2 \cdot e_3 = 0 \end{aligned} \tag{A.52}$$

Thus, a bilinear form that corresponds to inner product has a matrix:

$$\mathbf{A}_{\mathrm{B}} = \begin{pmatrix} 1 & 0 & 0 \\ 0 & 1 & 0 \\ 0 & 0 & 1 \end{pmatrix} \tag{A.53}$$

The inner product is defined as a symmetric bilinear form on the vector space:

$$a \cdot b = B(a, b) = \langle a, b \rangle \quad \forall a, b \in \mathcal{V} \tag{A.54}$$

### A.4.2 Quadratic Forms

Quadratic forms are just another way of representing symmetric bilinear forms. Thus, any quadratic form is equivalent to a symmetric bilinear form and vice-versa. A Clifford algebra is usually defined using a quadratic form hence the following are definitions and properties of quadratic forms.

- A quadratic form $\mathrm{Q}$ on a vector space $\mathcal{V}$ over a field $\mathcal{F}$ is a mapping $\mathrm{Q} : \mathcal{V} \to \mathcal{F}$ that corresponds to a symmetric bilinear form:

$$\begin{aligned} &\mathrm{Q} \text{ is a quadratic form on } (\mathcal{F}, \mathcal{V}, \oplus, \otimes) \text{ iff:} \\ &\mathrm{Q} : \mathcal{V} \to \mathcal{F} \text{ where} \\ &\mathrm{Q}(\lambda \otimes u) = \lambda^2\, \mathrm{Q}(u)\ \forall u \in \mathcal{V}, \lambda \in \mathcal{F} \text{ and} \\ &\mathrm{B}(u, v) = \mathrm{Q}(u + v) - \mathrm{Q}(u) - \mathrm{Q}(v) \text{ is a bilinear form} \end{aligned} \tag{A.55}$$

The bilinear form $\mathrm{B}$ is called the associated bilinear form to the quadratic form $\mathrm{Q}$ and is always symmetric.

The value of the quadratic form can be obtained from the matrix of its bilinear form as:

$$\begin{aligned}
&\text{If } \mathbf{E} = \langle e_1, e_2, \ldots, e_n \rangle^{\mathcal{V}}, u \in \mathcal{V} \text{ where} \\
&\mathrm{Rep}_{\mathbf{E}}(u) = (u_1, u_2, \ldots, u_n)^T = x \in \mathbb{R}^n \\
&\Rightarrow \mathrm{Q}(u) = \frac{1}{2}\mathrm{B}(u,u) = \frac{1}{2}x^T A_{\mathrm{B}} x
\end{aligned} \tag{A.56}$$

The signature of a quadratic form is the signature of its associated bilinear form.

## A.5 Clifford Algebra

Many different but consistent mathematical definitions for Clifford Algebra can be found in the literature (for example [10], [131], [132], [133], and [134]). That is mainly due to the universal applicability of that algebraic system to many scientific fields. Some definitions are based on vectors of abstract algebra [134] and tensor algebra [131], [130]. The more useful definitions for a computer scientist or engineer are, however, more directly based on simpler mathematical concepts like the definitions given in [132] and [133]. That makes the definition more related to the useful characteristics of Clifford Algebra from an application-oriented point of view. A definition of a Clifford algebra based on quadratic forms similar to the one in [133] is given below. The history of the development of Clifford algebra and geometric algebra can be found in several references. In [33] a powerful introduction to applications of GA in many fields of physics is presented. The introduction begins by a history of the development of GA. In addition, [135] contains a short historical background for GA. An extended historical introduction to the subject can also be found in [10].

Let $(\mathcal{F}, \mathcal{V}, \oplus, \otimes)$ be a vector space with an associated quadratic form $\mathrm{Q}$ having a signature $p,q$ where $n = p + q = \dim(\mathcal{V})$ and an associated symmetric bilinear form $\mathrm{B}(u,v) = \mathrm{Q}(u+v) - \mathrm{Q}(u) - \mathrm{Q}(v)\ \forall u,v \in \mathcal{V}$. The Clifford Algebra $C\ell_{p,q}$ is defined to be the unitary associative algebra over the field $\mathcal{F}$ of the non-degenerate quadratic form $\mathrm{Q}$ with signature $p,q$ on $\mathcal{V}$ which contains the sets $\mathcal{F}$ and $\mathcal{V}$ as distinct subspaces and satisfies the three conditions:

- $$uu = u^2 = \mathrm{Q}(u) \quad \forall u \in \mathcal{V} \qquad \text{(A.57)}$$

- $\mathcal{V}$ generates $C\ell_{p,q}$ as an algebra over $\mathcal{F}$
- $C\ell_{p,q}$ is not generated by any proper subspace of $\mathcal{V}$.

The first condition is equivalent to:

$$\mathrm{B}(u,v) = \langle u,v \rangle = uv + vu \quad \forall u,v \in \mathcal{V} \qquad \text{(A.58)}$$

In addition, having the orthonormal basis $\langle e_1, e_2, \ldots, e_n \rangle^{\mathcal{V}}$, the first condition can be expressed as:

$$\begin{aligned} e_k^2 &= 1 && ;k = 1,2,\ldots,p \\ e_m^2 &= -1 && ;m = p+1, p+2, \ldots, p+q \\ e_r e_s &= -e_s e_r && ;\forall r < s, s = 2,3,\ldots,p+q \end{aligned} \qquad \text{(A.59)}$$

Finally the third condition is only required for signatures satisfying: $p - q = 1,5,9,13,\ldots$ where $(e_1 e_2 \cdots e_{p+q})^2 = 1$.

The product operation of any unitary associative algebra $C\ell$ satisfies the following properties:

$$a(b+c) = ab + ac \quad \forall a,b,c \in \mathcal{V} \qquad \text{(A.60)}$$

$$(a+b)c = ac + bc \quad \forall a,b,c \in \mathcal{V} \qquad \text{(A.61)}$$

$$(\lambda a)b = a(\lambda b) = \lambda(ab) \quad \forall a,b \in \mathcal{V}, \lambda \in \mathcal{F} \qquad \text{(A.62)}$$

$$a(bc) = (ab)c \quad \forall a,b,c \in \mathcal{V} \qquad \text{(A.63)}$$

$$\exists 1 \in C\ell : A\,1 = 1\,A = A \quad \forall A \in C\ell \qquad \text{(A.64)}$$

If the quadratic form is degenerate with signature $p,q,r$ the corresponding Clifford algebra (denoted by $C\ell_{p,q,r}$) is also degenerate with signature $p,q,r$. If $p = \dim(\mathcal{V}), q = r = 0$ (all nonzero vectors have positive signatures) the Clifford algebra is said to be Euclidean (and is written as $C\ell_n$ as an abbreviation for $C\ell_{n,0,0}$). If $q = \dim(\mathcal{V}), p = r = 0$ (all nonzero vectors have negative signatures) the Clifford algebra is said to be anti-Euclidean (and is written as $C\ell_{0,n}$ as an abbreviation for $C\ell_{0,n,0}$). If $r > 0$ (nonzero null basis are present) it is a degenerate Clifford algebra. Finally if it is non-degenerate ( $r = 0$ ) it is written as $C\ell_{p,q}$ as an abbreviation for $C\ell_{p,q,0}$ .

Examples of quadratic forms and their associated Clifford Algebras as presented in [11] and [136] are:

- The geometric algebra of the 2D Euclidean space $C\ell_2$. Its even sub-algebra $C\ell_2^+$ is isomorphic to the complex numbers.
- The geometric algebra of the 3D Euclidean space $C\ell_3$. Its even sub-algebra $C\ell_3^+$ is isomorphic to the quaternion numbers.
- The geometric algebra $C\ell_{4,1}$ of the 5-dimensional quadratic vector space with a quadratic form having signature 4, 1. It is a very versatile algebraic model of the 3D Euclidean space called Conformal Geometric Algebra (CGA). Blades in $C\ell_{4,1}$ are in direct correspondence with points, lines, planes, circles, and spheres in 3D space. The operations of translation, join and intersection of these blades are performed using simple geometric product expressions.